\documentclass[seceq]{ptptex}
\usepackage{wrapft}
\notypesetlogo                       %comment in if to eliminate PTPTeX
\usepackage{ifpdf}
\usepackage{psfrag} % psfrag for using tex inside ps files.
\usepackage{graphicx}
\usepackage{epstopdf}
\usepackage{color}
\usepackage{dcolumn}
\usepackage{mathrsfs}
\usepackage{amsfonts}
\usepackage{amssymb}
\usepackage{amsmath}
\usepackage{dsfont}
\usepackage{bm}
\usepackage{bbm}
\usepackage{longtable}
\usepackage{fixmath}
\usepackage{upgreek}
\usepackage{bbding}
\usepackage{bold-extra}
\usepackage[T1]{fontenc}
\usepackage{ae,aecompl}
\usepackage[bbgreekl]{mathbbol}
\usepackage{slashed}
\usepackage[dvipsnames*,svgnames]{xcolor}
\usepackage[pdfauthor={Behnam Farid},%
            pdftitle={Many-body perturbation expansions without diagrams. I. Normal states},%
            plainpages=false,pdfpagelabels,linktocpage,%
            breaklinks=true,%
            hyperfootnotes=true,%
            bookmarks=true]{hyperref}
\hypersetup{
            colorlinks=true,%
            citecolor=DarkBlue,%
            filecolor=black,%
            linkcolor=DarkBlue,%
            urlcolor=blue,%pdftex,%
            pdfstartview={FitH}}
\usepackage{breakcites}
\usepackage{cite}
\usepackage{perpage} %the perpage package
\MakePerPage{footnote} %the perpage package command
\usepackage[perpage]{footmisc}
\renewcommand{\thefootnote}{\fnsymbol{footnote}}
\renewcommand{\thefootnote}{\ifcase\value{footnote}\or*\or \S \or \flat \or \natural \or \| \or \sharp\fi}
\usepackage{scrextend} % For using \footref{label}
\usepackage{accents}
\newlength{\dhatheight}

\def\wt#1{\widetilde{#1}}
\def\t#1{\tilde{#1}}
\def\wh#1{\widehat{#1}}
\def\h#1{\hat{#1}}
\def\b#1{\bar{#1}}

\usepackage{scalerel}
\usepackage[usestackEOL]{stackengine}
\def\wb#1{\ThisStyle{%
  \setbox0=\hbox{$\SavedStyle#1$}%
  \stackengine{1.0\LMpt}{$\SavedStyle#1$}{\rule{\wd0}{0.2\LMpt}}{O}{c}{F}{F}{S}%
}}
\DeclareMathOperator{\re}{Re}

\DeclareMathOperator{\e}{e}
\DeclareMathOperator{\rd}{d\!}

\DeclareMathOperator{\sign}{sign}
\DeclareMathOperator{\Hf}{Hf} % Hafnian
\DeclareMathOperator{\Pf}{Pf} % Pfaffian
\DeclareMathOperator{\1BZ}{1BZ}
\DeclareMathOperator{\fbz}{\scriptscriptstyle{1BZ}}
\DeclareMathOperator{\ii}{\hspace{-0.2pt}\mathsf{i}\hspace{-1.0pt}}
\DeclareMathOperator{\Tr}{Tr}

\usepackage{stackengine,wasysym}
\newcommand\Widetilde[1]{\ThisStyle{%
  \setbox0=\hbox{$\SavedStyle#1$}%
  \stackengine{-.1\LMpt}{$\SavedStyle#1$}{%
    \stretchto{\scaleto{\SavedStyle\mkern.2mu\AC}{.5150\wd0}}{.6\ht0}%
  }{O}{c}{F}{T}{S}%
}}
\newcommand{\X}[1]{
  \mathchoice
    {{\scriptstyle #1}}% \displaystyle
    {{\scriptstyle #1}}% \textstyle
    {{\scriptscriptstyle #1}}% \scriptstyle
    {\scalebox{.7}{$\scriptscriptstyle #1$}} }%\scriptscriptstyle
\newcommand\p[1]{\mathchoice{#1^\prime}{#1^\prime}{#1^\prime}%
  {#1^{\scalebox{.7}{$\scriptscriptstyle\prime$}}} } % \p{x} = x'
\makeatletter
\newcommand{\raisemath}[1]{\mathpalette{\raisem@th{#1}}}
\newcommand{\raisem@th}[3]{\raisebox{#1}{$#2#3$}}
\makeatother
\DeclareFontFamily{OT1}{pzc}{}
\DeclareFontShape{OT1}{pzc}{m}{it}{<-> s * [1.10] pzcmi7t}{}
\DeclareMathAlphabet{\mathpzc}{OT1}{pzc}{m}{it} %Zapf Chancery
\usepackage{marginnote}
\usepackage{geometry}
\usepackage[utf8]{inputenc}
\DeclareUnicodeCharacter{2014}{\dash}
\DeclareRobustCommand\dash{%
  \unskip\nobreak\thinspace\textemdash\allowbreak\thinspace\ignorespaces}
\makeatletter
 \renewcommand*\l@subsection{\@dottedtocline{2}{2.0em}{2.6em}}
\makeatother
\newcounter{dummy}
\newcounter{dummyX}
\usepackage[british]{babel}
\title{%        %You can use \\ for explicit line-break.
Many-body perturbation expansions without diagrams.\\ I. Normal states\,\footnote{Dedicated to the memory of Nicolaas Godfried van Kampen (22 June 1921--\hspace{1.2pt}6 October 2013). \protect\cite{NvK} }
}

\author{%       %Use \scshape for the family name.
Behnam \textsc{Farid}\,\footnote{\href{mailto:behnam.farid@btinternet.com}{behnam.farid@btinternet.com}}%
}

\abst{
On the basis of an exact perturbational expression for the interacting one-particle Green function $G(a,b)$ corresponding to bosons / fermions in terms of the bare two-body interaction potential $v$ and permanents / determinants of the non-interacting one-particle Green functions $\{G_{\protect\X{0}}(i,j)\| i, j\}$, we deduce four \textsl{recursive} perturbation series expansions for the proper self-energy $\Sigma(a,b)$. With $W$ denoting the dynamic screened two-body interaction potential, these four perturbation series expansions of $\Sigma(a,b)$ are formally identical to the expansions of this function in terms of (i) proper self-energy diagrams and $(v, G_{\protect\X{0}})$, (ii) $G$-skeleton self-energy diagrams and $(v, G)$ (singly `bold-line' diagrams), (iii) $W$-skeleton self-energy diagrams and $(W, G_{\protect\X{0}})$ (singly `bold-line' diagrams), and (iv) $G$- and $W$-skeleton self-energy diagrams and $(W,G)$ (doubly `bold-line' diagrams). For the calculation of $W$, we rely on a similar exact perturbational expression for the interacting two-particle Green function $G_{\protect\X{2}}(a,b;c,d)$ for bosons / fermions as for $G(a,b)$ in terms of $v$ and permanents / determinants of $\{G_{\protect\X{0}}(i,j)\| i, j\}$. From this expression, we deduce four \textsl{recursive} perturbation series expansions for the proper polarization function $P(a,b)$, necessary for the calculation of $W$, that are similar to those for the proper self-energy $\Sigma(a,b)$ specified above. Here $a$, $b$, $c$, $d$, $i$, and $j$ denote space-time-spin variables in the case of calculations at zero temperature (within the framework of the adiabatic approximation). At non-zero temperatures, they denote space-imaginary-time-spin variables when dealing with the imaginary-time formalism of Matsubara, and space-time-spin-$\mu$ variables when dealing with the real-time formalism of thermo-field dynamics (TFD), where $\mu$ denotes a binary variable marking the original and the tilde-conjugated fields. The doubling of the fields in this formalism brings about transformation of the trace over the thermal ensemble of states into an expectation value with respect to a thermal vacuum state. The finite-temperature TFD formalism is particularly advantageous in directly providing the \textsl{dynamic} correlation functions. By contrast, for the calculation of such correlation functions within Matsubara's imaginary-time formalism, analytic continuation of these functions towards the real-time axis is to be effected, which in general is nontrivial, if practicable at all, to accomplish. Although throughout this paper diagrams are often referred to, they do \textsl{not} explicitly feature in the above-mentioned series expansions. In an appendix, we explicitly apply the formalisms presented in this paper to the Hubbard Hamiltonian for spin-$\tfrac{1}{2}$ fermions on a lattice in arbitrary $d$ spatial dimensions. In two further appendices, we present methods and short programs for determining the $\nu$th-order diagrams corresponding to the perturbation series expansions of $G$ in terms of $(v,G_0)$ and $\Sigma$ in terms of $(v,G_0)$ and $(v,G)$ on the basis of the cycle decompositions of the elements of the symmetric group $S_{2\nu}$.
}
\begin{document}
\ifx\href\undefined\else\hypersetup{linktocpage=true}\fi

\maketitle
\renewcommand{\thefootnote}{\textrm{\alph{footnote}}}

{\scriptsize{\tableofcontents}}

\refstepcounter{dummyX}
\section{Introduction}
\phantomsection
\label{sec1}

\refstepcounter{dummyX}
\subsection{General considerations}
\phantomsection
\label{sec1a}
The conventional many-body perturbation series expansions for correlation functions of interacting systems \cite{RM69,MYS95,FW03,NO98,HBKF04,SvL13} are founded on the possibility of treating non-commuting (field) operators as commuting and anti-commuting functions\,\footnote{For respectively bosonic and fermionic field operators. Thus, similar to a bosonic field operator, a product of an \textsl{even} number of fermion field operators is treated like an ordinary function.} through introducing integral representations of these operators in terms of a time-like parameter, an `ordering parameter' \cite{RPF51}, in conjunction with a `time'-ordering operator\,\footnote{The \textsl{interaction} picture of operators \protect\cite{FJD49,FW03,RM69} is one such representation, which is a specific case of the more general representation introduced in Ref.\,\protect\citen{RPF51}. Concerning the underlying time-like parameter in the interaction picture, this is the physical time $t$, $t\in\mathds{R}$, when dealing with ground-state (GS) correlation functions within the framework of the adiabatic approximation, the imaginary time $t \equiv -\protect\ii\hspace{0.6pt}\tau$, $\tau\in\mathds{R}$, when dealing with equilibrium thermal ensemble of states at non-zero temperatures, \protect\cite{FW03,RM69,NO98,HBKF04,MS59,AGD75,TM55,ESF59} and a complex quantity parameterising the directed Konstantinov-Perel' \protect\cite{KP61} and the Keldysh \protect\cite{LVK65,RAC68,RS86,JR07} contours when considering non-equilibrium ensemble of states \protect\cite{SvL13,PD84,MW91}. In a special case, the contour relevant to the real-time formalism of thermo-field dynamics (TFD) \cite{UMT82,HU95} coincides with that in the Keldysh formalism, however in general the two contours are different: typically, the contour $\mathscr{C}$ in the TFD formalism is that given in Eq.\,(\protect\ref{es3}) below. For further relevant details, consult \S\S\,\protect\ref{sac01} and \protect\ref{s226}. The imaginary time $t \equiv -\protect\ii\hspace{0.6pt}\tau$ is inherent to Euclidean quantum field theories \protect\cite{RPF53,IZ80,HU95,GP98,NO98}, which form the basis for many pioneering calculations on quantum spin systems, on coupled boson-fermion as well as interacting boson and fermion systems over the course of the past several decades, \S\,\protect\ref{sec1c}. Canonical boson / fermion operators in the interaction picture have the important property that the commutation / anti-commutation of any pair of them is a $c$-number also for \textsl{unequal} time arguments of these, which in turn leads to the contractions of these operators, appendix \ref{saa}, to be similarly $c$-numbers. In contrast, the commutation / anti-commutation of any pair of canonical operators in the Heisenberg picture is a $c$-number \textsl{only} when the time arguments of these are equal.} that ultimately takes full account of the proper ordering of the operators in the perturbational expressions. In weak-coupling many-body perturbation expansions, the Wick \textsl{decomposition} theorem \cite{HK49,GCW50,PD84}, appendix \ref{saa}, forms a crucial link between complicated perturbational contributions, consisting of the expectation values or ensemble averages, as the case may be,\footnote{Depending on whether the correlation function of interest is defined as the expectation value with respect to the vacuum state of the problem, or an average over an ensemble of sates, in this paper generally the equilibrium thermal ensemble of states.} of `time'-ordered products of canonical operators (in the interaction picture) to superpositions of products of contractions of pairs of these operators.\footnote{The contractions of canonical (field) operators in the interaction picture are $c$-numbers, that is they are some complex-valued functions times $\h{1}$, the identity operator in the Fock space of the problem at hand.} By normalisation, the contributions of these terms to the relevant correlation function prove to be limited to those expressible as \textsl{connected} diagrams, with each such diagram representing a well-specified functional of the underlying interaction function $v$ and the non-interacting one-particle Green function $G_{\X{0}}$.

Despite their transparency and intuitive appeal, diagrammatic expansions are in general not efficient for high-order perturbational calculations in practice.\footnote{Building on the formalisms of appendices \protect\ref{sac} and \protect\ref{sad}, in Ref.\,\protect\citen{BF16a} we introduce a general symbolic-algebraic technique that can considerably simplify calculations based on these expansions.} This is rooted in the fact that the mathematical expressions associated with diagrammatic expansions can be more economically described in terms of permanents \cite{HM78,NW78,MA06} / determinants \cite{TM60,VD99} in the case of bosons / fermions.\footnote{Equivalently, the former can be expressed in terms of  Hafnians \protect\cite{DLRV70}, and the latter in terms of Pfaffians \protect\cite{TM60,VD99}, appendix \ref{saa}.} Diagrammatic series expansions for bosons / fermions explicitly rely on the \textsl{full} expansions of the relevant permanents / determinants (that is to say, on the \textsl{definitions} of these two mathematical objects), which, as we discuss below, prove to be of higher computational complexity \cite{PP13} than strictly necessary.

To clarify the above statement, we first note that the arithmetic complexity of the \textsl{full} expansion of a general $\mathpzc{n}$-permanent / -determinant is $\mathpzc{n} \times \mathpzc{n}!$.\footnote{The arithmetic complexity is $(\mathpzc{n}-1) \times \mathpzc{n}!$ if the arithmetic complexity of summation is neglected in comparison with that of multiplication. Approximating the total number of summations, that is $\mathpzc{n}! -1$, by $\mathpzc{n}!$, for notational convenience in this paper we opt for the value $\mathpzc{n} \times \mathpzc{n}!$.} Perturbational calculation of $G$ at the $\nu$th order in the bare interaction potential requires determination of at least one $(2\nu+1)$-permanent / -determinant. Thus the arithmetic complexity of \textsl{diagrammatic} calculations increases factorially with the order of the perturbation expansion (here for $G$). More specifically, from the asymptotic series expansion corresponding to $\nu\to\infty$ of the number of $\nu$th-order Green-function diagrams \cite{CLP78,PH07} one observes that to leading order this number is proportional to $(2\nu+1)!!$, where $(2\nu+1)!! \doteq 1\cdot 3 \dots \cdot (2\nu+1)$.\footnote{The numbers relevant to the present discussion are those presented under the heading `Exact electron propagator without Furry's theorem' in Table I of Ref.\,\protect\citen{CLP78}. The asymptotic expression, `Asymptotic', corresponding to these numbers is given in the same Table, with the relevant variables presented under the same heading, with $k = 2, 4, 6, \dots$ denoting the `Order', which is to be identified with $2\nu$. Our explicit calculations reveal that this asymptotic expression is to be multiplied by $\sqrt{\pi}$ in order to approximate the actual numbers accurately (in other words, the coefficient $C$ should be $\sqrt{2/\pi}$ instead of $\sqrt{2}/\pi$). Following this correction, the resulting expression coincides with the leading-order asymptotic expression for $(k+1)!! \equiv (2\nu+1)!!$. This result deviates from the leading-order asymptotic expression as presented in Table I of Ref.\,\protect\citen{PH07}, by a factor of $1/\mathrm{e}$, where $\mathrm{e}$ denotes the Euler number (that is $\gamma = 0.5772\dots$) in the latter reference. Interestingly, the exact numbers $1, 1, 4, 27, 248, \dots$ in Ref.\,\protect\citen{PH07} deviate from the exact numbers $2, 10, 74, 706, \dots$ (or $1, 4, 25, 208, \dots$, taking account of Furry's theorem). For completeness, we note that the work by Pavlyukh and H\"{u}bner \protect\cite{PH07} follows earlier relevant works by Molinari \protect\cite{LGM05}, and Molinari and Manini \protect\cite{MM06}.} From the equality $(2\nu+1)!! = 2^{\nu} (2\nu+1) \Gamma(\nu+1/2)/\sqrt{\pi}$ [pp.\,256 and 258 in Ref.\,\citen{AS72}]\,\footnote{The equality applies only for \textsl{integer} (positive, zero and negative) values of $\nu$.} one observes that, for sufficiently large $\nu$, to leading order the number of connected $\nu$th-order Green-function diagrams to be explicitly taken into account scales like $2^{\nu+1} \times \nu^{1/2} \times \nu!$ [\S\,6.1.37, p.\,257, in Ref.\,\citen{AS72}]. This amounts to a considerable reduction relative to $(2\nu+1) \times (2\nu +1)! \sim 2^{(2-1/\ln 2) \nu + 5/2} \times \nu^{\nu+2} \times \nu!$,\footnote{$2-1/\ln 2 \approx 0.5573$.} the reduction arising from a combination of two factors: firstly, not all $(2\nu+1)!$ terms resulting from the expansion of a $(2\nu+1)$-permanent / -determinant correspond to connected diagrams, and, secondly, by the permutation symmetry associated with $\nu$ interaction potentials, for each connected term (representable by a connected diagram) in the explicit expansion of a $(2\nu+1)$-permanent / -determinant, there are $\nu!-1$ other connected terms each of which makes exactly the same contribution to the Green function at the $\nu$th order of the perturbation theory; this permutation symmetry is explicitly taken account of in the diagrammatic expansion of the Green function [p.\,97 in Ref.\,\citen{FW03}].\footnote{See also appendix \protect\ref{sac}.}

In spite of the fact that in the \textsl{diagrammatic} series expansion of $G$, in terms of $(v,G_{\X{0}})$, one to leading order explicitly deals with of the order of $(2\nu+1)!!$ diagrams, instead of $(2\nu+1)!$ terms that the full expansion of a $(2\nu+1)$-permanent / -determinant gives rise to, it should be borne in mind that computational complexities of the processes of identifying disconnected diagrams and those related by permutation symmetry cannot be disregarded, appendices \ref{sac} and \ref{sad}; while these processes may not involve arithmetic floating-point operations, for sufficiently large values of $\nu$ they require extensive amount of data management and computer-memory access.

In view of the above observations, it is remarkable that the arithmetic complexity of evaluating a general $(2\nu+1)$-determinant is at most of the order of $(2\nu+1)^3 \sim 8 \nu^3$ [\S\,3.2, p.\,111, in Ref.\,\citen{GvL13}].\refstepcounter{dummy}\label{ByAtMost}\,\footnote{By `at most' we are here referring to the algorithm of Strassen \protect\cite{VS69}, according to which the power $3$ in $(2\nu+1)^3$ is reduced to $\log_2(7) \approx 2.807$, the method of Coppersmith and Winograd \protect\cite{CW90} that reduces this value to $2.376$, and the more recent methods reducing this value even further (for a review see Ref.\,\protect\citen{VVW11}). However, with $\upalpha \mathpzc{n}^{\omega}$ expressing the arithmetic complexity of these methods for dealing with general $\mathpzc{n}$-matrices, due to a rapid increase in $\upalpha$ for decreasing $\omega$, for $\omega < 2.7$ the number of multiplications must be in excess of $10^{23}$ before these methods can compete with the method of Strassen [\S\,4.6.4, p.\,501, in Ref.\,\protect\citen{DEK98}]. See also \S\,5.1, p.\,395, of Ref.\,\protect\citen{PP13}, and Ch.\,24, p.\,433, of Ref.\,\citen{NJH02}. \label{notec}} According to the algorithm of Ryser [\S\,73, p.\,124, in Ref.\,\citen{HM78}] [Ch.\,27, p.\,217, in Ref.\,\citen{NW78}], the arithmetic complexity of evaluating a general $(2\nu+1)$-permanent amounts to $(2\nu+1) \times 2^{2\nu +1} \sim \nu \times 4^{\nu+1}$. For increasing values of $\nu$, this arithmetic complexity becomes negligibly small in comparison with even $(2\nu+1)!!$. We remark that the computational complexity of the calculation of permanents is an NP-hard problem \cite{LGV79,PP13,MA06}.

\refstepcounter{dummyX}
\subsection{The considerations in this paper}
\phantomsection
\label{sec1b}
The considerations in this paper are based on two formally exact weak-coupling perturbational expressions for the one- and two-particle Green functions, respectively $G$ and $G_{\X{2}}$, in terms of the bare two-body interaction potential $v$ and permanents / determinants of the non-interacting one-particle Green function $G_{\X{0}}$. As we shall be more specific later in \S\,\ref{sec2}, in this paper we focus on the normal state of systems, as opposed to superfluid and superconductive states, which we shall consider in a separate publication \cite{BF16b}. In Ref.\,\citen{BF16b} we shall also deal with coupled fermion-boson systems, notably systems of electrons coupled with phonons. Regarding the bosons associated with charge and spin fluctuations \cite{MP934,DP97,DJS99,NP10}, they are, insofar as normal states are concerned, taken account of by the considerations of this paper.\footnote{Discussing \textsl{paramagnons}, Monien\protect\cite{HM02} emphasises the significance of the physics associated with the non-Gaussian order-parameter fluctuations in two-dimensional (cuprate) superconducting compounds. Accounting for these fluctuations amounts to the calculation of the screened interaction function $W$ beyond the random-phase approximation, RPA \protect\cite{JMZ69,PN89}. Such calculation is technically straightforward in the framework of the diagram-free formalisms of the present paper, \S\,\protect\ref{sec3}.}

We begin the main part of this paper by developing a \textsl{recursive} formalism for the calculation of the \textsl{ordered} sequence $\{ G^{\X{(\nu)}} \| \nu = 1,2,\dots\}$ of the terms in the perturbation series expansion of $G$ in terms of $(v,G_{\X{0}})$ to an arbitrary finite order $n$ in $v$, \S\,\ref{s22}. On the basis of this sequence, we deduce a recursive formalism for the calculation of the \textsl{ordered} sequence of the terms in the perturbation series expansion of the self-energy $\Sigma$ in terms of $(v,G_{\X{0}})$, \S\,\ref{s23}. The $\nu$th term of this sequence is identical to the total contribution of all $\nu$th-order proper self-energy diagrams \cite{FW03}\footnote{A (connected) self-energy diagram is \textsl{proper}, or \textsl{one-particle irreducible} (1PI), when it does not become disconnected on cutting a single line representing a one-particle Green function.} evaluated in terms of $(v,G_{\X{0}})$. This expansion describing $\Sigma$ as a functional of $v$ and $G_{\X{0}}$, for the systematic development of the formalisms to be presented in this paper it proves advantageous to denote the corresponding functional by $\Sigma_{\X{00}}$, where the first $\X{0}$ in the compound index $\X{00}$ refers to the \textsl{bare} interaction potential $v$, and the second $\X{0}$ to the \textsl{non-interacting} Green function $G_{\X{0}}$.\footnote{Depending on the nature of the interaction potential and whether the system under consideration is defined on a lattice embedded in $\mathds{R}^d$ or over a continuum subset of $\mathds{R}^d$, in particular the perturbational terms in the perturbation series expansion for $\Sigma_{\protect\X{00}}$ may not exist to an arbitrary order. Nonetheless, even though formal, this perturbation series plays a vital role in the construction of the perturbation series expansion for $\Sigma_{\protect\X{\varsigma\varsigma'}}$, with $\protect\X{\varsigma}, \protect\X{\varsigma'} \in\{\protect\X{0},\protect\X{1}\}$ not both equal to $\protect\X{0}$, in this paper.} Thus
\begin{equation}\label{e1}
\Sigma_{\X{00}}(a,b) \equiv \Sigma_{\X{00}}(a,b;[v,G_{\X{0}}]) \equiv \Upsigma(a,b) \equiv \Upsigma(a,b;[G_{\X{0}}]) \equiv \Upsigma(a,b;[v,G_{\X{0}}]),
\end{equation}
where the functional $\Upsigma[G_{\X{0}}]$ has been introduced and discussed in some detail in Ref.\,\protect\citen{BF13}.\footnote{Unfortunately, the symbol $\Upsigma$ coincides with the symbol for the self-energy within the framework of the TFD according to the notation adopted in the present paper. This will however cause no confusion.} Later in this paper, $\Sigma_{\X{00}}^{\X{(\nu)}}$ will denote the above-mentioned total contribution of \textsl{all} $\nu$th-order proper (or \textsl{one-particle irreducible}, 1PI) self-energy diagrams contributing to $\Sigma_{\X{00}}$.

The details underlying the recursive calculation of the functional $\Sigma_{\X{00}}$ directly lead us to recursive formalisms for the calculation of the other three perturbation series expansions for the self-energy $\Sigma$ indicated in the abstract of this paper. These define the self-energy $\Sigma$ as a functional of $v$ and $G$, to be denoted by $\Sigma_{\X{01}}$, \S\,\ref{s24}, of $W$ and $G_{\X{0}}$, to be denoted by $\Sigma_{\X{10}}$, \S\,\ref{s25}, and of $W$ and $G$, to be denoted by $\Sigma_{\X{11}}$, \S\,\ref{s26}, where $W$ stands for the dynamic screened interaction potential \cite{JH57}, to be considered in some detail in \S\,\ref{sec3}. In analogy with the identities in Eq.\,(\ref{e1}), one has
\begin{align}\label{e2}
\Sigma_{\X{01}}(a,b) &\equiv \Sigma_{\X{01}}(a,b;[v,G]),\nonumber\\
\Sigma_{\X{10}}(a,b) &\equiv \Sigma_{\X{10}}(a,b;[W,G_{\X{0}}]),\nonumber\\
\Sigma_{\X{11}}(a,b) &\equiv \Sigma_{\X{01}}(a,b;[W,G]).
\end{align}
For the \textsl{complete} perturbation series expansions, one formally \cite{BF13} has
\begin{equation}\label{e3}
\Sigma = \Sigma_{\X{00}}[v,G_{\X{0}}] = \Sigma_{\X{01}}[v,G] =  \Sigma_{\X{10}}[W,G_{\X{0}}] = \Sigma_{\X{11}}[W,G].
\end{equation}
For clarity, one can in principle calculate for instance the function $\Sigma_{\X{00}}(a,b;[v,G])$ (assuming that $G$ is given), which is distinct from the sought-after self-energy $\Sigma(a,b) \equiv \Sigma_{\X{00}}(a,b;[v,G_{\X{0}}])$. Similarly as regards the other functionals encountered in Eq.\,(\ref{e3}). \emph{Where in the following we suppress the arguments of the self-energy functionals that ordinarily signify their functional dependence on the relevant interaction function and the one-particle Green function, we implicitly assume that these have been evaluated in terms of the appropriate functions.}\footnote{This remark identically applies to other similar functionals encountered in this paper. For instance, $P_{\protect\X{01}}(a,b)$, \S\,\protect\ref{s33}, is equivalent to $P_{\protect\X{01}}(a,b;[v,G])$.} Thus, for instance, $\Sigma_{\X{01}}(a,b)$ is equivalent to the more extensive notation $\Sigma_{\X{01}}(a,b;[v,G])$.

For later reference, with $\mathscr{D}_{\X{\varsigma\p{\varsigma}}}$, $\X{\varsigma}, \X{\varsigma'} \in \{\X{0},\X{1}\}$, denoting the set of self-energy \textsl{diagrams} corresponding to the self-energy functional $\Sigma_{\X{\varsigma\p{\varsigma}}}$, $\mathscr{D}_{\X{00}}$ consists of all \textsl{proper} (or 1PI) self-energy diagrams \cite{FW03} (connected self-energy diagrams that remain connected on removing any single internal line representing a $G_{\X{0}}$), $\mathscr{D}_{\X{01}}$ of all $G$-skeleton (or \textsl{two-particle irreducible}, 2PI) self-energy diagrams \cite{LW60} (those proper self-energy diagrams from which no self-energy diagram, whether proper (\emph{i.e.} 1PI) or improper, can be excised by cutting two Green-function lines\,\footnote{See appendix \protect\ref{sad}.}), $\mathscr{D}_{\X{10}}$ of $W$-skeleton diagrams \cite{JH57} (those proper self-energy diagrams from which no polarization diagram \cite{FW03,JH57}, whether proper or improper, can be excised by cutting two interaction-function lines), and $\mathscr{D}_{\X{11}}$ of all $G$- and $W$-skeleton self-energy diagrams. One has\,\footnote{Similar relationships apply for the set of polarisation diagrams $\mathscr{P}_{\protect\X{\varsigma\varsigma'}}$, $\protect\X{\varsigma},\protect\X{\varsigma'} \in \{\protect\X{0}, \protect\X{1}\}$, to be encountered, however not explicitly discussed, later.}
\begin{align}\label{e4}
\mathscr{D}_{\X{11}} &\subset \mathscr{D}_{\X{01}} \subset \mathscr{D}_{\X{00}}, \nonumber\\
\mathscr{D}_{\X{11}} &\subset \mathscr{D}_{\X{10}} \subset \mathscr{D}_{\X{00}},
\end{align}
where $\subset$ signifies the set on the left as being a \textsl{proper} subset of the set on the right. Generalising the above notation, by $\mathscr{D}_{\X{\varsigma\p{\varsigma}}}^{\X{(\nu)}}$ we denote the subset of \textsl{all} $\nu$th-order elements of $\mathscr{D}_{\X{\varsigma\p{\varsigma}}}$. Similar relationships as in Eq.\,(\ref{e4}) apply to $\{\mathscr{D}_{\X{\varsigma\p{\varsigma}}}^{\X{(\nu)}} \| \X{\varsigma},\X{\varsigma'}\}$, $\forall\nu$, except that at the lowest order the $\subset$ are to be replaced by $\subseteq$, or, more sharply, $=$.

With reference to the above notations, we obtain the aforementioned series expansions for the self-energy functionals $\Sigma_{\X{01}}$, $\Sigma_{\X{10}}$, and $\Sigma_{\X{11}}$ from that for $\Sigma_{\X{00}}$ by introducing systematic subtraction schemes that \textsl{recursively} remove the contributions of the diagrams in the set $\{\mathscr{D}_{\X{00}}^{\X{(\nu)}} \| \nu = 1,2,\dots, n\}$ that do not feature in the set $\{\mathscr{D}_{\X{\varsigma\p{\varsigma}}}^{\X{(\nu)}} \| \nu = 1,2,\dots, n\}$, where $\X{\varsigma}$ and $\X{\varsigma'}$ do not simultaneously coincide with $\X{0}$, Eq.\,(\ref{e4}). \emph{We achieve the relevant subtractions without any explicit reliance on diagrams.}

Calculation of the self-energy functionals $\Sigma_{\X{10}}[W,G_{\X{0}}]$ and $\Sigma_{\X{11}}[W,G]$ is demanding of the calculation of the dynamic screened interaction potential $W$, \S\,\ref{sec3}. On this account, in this paper we also consider the two-particle Green function $G_{\X{2}}$, \S\S\,\ref{sec3}, \ref{s32}. On the basis of a formally exact weak-coupling perturbational expression for $G_{\X{2}}$ in terms of $(v,G_{\X{0}})$, analogous to that for $G$ in terms of $(v,G_{\X{0}})$, \S\,\ref{s212}, we develop a recursive scheme for the calculation of the \textsl{ordered} sequence $\{G_{\X{2}}^{\X{(\nu)}} \| \nu =1,2,\dots\}$ of the perturbational contributions to $G_{\X{2}}$, \S\,\ref{s32}. We note in passing that, in an approximate framework one may rely on an approximate calculation of $G_{\X{2}}$ based for instance on a conserving approximation of this function, as specified by Baym and Kadanoff \cite{BK61,KB62,PD84}, instead of relying on the just-mentioned systematic approach.

From the ordered perturbational sequences $\{G^{\X{(\nu)}} \| \nu\}$ and $\{G_{\X{2}}^{\X{(\nu)}} \| \nu\}$, \S\S\,\ref{s221} and \ref{s32}, we deduce a recursive formalism for the calculation of the terms in the perturbation series expansion of the polarisation function $P(a,b)$ in terms of $(v,G_{\X{0}})$, similar to those of $G$ and $\Sigma$ in terms of $(v,G_{\X{0}})$, \S\S\,\ref{s22}, \ref{s23}. In analogy with the case of the self-energy, we denote the thus-calculated functional by $P_{\X{00}}[v,G_{\X{0}}]$, and the underlying \textsl{ordered} sequence of the perturbational terms by $\{P_{\X{00}}^{\X{(\nu)}}[v,G_{\X{0}}] \| \nu \in \mathds{N}_0\}$, \S\,\ref{s33},\footnote{See Eqs\,(\protect\ref{e217}) -- (\protect\ref{e216c}) below.} where $\mathds{N}_0 \equiv \mathds{N} \cup \{0\}$. Introducing, in analogy with $\Sigma_{\X{\varsigma\varsigma'}}$, $\X{\varsigma}, \X{\varsigma'} \in \{\X{0},\X{1}\}$ (see above), the functional $P_{\X{\varsigma\varsigma'}}$, on the basis of the latter sequence of terms concerning the perturbation series expansion of $P_{\X{00}}[v,G_{\X{0}}]$, we deduce recursive formalisms for the calculation of the perturbation series expansions of $P_{\X{01}}[v,G]$, $P_{\X{10}}[W,G_{\X{0}}]$, and $P_{\X{11}}[W,G]$, \S\,\ref{s33}. In this way, we arrive at four distinct perturbational expressions for $W$, describing this function as functional of $(v,G_{\X{0}})$, $(v,G)$, $(W,G_{\X{0}})$, and $(W,G)$, to be denoted by respectively $W_{\X{00}}[v,G_{\X{0}}]$, $W_{\X{01}}[v,G]$, $W_{\X{10}}[W,G_{\X{0}}]$, and $W_{\X{11}}[W,G]$, \S\,\ref{s33}. The functionals $W_{\X{10}}[W,G_{\X{0}}]$ and $W_{\X{11}}[W,G]$ are to be used in the \textsl{self-consistent} calculation of respectively $\Sigma_{\X{10}}[W,G_{\X{0}}]$, and $\Sigma_{\X{11}}[W,G]$, \S\S\,\ref{s25} and \ref{s26}. In Ref.\,\citen{BF16a} we introduce, amongst others, a general and practicable formalism to be employed in the \textsl{self-consistent} calculations of the functionals encountered in this paper.

\refstepcounter{dummyX}
\subsection{A brief overview of related works}
\phantomsection
\label{sec1c}
The earliest form of the many-body perturbation expansion ``without use of Feynman graphs'' is due to Caianiello \cite{ERC53I,ERC59}. The formal perturbation series expansion of Dyson's \cite{FJD49} $S$-matrix in quantum electrodynamics in Ref.\,\citen{ERC53I} in essence coincides with that in the denominator of the expression for $G(a,b)$ in Eq.\,(\ref{e242}) below.\footnote{See Eqs\,(\protect\ref{e261}) and (\protect\ref{e260}) below. Regarding the $S$-matrix in the non-relativistic context of this paper, compare the expression in Eq.\,(\protect\ref{e242}) with for instance that in Eq.\,(8.9), p.\,85, of Ref.\,\protect\citen{FW03}.}

Concentrating on systems of interacting \textsl{fermions} for which the perturbational expressions for the one- and two-particle Green functions as adopted in this paper are described in terms of determinants, Eqs\,(\ref{e242}) and (\ref{e200}), determinantal (or determinant, or auxiliary-field) schemes \cite{FMPR81,SS81,BSS81,HSSB82,JEH83,JEH85,%
JEG85,SK86,LG92,BPS04,AE08} were developed in the early $1980$s and have since been extensively used in theoretical studies of correlated electron systems, as well as systems of conduction electrons coupled to (magnetic) impurities \cite{HF86,FHS87,GKKR96,KSHOPM06} and bosons \cite{SBS88}. The focus of these schemes is the grand partition function $\mathcal{Z}$,\footnote{The `projector' formalism \protect\cite{SK86,TC90,AE08} is suited for calculating the \textsl{GS} properties.} Eq.\,(\ref{e16}) below, from which various correlation functions, such as $G$ and $G_{\X{2}}$, can in principle be determined through functional differentiation with respect to auxiliary fields of vanishingly small amplitudes coupling to appropriate operators \cite{MS59,NO98}. We note in passing that the `worm' algorithm / updating scheme \cite{PST96,PST98,BPS06}, to be encountered later in this section, accommodates use of this procedure for the determination of general correlation functions (in particular the one-particle Green function and pair-correlation functions \cite{BPST06a,BPST06b}). When applying this algorithm, by allowing only for the relevant configurations in the underlying Monte Carlo simulations, one bypasses the need for dealing with auxiliary fields of small (ideally, infinitesimal) amplitude.\footnote{That is to say, configurations that depend linearly, or quadratically, \emph{etc.} (as the case may be), on the source fields. For illustration, consider the function $f(x) = a + b x + c x^2 + \dots\;$. Whereas one has $b = f'(0) \equiv \lim_{h\to 0} \{f(h) - f(0)\}/h$, the limit process $h\to 0$ can be bypassed by allowing only the contributions to $f(x)$ in the construction of this function that depend linearly on $x$.} This is advantageous, since accurate calculation of derivatives involves subtraction of similar numbers, imposing stringent demand on the accuracy with which the underlying calculations are to be carried out. To clarify, considering for transparency \textsl{functions} and the simplest approach for the numerical determination of their derivatives (\S\,5.7, p.\,180, in Ref.\,\citen{PTVF97}), for $\textrm{d}f(x)/\textrm{d}x \equiv f^{\X{(1)}}(x)$ to be accurate to $1$ part in $10^p$, the function $f(x)$ is to be calculated to an accuracy of at least $1$ part in $10^{2p}$. More generally, employing the same approach, for $\textrm{d}^mf(x)/\textrm{d}x^m \equiv f^{\X{(m)}}(x)$, $m\in \mathds{N}$, to be accurate to $1$ part in $10^p$, the function $f(x)$ is to be calculated to an accuracy of at least $1$ part in $10^{2mp}$.

With $\wh{A}$ denoting a quantum-mechanical operator,\footnote{An \textsl{observable} or otherwise.} its grand-canonical-ensemble average $\langle \wh{A} \rangle$ is equally obtained from the expression\,\footnote{Compare with Eqs\,(\protect\ref{e18}), (\protect\ref{e91}), and (\protect\ref{e95}) below.}
\begin{equation}\label{e350}
\langle \wh{A} \rangle = \Tr\!\big[\h{\varrho}\hspace{1.0pt}\wh{A}],
\end{equation}
where the statistical operator $\h{\varrho}$ is defined in Eq.\,(\ref{e17}) below. Describing the expression on the right-hand side (RHS) of Eq.\,(\ref{e350}) in terms of the path integral corresponding to the Euclidean action \cite{RPF53,IZ80,GP98,NO98} associated with the Hamiltonian $\wh{H}$ in Eq.\,(\ref{e15}) below, for systems of bosons one is to deal with commuting fields \cite{NO98}. In contrast, for systems of fermions one is to deal with anti-commuting Grassmann fields \cite{NO98}. To bypass use of the latter fields, in practice the direct interaction part of the fermions in the Euclidean action is dispensed with in exchange for a bosonic field (or \textsl{fields} in the case of fermions with spin) through the application of the Hubbard-Stratonovich \cite{JH59} transformation. This approach results in a determinant of non-interacting one-particle Green functions \cite{SS81,BSS81,LG92,AE08}, where the relevant Green function differs from the conventional Green function $\mathscr{G}_{\X{0}}$ (or $G_{\X{0}}$ in the zero-temperature limit) encountered in this paper\,\footnote{See Eqs\,(\protect\ref{e14}) and (\protect\ref{e26}), as well as Eqs\,(\protect\ref{e258}), (\protect\ref{e260}), and (\protect\ref{e203}). See also the sixth remark in \S\,\protect\ref{saa2b}, embedding Eq.\,(\protect\ref{ea1r}).} by not being defined in terms of `time'-ordered products of creation and annihilation field operators.\footnote{See in particular \S\S\,2.5, 2.6, and 2.10 (pp.\,189 and 194) in Ref.\,\citen{LG92}. Compare with the correlation functions $G^{\protect\X{<}}$ and $G^{\protect\X{>}}$ encountered in non-equilibrium formalisms \cite{PD84,RS86} (see in particular \S\,2 of Ref.\,\protect\citen{PD84}).}

Aside from the last observation, whereas the sizes of the matrices to be dealt with in the context of the considerations of this paper scale with the order of the perturbation theory, \S\S\,\ref{sec1a}, \ref{s212}, \ref{s32}, those of the matrices in a determinantal scheme cover a wide range of values that in general is unbounded.\footnote{For an approximate approach (amenable to being made arbitrarily accurate) bypassing this problem in the case of Hubbard-type lattice models, see Ref.\,\protect\citen{BPS04}.} An exception concerns models defined on finite lattices, with each lattice site potentially accommodating a finite number of particles. For illustration, consider such lattice model as the \textsl{single-band} Hubbard Hamiltonian \cite{PWA59,ThWR62,JH63} for spin-$\tfrac{1}{2}$ fermions defined on $N_{\textsc{s}}$ lattice sites, Eq.\,(\ref{e27}) below.\footnote{For the Hubbard Hamiltonian under discussion, in $d$ space dimensions the discrete Hubbard-Stratonovich transformations by Hirsch \cite{JEH83,GK97} enable one to deal with a discrete set of auxiliary variables, bypassing use of discretized Hubbard-Stratonovich fields in numerical calculations. For spin-$\tfrac{1}{2}$ fermions, at each Trotter \cite{HFT59,MS76} time slice the discrete Hubbard-Stratonovich variables have the form of the Ising spins, taking one of the two values $\pm 1$ at each site. For details of a Monte-Carlo calculation, corresponding to $d=2$, see \S\,3, p.\,195, in Ref.\,\protect\citen{LG92}. For an approximate, but $\mathrm{SU}(2)$-symmetry-preserving discrete Hubbard-Stratonovich transformation, see Ref.\,\protect\citen{FFA05} as well as Appendix 10.B, p.\,347, in Ref.\,\protect\citen{AE08}. The error in this transformation is of the order of $(\Delta\tau)^4$, where $\Delta\tau \equiv \hbar\beta/L$, with $L$ denoting the number of Trotter decompositions along the imaginary-time axis.} As the name indicates, in this model each lattice site can accommodate at most two particles of opposite spins, so that in the grand canonical ensemble one encounters $N$-particle states, with $N = \sum_{\sigma \X{\in\{\uparrow,\downarrow\}}} N_{\sigma}$, where $N_{\sigma}$, the number of particles with spin index $\sigma$, varies between $0$ and $N_{\textsc{s}}$, implying that in determinantal methods one in principle is to deal with determinants of matrices whose size can be as large as $N_{\textsc{s}} \times N_{\textsc{s}}$.\footnote{By symmetry, different spin species, signified by $\sigma \in \{\uparrow,\downarrow\}$, can be treated separately.} With $\b{N}$ denoting the ensemble average of the number of particles corresponding to fixed values of temperature and chemical potential, for general systems and $\b{N}_{\sigma} \simeq \b{N}/2$, $\forall\sigma$, in determinantal methods on average the relevant matrices are (approximately) of the size $\b{N}/2 \times \b{N}/2$.\footnote{The determinantal continuous-time Monte Carlo methods (to be discussed below), the category to which also the methods of this paper belong, are therefore suited for dealing with fermion systems in the thermodynamic limit.}

The determinantal approaches referred to above rely on the discretization of the integral of the above-mentioned Euclidean action along the imaginary-time axis, leading to inevitable inaccuracies that are difficult to overcome in practice. Since increasing the number of the Trotter slices \cite{HFT59,MS76} of the interval $[0,\hbar\beta]$, Eq.\,(\ref{e23}) below, is to be accompanied by increased accuracy in the underlying calculations \cite{AW82,DRdR83}, and since these approaches invariably rely on the quantum Monte Carlo sampling methods \cite{Note1,FS02,MRRTT53,JMT07,WF68,DCH62,WKH70,CF81,
AW13,GL10,HH75,NM02,JEG03,DMC95,DMC13,KW08,BH10,LB15,DRL85,MS87,VdL92,AWS97,FFA05}, the required accuracy can prove prohibitively difficult, if at all possible (in particular at sufficiently low temperatures),\footnote{Think of the \textsl{sign problem} \protect\cite{GBPF93,TW05,GMLRTW11}, discussed in the relevant references cited in Ref.\,\protect\citen{Note1}.} to achieve in practice.

The above-mentioned imaginary-time-discretization error can be avoided by employing the method of stochastic series expansion of $\exp(-\beta \wh{H})$ in powers of the Hamiltonian $\wh{H}$  \cite{SK91,RGM13},\footnote{See also \S\,4.2, p.\,614, in Ref.\,\protect\citen{TT06}, and \S\,10.3.2, p.\,301, in Ref.\,\protect\citen{AE08}.} Eqs\,(\ref{e15}) and (\ref{e16}) below, the continuous Euclidean-time loop algorithm \cite{BW967},\footnote{This algorithm is based on the functional-integral formalism of Farhi and Gutmann \protect\cite{FG92}, which in principle is applicable to systems based on a separable single-particle Hilbert space. Considering one-particle systems on a lattice, in Ref.\,\protect\citen{FG92} is has been shown that for non-relativistic particles the constructed functional integral is \textsl{not} well-defined in the continuum limit.} and the so-called continuous-integral methods. Before discussing the latter methods, we point out that the Trotter approximation applies to \textsl{bounded} operators \cite{HFT59,MS76}, implying that use of this approximation for general systems, \S\,\ref{s211}, is not warranted. Further, the computational complexity \cite{PP13} associated with the use of the  Hubbard-Stratonovich transformation \cite{JH59} is higher in the case of fermions interacting through a non-contact-type (or non-local) interaction potential than in the idealised case where the interaction potential is contact-type \cite{ANR03,RL04,RSL05}, as in the single-band Hubbard Hamiltonian, Eq.\,(\ref{e27}) below.

The continuous-integral methods indicated above are generally more completely referred to as continuous-time quantum \textsl{Monte Carlo} methods for their common applications in conjunction with the quantum Monte Carlo sampling method \cite{Note1}. These methods \cite{PST96,PST98,BPS06,RHJ98,ANR03,RL04,RSL05,WCMTM06,GWPT08,GT13,GMLRTW11} are invariably based on conventional perturbation series expansions, in particular of the grand partition function $\mathcal{Z}$, so that their novelty rests in the specific ways in which the underlying expressions are stochastically sampled, respecting detailed balance and ergodicity, as well as avoiding decline in the convergence rate arising from increased frequency of the rejection of the attempted Monte Carlo moves \cite{Note1}. The exact perturbational expression for the one-particle Green function $G$ that we employ in this paper, Eq.\,(\ref{e242}) below, in essence coincides with the perturbational expression for the $G$ underlying the continuous-time quantum Monte Carlo method by Rubtsov and collaborators \cite{ANR03,RL04,RSL05}, both expressions being the weak-coupling perturbation series expansion for $G$ (compare the expression in Eq.\,(\ref{e242}) below with for instance that in Eq.\,(6) of Ref.\,\citen{RSL05}).\footnote{As the details in \S\,\protect\ref{s222} make explicit, the \textsl{exact} division of the numerator by the denominator in the expression on the RHS of Eq.\,(\protect\ref{e242}), when both infinite sums herein are approximated by finite sums, gives rise to contributions corresponding \textsl{disconnected} Green-function diagrams. \emph{This source of uncontrolled error is absent in the schemes proposed in the present paper.}}

Considering the one-particle Green function $G$ for systems of interacting \textsl{fermions}, by formally expanding the determinants encountered in the formalism of Rubtsov and collaborators \cite{ANR03,RL04,RSL05} and discarding the perturbational contributions associated with disconnected Green-function diagrams, appendix \ref{sac}, one arrives at the diagrammatic Monte Carlo method \cite{PS98,PS08,vHKPS10,KvHGPPST10,ASM08,GWFSPT11,GMLRTW11,vHetal12,NP13} for $G$. In the framework of this method, the connected Green-function diagrams are stochastically sampled, using a Markov process \cite{WF68,Note1} that treats the order of the perturbation expansion and the variables associated with each order of the perturbation expansion on the same footing as the integrals and sums in terms of which the algebraic expressions associated with diagrams are described \cite{vHKPS10,KvHGPPST10,vHetal12,NP13}.

In applying the diagrammatic Monte Carlo method for calculating for instance $G$, in two different ways account is taken of the contributions of the relevant diagrams to an infinite order. These we describe in the next two paragraphs. A third approach, based on summation techniques and extrapolation of the calculated results associated with finite orders of perturbation theory to infinite order, has also been applied \cite{PS08,PS08R,vHKPS10,PPS10}. We shall not go into this approach here and relegate a detailed discussion of it to Ref.\,\citen{BF16a}.\footnote{\emph{Note added to \textsf{arXiv:1912.00474v2}}: To keep the extent of Ref.\,\protect\citen{BF16a} within reasonable bounds, in the final analysis we decided to relegate the `detailed discussion' as promised here to a separate publication. The text of Ref.\,\protect\citen{BF16a} that will be published at the same time as \textsf{arXiv:1912.00474v2}, contains a section [\S\,5] on a new method of (re-) summation of perturbation series that very deliberately avoids any `detailed discussion' of various summation techniques.}

The above-mentioned two approaches are referred to as `bold-line' methods, reflecting the fact that, in dealing with Feynman diagrams, solid \textsl{bold} lines are customarily used to represent $G$, to be contrasted with solid \textsl{thin} lines that customarily are used to represent $G_{\X{0}}$. In the first approach \cite{PS08,vHKPS10,KvHGPPST10,vHetal12,NP13}, the self-energy $\Sigma$ is calculated through performing diagrammatic Monte Carlo calculations on the set of skeleton self-energy diagrams,\footnote{Compare with the considerations in appendix \protect\ref{sad}.}\footnote{We note that Ref.\,\protect\citen{vHetal12} reports use of both $G$- and $W$-skeleton self-energy diagrams.} with the $G$ employed in the determination of the contributions of the Monte-Carlo-sampled skeleton self-energy diagrams being in principle self-consistently calculated on the basis of the Dyson equation.\footnote{The calculation reported in Ref.\,\protect\citen{PS08} concerns a polaron model, described by the Hamiltonian in Eq.\,(1) herein.} In practice, the bold-line method of this kind may be implemented partially \cite{vHKPS10},\footnote{See section `Bold propagators', p.\,102, in Ref.\,\protect\citen{vHKPS10}.} by for instance restricting the set of the self-energy diagrams to be considered to those without tadpole self-energy insertions, this in exchange for evaluating the self-energy diagrams in terms of $G_{\textsc{h}}$, the one-particle Green function corresponding to the Hartree Hamiltonian in which the static Hartree self-energy $\Sigma^{\textsc{h}}$ is to be calculated self-consistently\refstepcounter{dummy}\label{WhenDealingWith}\,\footnote{When dealing with the uniform ground states (GSs) of the single-band Hubbard Hamiltonian $\protect\wh{\mathcal{H}}$ for spin-$\tfrac{1}{2}$ fermions defined on a Bravais lattice \protect\cite{AM76}, Eq.\,(\protect\ref{e27}) below, with the interaction part $\protect\wh{\mathcal{H}}_{\X{1}}$ as represented in Eq.\,(\protect\ref{e32}) below, in the Fourier space (the $\bm{k}$ space, with $\bm{k}$ defined over the underlying first Brillouin zone \protect\cite{AM76}, $\protect\1BZ$), for the Hartree self-energy $\protect\h{\Sigma}^{\textsc{h}}$ one has \protect\cite{BF13} $\Sigma^{\textsc{h}}(\bm{k}) = Un/\hbar$, where $n = n_{\uparrow} + n_{\downarrow}$ is the total site occupation number. Because of the strict locality of the two-body interaction potential, in the case at hand the Fock (or the bare exchange) self-energy corresponding to spin-$\sigma$ particles, that is $\protect\h{\Sigma}_{\sigma}^{\textsc{f}}$, is similarly local, for which in the Fourier space one has \protect\cite{BF13} $\Sigma_{\sigma}^{\textsc{f}}(\bm{k}) = -U n_{\sigma}/\hbar$. Consequently, in the case at hand for the Hartree-Fock self-energy corresponding to spin-$\sigma$ particles, that is for $\protect\h{\Sigma}_{\sigma}^{\textsc{hf}} \equiv \protect\h{\Sigma}^{\textsc{h}} + \protect\h{\Sigma}_{\sigma}^{\textsc{f}}$, Eq.\,(\protect\ref{e250b}) below, in the Fourier space one has $\Sigma_{\sigma}^{\textsc{hf}}(\bm{k}) = U n_{\protect\b{\sigma}}/\hbar$, where $\protect\b{\sigma}$ denotes the spin index complementary to $\sigma$. In the paramagnetic state, where $n_{\sigma} = n_{\protect\b{\sigma}} = \tfrac{1}{2} n$, $\Sigma_{\sigma}^{\textsc{f}}(\bm{k})$ and $\Sigma_{\sigma}^{\textsc{hf}}(\bm{k})$ do not depend on $\sigma$. In this footnote, $\langle\bm{k}\vert \protect\h{A}\vert\bm{k}'\rangle = \langle\bm{k}\vert \protect\h{A}\vert\bm{k}\rangle \delta_{\bm{k},\bm{k}'} \equiv A(\bm{k}) \delta_{\bm{k},\bm{k}'}$, where $\h{A}$ stands for $\protect\h{\Sigma}^{\textsc{h}}$, $\protect\h{\Sigma}_{\sigma}^{\textsc{f}}$, and $\protect\h{\Sigma}_{\sigma}^{\textsc{hf}}$, and $\vert\bm{k}\rangle$ for the normalised eigenstate of the single-particle $\protect\h{\bm{k}}$ operator, subject to the box boundary condition, corresponding to the eigenvalue $\bm{k}$. For details, consult appendix A in Ref.\,\protect\citen{BF13}.} (see also the following paragraph).

In the second approach \cite{GRM10},\footnote{See also Refs\,\protect\citen{CGRM14} and \protect\citen{CRMG14}.} the `worm' algorithm / updating scheme \cite{PST96,PST98,BPS06} is used to sample a set of diagrams describing an \textsl{extended} partition function $\mathcal{Z}$ described in terms of $G$-skeleton self-energy diagrams.\footnote{As has been pointed out in Ref.\,\protect\citen{GRM10}, the complete set of Green-function diagrams in the bold-line expansion of this function does \textsl{not} coincide with the set of linked diagrams corresponding to all possible ways in which a single Green-function line can be \textsl{cut} in the bold-line diagrammatic expansion of the partition function $\mathcal{Z}$. To appreciate this observation, one should first consider the expression $\mathcal{Z} = \exp(-\beta \Omega)$, where $\Omega$ stands for the grand potential. Following this, one should consider the equality $\Omega(\lambda) = Y(\lambda)$ in Eq.\,(55) of Ref.\,\protect\citen{LW60}, where the bold-line perturbation expansion of the functional $Y(\lambda)$, for arbitrary coupling constant of interaction $\lambda$, is specified in Eqs\,(47) and (48) (for clarity, consult Table I, p.\,15, and Eqs\,(5.4), (5.7), and (5.8), p.\,24, of Ref.\,\protect\citen{BF07}). One observes that whereas the function $Y'(\lambda)$ is described in terms of bold-line skeleton self-energy diagrams [Eq.\,(48) in Ref.\,\protect\citen{LW60}], this is clearly \textsl{not} the case for the difference function $Y(\lambda) - Y'(\lambda)$.} The one-particle Green function determined in this way is proportional to the sought-after $G$, where the constant of proportionality is readily determined \cite{GRM10}. In Ref.\,\citen{GRM10} the Anderson impurity Hamiltonian has been considered, making use of the perturbation series expansion in powers of the hybridization potential, which couples the correlated impurity electrons with a bath of free conduction electrons, and two approximate diagrammatic schemes for the self-energy operator: the non-crossing approximation (NCA), and the one-crossing approximation (OCA) \cite{KK71,GK81,KM84,PG89,NEB87,ACH97,PF02}; depending on the approximate scheme adopted, the Monte-Carlo-sampled diagrams have been free from the relevant self-energy insertions.

Restricting ourselves to systems of fermions, we point out that the sign problem in Monte Carlo calculations \cite{Note1} is less severe\,\footnote{In some specific cases, such as the case of the single-band Hubbard Hamiltonian with attractive on-site interaction potential \protect\cite{GBPF93}, or repulsive on-site interaction potential at half-filling \protect\cite{ANR03,BPST06b,GMLRTW11}, the sign problem is absent.} in impurity problems than in other problems \cite{GMLRTW11}. The diagrammatic Monte Carlo methods, discussed above, can therefore be used with success for impurity problems. In other cases, \textsl{determinantal} diagrammatic methods are to be used instead, since dealing with the total contributions of subsets of diagrams, with each subset corresponding to the totality of the diagrams associated with a determinant, proves to ameliorate the sign problem \cite{BPST06b}.

\refstepcounter{dummyX}
\section{The formalism}
\phantomsection
\label{sec2}

\refstepcounter{dummyX}
\subsection{Preliminaries}
\phantomsection
\label{s21}
In preparation for the introduction of the diagram-free formalisms for the perturbation series expansions briefly described in \S\,\ref{sec1b}, in this section we present the specifics of the systems, of the (ensemble of) states and of the formalisms that we explicitly consider in this paper.

The considerations of this paper are applicable to both continuum models and lattice models. As regards continuum models, we restrict the considerations to systems in which particles interact through a \textsl{two-body} interaction potential. The two-body interaction potential that we explicitly consider is sufficiently general for many practical applications; to avoid unnecessary notational complication,\footnote{See the closing remark of \S\,\protect\ref{e211a}, p.\,\protect\pageref{TheSimplifiedNotation}.} however without loss of generality, we do not consider the most general two-body interaction potential in the spin space (see later). Regarding lattice models, we explicitly deal with the single-band Hubbard Hamiltonian \cite{PWA59,ThWR62,JH63} as a prominent representative of such models of interacting particles, with the interaction potential, the Hubbard $U$,\footnote{Similarly as in the case of the continuum models just described, $U$ corresponds to a \textsl{two-body} interaction potential.} operative only between the particles at the same lattice site.\footnote{In the Hubbard Hamiltonian to be discussed in \S\,\protect\ref{s224}, the bare interaction at the same site is further restricted between particles with different spin indices.}

The perturbation series expansions that we explicitly deal with in this paper are specific to \textsl{normal} states. With some modifications, these expansions can be made suitable for dealing with \textsl{superfluid} and \textsl{superconductive} states. The modifications may amount to the use of the Nambu-Gor'kov \cite{LPG58,YN60,AGD75,JRS99,KM69}\footnote{See also Ch.\,13, \S\,51, p.\,439, in Ref.\,\protect\citen{FW03}.} matrix formalism, relying on two-component spinor field operators \cite{PWA58}. For this, use of a general two-body interaction potential requires the underlying Hamiltonian to be appropriately Wick ordered\,\footnote{In appendix \protect\ref{saa} we briefly touch on this issue.} so as to avoid divergence at the lowest order of the perturbation theory. We shall not further touch upon these field operators in this paper,\footnote{See however Eq.\,(\protect\ref{es5}) below and the accompanying remark.} relegating the considerations with regard to superfluid and superconductive states to a future publication \cite{BF16b}. We only point out that the Wick theorem that underlies the considerations of this paper applies also in the framework of the weak-coupling perturbation expansions of the Nambu-Gor'kov Green functions \cite{AZ14}.

\refstepcounter{dummyX}
\subsection{Models and formalisms}
\phantomsection
\label{s211}
In this section we introduce two model Hamiltonians for systems of interacting fermions and bosons. Of these, one is a continuum model and the other a lattice model, explicitly, the single-band Hubbard Hamiltonian  \cite{PWA59,ThWR62,JH63}. We further introduce the one-particle Green functions corresponding to both zero temperature and non-zero temperatures. As regards non-zero temperatures, we explicitly consider the imaginary-time formalism of Matsubara \cite{FW03,NO98,MS59,AGD75,TM55,ESF59} and the real-time formalism of the thermo-field dynamics (TFD) \cite{UMT82,HU95}. The latter formalism shares aspects of the Keldysh formalism \cite{LVK65,RAC68,RS86,JR07}.\footnote{Consult for instance Ref.\,\protect\citen{MM86}, where this formalism is discussed under the general rubric of the \textsl{closed-time path} (CTP) formalism of Schwinger \protect\cite{JS61}, Keldysh \protect\cite{LVK65}, and Craig \protect\cite{RAC68}.}

\refstepcounter{dummyX}
\subsubsection{The continuum model}
\phantomsection
\label{e211a}
The continuum model that we consider is embedded in $\mathds{R}^d$ and is described by the Hamiltonian (in the Schr\"{o}dinger picture)
\begin{align}\label{e5}
\wh{H} &= \sum_{\sigma} \int \textrm{d}^d r\, \h{\psi}_{\sigma}^{\dag}(\bm{r}) \big(\uptau(\bm{r})+ v(\bm{r})\big) \h{\psi}_{\sigma}^{\phantom{\dag}}(\bm{r})\nonumber\\
&+ \hspace{-0.7pt} \frac{1}{2} \sum_{\sigma,\sigma'} \int \textrm{d}^d r \textrm{d}^d r'\, u_{\sigma,\sigma'}(\bm{r},\bm{r}')\hspace{0.6pt} \h{\psi}_{\sigma}^{\dag}(\bm{r}) \h{\psi}_{\sigma'}^{\dag}(\bm{r}') \h{\psi}_{\sigma'}^{\phantom{\dag}}(\bm{r}') \h{\psi}_{\sigma}^{\phantom{\dag}}(\bm{r})
\equiv \wh{H}_{\X{0}} + \wh{H}_{\X{1}},\hspace{0.2cm}
\end{align}
where $\h{\psi}_{\sigma}^{\phantom{\dag}}(\bm{r})$ and $\h{\psi}_{\sigma}^{\dag}(\bm{r})$ are respectively annihilation and creation field operators in the Schr\"{o}diger picture corresponding to particles with spin index $\sigma$ (see later), satisfying
\begin{equation}\label{e5a}
\big[\h{\psi}_{\sigma\hspace{-2.6pt}\phantom{'}}^{\phantom{\dag}}(\bm{r}), \h{\psi}_{\sigma'}^{\dag}(\bm{r}')\big]_{\mp} = \delta^d(\bm{r}-\bm{r}')\hspace{0.6pt} \delta_{\sigma,\sigma'}\hspace{0.6pt}\h{1},\;\; \big[\h{\psi}_{\sigma\hspace{-2.6pt}\phantom{'}}^{\phantom{\dag}}(\bm{r}), \h{\psi}_{\sigma'}^{\phantom{\dag}}(\bm{r}')\big]_{\mp} = [\h{\psi}_{\sigma\hspace{-2.6pt}\phantom{'}}^{\dag}(\bm{r}),
\h{\psi}_{\sigma'}^{\dag}(\bm{r}')]_{\mp} = \h{0},
\end{equation}
where $[\phantom{x},\phantom{x}]_{-/+}$ stands for commutation / anti-commutation,\footnote{See appendix \protect\ref{sae}.} depending on whether the particles under consideration are bosons / fermions, and $\delta^d$ for the $d$-dimensional Dirac $\delta$ function. Further, $\h{1}$ is the identity operator in the Fock space of the problem at hand, and $\h{0} \doteq 0 \times \h{1}$. The function $\uptau(\bm{r})$ on the RHS of Eq.\,(\ref{e5}) denotes the single-particle kinetic-energy operator, $v(\bm{r})$ the local external potential,\footnote{The considerations of this paper immediately apply to the cases where the external potential $v(\bm{r})$ is replaced by the more general spin-dependent potential $v_{\sigma}(\bm{r})$. This possibility is relevant for perturbational calculations in which the zeroth-order Hamiltonian is that encountered within the framework of spin density-functional theory \protect\cite{vBH72,RC73}. Naturally, with the $\protect\wh{H}_{\protect\X{0}}$ in such calculations deviating from that assumed in Eq.\,(\ref{e5}), the corresponding perturbation Hamiltonian $\protect\wh{H}_{\protect\X{1}}$ should be adjusted accordingly (see for instance Ref.\,\protect\citen{BF9799}). Because of the general form of the two-body interaction potential $u_{\sigma,\sigma'}(\bm{r},\bm{r}')$ considered here (specifically insofar as its dependence on $\sigma$ and $\sigma'$ is concerned), this is feasible. Insofar as the self-energy operator is concerned, the relevant details are similar to those encountered \S\,\protect\ref{s26} below, where the contribution of the \textsl{local} Hartree self-energy operator $\protect\h{\Sigma}^{\protect\textsc{h}}$ is isolated. In this connection, on using the identity $v(\bm{r}) \equiv v_{\sigma}(\bm{r}) + \lambda (v(\bm{r})-v_{\sigma}(\bm{r}))|_{\lambda=1}$, incorporation of the contribution of  $\lambda (v(\bm{r})-v_{\sigma}(\bm{r}))$ in $\protect\wh{H}_{\protect\X{1}}$ gives rise to a \textsl{local} self-energy contribution similar to $\protect\h{\Sigma}^{\protect\textsc{h}}$. As a result of this locality, any self-energy diagram of order $\nu \ge 2$ containing this self-energy cannot be $G$-skeleton / 2PI.} and $u_{\sigma,\sigma'}(\bm{r},\bm{r}')$ the bare two-body interaction potential. In first-principles calculations, one has
\begin{equation}\label{e6}
\uptau(\bm{r}) \equiv -\frac{\hbar^2}{2\mathsf{m}} \nabla_{\bm{r}}^2,
\end{equation}
where $\mathsf{m}$ denotes the bare particle mass, and the function $u_{\sigma,\sigma'}(\bm{r},\bm{r}')$ is identified with the spin-independent Coulomb potential $u_{\textsc{c}}(\bm{r}-\bm{r}')$, which is further a function of $\|\bm{r}-\bm{r}'\|$. The integrals in Eq.\,(\ref{e5}) are over the single-particle configuration space of the system under consideration, embedded in $\mathds{R}^d$.

\refstepcounter{dummy}\label{AsIsCommon}As is common to most condensed-matter applications, in this paper we assume that \textsl{irrespective} of the value of $d$ the spin of particles is associated with the rotation group specific to three-dimensional Euclidean space, that is $\mathrm{SO}(3)$, of which $\mathrm{SU}(2)$ is the universal covering group \cite{JF09,FS97}. Thus, the spin-$\mathsf{s}$ particles considered in this paper correspond to the $(2\mathsf{s}+1)$-dimensional unitary representation of the $\mathrm{SU}(2)$ group. Denoting the operators of the underlying Lie algebra $\mathfrak{su}(2)$ by $\{S^x, S^y, S^z\}$, the index $\sigma$, as encountered above and in the remaining part of this paper, stands in a one-to-one correspondence with an eigenvalue of the $(2\mathsf{s}+1)$-dimensional unitary representation of $S^z$. With
\begin{equation}\label{e6ax}
S^z  \vert\mathsf{s},m_z\rangle = \hbar\hspace{0.6pt}m_z \vert\mathsf{s},m_z\rangle,
\end{equation}
for instance for $\mathsf{s} = \frac{1}{2}$ the index $\sigma =\uparrow$ corresponds to $m_z = \tfrac{1}{2}$, and the index $\sigma = \downarrow$ to $m_z = -\tfrac{1}{2}$.

The interaction Hamiltonian $\wh{H}_{\X{1}}$ in Eq.\,(\ref{e5}) is a specific case of the following interaction Hamiltonian (\emph{cf.} Eq.\,(7.12), p.\,67, in Ref.\,\citen{FW03}):
\begin{equation}\label{e6a}
\wh{H}_{\X{1}}' =  \frac{1}{2} \sum_{\sigma,\sigma',\sigma'',\sigma'''} \int \textrm{d}^d r \textrm{d}^d r'\, \b{u}_{\sigma,\sigma''';\sigma',\sigma''}(\bm{r},\bm{r}')\hspace{0.6pt} \h{\psi}_{\sigma}^{\dag}(\bm{r}) \h{\psi}_{\sigma'}^{\dag}(\bm{r}') \h{\psi}_{\sigma''}^{\phantom{\dag}}(\bm{r}') \h{\psi}_{\sigma'''}^{\phantom{\dag}}(\bm{r}).
\end{equation}
The simpler interaction Hamiltonian $\wh{H}_{\X{1}}$ in Eq.\,(\ref{e5}) is recovered on effecting the substitution
\begin{equation}\label{e6b}
\b{u}_{\sigma,\sigma''';\sigma',\sigma''}(\bm{r},\bm{r}') \rightharpoonup u_{\sigma,\sigma'}(\bm{r},\bm{r}')\hspace{0.6pt} \delta_{\sigma,\sigma'''} \delta_{\sigma',\sigma''}.
\end{equation}
As we have indicated above, use of the two-body potential on the RHS of this substitution does not affect the generality of the formalisms introduced in this paper.

With $\wh{H}_{\X{1};\textsc{i}}(t)$ denoting the above $\wh{H}_{\X{1}}$ in the \textsl{interaction picture}, one has\,\footnote{In this paper $\protect\ii \equiv \sqrt{-1}$ is distinct from $i$, which we generally employ either as an integer-valued index, or a compound variable similar to $j$.} [p.\,54 in Ref.\,\citen{FW03}]
\begin{equation}\label{e7}
\wh{H}_{\X{1};\textsc{i}}(t) = \e^{\ii \wh{H}_{\X{0}} t/\hbar} \wh{H}_{\X{1}} \e^{-\ii \wh{H}_{\X{0}} t/\hbar} \equiv \frac{1}{2} \int \rd 1 \rd 2\; v(1,2) \hspace{0.6pt} \h{\psi}_{\textsc{i}}^{\dag}(1) \h{\psi}_{\textsc{i}}^{\dag}(2) \h{\psi}_{\textsc{i}}(2) \h{\psi}_{\textsc{i}}(1),
\end{equation}
where we have introduced the short-hand notation\,\footnote{Throughout this paper, we use the symbol $\rightleftharpoons$ to express a form of equivalence that cannot be expressed by the equality and identity signs.}
\begin{equation}\label{e8}
j \rightleftharpoons \bm{r}_j, t_j, \sigma_j \rightleftharpoons \bm{r}_j t_j \sigma_j,
\end{equation}
whereby\,\footnote{The field operator $\h{\psi}_{\textsc{i}}(j)$ is in the (real-time) \textsl{interaction} picture.}
\begin{equation}\label{e9}
\h{\psi}_{\textsc{i}}(j) \doteq \e^{\ii \wh{H}_{\X{0}} t_j/\hbar} \h{\psi}_{\sigma_j}(\bm{r}_j) \e^{-\ii \wh{H}_{\X{0}} t_j/\hbar}.
\end{equation}
Accordingly,
\begin{equation}\label{e10}
v(i,j) \doteq u_{\sigma_i,\sigma_j}(\bm{r}_i,\bm{r}_j)\hspace{0.6pt} \delta(t_i-t_j),
\end{equation}
and\,\footnote{As regards the integration with respect to $t_j$ over the interval $(-\infty,\infty)$, see Fig.\,4.5 (c), p.\,107, in Ref.\,\protect\citen{SvL13}, as well as \S\,5.4, p.\,140, herein. We are therefore implicitly relying on the \textsl{adiabatic approximation}, which can be relaxed.}
\begin{equation}\label{e11}
\int \rd j \rightleftharpoons \sum_{\sigma_j}\int_{-\infty}^{\infty} \rd t_j \int \textrm{d}^d r_j.
\end{equation}
In the following
\begin{equation}\label{e260b}
j^+ \rightleftharpoons \bm{r}_j,t_j^+,\sigma_j \rightleftharpoons \bm{r}_j t_j^+ \sigma_j,
\end{equation}
where $t_j^+ \doteq t_j + 0^+$. \refstepcounter{dummy}
\label{TheSimplifiedNotation}The simplified notation on the left in Eq.\,(\ref{e11}) is an immediate consequence of the specific assumption with regard to the two-body interaction function specified in Eq.\,(\ref{e6b}). Without this assumption, the number of summations with respect to spin indices in the defining expression for $\wh{H}_{\X{1}}$ would have been four, instead of two.\footnote{Additional summations with respect to spin indices would be somewhat similar to the summations with respect to $\{\mu_j\| j\}$ in \S\S\,\protect\ref{sac01} and \protect\ref{s226} below.}

\refstepcounter{dummyX}
\subsubsection{The one-particle Green functions for \texorpdfstring{$T=0$}{} and \texorpdfstring{$T>0$}{} (Matsubara formalism)}
\phantomsection
\label{s222x}
With \refstepcounter{dummy}\label{WithPsi}$\h{\psi}_{\sigma;\textsc{h}}(\bm{r}t)$ denoting the Heisenberg-picture \cite{FW03} counterpart of $\h{\psi}_{\sigma}(\bm{r})$, for the one-particle Green function $G_{\sigma,\sigma'}(\bm{r} t, \bm{r}' t')$ one has \cite{FW03}
\begin{equation}\label{e11a}
G_{\sigma,\sigma'}(\bm{r} t,\bm{r}'t') \equiv G(\bm{r} t \sigma,\bm{r}'t'\sigma')  \doteq -\ii\hspace{1.6pt} \langle\Psi_{N;0}\vert \mathcal{T}\big\{\h{\psi}_{\sigma\hspace{-2.4pt}\phantom{'};\textsc{h}}^{\phantom{\dag}}(\bm{r}t) \h{\psi}_{\sigma';\textsc{h}}^{\dag}(\bm{r}'t')\big\}\vert\Psi_{N;0}\rangle,
\end{equation}
where $\mathcal{T}$ denotes the boson / fermion \textsl{chronological} time-ordering operator\refstepcounter{dummy}\label{ToBeDistinguishedFrom}\,\footnote{To be distinguished from the \textsl{anti-chronological} time-ordering operator $\protect\b{\mathcal{T}}$ (or $T^{\textrm{a}}$ when $\mathcal{T}$ is denoted by $T^{\textrm{c}})$, which one encounters in the Keldysh formalism \protect\cite{LVK65,PD84,SvL13}.} (for the field operators satisfying the commutation / anti-commutation relations in Eq.\,(\ref{e5a})), and $\vert\Psi_{N;0}\rangle$ the \textsl{normalised} $N$-particle ground state (GS) of $\wh{H}$. The state $\vert\Psi_{N;0}\rangle$ is therefore in the Heisenberg picture.\footnote{Compare with Eqs\,(6.33) and (6.34) on p.\,59 of Ref.\,\protect\citen{FW03}.} With reference to Eq.\,(\ref{e8}), following the identifications
\begin{equation}\label{e12}
a \rightleftharpoons \bm{r} t \sigma,\;\; b \rightleftharpoons \bm{r}' t'\sigma',
\end{equation}
in the following we shall use the notation
\begin{equation}\label{e13}
G(a,b) \equiv G(\bm{r}t \sigma,\bm{r}'t'\sigma'),
\end{equation}
and similarly
\begin{equation}\label{e14}
G_{\X{0}}(a,b) \equiv G_{\X{0}}(\bm{r}t\sigma,\bm{r}'t'\sigma'),
\end{equation}
where $G_{\X{0}}$ is the one-particle Green function corresponding to $\wh{H}_{\X{0}}$, Eq.\,(\ref{e5}). For $\wh{H}$ and $\wh{H}_{\X{0}}$ time independent, the functions $G$ and $G_{\X{0}}$ in Eqs\,(\ref{e13}) and (\ref{e14}) depend on $t-t'$, rather than on $t$ and $t'$ separately.

\refstepcounter{dummy}
\label{SinceHIs}
For $\vert\Psi_{N;0}\rangle$ an eigenstate of the $z$ component of the total spin operator,\footnote{See p.\,\protect\pageref{AsIsCommon}.} $G_{\sigma,\sigma'}$ is diagonal in the spin space, that is
\begin{equation}\label{e11a1}
G_{\sigma,\sigma'}(\bm{r} t,\bm{r}'t') \equiv G_{\sigma}(\bm{r} t,\bm{r}'t')\hspace{0.7pt} \delta_{\sigma,\sigma'}.
\end{equation}
To clarify this observation, let $\mathds{S}^z = \hbar\hspace{0.6pt} \bbsigma^{z}(\mathsf{s})$ denote the $(2\mathsf{s}+1) \times (2\mathsf{s}+1)$ matrix representation of the single-particle operator $S^z$, referred to above, p.\,\pageref{AsIsCommon}. One has
\begin{equation}\label{e11b}
\bbsigma^{z}(\tfrac{1}{2}) =
\begin{pmatrix}
\tfrac{1}{2} & 0 \\
0 & \b{\tfrac{1}{2}}
\end{pmatrix}\hspace{-1.4pt},\;
\bbsigma^{z}(1) =
\begin{pmatrix}
1 & 0 & 0\\
0 & 0 & 0 \\
0 & 0 & \b{1}
\end{pmatrix}\hspace{-1.4pt},\;
\bbsigma^{z}(\tfrac{3}{2}) =
\begin{pmatrix}
\tfrac{3}{2} & 0 & 0 & 0\\
0 & \tfrac{1}{2} & 0 & 0\\
0 & 0 & \b{\tfrac{1}{2}} & 0 \\
0 & 0 & 0 & \b{\tfrac{3}{2}}
\end{pmatrix}\hspace{-1.4pt},\; \dots,
\end{equation}
where $\b{i} \equiv -i$. With these matrices at hand, for the $z$ component $\wh{S}^z$ of the total-spin operator $\hspace{2.6pt}\wh{\rule{-2.6pt}{-1.0ex}\bm{S}}$ one has
\begin{equation}\label{e11c}
\wh{S}^z = \hbar \sum_{\sigma,\sigma'} \int \textrm{d}^dr\; \h{\psi}_{\sigma}^{\dag}(\bm{r}) (\bbsigma^{z}(\mathsf{s}))_{\sigma,\sigma'}^{\phantom{\dag}} \h{\psi}_{\sigma'}^{\phantom{\dag}}(\bm{r}) \equiv \hbar \sum_{\sigma} (\bbsigma^{z}(\mathsf{s}))_{\sigma,\sigma} \wh{N}_{\sigma},
\end{equation}
where
\begin{equation}\label{e11d}
\wh{N}_{\sigma} \doteq \int \textrm{d}^dr\; \h{\psi}_{\sigma}^{\dag}(\bm{r}) \h{\psi}_{\sigma}^{\phantom{\dag}}(\bm{r})
\end{equation}
is the total-number operator corresponding to particles with spin index $\sigma$. In the particular case of spin-$\tfrac{1}{2}$ particles, one has
\begin{equation}\label{e11e}
\wh{S}^z = \frac{\hbar}{2} (\wh{N}_{\uparrow} - \wh{N}_{\downarrow}).\;\;\;\;\; \text{(For spin-$\tfrac{1}{2}$ particles)}
\end{equation}
One explicitly demonstrates that for the Hamiltonian in Eq.\,(\ref{e5})
\begin{equation}\label{e11f}
\big[\wh{H},\wh{N}_{\sigma}\big]_{-} = 0,
\end{equation}
so that
\begin{equation}\label{e11g}
\big[\wh{H},\wh{S}^{z}\big]_{-} = 0.
\end{equation}
Hence, $\vert\Psi_{N;0}\rangle$ can indeed be chosen as a simultaneous eigenstate of $\wh{H}$ and $\wh{S}^{z}$.

The one-particle Green function in Eq.\,(\ref{e11a}) is specific to zero temperature, $T=0$. To introduce the counterpart of this function corresponding to a non-zero temperature equilibrium ensemble of states, we begin with the grand canonical Hamiltonian $\wh{\mathcal{K}}$ corresponding to the Hamiltonian in Eq.\,(\ref{e5}) [Ch.\,7 in Ref.\,\citen{FW03}]:
\begin{equation}\label{e15}
\wh{\mathcal{K}} \doteq \wh{H} - \mu \wh{N} \equiv (\wh{H}_{\X{0}} - \mu \wh{N}) + \wh{H}_{\X{1}} \equiv \wh{\mathcal{K}}_{\X{0}} + \wh{\mathcal{K}}_{\X{1}},
\end{equation}
where $\mu$ is the chemical potential,\footnote{Not to be confused with the binary variable $\mu$ in \S\S\,\protect\ref{sac01} and \protect\ref{s226} below.} and $\wh{N} \equiv \sum_{\sigma} \wh{N}_{\sigma}$ the total-number operator, Eq.\,(\ref{e11d}). With
\begin{equation}\label{e16}
\mathcal{Z} \doteq \Tr\!\big[\e^{-\beta \wh{\mathcal{K}}}\big]
\end{equation}
the grand partition function, where $\beta \equiv 1/(k_{\textsc{b}} T)$, and\refstepcounter{dummy}\label{WithOmegaDenoting}\,\footnote{With $\Omega$ denoting the grand potential, one has $\mathcal{Z} = \exp(-\beta\Omega)$, so that $\h{\varrho} = \exp(\beta (\Omega \h{1}-\wh{\mathcal{K}}))$.}
\begin{equation}\label{e17}
\h{\varrho} \doteq \frac{1}{\mathcal{Z}} \e^{-\beta \wh{\mathcal{K}}},
\end{equation}
for the thermal one-particle Green function $\mathscr{G}$ in the Matsubara formalism \cite{FW03,NO98,MS59,AGD75,TM55,ESF59} one has (\emph{cf.} Eq.\,(\ref{e11a}))
\begin{equation}\label{e18}
\mathscr{G}_{\sigma,\sigma'}(\bm{r}\tau,\bm{r}'\tau') \equiv \mathscr{G}(\bm{r}\tau\sigma,\bm{r}'\tau'\sigma') \doteq -\Tr\!\big[\h{\varrho}\hspace{0.8pt} \mathcal{T}_{\X{\uptau}}\big\{\h{\psi}_{\sigma;\textsc{k}}^{\phantom{\dag}}(\bm{r}\tau) \h{\psi}_{\sigma';\textsc{k}}^{\dag}(\bm{r}'\tau')\big\}\big],
\end{equation}
where $\tau \equiv \ii t$ and $\tau' \equiv \ii t'$ correspond to imaginary times $t$ and $t'$ for $\tau, \tau' \in\mathds{R}$, and $\mathcal{T}_{\X{\uptau}}$ the boson / fermion imaginary-time-ordering operator (compare with the time-ordering operator $\mathcal{T}$ in Eq.\,(\ref{e11a})). \emph{Unless we indicate otherwise, in this paper} $\tau, \tau' \in \mathds{R}$ (\emph{cf.} Eq.\,(\ref{e23}) below). The field operators $\h{\psi}_{\sigma;\textsc{k}}^{\phantom{\dag}}(\bm{r}\tau)$ and $\h{\psi}_{\sigma;\textsc{k}}^{\dag}(\bm{r}\tau)$ are the imaginary-time Heisenberg pictures of respectively $\h{\psi}_{\sigma}^{\phantom{\dag}}(\bm{r})$ and $\h{\psi}_{\sigma}^{\dag}(\bm{r})$. In contrast to $\h{\psi}_{\sigma;\textsc{h}}^{\dag}(\bm{r}t)$ which is the Hermitian conjugate of $\h{\psi}_{\sigma;\textsc{h}}^{\phantom{\dag}}(\bm{r}t)$ for $t\in \mathds{R}$, $\h{\psi}_{\sigma;\textsc{k}}^{\dag}(\bm{r}\tau)$ is clearly \textsl{not} the Hermitian conjugate of $\h{\psi}_{\sigma;\textsc{k}}^{\phantom{\dag}}(\bm{r}\tau)$ for $\tau \in \mathds{R}\backslash \{0\}$.\footnote{For this reason, it may be preferable to use the notation $\b{\h{\psi}}_{\sigma;\textsc{k}}(\bm{r}\tau)$, or simply $\b{\psi}_{\sigma;\textsc{k}}(\bm{r}\tau)$.} We shall have occasion (for instance in appendix \protect\ref{saa}) to refer to the non-interacting counterpart of $\h{\varrho}$, that is (\emph{cf.} Eqs\,(\ref{e16}) and (\ref{e17}))
\begin{equation}\label{e17a}
\h{\varrho}_{\X{0}} \doteq \frac{1}{\mathcal{Z}_{\X{0}}} \e^{-\beta \wh{\mathcal{K}}_{\X{0}}},
\;\;\text{where}\;\; \mathcal{Z}_{\X{0}} \doteq \Tr[\protect\e^{-\beta\protect\wh{\mathcal{K}}_{\X{0}}}].
\end{equation}

With $\wh{\mathcal{K}}_{\X{1};\textsc{i}}(\tau)$ denoting $\wh{\mathcal{K}}_{\X{1}} \equiv \wh{H}_{\X{1}}$ in the imaginary-time \textsl{interaction picture}, one has (\emph{cf.} Eq.\,(\ref{e7})) [p.\,235 in Ref.\,\citen{FW03}]
\begin{equation}\label{e19}
\wh{\mathcal{K}}_{\X{1};\textsc{i}}(\tau) = \e^{\wh{\mathcal{K}}_{\X{0}} \tau/\hbar} \wh{\mathcal{K}}_{\X{1}} \e^{-\wh{\mathcal{K}}_{\X{0}} \tau/\hbar} \equiv \frac{1}{2} \int \rd 1 \rd 2\; v(1,2) \hspace{0.6pt} \h{\psi}_{\textsc{i}}^{\dag}(1) \h{\psi}_{\textsc{i}}^{\dag}(2) \h{\psi}_{\textsc{i}}(2) \h{\psi}_{\textsc{i}}(1),
\end{equation}
where\,\footnote{The field operator $\h{\psi}_{\textsc{i}}(j)$ is in the imaginary-time \textsl{interaction} picture.} (\emph{cf.} Eqs\,(\ref{e8}) -- (\ref{e11}))
\begin{equation}\label{e20}
j \rightleftharpoons \bm{r}_j, \tau_j, \sigma_j \rightleftharpoons \bm{r}_j\tau_j\sigma_j,
\end{equation}
\begin{equation}\label{e21}
\h{\psi}_{\textsc{i}}(j) \doteq \e^{\wh{\mathcal{K}}_{\X{0}}\tau/\hbar} \h{\psi}_{\sigma_j}^{\phantom{\dag}}(\bm{r}_j) \e^{-\wh{\mathcal{K}}_{\X{0}} \tau/\hbar},\;\; \h{\psi}_{\textsc{i}}^{\dag}(j) \doteq \e^{\wh{\mathcal{K}}_{\X{0}}\tau/\hbar} \h{\psi}_{\sigma_j}^{\dag}(\bm{r}_j) \e^{-\wh{\mathcal{K}}_{\X{0}} \tau/\hbar},
\end{equation}
\begin{equation}\label{e22}
v(i,j) \doteq u_{\sigma_i,\sigma_j}(\bm{r}_i,\bm{r}_j) \hspace{0.6pt} \delta(\tau_i -\tau_j),
\end{equation}
\begin{equation}\label{e23}
\int \rd j \rightleftharpoons \sum_{\sigma_j} \int_{0}^{\hbar\beta} \rd\tau_j \int \textrm{d}^d r_j.
\end{equation}
Clearly, for $\tau \in \mathds{R}\backslash \{0\}$ the operators $\h{\psi}_{\textsc{i}}^{\phantom{\dag}}(j)$ and $\h{\psi}_{\textsc{i}}^{\dag}(j)$ in Eq.\,(\ref{e21}) are \textsl{not} each other's Hermitian conjugates. Similarly as in Eq.\,(\ref{e260b}),
\begin{equation}\label{e23a}
j^+ \rightleftharpoons \bm{r}_j, \tau_j^+,\sigma_j \rightleftharpoons \bm{r}_j \tau_j^+ \sigma_j,
\end{equation}
where, with $\tau_j \in \mathds{R}$, $\tau_j^+ \doteq \tau_j + 0^+$ [p.\,229 in Ref.\,\citen{FW03}].\footnote{Note that the integral with respect to $\tau_j$ in Eq.\,(\protect\ref{e23}) is over the \textsl{real} interval $[0,\hbar\beta]$. See Fig.\,4.5 (a), p.\,107, in Ref.\,\protect\citen{SvL13}, as well as \S\,5.4, p.\,140, herein.}

With (\emph{cf.} Eq.\,(\ref{e12}))
\begin{equation}\label{e24}
a \rightleftharpoons \bm{r}\tau\sigma,\;\; b \rightleftharpoons \bm{r}'\tau'\sigma',
\end{equation}
we introduce the notation
\begin{equation}\label{e25}
\mathscr{G}(a,b) \equiv \mathscr{G}(\bm{r}\tau\sigma,\bm{r}'\tau'\sigma'),
\end{equation}
and similarly
\begin{equation}\label{e26}
\mathscr{G}_{\X{0}}(a,b) \equiv \mathscr{G}_{\X{0}}(\bm{r}\tau\sigma,\bm{r}'\tau'\sigma')
\end{equation}
for the non-interacting counterpart of $\mathscr{G}(a,b)$. For $\wh{H}$ and $\wh{H}_{\X{0}}$ time independent, the functions $\mathscr{G}$ and $\mathscr{G}_{\X{0}}$ in Eqs\,(\ref{e25}) and (\ref{e26}) depend on $\tau-\tau'$, rather than on $\tau$ and $\tau'$ separately. Further, for $\tau-\tau' < 0$ one explicitly shows that [Eqs\,(24.14) and (24.15), p.\,236, in Ref.\,\citen{FW03}]
\begin{equation}\label{e26a}
\mathscr{G}(\bm{r}\tau\sigma,\bm{r}'\tau'\sigma') = \pm
\mathscr{G}(\bm{r}\X{(}\tau+\hbar\beta\X{)}\sigma,\bm{r}'\tau'\sigma'),\;\;\text{for bosons / fermions},
\end{equation}
that is $\mathscr{G}(\bm{r}\tau\sigma,\bm{r}'\tau'\sigma')$ is a periodic / anti-periodic function of $\tau-\tau'$.
A similar equality as in Eq.\,(\ref{e26a}) applies for $\mathscr{G}_{\X{0}}$. The equality in Eq.\,(\ref{e26a}) is referred to as the Kubo-Martin-Schwinger (KMS)\cite{RK57,MS59} relation.

Since $\h{\mathcal{K}}_{\X{1}} \equiv \wh{H}_{\X{1}}$, Eq.\,(\ref{e15}), it immediately follows that the weak-coupling perturbation series expansion of $\mathscr{G}(a,b)$ is functionally identical to that of $G(a,b)$, with $\mathscr{G}_{\X{0}}$ taking the place of $G_{\X{0}}$ (see Eqs\,(\ref{e242}) -- (\ref{e260}) below). Correspondingly, the $\mathsf{v}(i,j)$ in Eqs\,(\ref{e260a}), (\ref{e257}), and (\ref{e261}) below is related to the two-body interaction function $v(i,j)$ in Eq.\,(\ref{e22}), and the integrals with respect to $1, 2,\dots, 2\nu$ in Eqs\,(\ref{e257}) and (\ref{e261}) below are defined in accordance with the prescription in Eq.\,(\ref{e23}), instead of that in Eq.\,(\ref{e11}).

\refstepcounter{dummyX}
\subsubsection{The one-particle Green function for \texorpdfstring{$T>0$}{}
(the \textsl{real-time} thermo-field dynamics, TFD)}
\phantomsection
\label{sac01}
In this section we consider the TFD formalism \cite{UMT82,HU95}.\footnote{We shall consider this formalism also in Ref.\,\protect\citen{BF16b}.} Conform conventional notation, in this section we suppress carets on the symbols that in other sections of this paper denote second-quantised operators,\footnote{In this paper we encounter some single-particle operators, such as the Green operator $\protect\h{G}$, that are furnished with caret but are not \textsl{second-quantised} operators. To underline this fact, we generally qualify these operators with the adjective \textsl{single-particle}.} as within the framework of the TFD caret on a symbol generally signifies the \textsl{difference} of two second-quantized operators that share the same basic symbol (see Eqs\,(\ref{es1}) and (\ref{es2}) below). Thus, in the this section $H_{\X{0}}$ and $H_{\X{1}}$ denote the operators $\wh{H}_{\X{0}}$ and $\wh{H}_{\X{1}}$ of the previous sections of this paper. Similarly as regards the field operators $\h{\psi}$ and $\h{\psi}^{\dag}$, except that in this case at places we additionally employ the symbol $\b{\psi}$ for $\psi^{\dag}$, this partly on account of the fact that on complex time contours $\psi^{\dag}$ is not the Hermitian conjugate of $\psi$. In this connection, we recall that also within the finite-temperature formalism of Matsubara, \S\,\ref{s211}, for $\tau \in \mathds{R}\backslash \{0\}$ the creation operator $\h{\psi}^{\dag}$ is \textsl{not} the Hermitian conjugate of the annihilation operator $\h{\psi}$.

Before proceeding with details, we point out that calculation of the dynamical correlation functions within Matsubara's  imaginary-time formalism \cite{TM55,ESF59,MS59,FW03,AGD75} is generally \textsl{not} straightforward, it requiring the analytic continuation of these functions from along the imaginary-time axis to along the real-time axis. Alternatively, and considering for concreteness the interacting one-particle Green function $\mathscr{G}(\bm{r}\tau\sigma,\bm{r}'\sigma'\tau')$,\footnote{Following the periodicity / anti-periodicity of $\mathscr{G}(\bm{r}\tau\sigma,\bm{r}'\sigma'\tau')$ as function of $\tau-\tau'$ for boson / fermion systems, Eq.\,(\protect\ref{e26a}), the imaginary-time Fourier transform of this function is discrete, defined over the discrete set of Matsubara energies (or frequencies).} Eq.\,(\ref{e18}), while determination of the time-Fourier transform of this function at an arbitrary complex energy $z$, specifically for $z = \varepsilon \pm \ii 0^+$, with $\varepsilon \in \mathds{R}$, is in principle possible \cite{BM61}, in practice this determination is generally non-trivial \cite{MW58,AGD59,JSL61}.\footnote{Recent progress in this area, under the heading of `algorithmic Matsubara integration' (AMI), for Hubbard-like models has been reported \protect\cite{TCL19} and implemented \protect\cite{VF19}. A comparable approach based on \textsl{time-ordered} diagrams (\S\,3.2, p.\,157, in Ref.\,\protect\citen{NO98}) is conceivable.} The complexity of the process of analytic continuation over the complex energy plane increases with the order of the dynamical correlation function, the imaginary-time-Fourier transform of higher-order dynamical correlations depending on multiple discrete Matsubara energies (or frequencies).\footnote{\emph{Note added to \textsf{arXiv:1912.00474v2}}: In Ref.\,\protect\citen{BF16a} we show that calculation of thermal correlation functions within the Matsubara formalism is generally unsafe when carried out in the frequency / energy domain and that for reliable calculations these have to be carried out in the imaginary-time domain; the sought-after correlation functions at the relevant Matsubara frequencies / energies are thus to be determined through the explicit Fourier transformations of these along the imaginary-time axis. The analysis in Ref.\,\protect\citen{BF16a} reveals that the intermediate functions contributing to a many-body correlation function can contain vital information that is irrecoverably lost on being evaluated at the Matsubara frequencies. This is however not the case when these intermediate functions are calculated in the imaginary-time domain.}

The above-mentioned problem associated with the process of analytic continuation of finite-temperature correlation functions is fully overcome within the real-time formalism\,\footnote{Relativistic as well as non-relativistic field theories in which the time path is entirely along the real (imaginary) axis are commonly qualified as Minkowskian (Euclidean). The TFD formalism does not fall into either of the two categories.} of thermo-field dynamics (TFD) \cite{UMT82,HU95}.\footnote{As regards relevant original publications, we refer the reader to Refs\,\citen{TU75,IO81}. The framework of the TFD has been expanded for dealing with non-equilibrium ensemble of states \cite{AU85,MS85,JPW85,APU86,UA86,AU87}. For a review, consult Ref.\,\protect\citen{LvW87}.}\footnote{We note that the super-operators acting on a Liouville space of a system of fermions / bosons with a given number of degrees of freedom constitute an algebra corresponding to a system of super-fermions / super-bosons with doubled degrees of freedom \cite{MS78}. This doubling of degrees of freedom coincides with that in the framework of the TFD through the process of `tilde substitution' \cite{TU75,IO81,MNU85,MS78}. For reviews, consult Refs\,\protect\citen{LvW87,PAH95}. The review by Landsman and van Weert \protect\cite{LvW87} provides amongst others also a comprehensive overview of the operator structure of the TFD in a $C^*$-algebraic context, tracing the roots of it to the classic work by Haag, Hugenholtz, and Winnink \protect\cite{HHW67,RH96,KL17} on the equilibrium states of quantum statistical mechanics.} The \textsl{structure} of the perturbation series expansion of the Green functions in the framework of the TFD is identical to that of the zero-temperature formalism \cite{HM77,HM86,MNUMM83,SU83,MOU84,RB94,HM88,CU93} that we consider in detail in this paper. Technically, in the TFD formalism the role of the (causal) non-interacting Green function $G_{\X{0}}$ of the zero-temperature formalism is played by the $2\times 2$ (causal) non-interacting one-particle Green matrix $\mathbb{G}_{\X{0}}$, Eqs\,(\ref{es9}) and (\ref{es16fa}) below. Similar to $\mathbb{G}_{\X{0}}$, in this formalism the interacting one-particle Green function, Eqs\,(\ref{es7}) and (\ref{es16f}) below, as well as the self-energy operator, the polarisation function, the dielectric function and the screened interaction potential, \S\,\ref{sec3}, are $2\times 2$ matrices \cite{UMT82,HM88,HU95}. We note that there exists a direct formal association between the TFD \cite{MNUMM83,MM86} and the Keldysh formalism \cite{RS86}.

Within the framework of the TFD the role of the second-quantised Hamiltonian operator $\wh{H}$ in the previous sections of this paper is played by the operator
\begin{equation}\label{es1}
\wh{H} \doteq H - \wt{H},
\end{equation}
where the $H$ on the RHS is the interacting Hamiltonian as defined in Eq.\,(\ref{e5}),\footnote{Or the Hubbard Hamiltonian $\wh{\mathcal{H}}$ in Eq.\,(\protect\ref{e27}) below.} and $\wt{H}$ its tilde conjugation.\footnote{For operators $\mathcal{O}_1$ and $\mathcal{O}_2$, and complex $c$-numbers $c_1$ and $c_2$, one has $\widetilde{\mathcal{O}_1 \mathcal{O}_2} = \protect\wt{\mathcal{O}}_1 \widetilde{\mathcal{O}}_2$, and $\Widetilde{c_1 \mathcal{O}_1 + c_2 \mathcal{O}_2} = c_1^* \protect\wt{\mathcal{O}}_1 + c_2^* \protect\wt{\mathcal{O}}_2$, where $c_i^*$, $i=1,2$, is the complex conjugate of $c_i$. One further has $(\mathcal{O}^{\dag})\phantom{.}\wt{{}} = (\wt{\mathcal{O}})^{\dag}$, and $(\wt{\mathcal{O}})\phantom{.}\wt{{}} = \pm \mathcal{O}$ for boson / fermion operators (see for instance Eqs\,(2.5a) and (2.5b), p.\,350, in Ref.\,\protect\citen{MOU84}, and Eqs\,(7.40) -- (7.45), p.\,145, in Ref.\,\protect\citen{HU95}).} Correspondingly, one has
\begin{equation}\label{es2}
\wh{H}_{\X{0}} \doteq H_{\X{0}} - \wt{H}_{\X{0}},\;\;\;\; \wh{H}_{\X{1}} \doteq H_{\X{1}} - \wt{H}_{\X{1}},
\end{equation}
where $H_{\X{0}}$ ($H_{\X{1}}$) is the non-interacting (interaction) Hamiltonian $\wh{H}_{\X{0}}$ ($\wh{H}_{\X{1}}$) in Eq.\,(\ref{e5}) and $\wt{H}_{\X{0}}$ ($\wt{H}_{\X{1}}$) its tilde conjugation.

In order to be capable of calculating correlation functions for real times within the framework of the TFD, the time contour on which the Heisenberg- and interaction-picture operators are defined must consist of a part that covers the relevant interval of the real time axis. Conventionally, within this formalism one adopts the following \textsl{directed} time contour:
\begin{equation}\label{es3}
\mathscr{C} = \sum_{i=1}^4 \mathscr{C}_i,
\end{equation}
where\,\footnote{Here, the direction of the contour segment $[a,b]$ is from $a$ to $b$.}
\begin{align}\label{es4}
\mathscr{C}_{\X{1}} &= [t_{\textrm{i}},t_{\textrm{f}}],\nonumber\\
\mathscr{C}_{\X{2}} &= [t_{\textrm{f}}-\ii\hspace{1.0pt} (1-\alpha)\hbar\beta,t_{\textrm{i}}-\ii\hspace{1.0pt} (1-\alpha)\hbar\beta],\nonumber\\
\mathscr{C}_{\X{3}} &= [t_{\textrm{f}},t_{\textrm{f}}-\ii\hspace{1.0pt} (1-\alpha)\hbar\beta],\nonumber\\
\mathscr{C}_{\X{4}} &= [t_{\textrm{i}}-\ii\hspace{1.0pt} (1-\alpha)\hbar\beta,t_{\textrm{i}}-\ii \hbar\beta],\;\;
\text{where}\;\; \alpha \in (0,1],
\end{align}
with $\alpha = \tfrac{1}{2}$ corresponding to the Hermitian representation \cite{EHUY92};\footnote{See the second footnote associated with Eq.\,(\protect\ref{es8}) below.} the initial and final times, $t_{\textrm{i}}$ and $t_{\textrm{f}}$ (both real), can be identified with respectively $-\infty$ and $+\infty$, in which case the contributions arising from $\mathscr{C}_{\X{3}}$ and $\mathscr{C}_{\X{4}}$ can be discarded.\footnote{For the contours in the TFD, see in particular Ref.\,\protect\citen{MNUMM83,MNU84,NS84a,NS84b,HU95}. For the implications of various choices of the parameter $\alpha$ in Eq.\,(\protect\ref{es4}), consult Ref.\,\protect\citen{EHUY92}.}

With\,\footnote{The two-component field operator $\uppsi$ in Eq.\,(\protect\ref{es5}) is different from the Nambu \protect\cite{YN60,PWA58} two-component field operator to which we have referred earlier in this section. The difference lies in the fact that the first component of the Nambu two-component annihilation field operator consists of an annihilation field operator and its second component of the time-reversed creation field operator. In contrast, the second component of $\uppsi$ consists of the tilde-conjugated creation field operator.}
\begin{equation}\label{es5}
\uppsi \doteq
\begin{pmatrix}
\psi^1 \\ \psi^2
\end{pmatrix}\equiv
\begin{pmatrix}
\psi \\ \t{\psi}^{\dag}
\end{pmatrix}, \;\;
\b{\uppsi} \doteq
\begin{pmatrix}
\b{\psi}^1, &\hspace{-0.32cm} \b{\psi}^2
\end{pmatrix}\equiv
\begin{pmatrix}
\psi^{\dag}, &\hspace{-0.32cm} \mp\t{\psi}
\end{pmatrix}
\;\;\;\text{for}\;\; \text{bosons / fermions},
\end{equation}
and assuming that $u_{\sigma,\sigma'}(\bm{r},\bm{r}') \in \mathds{R}$, from the expression in Eq.\,(\ref{e5}) one obtains
\begin{align}\label{es6a}
&\wh{H}_{\X{1}} = \frac{1}{2}\sum_{\sigma,\sigma'} \int \mathrm{d}^dr\mathrm{d}^dr'\, u_{\sigma,\sigma'}(\bm{r},\bm{r}')
\big(\b{\psi}_{\sigma}^1(\bm{r}) \b{\psi}_{\sigma'}^1(\bm{r}') \psi_{\sigma'}^1(\bm{r}')
\psi_{\sigma}^1(\bm{r})\nonumber\\
&\hspace{5.3cm}-\psi_{\sigma}^2(\bm{r}) \psi_{\sigma'}^2(\bm{r}') \b{\psi}_{\sigma'}^2(\bm{r}') \b{\psi}_{\sigma}^2(\bm{r})\big).
\end{align}
With $\wh{H}_{\X{1};\textsc{i}}(t)$, $t\in \mathscr{C}$, denoting $\wh{H}_{\X{1}}$ in the interaction picture,\footnote{The relationship between the interaction-picture and the Heisenberg-picture operators within the TFD formalism has been discussed in Refs\,\protect\citen{MNUMM83,MM86,EHUY92,EHUY93}. For a comprehensive discussion of `the contour idea', consult Ch.\,4, p.\,95, of Ref.\,\protect\citen{SvL13}, and for the equations of motion on contours, \S\,4.4, p.\,110, herein. Note that the $z_i$ in the latter reference is to be identified with the initial time $t_{\textrm{i}}$ in Eq.\,(\protect\ref{es4}).} the associated expression is obtained by replacing the field operators on the RHS of Eq.\,(\ref{es6a}) by their interaction-picture counterparts. With\,\footnote{The binary variable $\mu$ is not to be confused with the \textsl{chemical potential} introduced in \S\,\protect\ref{s222x}.}
\begin{equation}\label{es7a}
\psi_{\textsc{i}}^{\mu}(\bm{r}t\sigma) \equiv \psi_{\sigma;\textsc{i}}^{\mu}(\bm{r}t),\;\;
\b{\psi}_{\textsc{i}}^{\mu}(\bm{r}t\sigma) \equiv \b{\psi}_{\sigma;\textsc{i}}^{\mu}(\bm{r}t),\;\; \mu \in \{1,2\},
\end{equation}
one thus has
\begin{align}\label{es6}
&\hspace{-0.2cm}\wh{H}_{\X{1};\textsc{i}}(t) = \frac{1}{2}\sum_{\sigma,\sigma'} \int \mathrm{d}^dr\mathrm{d}^dr'
\mathrm{d}t'\, u_{\sigma,\sigma'}(\bm{r},\bm{r}') \delta(t,t') \nonumber\\
&\hspace{0.2cm}\times \big(\b{\psi}_{\textsc{i}}^1(\bm{r}t\sigma) \b{\psi}_{\textsc{i}}^1(\bm{r}'t'\sigma')
\psi_{\textsc{i}}^1(\bm{r}'t'\sigma')\psi_{\textsc{i}}^1(\bm{r}t\sigma)
-\psi_{\textsc{i}}^2(\bm{r}t\sigma) \psi_{\textsc{i}}^2(\bm{r}'t'\sigma') \b{\psi}_{\textsc{i}}^2(\bm{r}'t'\sigma')
\b{\psi}_{\textsc{i}}^2(\bm{r}t\sigma)\big),\nonumber\\
\end{align}
where we have introduced the integral with respect to $t'$ in the usual manner (\emph{cf}. Eqs\,(\ref{e10}) and (\ref{e22})). The distribution $\delta(t,t')$ is defined for $t, t' \in \mathscr{C}$ and coincides with $\delta(t-t')$ for $t,t'\in \mathds{R}$.\footnote{For the relevant details, consult for instance \S\,4.5, p.\,114, in Ref.\,\protect\citen{SvL13}.} For ease of notation, \emph{where in the remaining part of this paper the TFD formalism is concerned, the distribution $\delta(t-t')$ is to be understood as denoting $\delta(t,t')$.}

With $\vert 0(\beta)\rangle$ denoting the temperature-dependent vacuum state in the Heisenberg picture, the $(\mu,\mu')$ element of the interacting one-particle Green matrix $\mathbb{G}$ is defined as \cite{HU95}\footnote{See Eq.\,(7.212), p.\,165, in Ref.\,\protect\citen{HU95}. See also Eq.\,(2.25a) in Ref.\,\protect\citen{HM77} and note that the $-\protect\ii\,$ in the definition in Eq.\,(\protect\ref{es7}) is only a matter of convention, relevant only when applying the rules based on the definition in Eq.\,(\protect\ref{e11a}) for evaluating the contributions of Feynman diagrams.} (\emph{cf.} Eq.\,(\ref{e11a}))
\begin{equation}\label{es7}
\mathsf{G}^{\mu\mu'}\hspace{-2.4pt}(a,b) \doteq -\ii\hspace{1.6pt} \langle 0(\beta)\vert \mathcal{T}_{\hspace{-1.2pt}{}_{\mathscr{C}}}\hspace{-1.0pt}\big\{\psi_{\textsc{h}}^{\mu}(a) \b{\psi}_{\textsc{h}}^{\mu'}(b)\big\}\vert 0(\beta)\rangle,\;\; \mu,\mu' \in \{1,2\},
\end{equation}
where $\mathcal{T}_{\hspace{-1.2pt}{}_{\mathscr{C}}}$ is the chronological time-ordering operator on $\mathscr{C}$, and the subscript $\textsc{h}$ attached to field operators marks these as being in the Heisenberg picture. We note that with $\wh{A}$ denoting an observable, by definition\,\footnote{See for instance Eqs\,(1.1) -- (1.5) in Ref.\,\protect\citen{TU75}.}\footnote{We note in passing that the $\alpha$ in Eq.\,(\protect\ref{es4}) is tied to employing the identity $\protect\h{\varrho} \equiv \protect\h{\varrho}^{\alpha} \protect\h{\varrho}^{1-\alpha}$ and, on the basis of the invariance of the trace of a product of operators under their cyclic permutations, expressing $\protect\Tr\!\big[\protect\h{\varrho}\hspace{0.6pt} \wh{A}\hspace{1.0pt}\big]$ as $\protect\Tr\!\big[\protect\h{\varrho}^{1-\alpha}\hspace{0.6pt} \wh{A}\hspace{0.6pt}\protect\h{\varrho}^{\alpha}\big]$ (for the relevance of the latter expression, consult Ref.\,\protect\citen{EHUY92}). The latter expression makes evident the way in which $\alpha = \tfrac{1}{2}$ is special.} (see Eqs\,(\ref{e16}) and (\ref{e17}))
\begin{equation}\label{es8}
\langle 0(\beta)\vert\wh{A}\vert 0(\beta)\rangle \equiv
\Tr\!\big[\h{\varrho}\hspace{0.6pt} \wh{A}\hspace{1.0pt}\big].
\end{equation}
For the $(\mu,\mu')$ element of the non-interacting one-particle Green matrix $\mathbb{G}_{\X{0}}$, one has\,\footnote{For the elements of the (causal) one-particle Green matrix $\mathbb{G}_{\protect\X{0}}(a,b)$, see for instance \S\,7.2.6, p.\,153, in Ref.\,\protect\citen{HU95}. For some explicit diagrammatic calculations, see Refs\,\protect\citen{FG85,TSE87,FY88,KS88}. For the application of the TFD formalism to the problem of time evolution of large systems, see  Ref.\,\protect\citen{KN74}, and to the problem of the particle-antiparticle symmetry in nuclear physics, Ref.\,\protect\citen{KK76}.}
\begin{equation}\label{es9}
\mathsf{G}_{\X{0}}^{\mu\mu'}\hspace{-2.4pt}(a,b) \doteq -\ii\hspace{1.6pt} \langle\X{0}(\beta)\vert \mathcal{T}_{\hspace{-1.2pt}{}_{\mathscr{C}}}\hspace{-1.0pt}\big\{\psi_{\textsc{i}}^{\mu}(a) \b{\psi}_{\textsc{i}}^{\mu'}(b)\big\}\vert\X{0}(\beta)\rangle,\;\; \mu,\mu' \in \{1,2\},
\end{equation}
where $\vert\X{0}(\beta)\rangle$ (not to be confused with $\vert 0(\beta)\rangle$) is the temperature-dependent vacuum state in the interaction picture, for which one has (\emph{cf.} Eq.\,(\ref{es8}))
\begin{equation}\label{es10}
\langle \X{0}(\beta)\vert\wh{A}\vert \X{0}(\beta)\rangle \equiv
\Tr\!\big[\h{\varrho}_{\X{0}}\hspace{0.6pt} \wh{A}\hspace{1.0pt}\big],
\end{equation}
where $\h{\varrho}_{\X{0}}$ is the statistical operator defined in Eq.\,(\ref{e17a}).

The many-body perturbation expansion of $\mathsf{G}^{\mu\mu'}$ in terms of $\{\mathsf{G}_{\X{0}}^{\mu\mu'} \| \mu,\mu'\}$,\footnote{Presented in \S\S\,\protect\ref{s212} and \protect\ref{s226} below.} deduced by applying the Gell-Mann and Low theorem \cite{GL51},\footnote{For a pedagogical exposition of this theorem in relation to the expansion of the $N$-particle GS $\vert\Psi_{N;0}\rangle$ of $\wh{H}$ in terms of the $N$-particle GS $\vert\Phi_{N;0}\rangle$ of $\wh{H}_{\X{0}}$, under the adiabatic assumption, see Ref.\,\protect\citen{FW03}, p.\,61.}\footnote{Within the framework of the TFD, use of the Gell-Mann and Law theorem can be bypassed through a judicious choice of the parameter $\alpha$ in the definition of the contour $\mathscr{C}$, Eqs\,(\protect\ref{es3}) and (\protect\ref{es4}). For details, the reader is referred to Ref.\,\protect\citen{EHUY92}, as well as \S\,7.5, p.\,164, of Ref.\,\protect\citen{HU95}.} involves the expressions\,\footnote{See for instance Eq.\,(2.9) in Ref.\,\protect\citen{MM86} and compare this with Eq.\,(8.9), p.\,85, in Ref.\,\protect\citen{FW03}.}
\begin{equation}\label{es11}
\langle\X{0}(\beta)\vert \mathcal{T}_{\hspace{-1.2pt}{}_{\mathscr{C}}}\hspace{-1.0pt}\big\{
\wh{H}_{\X{1};\textsc{i}}(t_1) \wh{H}_{\X{1};\textsc{i}}(t_2)\dots \wh{H}_{\X{1};\textsc{i}}(t_{\nu})
\psi_{\textsc{i}}^{\mu}(a) \b{\psi}_{\textsc{i}}^{\mu'}(b)\big\}\vert\X{0}(\beta)\rangle,
\end{equation}
and
\begin{equation}\label{es12}
\langle\X{0}(\beta)\vert \mathcal{T}_{\hspace{-1.2pt}{}_{\mathscr{C}}}\hspace{-1.0pt}\big\{
\wh{H}_{\X{1};\textsc{i}}(t_1) \wh{H}_{\X{1};\textsc{i}}(t_2)\dots \wh{H}_{\X{1};\textsc{i}}(t_{\nu})
\big\}\vert\X{0}(\beta)\rangle.
\end{equation}
Since
\begin{equation}\label{es13}
\mathcal{T}_{\hspace{-1.2pt}{}_{\mathscr{C}}}\hspace{-1.0pt}\big\{
\psi_{\textsc{i}}^2(\bm{r}t\sigma) \psi_{\textsc{i}}^2(\bm{r}'t'\sigma') \b{\psi}_{\textsc{i}}^2(\bm{r}'t'\sigma')
\b{\psi}_{\textsc{i}}^2(\bm{r}t\sigma)\big\} \equiv
\mathcal{T}_{\hspace{-1.2pt}{}_{\mathscr{C}}}\hspace{-1.0pt}\big\{
\b{\psi}_{\textsc{i}}^2(\bm{r}t\sigma) \b{\psi}_{\textsc{i}}^2(\bm{r}'t'\sigma') \psi_{\textsc{i}}^2(\bm{r}'t'\sigma')
\psi_{\textsc{i}}^2(\bm{r}t\sigma)\big\},
\end{equation}
it follows that under the path-ordering operation $\mathcal{T}_{\hspace{-1.2pt}{}_{\mathscr{C}}}$ the operator $\wh{H}_{\X{1};\textsc{i}}(t_j)$, $j=1,2,\dots, \nu$, can be expressed as
\begin{equation}\label{es14}
\wh{H}_{\X{1};\textsc{i}}(t_j) = \wh{H}_{\X{1};\textsc{i}}^1(t_j) + \wh{H}_{\X{1};\textsc{i}}^2(t_j),\;\;\;\;\;
\text{(Under the path ordering $\mathcal{T}_{\hspace{-1.2pt}{}_{\mathscr{C}}}$)}
\end{equation}
where
\begin{align}\label{es15}
&\hspace{-0.0cm}\wh{H}_{\X{1};\textsc{i}}^{\mu_j}(t_j) \doteq
\frac{1}{2}\sum_{\sigma_j\phantom{'}\hspace{-2.7pt},\sigma_j'} \sum_{\mu_j' \in \{1,2\}}
\int_{\mathscr{C}} \mathrm{d}t_j'
\int \mathrm{d}^dr_j\mathrm{d}^dr_j'\, (-1)^{\lfloor \mu_j/2\rfloor} u_{\sigma_j,\sigma_j'}(\bm{r}_j,\bm{r}_j') \delta(t_j,t_j') \delta_{\mu_j,\mu_j'}\nonumber\\
&\hspace{1.4cm}\times \b{\psi}_{\textsc{i}}^{\mu_j}(\bm{r}_jt_j\sigma_j) \b{\psi}_{\textsc{i}}^{\mu_j'}(\bm{r}_j't_j'\sigma_j')
\psi_{\textsc{i}}^{\mu_j'}(\bm{r}_j't_j'\sigma_j')\psi_{\textsc{i}}^{\mu_j}(\bm{r}_jt_j\sigma_j),\;\;\;
\mu_j \in\{1,2\},
\end{align}
in which $\lfloor x\rfloor$ is the floor function, yielding the greatest integer less than or equal to $x$. The summation with respect to $\mu_j'$ as introduced on the RHS of Eq.\,(\ref{es15}) serves a similar purpose as the integration with respect to $t'$ on the RHS of Eq.\,(\ref{es6}). Thus, under the path-ordering operation $\mathcal{T}_{\hspace{-1.2pt}{}_{\mathscr{C}}}$ one can write
\begin{align}\label{es16}
&\hspace{-0.6cm}\wh{H}_{\X{1};\textsc{i}}(t_1) \wh{H}_{\X{1};\textsc{i}}(t_2)\dots \wh{H}_{\X{1};\textsc{i}}(t_{\nu})\nonumber\\
&\hspace{0.4cm} =\sum_{\mu_1,\mu_2,\dots,\mu_{\nu} \in \{1,2\}}
[\wh{H}_{\X{1};\textsc{i}}^{1}(t_1)]^{1-\lfloor\mu_1/2\rfloor}
[\wh{H}_{\X{1};\textsc{i}}^{1}(t_2)]^{1-\lfloor\mu_2/2\rfloor}\dots [\wh{H}_{\X{1};\textsc{i}}^{1}(t_{\nu})]^{1-\lfloor\mu_{\nu}/2\rfloor}\nonumber\\
&\hspace{2.9cm}\times
[\wh{H}_{\X{1};\textsc{i}}^{2}(t_1)]^{\lfloor\mu_1/2\rfloor}
[\wh{H}_{\X{1};\textsc{i}}^{2}(t_2)]^{\lfloor\mu_2/2\rfloor}\dots  [\wh{H}_{\X{1};\textsc{i}}^{2}(t_{\nu})]^{\lfloor\mu_{\nu}/2\rfloor},\nonumber\\
&\hspace{6.6cm}
\text{(Under the path ordering $\mathcal{T}_{\hspace{-1.2pt}{}_{\mathscr{C}}}$)}
\end{align}
where $[\wh{H}_{\X{1};\textsc{i}}^{\mu}(t_j)]^0 \equiv \h{1}$, the identity operator in the Fock space, and $[\wh{H}_{\X{1};\textsc{i}}^{\mu}(t_j)]^1 \equiv \wh{H}_{\X{1};\textsc{i}}^{\mu}(t_j)$, $\mu \in \{1,2\}$. The RHS of Eq.\,(\ref{es16}) consists of a superposition of $2^{\nu}$ terms, each of which is of the $\nu$th order in the two-body interaction potential.

As in earlier sections, it proves advantageous to make the following identifications within the framework of the TFD (\emph{cf.} Eqs\,(\ref{e12}), (\ref{e8}), (\ref{e260b}), and (\ref{e11})):
\begin{equation}\label{es16a}
a \rightleftharpoons \bm{r} t \sigma \mu,\;\; b \rightleftharpoons \bm{r}' t'\sigma'\mu',
\end{equation}
\begin{equation}\label{es16b}
j \rightleftharpoons \bm{r}_j, t_j, \sigma_j, \mu_j \rightleftharpoons \bm{r}_j t_j \sigma_j \mu_j,
\end{equation}
\begin{equation}\label{es16e}
j^+ \rightleftharpoons \bm{r}_j,t_j^+,\sigma_j,\mu_j \rightleftharpoons \bm{r}_j t_j^+ \sigma_j\mu_j.
\end{equation}
\begin{equation}\label{es16d}
\int \rd j \rightleftharpoons \sum_{\sigma_j}\sum_{\mu_j \in\{1,2\}}\int_{\mathscr{C}} \rd t_j \int \textrm{d}^d r_j,
\end{equation}
and further to define $v(i,j)$ as (\emph{cf.} Eq.\,(\ref{e10}))
\begin{equation}\label{es16c}
v(i,j) \doteq (-1)^{\lfloor\mu_i/2\rfloor} u_{\sigma_i,\sigma_j}(\bm{r}_i,\bm{r}_j)\hspace{0.6pt} \delta(t_i,t_j) \delta_{\mu_i,\mu_j}.
\end{equation}
Thus, in the light of the identifications in Eq.\,(\ref{es16a}), in the following (\emph{cf.} Eq.\,(\ref{e13}))
\begin{equation}\label{es16f}
\mathsf{G}(a,b) \equiv \mathsf{G}^{\mu\mu'}\hspace{-2.4pt}(\bm{r}t\sigma,\bm{r}'t'\sigma')
\equiv (\mathbb{G}(\bm{r}t\sigma,\bm{r}'t'\sigma'))_{\mu,\mu'},
\end{equation}
and (\emph{cf.} Eq.\,(\ref{e14}))
\begin{equation}\label{es16fa}
\mathsf{G}_{\X{0}}(a,b) \equiv \mathsf{G}_{\X{0}}^{\mu\mu'}\hspace{-2.4pt}(\bm{r}t\sigma,\bm{r}'t'\sigma')
\equiv (\mathbb{G}_{\X{0}}(\bm{r}t\sigma,\bm{r}'t'\sigma'))_{\mu,\mu'}.
\end{equation}
Taking account of the above specifications, the structure of the perturbation series expansion of $\mathsf{G}(a,b)$ in terms of $\{\mathsf{G}_{\X{0}}(i,j)\| i,j\}$ is identical to that of $G(a,b)$ in terms of $\{G_{\X{0}}(i,j)\| i,j\}$ corresponding to the zero-temperature formalism, \S\,\ref{s222x}.

\refstepcounter{dummyX}
\subsubsection{The lattice model: the single-band Hubbard Hamiltonian}
\phantomsection
\label{s224}
Insofar as lattice models are concerned, we restrict the explicit considerations in this paper to the single-band Hubbard Hamiltonian $\wh{\mathcal{H}}$ for spin-$\tfrac{1}{2}$ fermions \cite{PWA59,ThWR62,JH63}, for which one has
\begin{equation}\label{e27}
\wh{\mathcal{H}} = \sum_{\sigma} \sum_{l,l'=1}^{N_{\textsc{s}}} T_{l,l'}\hspace{0.7pt} \h{c}_{l;\sigma}^{\dag} \h{c}_{l';\sigma}^{\phantom{\dag}} + \frac{U}{2}\sum_{\sigma}\sum_{l=1}^{N_{\textsc{s}}} \h{n}_{l;\sigma}  \h{n}_{l;\b{\sigma}} \equiv \wh{\mathcal{H}}_{\X{0}} + \wh{\mathcal{H}}_{\X{1}},
\end{equation}
where $\{T_{l,l'} \| l,l'\}$ are hopping matrix elements, $U$ the on-site interaction energy, $\b{\sigma}$ the spin index complementary to $\sigma$,\footnote{With $\sigma = \uparrow$, one has $\b{\sigma} = \downarrow$, and \emph{vice versa}. See p.\,\protect\pageref{AsIsCommon}.} and
\begin{equation}\label{e28}
\h{n}_{l;\sigma} \doteq \h{c}_{l;\sigma}^{\dag} \h{c}_{l;\sigma}^{\phantom{\dag}},
\end{equation}
the site-occupation-number operator. The indices $l$ and $l'$ mark the vectors $\{\bm{R}_l \| l =1,2,\dots,N_{\textsc{s}}\}$ spanning the lattice on which $\wh{\mathcal{H}}$ is defined. Assuming the latter to be a Bravais lattice \cite{AM76}, with $\1BZ$ denoting the first Brillouin zone in the corresponding reciprocal space, for $T_{l,l'}$ one has
\begin{equation}\label{e29}
T_{l,l'} = \frac{1}{N_{\textsc{s}}} \sum_{\bm{k} \in \fbz} \varepsilon_{\bm{k}} \e^{\ii \bm{k}\cdot (\bm{R}_l -\bm{R}_{l'})},
\end{equation}
where $\varepsilon_{\bm{k}}$ denotes the underlying non-interacting single-particle energy dispersion. Customarily, one adjusts the energy dispersion $\varepsilon_{\bm{k}}$ by a constant shift so that $T_{l,l} = 0$ for all $l$.

The Hubbard Hamiltonian for bosons \cite{PWA64,KBE80,FG88,FWGF89,ES94,FM94,KR97,JBCGZ98,OSS03,WSGY94,%
PS04,PST98,SS11},\footnote{Ch.\,9, p.\,117, in Ref.\,\protect\citen{SS11}.} often referred to as the Bose-Hubbard and the Boson Hubbard model, can be treated along the same lines as the Hubbard Hamiltonian for fermions. We do not explicitly deal with this Hamiltonian in this paper for two reasons. Firstly, in the applications of contemporary interest the strong-coupling perturbation expansion \cite{FM94} turns out to be the appropriate choice in dealing with this Hamiltonian, to be contrasted with the weak-coupling perturbation expansions dealt with in this paper, which crucially rely on the Wick decomposition theorem, appendix \ref{saa}. Secondly, the considerations of this model in many applications relate to both the normal \textsl{and} superfluid \cite{FM94,KR97,JBCGZ98,OSS03,SS11} as well as the normal \textsl{and} superconductive states in granulated material \cite{PWA64,KBE80,FG88,FWGF89,ES94}. In particular, determination of the boundary between the normal and superfluid / superconductive phases of the systems under consideration is of prime interest. As we have indicated earlier, in this paper we focus on the normal states of systems and relegate considerations of superconductive and superfluid states to a future publication \cite{BF16b}.

Since $\wh{\mathcal{H}}_{\X{0}}$, the non-interacting part of the Hubbard Hamiltonian $\wh{\mathcal{H}}$, is accounted for by the non-interacting Green function, $G_{\X{0}}$, $\mathscr{G}_{\X{0}}$ or $\mathsf{G}_{\X{0}}$,\footnote{Associated with respectively the $G$, $\mathscr{G}$, and $\mathsf{G}$ in Eqs\,(\protect\ref{e11a}), (\protect\ref{e18}), and (\protect\ref{es7}).} in what follows we focus on the interaction part $\wh{\mathcal{H}}_{\X{1}}$ of $\wh{\mathcal{H}}$. Before proceeding, however, with reference to the \textsl{double sum} (with respect to $l$ and $l'$) on the RHS of Eq.\,(\ref{e27}), we note that we could have defined the $\wh{H}_{\X{0}}$ in Eq.\,(\ref{e5}) equivalently as follows:
\begin{equation}\label{e30}
\wh{H}_{\X{0}} = \sum_{\sigma} \int \textrm{d}^dr \textrm{d}^dr'\; \h{\psi}_{\sigma}^{\dag}(\bm{r}) \big(\t{\uptau}(\bm{r},\bm{r}') + \t{v}(\bm{r},\bm{r}') \big) \h{\psi}_{\sigma}^{\phantom{\dag}}(\bm{r}'),
\end{equation}
where $\t{v}(\bm{r},\bm{r}')$ denotes a non-local external potential, which we do not specify further here except that the external potential $v(\bm{r})$ in Eq.\,(\ref{e5}) corresponds to the specific case where
\begin{equation}\label{e30a}
\t{v}(\bm{r},\bm{r}') = \delta(\bm{r}-\bm{r}')\hspace{0.7pt} v(\bm{r}').
\end{equation}
Similarly, the kinetic-energy operator $\uptau(\bm{r})$ in Eqs\,(\ref{e5}) and (\ref{e6}) corresponds to the case where
\begin{equation}\label{e31}
\t{\uptau}(\bm{r},\bm{r}') = \delta(\bm{r}-\bm{r}')\hspace{0.7pt}\uptau(\bm{r}') \equiv  +\frac{\hbar^2}{2\mathsf{m}}\hspace{0.6pt}\delta(\bm{r}-\bm{r}')\hspace{0.6pt} \bm{\nabla}_{\bm{r}} \cdot \bm{\nabla}_{\bm{r}'}.
\end{equation}
From the perspective of the considerations of this paper, the \textsl{double integral} on the RHS of Eq.\,(\ref{e30}) is the equivalent of the \textsl{double sum} in the expression for $\wh{\mathcal{H}}_{\X{0}}$ in Eq.\,(\ref{e27}).

To make contact with the details bearing on the Hamiltonian $\wh{H}$ in Eq.\,(\ref{e5}), we express the interaction Hamiltonian $\wh{\mathcal{H}}_{\X{1}}$ in Eq.\,(\ref{e27}) as\,\footnote{In Ref.\,\protect\citen{BF16a} we discuss different, but equivalent, representations of the interacting part of the Hubbard Hamiltonian $\wh{\mathcal{H}}$ and their associated distinct diagrammatic expansions. The representation of $\wh{\mathcal{H}}_{\X{1}}$ in Eq.\,(\protect\ref{e32}) is the most suitable one for the considerations of the present paper.}
\begin{equation}\label{e32}
\wh{\mathcal{H}}_{\X{1}} = \frac{1}{2} \sum_{\sigma,\sigma'} \sum_{l,l' = 1}^{N_{\textsc{s}}} U_{\sigma,\sigma'}(l,l')\hspace{0.8pt} \h{c}_{l;\sigma}^{\dag} \h{c}_{l';\sigma'}^{\dag} \h{c}_{l';\sigma'}^{\phantom{\dag}} \h{c}_{l;\sigma}^{\phantom{\dag}},
\end{equation}
where
\begin{equation}\label{e33}
U_{\sigma,\sigma'}(l,l') \doteq U (1-\delta_{\sigma,\sigma'})\hspace{0.7pt} \delta_{l,l'}.
\end{equation}
One observes that through the identifications
\begin{equation}\label{e34}
\bm{r}\rightleftharpoons l,\;\; u_{\sigma,\sigma'}(\bm{r},\bm{r}') \rightleftharpoons U_{\sigma,\sigma'}(l,l'),\;\; \h{\psi}_{\sigma}(\bm{r}) \rightleftharpoons \h{c}_{l;\sigma},\;\;
\int \textrm{d}^d r \rightleftharpoons \sum_{l =1}^{N_{\textsc{s}}},
\end{equation}
one has
\begin{equation}\label{e35}
\wh{H} \rightleftharpoons \wh{\mathcal{H}}.
\end{equation}
Consequently, with (\emph{cf.} Eq.\,(\ref{e12}))
\begin{equation}\label{e36}
a \rightleftharpoons l t \sigma,\;\; b \rightleftharpoons l' t' \sigma',
\end{equation}
the relevant perturbational expression for the Green function $G(a,b)$ corresponding to $\wh{\mathcal{H}}$ coincides with that corresponding to the Hamiltonian $\wh{H}$ in Eq.\,(\ref{e5}), provided that
\begin{equation}\label{e37}
j \rightleftharpoons l_j, t_j, \sigma_j \rightleftharpoons l_j t_j\sigma_j,
\end{equation}
\begin{equation}\label{e38}
v(i,j) \doteq U_{\sigma_i,\sigma_j}(l_i,l_j)\hspace{0.7pt} \delta(t_i-t_j) \equiv U (1-\delta_{\sigma_i,\sigma_j}) \delta_{l_i,l_j}\hspace{0.7pt} \delta(t_i-t_j),
\end{equation}
\begin{equation}\label{e39}
\int \rd j \rightleftharpoons \sum_{\sigma_j} \int_{-\infty}^{\infty} \rd t_j \sum_{l_j=1}^{N_{\textsc{s}}}.
\end{equation}
Similarly as regards $\mathscr{G}(a,b)$ and $\mathsf{G}(a,b)$; for the calculation of $\mathscr{G}(a,b)$ one has to adopt the following conventions (\emph{cf.} Eqs\,(\ref{e24}), (\ref{e20}), (\ref{e22}), and (\ref{e23})):
\begin{equation}\label{e24a}
a \rightleftharpoons l \tau \sigma,\;\; b \rightleftharpoons l' \tau'\sigma',
\end{equation}
\begin{equation}\label{e20a}
j \rightleftharpoons l_j, \tau_j, \sigma_j \rightleftharpoons l_j\tau_j\sigma_j,
\end{equation}
\begin{equation}\label{e41}
v(i,j) \doteq U_{\sigma_i,\sigma_j}(l_i,l_j)\hspace{0.7pt} \delta(\tau_i-\tau_j) \equiv U (1-\delta_{\sigma_i,\sigma_j}) \delta_{l_i,l_j}\hspace{0.7pt} \delta(\tau_i-\tau_j),
\end{equation}
\begin{equation}\label{e42}
\int \rd j \rightleftharpoons \sum_{\sigma_j} \int_0^{\hbar\beta} \rd \tau_j \sum_{l_j=1}^{N_{\textsc{s}}},
\end{equation}
and for the calculation of $\mathsf{G}(a,b)$ the following conventions (\emph{cf.} Eqs\,(\ref{es16a}), (\ref{es16b}), (\ref{es16c}), and (\ref{es16d})):
\begin{equation}\label{e42c}
a \rightleftharpoons l t \sigma \mu,\;\; b \rightleftharpoons l' t'\sigma'\mu',
\end{equation}
\begin{equation}\label{e42d}
j \rightleftharpoons l_j, t_j, \sigma_j, \mu_j \rightleftharpoons l_j t_j \sigma_j \mu_j,
\end{equation}
\begin{align}\label{e41a}
v(i,j) &\doteq (-1)^{\lfloor\mu_i/2\rfloor} U_{\sigma_i,\sigma_j}(l_i,l_j)\hspace{0.7pt} \delta(t_i,t_j) \delta_{\mu_i,\mu_j} \nonumber\\
&\equiv (-1)^{\lfloor\mu_i/2\rfloor} U (1-\delta_{\sigma_i,\sigma_j}) \delta_{l_i,l_j}\hspace{0.7pt} \delta(t_i,t_j)\delta_{\mu_i,\mu_j},
\end{align}
\begin{equation}\label{e42b}
\int \rd j \rightleftharpoons \sum_{\sigma_j} \sum_{\mu_j \in \{1,2\}}\int_{\mathscr{C}} \rd t_j \sum_{l_j=1}^{N_{\textsc{s}}}.
\end{equation}

\refstepcounter{dummyX}
\subsubsection{The one-particle Green function (General)}
\phantomsection
\label{s212}
For what follows, it proves convenient to express the two-body interaction potential $v(i,j)$, Eqs\,(\ref{e10}), (\ref{e22}), (\ref{es16c}), (\ref{e38}), (\ref{e41}), and (\ref{e41a}), as
\begin{equation}\label{e260a}
v(i,j) \equiv \lambda\hspace{0.6pt} \mathsf{v}(i,j),
\end{equation}
where $\lambda$, the dimensionless coupling constant of interaction, serves as a book-keeping devise that in the actual calculations is to be identified with unity.

Bearing in mind that $G(a,b)$ and $G_{\X{0}}(a,b)$ denote respectively the interacting and non-interacting one-particle Green functions, specific to both $T=0$ and $T > 0$, corresponding to the Hamiltonians discussed in \S\,\ref{s211},\footnote{Thus, in the case of $T>0$ the functions $G(a,b)$ and $G_{\protect\X{0}}(a,b)$ stand for $\mathscr{G}(a,b)$ and $\mathscr{G}_{\protect\X{0}}(a,b)$ ($\mathsf{G}(a,b)$ and $\mathsf{G}_{\protect\X{0}}(a,b)$) when dealing with the Matsubara (TFD) formalism.} for $G(a,b)$ one has the following exact weak-coupling perturbational expression [Eq.\,(13.12), p.\,228, in Ref.\,\citen{HBKF04}] [Eq.\,(5.32), p.\,138, in Ref.\,\citen{SvL13}]:\,\refstepcounter{dummy}\label{SeeAlsoRef}\footnote{We note that whereas the functions $G$, $G_2$, and $G_{\protect\X{0}}$ as introduced in this paper coincide with those introduced in Ref.\,\protect\citen{SvL13} (the same applies to $\mathsf{G}$ and $\mathsf{G}_{\X{0}}$, which however have no counterparts in Ref.\,\protect\citen{SvL13}), this is \textsl{not} the case as regards $\mathscr{G}$, $\mathscr{G}_{\protect\X{2}}$, and $\mathscr{G}_{\protect\X{0}}$: the prefactor of \textsl{all} $n$-particle Green functions in Ref.\,\protect\citen{SvL13} is $1/\protect\ii\hspace{0.2pt}^n \equiv (-\protect\ii\hspace{1.0pt})^n$, while in contrast the prefactor of the $n$-particle \textsl{Matsubara} Green functions as defined in this paper is $(-1)^n$, Eq.\,(\protect\ref{e18}) and Eq.\,(\protect\ref{e91}) below. Nonetheless, the expression in Eq.\,(\ref{e242}) applies equally to $\mathscr{G}(a,b)$. This follows from the fact that, insofar as the Matsubara formalism is concerned, whereas in this paper the internal variables $\tau_j$ are integrated over the real interval $[0,\hbar\beta]$ (see Eqs\,(\protect\ref{e23}) and (\protect\ref{e42})), the relevant internal variables $\protect\b{z}_j$ (or $t_j$) in Ref.\,\protect\citen{SvL13} (in particular those in Eq.\,(5.32) herein) are integrated over the interval $[t_0,t_0-\protect\ii\beta]$, from $t_0$ towards $t_0 -\protect\ii\beta$ (using the units in which $\hbar =1$); see \S\,5.4, p.\,140, in Ref.\,\protect\citen{SvL13}. Identifying $t_0$ with zero, with $\protect\b{z} = -\protect\ii\hspace{0.6pt}\tau$, one has $\mathrm{d}\protect\b{z} = -\protect\ii\hspace{0.6pt}\mathrm{d}\tau$, so that $\prod_{j=1}^{2\nu}\mathrm{d}\protect\b{z}_j = (-\protect\ii\hspace{1.0pt})^{2\nu} \prod_{j=1}^{2\nu}\mathrm{d}\tau_j$. Further, since \textsl{our} $\mathscr{G}_0$ is identical to $\protect\ii\hspace{1.0pt}$ times the corresponding Green function in Ref.\,\protect\citen{SvL13}, the $A_{2\nu+1}^{\textsc{b}}$ and $A_{2\nu}$, Eqs\,(\ref{e258}) and (\ref{e260}), as expressed in terms of the latter Green function are respectively $\protect\ii\hspace{0.2pt}^{2\nu+1}$ and $\protect\ii\hspace{0.2pt}^{2\nu}$ times the $A_{2\nu+1}^{\textsc{b}}$ and $A_{2\nu}$ as expressed in terms of the former Green function. Combining these two observations, the validity of our above assertion is established. A similar reasoning establishes the applicability of the expression in Eq.\,(\ref{e200}) below to $\mathscr{G}_{\protect\X{2}}(a,b;c,d)$.}
\begin{equation}\label{e242}
G(a,b) = \frac{G_{\X{0}}(a,b) + \sum_{\nu=1}^{\infty}\lambda^{\nu} N_{\nu}(a,b)}{1 + \sum_{\nu=1}^{\infty}\lambda^{\nu} D_{\nu}},
\end{equation}
where
\begin{equation}\label{e257}
N_{\nu}(a,b) \doteq \frac{1}{\nu !} \Big(\frac{\ii}{2\hbar}\Big)^{\nu} \int \prod_{j=1}^{2\nu} \rd j\; \mathsf{v}(1,2) \dots \mathsf{v}(2\nu-1,2\nu) \hspace{0.6pt} A_{2\nu+1}^{\textsc{b}}(a,b;1,2,\dots,2\nu-1,2\nu),
\end{equation}
\begin{equation}\label{e261}
D_{\nu} \doteq \frac{1}{\nu !} \Big(\frac{\ii}{2\hbar}\Big)^{\nu} \int \prod_{j=1}^{2\nu} \rd j\; \mathsf{v}(1,2) \dots \mathsf{v}(2\nu-1,2\nu) \hspace{0.6pt} A_{2\nu}(1,2,\dots,2\nu-1,2\nu),
\end{equation}
in which\,\footnote{For the superscripts $+$, see Eqs\,(\ref{e260b}), (\ref{e23a}), and (\ref{es16e}).}
\begin{equation}\label{e258}
A_{2\nu+1}^{\textsc{b}}(a,b;1,2,\dots,2\nu) \equiv
\begin{vmatrix}
G_{\X{0}}(a,b)& G_{\X{0}}(a,1^+) & G_{\X{0}}(a,2^+) & \dots & G_{\X{0}}(a,2\nu^+) \\
G_{\X{0}}(1,b) & G_{\X{0}}(1,1^+) & G_{\X{0}}(1,2^+) & \dots & G_{\X{0}}(1,2\nu^+) \\
G_{\X{0}}(2,b) & G_{\X{0}}(2,1^+) & G_{\X{0}}(2,2^+) & \dots & G_{\X{0}}(2,2\nu^+) \\
\vdots & \vdots & \vdots & \ddots & \vdots \\
G_{\X{0}}(2\nu,b) & G_{\X{0}}(2\nu,1^+) & G_{\X{0}}(2\nu,2^+) & \dots & G_{\X{0}}(2\nu,2\nu^+)
\end{vmatrix}_{\pm},
\end{equation}
\begin{equation}\label{e260}
A_{2\nu}(1,2,\dots,2\nu) \equiv
\begin{vmatrix}
G_{\X{0}}(1,1^+) & G_{\X{0}}(1,2^+) & \dots & G_{\X{0}}(1,2\nu^+) \\
G_{\X{0}}(2,1^+) & G_{\X{0}}(2,2^+) & \dots & G_{\X{0}}(2,2\nu^+) \\
\vdots & \vdots & \ddots & \vdots \\
G_{\X{0}}(2\nu,1^+) & G_{\X{0}}(2\nu,2^+) & \dots & G_{\X{0}}(2\nu,2\nu^+)
\end{vmatrix}_{\pm}.
\end{equation}
The functions $A_{2\nu+1}^{\textsc{b}} \doteq \vert \mathbb{A}_{2\nu+1}^{\textsc{b}}\vert_{\X{+/-}}$ and $A_{2\nu} \doteq \vert \mathbb{A}_{2\nu}\vert_{\X{+/-}}$ are permanents \cite{HM78,NW78} / determinants\cite{TM60,VD99}, specific to bosons / fermions, associated with the $(2\nu+1)\times (2\nu+1)$ matrix $\mathbb{A}_{2\nu+1}^{\textsc{b}}$ and the $2\nu\times 2\nu$ matrix $\mathbb{A}_{2\nu}$. One observes that $\mathbb{A}_{2\nu+1}^{\textsc{b}}$ is a \textsl{bordered} matrix associated with $\mathbb{A}_{2\nu}$. For fermions, $A_{2\nu+1}^{\textsc{b}}$ is a \textsl{bordered} determinant \cite{TM60,VD99} associated with the determinant $A_{2\nu}$.

We note that since each of the diagonal elements $\{G_{\X{0}}(j,j^+)\| j=1,2,\dots,2\nu\}$ in the above permanents / determinants, which is independent of time (imaginary time),\footnote{See p.\,\protect\pageref{SinceHIs}.} corresponds to a tadpole diagram, appendices \ref{sac} and \ref{sad}, they can be suppressed by including the contribution of the static Hartree potential in the non-interacting Hamiltonian, Eqs\,(\ref{e5}) and (\ref{e27}).\footnote{See p.\,\protect\pageref{WhenDealingWith}, where we discuss $G_{\textsc{h}}$, $\protect\h{\Sigma}^{\textsc{h}}$, $\protect\h{\Sigma}_{\sigma}^{\textsc{f}}$, and $\protect\h{\Sigma}_{\sigma}^{\textsc{hf}} \equiv \protect\h{\Sigma}^{\textsc{h}} + \protect\h{\Sigma}_{\sigma}^{\textsc{F}}$. See also Eq.\,(\ref{e250b}) below.} In the case of uniform GSs (or uniform ensembles of states), where $\{G_{\X{0}}(j,j^+)\| j=1,2,\dots,2\nu\}$ are fully independent of the spatial coordinate associated with $j$, the task is straightforwardly achieved.\footnote{Compare for instance with the expressions in Eqs\,(3)--(5), (10), (11), and (14) of Ref.\,\protect\citen{YY70I}.}\footnote{Unless $G_{0;\sigma_j} = G_{0;\protect\b{\sigma}_j}$ for all $j$, even for uniform ground states (thermal ensemble of states, within the Matsubara formalism) $G_{\X{0}}(j,j^+)$ retains a dependence on $j$. This applies the stronger within the TFD formalism, where $\mathsf{G}_{\X{0}}(j,j^+)$ depends additionally on $\mu_j$, Eqs\,(\protect\ref{es16b}) and (\protect\ref{e42d}).} In fact, this procedure amounts to a special case of a more general one encountered in for instance Refs\,\citen{ANR03,RL04,RSL05} and \citen{BF9799}. In Refs\,\citen{ANR03,RL04,RSL05}, the functions $\{ \upalpha_{\X{j}}^{\X{j'}}\| j,j'\}$ are varied to minimise the impact of the sign problem in the underlying Monte Carlo calculations.

\refstepcounter{dummyX}
\subsubsection{The TFD revisited}
\phantomsection
\label{s226}
In this section we revert to the definitions for $a$, $b$, $j$, $v(i,j)$, and $\int \mathrm{d}j$ as in Eqs\,(\ref{e12}), (\ref{e8}), (\ref{e10}), and (\ref{e11}), respectively, to be distinguished from their counterparts in \S\,\ref{sac01}.

In the light of the above observations, within the framework of the TFD for the exact weak-coupling perturbational expression of $\mathsf{G}^{\mu\mu'}\hspace{-2.4pt}(a,b)$ one has (\emph{cf.} Eq.\,(\ref{e242}))
\begin{equation}\label{es17}
\mathsf{G}^{\mu\mu'}\hspace{-2.4pt}(a,b) = \frac{\mathsf{G}_{\X{0}}^{\mu\mu'}\hspace{-2.4pt}(a,b) +
\sum_{\nu=1}^{\infty}\lambda^{\nu} \mathsf{N}_{\nu}^{\mu\mu'}\hspace{-2.4pt}(a,b)}{1 +
\sum_{\nu=1}^{\infty}\lambda^{\nu} \mathsf{D}_{\nu}},
\end{equation}
where $\mathsf{N}_{\nu}^{\mu\mu'}\hspace{-2.4pt}(a,b)$ and $\mathsf{D}_{\nu}$ are determined according to the expressions in respectively Eq.\,(\ref{e257}) and Eq.\,(\ref{e261}), however in terms of the functions $\mathsf{A}_{2\nu+1}^{\textsc{b},\mu\mu'}$ and $\mathsf{A}_{2\nu}$, defined according to (\emph{cf.} Eq.\,(\ref{e258}))
\begin{align}\label{es18}
&\hspace{0.2cm}\mathsf{A}_{2\nu+1}^{\textsc{b},\mu\mu'}\hspace{-1.2pt}(a,b;1,2,\dots,2\nu) \equiv
\sum_{\mu_1,\mu_3\dots,\mu_{2\nu-1} \in \{1,2\}}
(-1)^{\lfloor\mu_1/2\rfloor +\lfloor\mu_3/2\rfloor +\dots +\lfloor\mu_{2\nu-1}/2\rfloor}
\nonumber\\
&\hspace{1.0cm}\times
\begin{vmatrix}
\mathsf{G}_{\X{0}}^{\mu\mu'}\hspace{-2.4pt}(a,b)&
\mathsf{G}_{\X{0}}^{\mu\mu_{\eta_1}}\hspace{-2.0pt}(a,1^+) &
\mathsf{G}_{\X{0}}^{\mu\mu_{\eta_2}}\hspace{-2.0pt}(a,2^+) & \dots &
\mathsf{G}_{\X{0}}^{\mu\mu_{\eta_{2\nu}}}\hspace{-2.0pt}(a,2\nu^+) \\
\mathsf{G}_{\X{0}}^{\mu_{\eta_1}\mu'}\hspace{-2.4pt}(1,b) &
\mathsf{G}_{\X{0}}^{\mu_{\eta_1}\mu_{\eta_1}}\hspace{-2.0pt}(1,1^+) &
\mathsf{G}_{\X{0}}^{\mu_{\eta_1}\mu_{\eta_2}}\hspace{-2.0pt}(1,2^+) & \dots &
\mathsf{G}_{\X{0}}^{\mu_{\eta_1}\mu_{\eta_{2\nu}}}\hspace{-2.0pt}(1,2\nu^+) \\
\mathsf{G}_{\X{0}}^{\mu_{\eta_{2}}\mu'}\hspace{-2.0pt}(2,b) &
\mathsf{G}_{\X{0}}^{\mu_{\eta_2}\mu_{\eta_1}}\hspace{-2.0pt}(2,1^+) &
\mathsf{G}_{\X{0}}^{\mu_{\eta_2}\mu_{\eta_2}}\hspace{-1.8pt}(2,2^+) & \dots &
\mathsf{G}_{\X{0}}^{\mu_{\eta_2}\mu_{\eta_{2\nu}}}\hspace{-2.0pt}(2,2\nu^+) \\
\vdots & \vdots & \vdots & \ddots & \vdots \\
\mathsf{G}_{\X{0}}^{\mu_{\eta_{2\nu}}\mu'}\hspace{-2.4pt}(2\nu,b) &
\mathsf{G}_{\X{0}}^{\mu_{\eta_{2\nu}}\mu_{\eta_1}}\hspace{-2.0pt}(2\nu,1^+) &
\mathsf{G}_{\X{0}}^{\mu_{\eta_{2\nu}}\mu_{\eta_2}}\hspace{-1.8pt}(2\nu,2^+) & \dots &
\mathsf{G}_{\X{0}}^{\mu_{\eta_{2\nu}}\mu_{\eta_{2\nu}}}\hspace{-1.8pt}(2\nu,2\nu^+)
\end{vmatrix}_{\pm},\nonumber\\
\end{align}
and (\emph{cf.} Eq.\,(\ref{e260}))
\begin{align}\label{es19}
&\mathsf{A}_{2\nu}(1,2,\dots,2\nu) \equiv
\sum_{\mu_1,\mu_3,\dots,\mu_{2\nu-1} \in \{1,2\}}
(-1)^{\lfloor\mu_1/2\rfloor +\lfloor\mu_3/2\rfloor +\dots +\lfloor\mu_{2\nu-1}/2\rfloor}
\nonumber\\
&\hspace{1.0cm}\times
\begin{vmatrix}
\mathsf{G}_{\X{0}}^{\mu_{\eta_1}\mu_{\eta_1}}\hspace{-2.0pt}(1,1^+) &
\mathsf{G}_{\X{0}}^{\mu_{\eta_1}\mu_{\eta_2}}\hspace{-2.0pt}(1,2^+) & \dots &
\mathsf{G}_{\X{0}}^{\mu_{\eta_1}\mu_{\eta_{2\nu}}}\hspace{-2.0pt}(1,2\nu^+) \\
\mathsf{G}_{\X{0}}^{\mu_{\eta_2}\mu_{\eta_1}}\hspace{-2.0pt}(2,1^+) &
\mathsf{G}_{\X{0}}^{\mu_{\eta_2}\mu_{\eta_2}}\hspace{-1.8pt}(2,2^+) & \dots &
\mathsf{G}_{\X{0}}^{\mu_{\eta_2}\mu_{\eta_{2\nu}}}\hspace{-2.0pt}(2,2\nu^+) \\
\vdots & \vdots & \ddots & \vdots \\
\mathsf{G}_{\X{0}}^{\mu_{\eta_{2\nu}}\mu_{\eta_1}}\hspace{-2.0pt}(2\nu,1^+) &
\mathsf{G}_{\X{0}}^{\mu_{\eta_{2\nu}}\mu_{\eta_2}}\hspace{-1.8pt}(2\nu,2^+) & \dots &
\mathsf{G}_{\X{0}}^{\mu_{\eta_{2\nu}}\mu_{\eta_{2\nu}}}\hspace{-1.8pt}(2\nu,2\nu^+)
\end{vmatrix}_{\pm},\nonumber\\
\end{align}
where\,\footnote{$\eta_{2k-1} = \eta_{2k} = 2k -1$, $\forall k \in \mathds{N}$.}
\begin{equation}\label{es20a}
\eta_j \doteq 2\Big\lfloor \frac{j+1}{2} \Big\rfloor -1.
\end{equation}
$\phantom{X}$

\refstepcounter{dummyX}
\subsubsection{The Hubbard Hamiltonian revisited}
\phantomsection
\label{s227}
As we have indicated earlier, expressions in Eqs\,(\ref{e242}) -- (\ref{e260}) equally apply to the Hubbard Hamiltonian $\wh{\mathcal{H}}$ discussed in \S\,\ref{s224}. On making explicit use of the two-body interaction potential in Eq.\,(\ref{e38}), from the expressions in Eqs\,(\ref{e257}) and (\ref{e261}) for the $N_{\nu}(a,b)$ and $D_{\nu}$, $\nu\in\mathds{N}$, specific to $\wh{\mathcal{H}}$ one obtains
\begin{align}\label{e45}
&\hspace{-1.0cm} N_{\nu}(a,b) = \frac{1}{\nu!} \Big(\frac{\ii U}{2\hbar}\Big)^{\nu} \sum_{\sigma_1,\dots, \sigma_{\nu}}\; \sum_{l_1,\dots,l_{\nu} =1}^{N_{\textsc{s}}}\nonumber\\
&\hspace{0.6cm} \times\int \prod_{j=1}^{\nu} \rd t_j\; A_{2\nu+1}^{\textsc{b}}(a,b;l_1 t_1\sigma_1,l_1 t_1\b{\sigma}_1,\dots,l_{\nu} t_{\nu}\sigma_{\nu},
l_{\nu} t_{\nu}\b{\sigma}_{\nu}),
\end{align}
\begin{align}\label{e46}
&\hspace{-1.5cm} D_{\nu} = \frac{1}{\nu!} \Big(\frac{\ii U}{2\hbar}\Big)^{\nu} \sum_{\sigma_1,\dots, \sigma_{\nu}}\; \sum_{l_1,\dots,l_{\nu} =1}^{N_{\textsc{s}}} \nonumber\\
&\hspace{0.1cm} \times\int \prod_{j=1}^{\nu} \rd t_j\; A_{2\nu}(l_1 t_1\sigma_1,l_1 t_1\b{\sigma}_1,\dots,l_{\nu}t_{\nu}\sigma_{\nu},
l_{\nu}t_{\nu}\b{\sigma}_{\nu}).
\end{align}
These expressions are explicitly applicable to the $T=0$ formalism. They equally apply to the case of Matsubara's $T>0$ formalism, provided that $t_i$, $t_j$, and $\int \rd t_j$ be understood as representing respectively $\tau_i$, $\tau_j$, and $\int_0^{\hbar\beta} \mathrm{d}\tau_j$. With reference to the considerations in \S\,\ref{s226}, the relevant expressions for $\mathsf{N}_{\nu}^{\mu\mu'}(a,b)$ and $\mathsf{D}_{\nu}$, specific to the TFD formalism, are obtained from those in Eqs\,(\ref{e45}) and (\ref{e46}) through replacing the $A_{2\nu+1}^{\textsc{b}}$ and $A_{2\nu}$ herein by respectively the $\mathsf{A}_{2\nu+1}^{\textsc{b},\mu\mu'}$ and $\mathsf{A}_{2\nu}$ as defined in Eqs\,(\ref{es18}) and (\ref{es19}).

\refstepcounter{dummy}\label{InAppendixD}In appendix \ref{sab} we consider the Hubbard Hamiltonian for spin-$\tfrac{1}{2}$ fermions in some detail. The lattice on which this Hamiltonian is defined is embedded in a $d$-dimensional space, with $d$ arbitrary, including $d=\infty$ \cite{EMH89b,MV89,EMH89a,GKKR96,PF03}, which corresponds to the framework of the dynamical mean-field theory \cite{GKKR96} (DMFT). We should emphasise that \emph{the simplified expressions in Eqs\,(\ref{e45}) and (\ref{e46}), as well as those presented in appendix \ref{sab}, are suited for the calculation of the operators $\h{\Sigma}_{\X{00}}[v,G_{\X{0}}]$ and $\h{\Sigma}_{\X{01}}[v,G]$, but \textsl{not} for that of $\h{\Sigma}_{\X{10}}[W,G_{\X{0}}]$ and $\h{\Sigma}_{\X{11}}[W,G]$.} This aspect will be clarified later in this paper, where we explicitly deal with the latter two self-energy operators.

\refstepcounter{dummyX}
\subsection{The perturbation series expansion of \texorpdfstring{$G$}{} in terms of
\texorpdfstring{$(v,G_{\protect\X{0}})$}{} }
\phantomsection
\label{s22}
The discussions in the preceding section have made explicit that the diagram-free formalism of the perturbation series expansion for the one-particle Green function, to be discussed in detail in this section, is structurally the same irrespective of whether one deals with this function as corresponding to a GS or to a non-zero-temperature equilibrium ensemble of states, or whether the underlying system is defined over a continuum subset of $\mathds{R}^d$ or on a lattice embedded in this space. Depending on the specifics of the case at hand, one is consistently to adopt one of the conventions in
Eqs\,(\ref{e8}), (\ref{e10}), (\ref{e11}), and (\ref{e12}), in Eqs\,(\ref{e20}), (\ref{e22}), (\ref{e23}), and (\ref{e24}), in Eqs\,(\ref{es16a}), (\ref{es16b}), (\ref{es16d}), and (\ref{es16c}), in Eqs\,(\ref{e36}), (\ref{e37}), (\ref{e38}), and (\ref{e39}), in Eqs\,(\ref{e24a}), (\ref{e20a}), (\ref{e41}), and (\ref{e42}), and in Eqs\,(\ref{e42c}), (\ref{e42d}), (\ref{e41a}), and (\ref{e42b}). \emph{In the following, we shall therefore employ the symbols $G$ and $G_{\X{0}}$ for respectively the interacting and non-interacting one-particle Green functions irrespective of whether $T=0$ or $T>0$.}\footnote{For $T>0$, irrespective of whether the Matsubara formalism is concerned or the TFD one.} By the same reasoning, \emph{in the following $\Sigma$ will similarly denote the self-energy for both cases of $T=0$ and $T> 0$.}

The diagram-free perturbation series expansion of $G$ in terms of $G_{\X{0}}$ and the bare two-body interaction function $v$ is relevant both in its own right and from the perspective of developing the perturbation series expansions for the self-energy functionals $\h{\Sigma}_{\X{00}}[v,G_{\X{0}}]$, $\h{\Sigma}_{\X{01}}[v,G]$, $\h{\Sigma}_{\X{10}}[W,G_{\X{0}}]$, and $\h{\Sigma}_{\X{11}}[W,G]$, briefly described in \S\,\ref{sec1}. Calculation of $\h{\Sigma}_{\X{00}}[v,G_{\X{0}}]$ from the perturbation series expansion to be discussed in this section is immediate. This is however not the case as regards the perturbation series expansions of the remaining three functionals for the self-energy operator. Nonetheless, an insight gained from the considerations of this section proves crucial for the construction of these perturbation series expansions.

We point out that for long-range interaction potentials and systems defined on a \textsl{continuum} subset of $\mathds{R}^d$, the perturbation series expansions of $G$, $\h{\Sigma}_{\X{00}}[v,G_{\X{0}}]$, and $\h{\Sigma}_{\X{01}}[v,G]$ to arbitrary order $n$ are strictly ill-defined, this on account of the fact that beyond a certain order, which depends on the nature of $v$ and the value of $d$, for long-range interaction potentials the terms of these series are infrared divergent, and for the mentioned systems these terms are ultraviolet divergent. In these cases, the relevant expansions in terms of the screened interaction potential $W$ are to be employed. As will become evident, although the perturbation series expansions in terms of $W$ (to be considered in \S\S\, \ref{s25} and \ref{s26}) rely on those in terms of $v$ (to be considered in \S\S\,\ref{s221}, \ref{s23}, and \ref{s24}), since the functionals to be relied upon in \S\S\,\ref{s25} and \ref{s26} are evaluated in terms of $W$, such reliance does not entail any practical or fundamental limitations.

\refstepcounter{dummyX}
\subsubsection{Details}
\phantomsection
\label{s221}
With $A_{r,s}^{\X{(2\nu-1)}}$ denoting the $(r,s)$ first cofactor [\S\,2.3.3, p.\,12, in Ref.\,\citen{VD99}]\,\footnote{Here we adopt the notation of Ref.\,\protect\citen{VD99}, with the superscript $(2\nu-1)$ signifying $A_{r,s}^{\protect\X{(2\nu-1)}}$ as being a $(2\nu-1)$-permanent / -determinant.} associated with $A_{2\nu}$, Eq.\,(\ref{e260}), for the function $A_{2\nu+1}^{\textsc{b}}$ in Eq.\,(\ref{e258}) one has [Theorem 3.9, p.\,47, in Ref.\,\citen{VD99}] [\S\,7.1, p.\,198, in Ref.\,\citen{BR91}]\,\refstepcounter{dummy}\label{InRefVD99}\footnote{In Ref.\,\protect\citen{BR91}, the term `cofactor' is used only in connection with determinants, however comparison of the expression in Eq.\,(7.2), p.\,199, herein (see also Eq.\,(1.2), p.\,16, in Ref.\,\protect\citen{HM78}) with that in Eq.\,(1.3.16) of Ref.\,\protect\citen{VD99} clearly shows that use of this term is justified also in dealing with permanents. Since in the case of determinants cofactor is defined as a \textsl{signed} minor [Eq.\,(1.3.12) in Ref.\,\protect\citen{VD99}], the sign in the case of permanents is to be identified with $+$ and the \textsl{minor} itself with a permanent.}
\begin{equation}\label{e259}
A_{2\nu+1}^{\textsc{b}} = A_{2\nu}\hspace{0.6pt} G_{\X{0}}(a,b) \pm \sum_{r,s=1}^{2\nu}
A_{r,s}^{\X{(2\nu-1)}} G_{\X{0}}(a,s^+) G_{\X{0}}(r,b),
\end{equation}
where $+/-$ corresponds to bosons / fermions. Hence from Eq.\,(\ref{e257}) one obtains
\begin{equation}\label{e259a}
N_{\nu}(a,b) = D_{\nu}\hspace{0.6pt} G_{\X{0}}(a,b) + M_{\nu}(a,b),
\end{equation}
where
\begin{align}\label{e259b}
&\hspace{-0.8cm} M_{\nu}(a,b) \doteq \pm\frac{1}{\nu !} \Big(\frac{\ii}{2\hbar}\Big)^{\nu} \sum_{r,s=1}^{2\nu}\int \prod_{j=1}^{2\nu} \rd j\; \mathsf{v}(1,2) \dots \mathsf{v}(2\nu-1,2\nu) \nonumber\\
&\hspace{3.0cm}\times A_{r,s}^{\X{(2\nu-1)}}(1,2,\dots,2\nu-1,2\nu)\hspace{0.8pt} G_{\X{0}}(a,s^+) G_{\X{0}}(r,b),
\end{align}
in which $+/-$ corresponds to bosons / fermions. We note in passing that [Eq.\,(2.3.10), p.\,12, in Ref.\,\citen{VD99}]:\,\footnote{This equality applies also to permanents.}
\begin{equation}\label{e254x}
A_{r,s}^{\X{(2\nu-1)}} = \frac{\partial A_{2\nu}}{\partial G_{\X{0}}(r,s^+)}.
\end{equation}

To calculate the $n$th-order perturbation series expansion of $G$, we begin with the calculation of the elements of the sequence $\{F_{\nu} \| \nu=1,\dots,n\}$ as encountered in the expression
\begin{equation}\label{e243}
\frac{1}{1+ \sum_{\nu=1}^{\infty}\lambda^{\nu} D_{\nu}} = 1 -\sum_{\nu=1}^{n} \lambda^{\nu} F_{\nu} + O(\lambda^{n+1}).
\end{equation}
Making use of the equality
\begin{equation}\label{e246}
\big(1 + \sum_{\nu=1}^{n} \lambda^n D_{\nu} \big) \big(1 -\sum_{\nu=1}^{n} \lambda^{\nu} F_{\nu}\big) = 1 + O(\lambda^{n+1}),
\end{equation}
followed by equating the coefficient of $\lambda^j$, $j = 1,\dots,n$, in the polynomial on the left-hand side (LHS) of Eq.\,(\ref{e246}) with zero, one arrives at\,\footnote{In the present expression, as well as in similar later expressions (unless we indicate otherwise), the condition $\nu \ge 2$ is a substitute for the more accurate pair of conditions $2 \le \nu \le n$, which implies $n \ge 2$. In this connection, $n=0$ corresponds to the trivial case where $G(a,b) \equiv G_{\X{0}}(a,b)$. For the case of $n=1$, one has $F_1 = D_1$ and the second equality is redundant.}
\begin{equation}\label{e247}
F_1 = D_1,\;\; F_{\nu} = D_{\nu} -\sum_{\nu'=1}^{\nu-1} D_{\nu-\nu'} F_{\nu'},\;\; \nu \ge 2,
\end{equation}
from which one recursively determines the elements of the sequence $\{F_{\nu} \| \nu=1,\dots,n\}$.

From Eqs\,(\ref{e242}) and (\ref{e243}), one has
\begin{equation}\label{e244}
G(a,b) = \big(1 -\sum_{\nu=1}^{n} \lambda^{\nu} F_{\nu} \big) \big(G_{\X{0}}(a,b) + \sum_{\nu=1}^{n} \lambda^{\nu} N_{\nu}(a,b) \big)  + O(\lambda^{n+1}),
\end{equation}
leading to the perturbation series expansion\,\footnote{For a diagrammatic determination of $G^{\protect\X{(\nu)}}(a,b) \equiv G^{\protect\X{(\nu)}}(a,b;[v,G_{\protect\X{0}}])$, see appendix \protect\ref{sac}.}
\begin{equation}\label{e245}
G(a,b) = G_{\X{0}}(a,b) + \sum_{\nu=1}^{n} \lambda^{\nu} G^{\X{(\nu)}}(a,b) + O(\lambda^{n+1}),
\end{equation}
where $G^{\X{(\nu)}}(a,b)$, $\nu \ge 1$, denotes the $\nu$th-order perturbational contribution to the interacting Green function $G(a,b)$. In a similar manner as in the case of $\{F_{\nu}\| \nu\}$, one obtains
\begin{align}\label{e248}
G^{\X{(1)}}(a,b) &= N_{1}(a,b) -  F_{1}\hspace{0.8pt}  G_{\X{0}}(a,b),\nonumber\\
G^{\X{(\nu)}}(a,b) &= N_{\nu}(a,b) - F_{\nu}\hspace{0.8pt} G_{\X{0}}(a,b) -\sum_{\nu'=1}^{\nu-1}\hspace{0.8pt} F_{\nu-\nu'} N_{\nu'}(a,b),\;\; \nu \ge 2.
\end{align}
The expression Eq.\,(\ref{e245}), in conjunction with the expressions in Eq.\,(\ref{e248}), is equivalent to that in Eq.\,(26), p.\,5, of Ref.\,\citen{RSL05}. The two expressions are however deduced along different lines.

For the considerations of appendices \ref{sac} and \ref{sad} it will prove significant to simplify the expressions in Eq.\,(\ref{e248}). One verifies that
\begin{align}\label{e248a}
G^{\X{(1)}}(a,b) &= M_1(a,b),\nonumber\\
G^{\X{(\nu)}}(a,b) &= M_{\nu}(a,b) -\sum_{\nu'=1}^{\nu-1} F_{\nu-\nu'}\hspace{0.8pt} M_{\nu'}(a,b),\;\; \nu \ge 2,
\end{align}
where $M_{\nu}(a,b)$, $\nu \in\mathds{N}$, is defined in Eq.\,(\ref{e259b}). The \textsl{ordered} sequence $\{ G^{\X{(\nu)}}(a,b) \| \nu \in \mathds{N}\}$ thus obtained corresponds to \textsl{connected} diagrams, \S\,\ref{s222}. Thus, neglecting the contribution $O(\lambda^{n+1})$ on the RHS of Eq.\,(\ref{e245}), one obtains the expression for the exact $n$th-order perturbation series expansion of $G(a,b)$ in terms of the contributions of the \textsl{connected} Green-function diagrams \cite{FW03} determined in terms of the non-interacting Green function $G_{\X{0}}$. This is \textsl{not} the case for the $G(a,b)$ obtained by merely replacing the $\infty$ by $n$ in the numerator and the denominator of the expression on the RHS of Eq.\,(\ref{e242}), which in addition takes account of contributions arising from disconnected Green-function diagrams. We shall discuss this observation below, \S\,\ref{s222}.

In appendix \ref{sac} we describe a practical approach whereby contributions to $G$ corresponding to disconnected Green-function diagrams are explicitly discarded, leading to the standard diagrammatic expansion of $G$ in terms of $(v,G_{\X{0}})$. The approach of appendix \ref{sac} serves as a stepping stone for devising a similar practical approach, to be described in appendix \ref{sad}, for determining the perturbational contributions corresponding to $G$-skeleton (\emph{i.e.} 2PI) self-energy diagrams on the basis of the perturbational contributions corresponding to the proper (\emph{i.e.} 1PI) self-energy diagrams.

\refstepcounter{dummyX}
\subsubsection{Discussion}
\phantomsection
\label{s222}
Following Eq.\,(\ref{e248a}), we indicated that the sequence $\{G^{\X{(\nu)}}(a,b) \| \nu\}$ corresponds to \textsl{connected} Green-function diagrams. This assertion can be appreciated directly from the expressions in Eq.\,(\ref{e248}), where $N_{\nu}(a,b)$, $\nu \in \{1,2,\dots,n\}$, is due to \textsl{all} $\nu$th-order Green-function diagrams, and the functions $F_{\nu} G_{\X{0}}$, $F_{\nu-\nu'} N_{\nu'}$, $\nu' \in \{1,2,\dots,\nu-1\}$, due to \textsl{all} $\nu$th-order \textsl{disconnected} diagrams, a fact clearly reflected in the multiplicative nature of these functions.\footnote{Note that the sum of the subscripts $\nu-\nu'$ and $\nu'$ is equal to $\nu$ for all relevant value of $\nu'$.} Note that, following Eqs\,(\ref{e259a}) and (\ref{e247}), depending on the values of $\nu$ and $\nu'$ the functions $F_{\nu}$, $F_{\nu-\nu'}$ and $N_{\nu'}$ in turn describe contributions corresponding to disconnected diagrams.

With reference to Eq.\,(\ref{e242}), let
\begin{equation}\label{e301}
G^{[m/m']}(a,b) \doteq \frac{G_{\X{0}}(a,b) + \sum_{\nu=1}^{m}\lambda^{\nu} N_{\nu}(a,b)}{1 + \sum_{\nu=1}^{m'}\lambda^{\nu} D_{\nu}}.
\end{equation}
Clearly, up to an error of order $\lambda^{n+1}$ the series expansion in Eq.\,(\ref{e245}) equally describes the function $G^{\X{[n/n]}}(a,b)$. This observation is interesting, since the function in the denominator of the expression on the RHS Eq.\,(\ref{e242}) is an \textsl{exact} divisor of the function in the numerator [Fig.\,9.2, p.\,95, and Eq.\,(9.4), p.\,96, in Ref.\,\citen{FW03}] [\S\,8.3 in Ref.\,\citen{AGD75}]. On this account, it is tempting to suspect that the function $G^{\X{[2n/n]}}(a,b)$ were identically equal to an $n$th-order polynomial of $\lambda$ and therefore the appropriate perturbational description of $G(a,b)$ up to and including the $n$th order in the interaction potential. Below we show that only in the limit $n =\infty$ is the function in the denominator of the expression for $G^{\X{[2n/n]}}(a,b)$ an exact divisor of the function in the numerator. Defining the function $\mathpzc{G}^{\X{[2n/n]}}(a,b)$ as consisting of the ratio of a $(2n)$th-order polynomial of $\lambda$ and the $n$th-order polynomial that comprises the denominator of the function $G^{\X{[2n/n]}}(a,b)$ in such a way that $\mathpzc{G}^{\X{[2n/n]}}(a,b)$ is identically equal to an $n$th-order polynomial of $\lambda$, one explicitly shows that $\mathpzc{G}^{\X{[2n/n]}}(a,b) \not\equiv G^{\X{[2n/n]}}(a,b)$ for $n <\infty$, Eq.\,(\ref{e316}) below. This follows from the fact that \textsl{only} up to and including the $n$th order in $\lambda$ are the polynomials in the numerators of $\mathpzc{G}^{\X{[2n/n]}}(a,b)$ and $G^{\X{[2n/n]}}(a,b)$ identical. As a result, the requirement of the function $\mathpzc{G}^{\X{[2n/n]}}(a,b)$ being an $n$th-order polynomial of $\lambda$ is only satisfied by leaving out some perturbational contributions to the Green function beyond the $n$th order in $\lambda$. Had this not been the case, one would be able to construct a recursive scheme for the calculation of $G(a,b)$ to arbitrary order in $\lambda$ on the basis of the knowledge of the function $N_1(a,b)$ and the constants $\{D_{\nu} \| \nu\in \mathds{N}\}$.

To proceed, with reference to the expression in Eq.\,(\ref{e242}) we consider the equality
\begin{equation}\label{e302}
\frac{1 + \sum_{\nu=1}^{2n} \lambda^{\nu} \uppi_{\nu}(a,b)}{1 + \sum_{\nu=1}^{n} \lambda^{\nu} D_{\nu}} = 1 + \sum_{\nu=1}^{n} \lambda^{\nu} \uprho_{\nu}(a,b),
\end{equation}
where
\begin{equation}\label{e303}
\uprho_{\nu}(a,b) \doteq \frac{G^{\X{(\nu)}}(a,b)}{G_{\X{0}}(a,b)}.
\end{equation}
In the light of the equality in Eq.\,(\ref{e245}), one expects that
\begin{equation}\label{e304}
\uppi_{\nu}(a,b) \equiv \frac{N_{\nu}(a,b)}{G_{\X{0}}(a,b)}\;\;\; \text{for}\;\;\; \nu \in \{1,2,\dots,n\}.
\end{equation}
With\,\footnote{From Eq.\,(\protect\ref{e302}) one has: $\sum_{\nu=1}^{2n} \lambda^{\nu} \uppi_{\nu}(a,b) = \sum_{\nu=1}^{n} \lambda^{\nu} (D_{\nu} + \uprho_{\nu}(a,b)) + \mathscr{F}_{2n}(\lambda\vert a,b)$. For the specific case of $n=1$, from this equality one in particular obtains $\uppi_{2}(a,b) = \upphi_2(a,b)$.}
\begin{equation}\label{e305}
\mathscr{F}_{2n}(\lambda\vert a,b) \doteq \sum_{\nu=1}^{n} \sum_{\nu'=1}^{n} \lambda^{\nu+\nu'} D_{\nu} \uprho_{\nu'}(a,b) \equiv \sum_{\nu=2}^{2n} \lambda^{\nu} \upphi_{\nu}(a,b),
\end{equation}
since
\begin{equation}\label{e306y}
\upphi_{\nu}(a,b) =\left. \frac{1}{\nu!} \frac{\partial^{\nu}}{\partial\lambda^{\nu}} \mathscr{F}_{2n}(\lambda\vert a,b)\right|_{\lambda=0},
\end{equation}
one readily obtains that
\begin{equation}\label{e307y}
\upphi_{\nu}(a,b) = \sum_{\nu'=1}^{n}\Delta_{\nu-\nu'}(n) D_{\nu'} \uprho_{\nu-\nu'}(a,b)  \equiv  \sum_{\nu'=1}^{n}  \Delta_{\nu-\nu'}(n) D_{\nu-\nu'} \uprho_{\nu'}(a,b),
\end{equation}
where
\begin{equation}\label{e308y}
\Delta_{\nu}(n) \doteq \sum_{\nu'=1}^{n}  \delta_{\nu,\nu'} \equiv \left\{
\begin{array}{cc}
1, & 1\le \nu \le n, \\ \\
0, & \nu < 1 \vee \nu >  n.
\end{array} \right.
\end{equation}
It follows that, for $n=1$,
\begin{equation}\label{e309a}
\upphi_2(a,b) = D_1 \uprho_1(a,b),
\end{equation}
and, for $n\ge 2$,
\begin{equation}\label{e309y}
\upphi_{\nu}(a,b) = \left\{ \begin{array}{lc}
\sum_{\nu'=1}^{\nu-1} D_{\nu-\nu'} \uprho_{\nu'}(a,b), & 2 \le \nu \le n, \\ \\
\sum_{\nu' = \nu-n}^{n} D_{\nu-\nu'} \uprho_{\nu'}(a,b), & n < \nu \le 2n.
\end{array} \right.
\end{equation}
Evidently, $\upphi_{\nu}(a,b) \equiv 0$ for $\nu < 2$.

Multiplying both sides of the equality in Eq.\,(\ref{e302}) by the denominator of the function on the LHS, expressing the resulting expression as a $(2n)$th-order polynomial of $\lambda$, one obtains
\begin{equation}\label{e310}
\uppi_{\nu}(a,b) = \left\{\begin{array}{lc}
\uprho_{1}(a,b) + D_{1}, & \nu = 1, \\ \\
\uprho_{\nu}(a,b) + D_{\nu} +\sum_{\nu'=1}^{\nu-1} D_{\nu-\nu'} \uprho_{\nu'}(a,b), & 2 \le \nu \le n, \\ \\
\sum_{\nu'=\nu-n}^{n}  D_{\nu-\nu'} \uprho_{\nu'}(a,b), & n < \nu \le 2n.
\end{array} \right.
\end{equation}
Note that, for $n=1$ one indeed has
\begin{align}\label{e310a}
\uppi_1(a,b) &= \uprho_1(a,b) + D_1, \nonumber\\
\uppi_2(a,b) &= D_1 \uprho_1(a,b).
\end{align}
With reference to the equalities in Eqs\,(\ref{e303}) and (\ref{e304}), the above results corresponding to $\nu=1$ and $\nu \in \{2,\dots, n\}$ are seen to coincide with those in Eq.\,(\ref{e248}), with the constants $\{F_{\nu}\| \nu\}$ in the latter equation determined from the recursive expression in Eq.\,(\ref{e247}). The equality in Eq.\,(\ref{e310}) makes explicit that for $\nu \in \{n+1,\dots,2n\}$ the function $\uppi_{\nu}(a,b)$ only takes account of the contributions of $\nu$th-order \textsl{disconnected} diagrams (see the opening remarks of this section, p.\,\pageref{s222}).

It is useful to denote the function $\uppi_{\nu}(a,b)$ as specified in Eq.\,(\ref{e310}) by $\uppi_{\nu}^{\X{(n)}}(a,b)$. Since however $\uppi_{\nu}^{\X{(n)}}(a,b)$ has \textsl{no} explicit dependence on $n$ for $1 \le\nu \le n$, in contrast to the cases corresponding to $n < \nu \le 2n$, one can suppress the superscript $(n)$ for $1 \le \nu \le n$, thus emphasising the universality of the relevant functions. \emph{When appropriate, in the following we shall follow this convention.}

For illustration, let us consider the case of $n=2$, for which one has
\begin{align}\label{e311}
\uppi_1(a,b) &= \uprho_1(a,b) + D_1,\;\; \uppi_2(a,b) = \uprho_2(a,b) + D_2 + D_1 \uprho_1(a,b),\nonumber\\
\uppi_3^{\X{(2)}}(a,b) &= D_2 \uprho_1(a,b) + D_1 \uprho_2(a,b),\;\; \uppi_4^{\X{(2)}}(a,b) = D_2 \uprho_2(a,b),
\end{align}
whereby
\begin{equation}\label{e312}
1 + \sum_{\nu=1}^{4} \lambda^{\nu} \uppi_{\nu}^{\X{(2)}}(a,b) = (1 + \lambda D_1 + \lambda^2 D_2) \big(1 + \lambda \uprho_1(a,b) + \lambda^2 \uprho_2(a,b)\big).
\end{equation}
One thus has
\begin{equation}\label{e313}
\frac{\mathpzc{G}^{[4/2]}(a,b)}{G_{\X{0}}(a,b)} \equiv
\frac{1+ \sum_{\nu=1}^{4} \lambda^{\nu} \uppi_{\nu}^{\X{(2)}}(a,b)}{1 + \sum_{\nu=1}^{2} \lambda^{\nu} D_{\nu}} = 1 + \sum_{\nu=1}^{2} \lambda^{\nu} \uprho_{\nu}(a,b).
\end{equation}
With reference to Eq.\,(\ref{e303}), this is equivalent to the perturbation series in Eq.\,(\ref{e245}) up to and including the second order in $\lambda$. Note that for the functions $\uppi_3^{\X{(3)}}(a,b) \equiv \uppi_{3}(a,b)$ and $\uppi_4^{\X{(3)}}(a,b)$ one has (\emph{cf.} Eq.\,(\ref{e311}))
\begin{align}\label{e314}
\uppi_3(a,b) &= \uprho_3(a,b) + D_3 + D_2 \uprho_1(a,b) + D_1 \uprho_2(a,b), \nonumber\\
\uppi_4^{\X{(3)}}(a,b) &= D_3 \uprho_1(a,b) + D_2 \uprho_2(a,b) + D_1 \uprho_3(a,b),
\end{align}
where the function $\uppi_4^{\X{(3)}}(a,b)$ is to be contrasted with
\begin{equation}\label{e315}
\uppi_4(a,b) = \uprho_4(a,b) + D_4 + D_3 \uprho_1(a,b) + D_2 \uprho_2(a,b) + D_1 \uprho_3(a,b).
\end{equation}

Lastly, with reference to Eqs\,(\ref{e301}), (\ref{e304}), and (\ref{e313}), one has
\begin{align}\label{e316}
\frac{G^{[2n/n]}(a,b)}{G_{\X{0}}(a,b)} &\equiv \frac{1+ \sum_{\nu=1}^{2n} \lambda^{\nu} \uppi_{\nu}(a,b)}{1 + \sum_{\nu=1}^{n} \lambda^{\nu} D_{\nu}}\nonumber\\
&= \frac{\mathpzc{G}^{[2n/n]}(a,b)}{G_{\X{0}}(a,b)}
+\frac{\sum_{\nu=n+1}^{2n} \lambda^{\nu} (\uppi_{\nu}(a,b)-\uppi_{\nu}^{\X{(n)}}(a,b))}{1 + \sum_{\nu=1}^{n} \lambda^{\nu} D_{\nu}},
\end{align}
where (\emph{cf.} Eqs\,(\ref{e245}), (\ref{e303}), and (\ref{e313}))
\begin{equation}\label{e317}
\frac{\mathpzc{G}^{[2n/n]}(a,b)}{G_{\X{0}}(a,b)} \doteq \frac{\sum_{\nu=1}^{2n} \lambda^{\nu} \uppi_{\nu}^{\X{(n)}}(a,b)}{1 + \sum_{\nu=1}^{n} \lambda^{\nu} D_{\nu}}
= 1 + \sum_{\nu=1}^{n} \lambda^{\nu} \uprho_{\nu}(a,b),
\end{equation}
and the last term on the RHS of Eq.\,(\ref{e316}) is non-vanishing and clearly of the order of $\lambda^{n+1}$. In the particular case of $n=2$, one has (\emph{cf.} Eqs\,(\ref{e311}), (\ref{e314}), and (\ref{e315}))
\begin{align}\label{e318}
\uppi_3(a,b) -\uppi_3^{\X{(2)}}(1,b) &= \uprho_3(a,b) +  D_3,\nonumber\\
\uppi_4(a,b) -\uppi_4^{\X{(2)}}(a,b) &= \uprho_4(a,b) + D_4 + D_3 \uprho_1(a,b) + D_1 \uprho_3(a,b).
\end{align}

The above observations make explicit that, for any \textsl{finite} value of $n$, use of the expression $G^{\X{[n/n]}}(a,b)$, Eq.\,(\ref{e301}), in place of the expression in Eq.\,(\ref{e245}) wherein the term $O(\lambda^{n+1})$ has been discarded, takes undue account of \textsl{disconnected} Green-function diagrams of order $n+1$ and higher; the contributions of these diagrams corresponding to no physical processes, they are not to be taken into account. To bypass this problem, one has to deal with the function $\mathcal{G}^{\X{[2n/n]}}(a,b)$, specified following Eq.\,(\ref{e301}) above.\footnote{The function $\mathcal{G}^{\protect\X{[2n/n]}}(a,b)$ coincides with that on the LHS of Eq.\,(\protect\ref{e302}) times $G_{\X{0}}(a,b)$.}

\refstepcounter{dummyX}
\subsection{Fredholm integral equations and perturbation series expansions
\texorpdfstring{\dash}{} a digression}
\phantomsection
\label{sac1}
In this section we establish a link between the exact perturbational expression for $G(a,b)$ in Eq.\,(\ref{e242}) and the exact solution of the Dyson equation, which, for the exact self-energy $\Sigma$ assumed as given, can be viewed as a Fredholm integral equation \cite{MF53,WW62,SG09} for $G(a,b)$. The observations of this section are of relevance to some fundamental analytic properties of the perturbation series expansions of $G(a,b)$ and $\Sigma(a,b)$, to be discussed in detail in Ref.\,\citen{BF16a}.\footnote{\emph{Note added to \textsf{arXiv:1912.00474v2}}: To keep the extent of Ref.\,\protect\citen{BF16a} within reasonable bounds, ultimately we decided to relegate the discussion as promised here to a separate publication. The  text of Ref.\,\protect\citen{BF16a} will be published at the same time as \textsf{arXiv:1912.00474v2}.}

Considering the self-energy $\Sigma$ as given, by introducing the function
\begin{equation}\label{e271}
K(a,b) \doteq \int \rd r\; G_{\X{0}}(a,r) \b{\Sigma}(r,b)
\end{equation}
where (\emph{cf.} Eq.\,(\ref{e260a}))
\begin{equation}\label{e272}
\b{\Sigma}(r,b) \doteq \frac{1}{\lambda}\hspace{0.6pt} \Sigma(r,b),
\end{equation}
and the functions
\begin{equation}\label{e273}
f_b(a) \doteq G_{\X{0}}(a,b),\;\; \phi_b(a) \doteq G(a,b),
\end{equation}
the Dyson equation
\begin{equation}\label{e254a}
G(a,b) = G_{\X{0}}(a,b) + \int \rd r\rd s\; G_{\X{0}}(a,r) \Sigma(r,s) G(s,b)
\end{equation}
can be equivalently written as\,\footnote{To keep the discussions of this section general, in the light of the equalities in Eq.\,(\protect\ref{e3}), here we identify $\Sigma_{\protect\X{01}}(a,b)$ with $\Sigma(a,b)$. Later in this section we identify $\Sigma(a,b)$ with $\Sigma_{\protect\X{00}}(a,b)$.}
\begin{equation}\label{e274}
\phi_{b}(a) = f_b(a) + \lambda \int \rd r\; K(a,r) \phi_{b}(r),
\end{equation}
which is the standard form for the Fredholm integral equation (of the second kind) [\S\,11.2, p.\,213, Ref.\,\citen{WW62}] [Ch.\,5, p.\,140, Ref.\,\citen{SG09}]. As in other similar cases considered elsewhere in this paper, here $\lambda$ serves mainly, but not entirely, as a book-keeping devise. In particular, by identifying $\Sigma$ with $\Sigma_{\X{01}}[v,G]$, one should note that $\b{\Sigma}_{\X{01}}[v,G]$ is a function of $\lambda$, depending explicitly (implicitly) on $\lambda$ through dependence of  $\Sigma_{\X{01}}[v,G]$ on $v$ ($G$), Eq.\,(\ref{e70}) below. According to the notation adopted above, $G_{\X{0}}(a,b)$ and $G(a,b)$ are functions of $a$ that parametrically depend on $b$.

The integral in Eq.\,(\ref{e274}) is the short-hand notation for one of the compound operations specified in Eqs\,(\ref{e11}), (\ref{e23}), (\ref{es16d}), (\ref{e39}), (\ref{e42}), and (\ref{e42b}). This deviation from the convention regarding integral equations is no bar to identifying the equation in Eq.\,(\ref{e274}) as an integral equation (for $\Sigma(a,b)$ considered as given).\footnote{To stay close to the considerations in Ref.\,\protect\citen{WW62}, one should deal with the $T> 0$ formalisms: within the Matsubara formalism the continuous integral with respect to $\tau_r$ as implied by $\int \rd r$,  Eqs\,(\protect\ref{e23}) and (\protect\ref{e42}), is over the \textsl{finite} interval $[0,\hbar\beta]$; within the TFD formalism, the $t_{\mathrm{i}}$ and $t_{\mathrm{f}}$ in Eq.\,(\ref{es4}) may be identified with \textsl{finite} real values.}

For $\Sigma(a,b)$ given, the solution of the integral equation in Eq.\,(\ref{e274}) has the form [p.\,214 in Ref.\,\citen{WW62}]
\begin{equation}\label{e275}
\phi_b(a) = f_b(a) + \frac{1}{\mathcal{D}(\lambda)} \int \rd r\; \mathcal{D}(a,r;\lambda) f_b(r),
\end{equation}
or, equivalently,\footnote{With $\mathcal{D}(a,b;\lambda)/\mathcal{D}(\lambda) \equiv \langle a\vert \h{\mathscr{D}}(\lambda)\vert b\rangle$, from the Dyson equation $\h{G} = \h{G}_{\X{0}} + \h{G} \h{\Sigma} \h{G}_{\X{0}}$ one infers that $\h{\mathscr{D}}(\lambda) = \h{G} \h{\Sigma} \Leftrightarrow \h{\Sigma} = \h{G}_{\X{0}}^{-1} \h{\mathscr{D}}(\lambda) (\h{1} + \h{\mathscr{D}}(\lambda))^{-1}$, which amounts to a self-consistent equation for $\h{\Sigma}$.}
\begin{equation}\label{e276}
G(a,b) = G_{\X{0}}(a,b) + \frac{1}{\mathcal{D}(\lambda)} \int \rd r\; \mathcal{D}(a,r;\lambda)\hspace{0.6pt} G_{\X{0}}(r,b),
\end{equation}
where $\mathcal{D}(\lambda)$ and $\mathcal{D}(a,b;\lambda)$ are the $\mathpzc{n} =\infty$ limits of respectively the Fredholm determinant $\mathcal{D}_{\hspace{-0.6pt}\mathpzc{n}}(\lambda)$ and its corresponding $(a,b)$ cofactor $\mathcal{D}_{\hspace{-0.6pt}\mathpzc{n}}(a,b;\lambda)$ (\emph{cf.} Eqs\,(\ref{e261}), (\ref{e260}), and (\ref{e254x})). One has [p.\,214 in Ref.\,\citen{WW62}] [Eqs\,(9.161) and (9.163), p.\,341, in Ref.\,\citen{SG09}]
\begin{equation}\label{e276b}
\mathcal{D}(\lambda) = 1 + \sum_{\nu=1}^{\infty} \frac{(-\lambda)^{\nu}}{\nu!} \int \prod_{j=1}^{\nu} \rd j\; B_{\nu}(1,\dots,\nu),
\end{equation}
and
\begin{equation}\label{e276c}
\mathcal{D}(a,b;\lambda) = \lambda K(a,b) + \lambda \sum_{\nu=1}^{\infty} \frac{(-\lambda)^{\nu}}{\nu!} \int \prod_{j=1}^{\nu} \rd j\; B_{\nu+1}^{\textsc{b}}(a,b;1,\dots,\nu),
\end{equation}
where (\emph{cf.} Eq.\,(\ref{e260}))
\begin{equation}\label{e276d}
B_{\nu}(1,\dots,\nu) \doteq
\begin{vmatrix}
K(1,1) & K(1,2) & \dots & K(1,\nu) \\
K(2,1) & K(2,2) & \dots & K(2,\nu) \\
\vdots & \vdots & \ddots & \vdots \\
K(\nu,1) & K(\nu,2) & \dots & K(\nu,\nu)
\end{vmatrix}_{-},
\end{equation}
and (\emph{cf.} Eq.\,(\ref{e258}))
\begin{equation}\label{e276e}
B_{\nu+1}^{\textsc{b}}(a,b;1,\dots,\nu) \doteq
\begin{vmatrix}
K(a,b) & K(a,1)  & K(a,2)  & \dots & K(a,\nu) \\
K(1,b) & K(1,1) & K(1,2) & \dots & K(1,\nu) \\
K(2,b) & K(2,1) & K(2,2) & \dots & K(2,\nu) \\
\vdots & \vdots & \vdots & \ddots & \vdots \\
K(\nu,b) & K(\nu,1) & K(\nu,2) & \dots & K(\nu,\nu)
\end{vmatrix}_{-}.
\end{equation}

From the above expressions for $\mathcal{D}(\lambda)$ and $\mathcal{D}(a,b;\lambda)$, one notices a close similarity between the expression in Eq.\,(\ref{e276}) and that in Eq.\,(\ref{e242}). This is in particular the case for systems of fermions for which the functions in the expression on the RHS of Eq.\,(\ref{e242}) are described in terms of determinants. In contrast to $\{N_{\nu}(a,b)\| \nu\}$ and $\{D_{\nu}\| \nu\}$, which are functionals of $(v,G_{\X{0}})$, the terms in the power series expansions of $\mathcal{D}(\lambda)$ and $\mathcal{D}(a,b;\lambda)$ are functionals of $(\Sigma,G_{\X{0}})$, as apparent from the defining expression for $K(a,b)$ in Eq.\,(\ref{e271}). To deduce the expression in Eq.\,(\ref{e242}) from that in Eq.\,(\ref{e276}), it is therefore required to identify the $\Sigma$ in Eq.\,(\ref{e272}) with $\Sigma_{\X{00}}[v,G_{\X{0}}]$, Eq.\,(\ref{e3}), and employ the perturbation series expansion for the latter functional as presented in Eq.\,(\ref{e60d}) below. Following this, the series expansions for $\mathcal{D}(\lambda)$ and $\mathcal{D}(a,b;\lambda)$ in Eqs\,(\ref{e276b}) and (\ref{e276c}) are to be cast into power series of $\lambda$, with the corresponding coefficients independent of $\lambda$. This task is simplified by making use of the general properties of determinants, in particular those under \textbf{g} on pages 9 and 10 of Ref.\,\citen{VD99}. In this way, the power series expansions in the numerator and the denominator of the expression on the RHS of Eq.\,(\ref{e242}) are recovered. We note that it is through the above-mentioned perturbation series expansion of $\Sigma_{\X{00}}[v,G_{\X{0}}]$ in terms of $(v,G_{\X{0}})$ that for bosons the series expansions of $\mathcal{D}(\lambda)$ and $\mathcal{D}(a,b;\lambda)$ in terms of determinants transform into expressions in terms of permanents. This can be surmised by recalling that each particle loop\,\footnote{The direct link between particle loops and \textsl{cycles} of permutations is highlighted in appendices \protect\ref{sac} and \protect\ref{sad}.}  in a self-energy diagram (as well as other Feynman diagrams) contributes a multiplicative factor $\upzeta$ to the analytic expression associated with that diagram, where $\upzeta = \pm 1$ in the case of bosons / fermions [p.\,81 in Ref.\,\citen{NO98}].\footnote{For illustration, from Eqs\,(\protect\ref{e271}), (\ref{e272}), (\protect\ref{e276b}), and Eq.\,(\protect\ref{e60d}) below, to leading order one has $\mathcal{D}(\lambda) \sim 1 - \int \protect\rd 1 \protect\rd 2\; G_{\X{0}}(1,2) \Sigma_{\protect\X{00}}^{\protect\X{(1)}}(2,1)$, which for \textsl{fermions} amounts to the analytic expression associated with the diagrams displayed in Fig.\,9.3, p.\,95, of Ref.\,\protect\citen{FW03}. The minus sign in the above expression (to be contrasted with the plus signs in the just-indicated figure) accounts for the multiplicative factor $\upzeta = -1$ associated with the fermion loops that the expression $\int \protect\rd 1 \protect\rd 2\; G_{\X{0}}(1,2) \Sigma_{\protect\X{00}}^{\protect\X{(1)}}(2,1)$ brings about.}

We note in passing that [pp.\,217 and 220 in Ref.\,\citen{WW62}]
\begin{equation}\label{e276a}
\int \rd r\; \mathcal{D}(r,r;\lambda) = -\lambda\hspace{0.6pt} \frac{\partial \mathcal{D}(\lambda)}{\partial\lambda},
\end{equation}
where by the \textsl{partial} derivative on the RHS\,\footnote{To be contrasted with the \textsl{total} derivative in the relevant expressions in Ref.\,\protect\citen{WW62}.} we emphasise that the dependence of $\mathcal{D}(\lambda)$ on $\lambda$ as arising from the dependence of $K(a,b)$ on $\lambda$, through that of $\b{\Sigma}(r,b)$ (to be distinguished from $\Sigma(r,b)$), Eq.\,(\ref{e272}), is to be neglected.

The Fredholm integral equation in Eq.\,(\ref{e274}) leads in a natural way to the notion of Volterra's \textsl{reciprocal functions} [\S\,11.22, p.\,218, in Ref.\,\citen{WW62}]. The functions $K(a,b)$ and $k(a,b;\lambda)$ are defined as being \textsl{reciprocal} provided that
\begin{equation}\label{e277}
K(a,b) + k(a,b;\lambda) = \lambda \int \rd r\; k(a,r;\lambda) K(r,b).
\end{equation}
One thus has [p.\,218 in Ref.\,\citen{WW62}]
\begin{equation}\label{e278}
k(a,b;\lambda) = -\frac{1}{\lambda\mathcal{D}(\lambda)}\hspace{0.6pt} \mathcal{D}(a,b;\lambda),
\end{equation}
so that, with reference to Eq.\,(\ref{e274}), [p.\,219 in Ref.\,\citen{WW62}]
\begin{equation}\label{e279}
f_b(a) = \phi_b(a) + \lambda \int \rd r\; k(a,r;\lambda) f_b(r),
\end{equation}
or, equivalently,
\begin{equation}\label{e280}
G_{\X{0}}(a,b) = G(a,b) + \lambda \int \rd r\; k(a,r;\lambda) G_{\X{0}}(r,b).
\end{equation}
For the function $k(a,r;\lambda)$ given, one in analogy with the result in Eq.\,(\ref{e275}) has\,\footnote{The symbol $\mathpzc{D}$ is not to be confused with the symbol $\mathcal{D}$.}
\begin{equation}\label{e281}
f_b(a) = \phi_b(a) + \frac{1}{\mathpzc{D}(\lambda)} \int \rd r\; \mathpzc{D}(a,r;\lambda)\hspace{0.6pt} \phi_b(r),
\end{equation}
or, equivalently,
\begin{equation}\label{e282}
G_{\X{0}}(a,b) = G(a,b) + \frac{1}{\mathpzc{D}(\lambda)} \int \rd r\; \mathpzc{D}(a,r;\lambda)\hspace{0.6pt} G(r,b),
\end{equation}
where the functions $\mathpzc{D}(\lambda)$ and $\mathpzc{D}(a,r;\lambda)$ have the same functional form as respectively $\mathcal{D}(\lambda)$ and $\mathcal{D}(a,b;\lambda)$, Eqs\,(\ref{e276b}) and (\ref{e276c}), with the function $k(r,s;\lambda)$ taking the place of the function $K(r,s)$ in the expressions in Eqs\,(\ref{e276d}) and (\ref{e276e}).

We shall not go into further details regarding the expression in Eq.\,(\ref{e282}) and suffice to mention that this expression is related to the series expansion in Eq.\,(\ref{e72}) below.

Similar considerations as discussed above regarding the functions $G$ and $G_{\X{0}}$ apply to $W$ and $v$, respectively the dynamic screened and the bare interaction potential, \S\,\ref{sec3}. This follows from the similarity between the Dyson equation with which the equation in Eq.\,(\ref{e274}) is equivalent, and the Dyson-type equation in Eq.\,(\ref{e78}) below. In particular, the counterpart of the equality in Eq.\,(\ref{e281}), or equivalently Eq.\,(\ref{e282}), expressing $v$ in terms of $W$ (assuming the polarisation function $P$ as given), is related to the expansion in Eq.\,(\ref{e84}) below.

\refstepcounter{dummyX}
\subsection{The self-energy operator
\texorpdfstring{$\h{\Sigma}_{\protect\X{00}}[v,G_{\protect\X{0}}]$}{}}
\phantomsection
\label{s23}
Knowledge of the perturbation series expansion for $G$ in terms of $v$ and $G_{\X{0}}$ is sufficient for directly determining the perturbation series expansion of the self-energy operator $\h{\Sigma}_{\X{00}}[v,G_{\X{0}}]$ via the Dyson equation \cite{FW03}. To this end, let $\h{G}$, $\h{G}_{\X{0}}$, and $\h{\Sigma}$ denote the single-particle operators corresponding to respectively the interacting Green function, the non-interacting Green function, and the self-energy in the single-particle Hilbert space of the system at hand.\refstepcounter{dummy}\label{ThusForInstance}\footnote{Thus, for instance, $G(a,b) \equiv \langle a\vert\protect\h{G}\vert b\rangle$, where $\vert a\rangle$ and $\vert b\rangle$ are normalised single-particle states in the underlying single-particle Hilbert space. The same applies to, for instance, the operator $\protect\h{G}^{\protect\X{(\nu)}}$ in Eq.\,(\protect\ref{e60a}) below. For the completeness relation in this Hilbert space one has $\int \mathrm{d} j\, \vert j\rangle \langle j\vert = \protect\h{I}$, where $\protect\h{I}$ denotes the identity operator in this space, and $\int \mathrm{d} j$ coincides with one of those in Eqs\,(\protect\ref{e11}), (\protect\ref{e23}), (\protect\ref{es16d}), (\protect\ref{e39}), (\protect\ref{e42}), and (\protect\ref{e42b}), depending on the system and the framework under consideration. This definition of $\int \mathrm{d} j$ applies to all integrals encountered in \S\,\protect\ref{sac1}. As regards $\protect\h{I}$, for instance in the case of $T=0$, and with reference to Eq.\,(\ref{e8}), one has $\langle j\vert \protect\h{I}\vert j'\rangle \equiv  \langle j\vert j'\rangle= \delta^d(\bm{r}_j-\bm{r}_{j'}) \delta(t_j-t_{j'}) \delta_{\sigma_j,\sigma_{j'}}$. In this connection, disregarding $\delta(t_j-t_{j'})$ (and therefore the time) for a moment, the latter equality would amount to the normalisation condition required of the simultaneous eigenstates of the single particle operators $\protect\h{\bm{r}}$ and $\protect\h{\sigma}^{\X{z}}$ (for the matrix representations of $\protect\h{\sigma}^{\X{z}}$, see Eq.\,(\protect\ref{e11b})). Since it is \textsl{not} possible to define a single-particle time operator $\protect\h{t}$, conjugate to the energy operator (for instance the single-particle Hamiltonian $\protect\h{h}$ to which the $\protect\wh{H}_0$ in Eq.\,(\protect\ref{e5}) corresponds), without extending the underlying (single-particle) Hilbert space \protect\cite{RGN80,MB83}, it is ruled out to associate $\vert j\rangle$ with a simultaneous eigenstate of $\protect\h{\bm{r}}$, $\protect\h{\sigma}^{\X{z}}$, and $\protect\h{t}$ (with eigenvalues $\bm{r}_j$, $\sigma_j$, and $t_j$). Here we are therefore relying on the completeness relations of the time- and energy-Fourier transforms in the space of the single-variable functions considered here. This enables us to associate a function of $t$, say $f(t)$, with the abstract ket vectors $\vert t\rangle$ and $\vert f\rangle$ according to the relation $f(t) = \langle t \vert f\rangle$, and its time-Fourier transform $\protect\t{f}(\varepsilon)$ with the additional abstract ket vector $\vert\varepsilon\rangle$, according to the relation $\protect\t{f}(\varepsilon) = \langle\varepsilon \vert f\rangle$. With $\langle t\vert\varepsilon\rangle = \protect\e^{-\protect\ii\hspace{0.4pt} \varepsilon t/\hbar}/\sqrt{2\pi\hbar}$, one clearly recovers the above-mentioned completeness relations $\langle t\vert t'\rangle = \delta(t-t')$ and $\langle \varepsilon \vert \varepsilon'\rangle = \delta(\varepsilon-\varepsilon')$. \label{noteb}} With $\h{G}^{-1}$ and $\h{G}_{\X{0}}^{-1}$ denoting the inverses of the relevant operators, following the Dyson equation \cite{FW03} one has
\begin{equation}\label{e60}
\h{\Sigma} = \h{G}_{\X{0}}^{-1} - \h{G}^{-1}.
\end{equation}
On the basis of the series expansion in Eq.\,(\ref{e245}), we write
\begin{equation}\label{e60a}
\h{G} = \h{G}_{\X{0}} + \sum_{\nu=1}^{n} \lambda^{\nu} \h{G}^{\X{(\nu)}} + O(\lambda^{n+1}),
\end{equation}
where $\h{G}^{\X{(\nu)}}$ denotes the single-particle operator associated with the function $G^{\X{(\nu)}}(a,b)$ in Eq.\,(\ref{e245}). From the equality in Eq.\,(\ref{e60a}), one obtains (\emph{cf.} Eq.\,(\ref{e243}))
\begin{align}\label{e60b}
\h{G}^{-1} &= \big(\h{I} + \h{G}_{\X{0}}^{-1} \sum_{\nu=1}^{n} \lambda^{\nu}
\h{G}^{\X{(\nu)}}\big)^{-1} \h{G}_{\X{0}}^{-1} + O(\lambda^{n+1}) \nonumber\\
&\equiv \h{G}_{\X{0}}^{-1} - \sum_{\nu=1}^{n} \lambda^{\nu} \h{\Sigma}_{\X{00}}^{\X{(\nu)}}[\mathsf{v},G_{\X{0}}] + O(\lambda^{n+1}),
\end{align}
where $\h{I}$ denotes the identity operator in the single-particle Hilbert space at hand, and (\emph{cf.} Eq.\,(\ref{e247}))
\begin{align}\label{e60c}
\h{\Sigma}_{\X{00}}^{\X{(1)}}[\mathsf{v},G_{\X{0}}] &= \h{G}_{\X{0}}^{-1} \h{G}^{\X{(1)}} \h{G}_{\X{0}}^{-1}, \nonumber\\
\h{\Sigma}_{\X{00}}^{\X{(\nu)}}[\mathsf{v},G_{\X{0}}] &= \h{G}_{\X{0}}^{-1} \h{G}^{\X{(\nu)}} \h{G}_{\X{0}}^{-1} - \h{G}_{\X{0}}^{-1} \sum_{\nu'=1}^{\nu-1} \h{G}^{\X{(\nu-\nu')}} \h{\Sigma}_{\X{00}}^{\X{(\nu')}}[\mathsf{v},G_{\X{0}}],\;\; \nu \ge 2.
\end{align}
From the equality in Eq.\,(\ref{e60}) and the last equality in Eq.\,(\ref{e60c}), one clearly observes that
\begin{equation}\label{e60d}
\h{\Sigma}_{\X{00}}[v,G_{\X{0}}] \equiv \sum_{\nu=1}^{n} \lambda^{\nu} \h{\Sigma}_{\X{00}}^{\X{(\nu)}}[\mathsf{v},G_{\X{0}}] + O(\lambda^{n+1})
\end{equation}
amounts to the $n$th-order perturbation series expansion of the self-energy operator $\h{\Sigma}$ in terms of $(v,G_{\X{0}})$. We note that the pre- and post-multiplications by $\h{G}_{\X{0}}^{-1}$ of the operator $\h{G}^{\X{(\nu)}}$, $\forall\nu \in\mathds{N}$, in the above expressions reflect the process of amputating the two \textsl{external} Green-function lines in obtaining the diagrammatic series expansion of the self-energy from that of the interacting Green function.

Introducing the single-particle operator\refstepcounter{dummy}\label{FNStar}\,\footnote{This notation does not accord with that in Ref.\,\protect\citen{FW03}, where $\star$ (not to be confused with $*$, which conventionally denotes complex conjugation) marks the \textsl{proper} self-energy.}
\begin{equation}\label{e60e}
\h{\Sigma}_{\X{00}}^{\star\hspace{0.5pt}\X{(\nu)}}[\mathsf{v},G_{\X{0}}] \doteq \h{G}_{\X{0}}^{-1}\hspace{0.6pt} \h{G}^{\X{(\nu)}}[\mathsf{v},G_{\X{0}}]\hspace{0.6pt} \h{G}_{\X{0}}^{-1},
\end{equation}
the expressions in Eq.\,(\ref{e60c}) can be written as
\begin{align}\label{e60f}
\h{\Sigma}_{\X{00}}^{\X{(1)}}[\mathsf{v},G_{\X{0}}] &= \h{\Sigma}_{\X{00}}^{\star \X{(1)}}[\mathsf{v},G_{\X{0}}], \nonumber\\
\h{\Sigma}_{\X{00}}^{\X{(\nu)}}[\mathsf{v},G_{\X{0}}] &= \h{\Sigma}_{\X{00}}^{\star (\nu)}[\mathsf{v},G_{\X{0}}] -\sum_{\nu'=1}^{\nu-1} \h{\Sigma}_{\X{00}}^{\star (\nu-\nu')}[\mathsf{v},G_{\X{0}}] \h{G}_{\X{0}}
\h{\Sigma}_{\X{00}}^{\X{(\nu')}}[\mathsf{v},G_{\X{0}}],\;\; \nu \ge 2.\;\;\;\;\;\;\;
\end{align}
Analogously to the second equality in Eq.\,(\ref{e248a}) where the second term on the RHS removes the contributions of the disconnected Green-function diagrams from $G^{\X{(\nu)}}(a,b)$, \S\,\ref{s221}, the second term on the RHS of Eq.\,(\ref{e60f}) removes the contributions of the \textsl{improper} self-energy diagrams from $\h{\Sigma}_{\X{00}}^{\star\X{(\nu)}}$ (compare with Fig.\,9.13, p.\,106, in Ref.\,\citen{FW03}). For illustration, for $\nu=2$, from the second equality in Eq.\,(\ref{e60f}), one has
\begin{equation}\label{e60g}
\h{\Sigma}_{\X{00}}^{\X{(2)}}[\mathsf{v},G_{\X{0}}] = \h{\Sigma}_{\X{00}}^{\star \X{(2)}}[\mathsf{v},G_{\X{0}}] -\h{\Sigma}_{\X{00}}^{\X{(1)}}[\mathsf{v},G_{\X{0}}]\hspace{0.7pt} \h{G}_{\X{0}}\hspace{0.7pt}\h{\Sigma}_{\X{00}}^{\X{(1)}}[\mathsf{v},G_{\X{0}}],
\end{equation}
where the second term on the RHS is clearly a second-order \textsl{improper} self-energy contribution. The first term on the RHS of Eq.\,(\ref{e60g}) is described in terms of ten connected self-energy diagrams,\footnote{As is evident from Eq.\,(\protect\ref{e60e}), the total number of $\nu$th-order diagrams describing $\protect\h{\Sigma}_{\X{00}}^{\star \X{(\nu)}}[\mathsf{v},G_{\X{0}}]$ is \textsl{exactly} equal to the number of \textsl{connected} diagrams describing $\protect\h{G}^{\protect\X{(\nu)}}[\mathsf{v},G_{\protect\X{0}}]$. The number of the latter diagrams corresponding to $\nu \equiv k/2 \in\{ 1,2,\dots,10\}$ is presented in Table I of Ref.\,\protect\citen{CLP78}. This number is calculated, for in principle an arbitrary value of $\nu \in \mathds{N}$, by the program \texttt{Gnu}, p.\,\protect\pageref{Gnu}, in appendix \protect\ref{sac}.} of which four are improper. The latter are indeed fully removed by the four improper self-energy diagrams associated with the second term on the RHS of Eq.\,(\ref{e60g}), thus correctly resulting in six \textsl{proper} (\emph{i.e.} 1PI) diagrams in terms of which $\h{\Sigma}_{\X{00}}^{\X{(2)}}[\mathsf{v},G_{\X{0}}]$ is described.\footnote{See Fig.\,\protect\ref{f10}, p.\,\protect\pageref{CompleteSet}.}

We note that for uniform GSs or thermal ensemble of states, the formalism of this section greatly simplifies on employing the energy-momentum representation of the single-particle operators encountered above. \emph{Similarly as regards the formalisms to be introduced in the following sections.}

\refstepcounter{dummyX}
\subsection{The self-energy operator \texorpdfstring{$\h{\Sigma}_{\protect\X{01}}[v,G]$}{}}
\phantomsection
\label{s24}
For constructing the diagram-free perturbation series expansion of the self-energy operator that formally coincides with the diagrammatic series expansion of this operator in terms of $G$-skeleton self-energy diagrams and $(v,G)$, we first note that this perturbation series expansion is included in that of $\h{\Sigma}_{\X{00}}[v,G]$ (note the $G$ taking the place of $G_{\X{0}}$), Eq.\,(\ref{e4}). The single-particle operator $\h{\Sigma}_{\X{01}}[v,G]$ can therefore be obtained from $\h{\Sigma}_{\X{00}}[v,G]$ by subtracting the contributions of non-skeleton self-energy diagrams determined in terms of $(v,G)$. To do so, we take our cue from the observations in \S\,\ref{s221}, where the second equality in Eq.\,(\ref{e248}), or that in Eq.\,(\ref{e248a}), systematically (that is, order-by-order) removes the contributions of the \textsl{disconnected} Green-function diagrams from the set of all Green-function diagrams. This suggests the possibility of constructing a formalism whereby the contributions of non-skeleton self-energy diagrams are removed from those of all proper (\emph{i.e.} 1PI) self-energy diagrams determined in terms of $(v,G)$. Below we construct such formalism.

We begin with the perturbational expression for the self-energy operator,\footnote{For a diagrammatic determination of $\protect\h{\Sigma}_{\protect\X{01}}^{\protect\X{(\nu)}}[v,G]$, see appendix \protect\ref{sad}. The relevant diagrams and their multiplicity are determined by the program \texttt{Snu}, p.\,\protect\pageref{Snu}.}
\begin{equation}\label{e70}
\h{\Sigma}_{\X{01}}[v,G] = \sum_{\nu=1}^{n} \lambda^{\nu}
\h{\Sigma}_{\X{01}}^{\X{(\nu)}}[\mathsf{v},G] + O(\lambda^{n+1}),
\end{equation}
where $\lambda^{\nu} \h{\Sigma}_{\X{01}}^{\X{(\nu)}}[\mathsf{v},G] \equiv  \h{\Sigma}_{\X{01}}^{\X{(\nu)}}[v,G]$ (note the $\mathsf{v}$ and $v$, Eq.\,(\ref{e260a})) denotes the total contribution of the $\nu$th-order $G$-skeleton (\emph{i.e.} 2PI) self-energy diagrams in terms of $v$ and $G$ to $\h{\Sigma}_{\X{01}}[v,G]$. The factor $\lambda^{\nu}$ accounts for the \textsl{explicit} dependence of the $\nu$th-order self-energy diagrams on the coupling constant $\lambda$ of the interaction potential $v$, Eq.\,(\ref{e260a}). Making use of the expression on the RHS of Eq.\,(\ref{e70}), from the Dyson equation \cite{FW03} one obtains the following expression for the non-interacting Green function:
\begin{equation}\label{e71}
\h{G}_{\X{0}} = \h{G} \big(\h{I} + \sum_{\nu=1}^{n} \lambda^{\nu}
\h{\Sigma}_{\X{01}}^{\X{(\nu)}}[\mathsf{v},G] \hspace{0.7pt} \h{G}\big)^{-1} + O(\lambda^{n+1}).
\end{equation}
On expressing the exact equality in Eq.\,(\ref{e71}) as\,\footnote{\emph{Cf.} Eq.\,(\protect\ref{e280}).}
\begin{equation}\label{e72}
\h{G}_{\X{0}} = \h{G} - \sum_{\nu=1}^{n} \lambda^{\nu}
\h{\mathcal{G}}_{\X{01}}^{\X{(\nu)}}[\mathsf{v},G] + O(\lambda^{n+1}),
\end{equation}
along the same lines as arriving at the recursive expression in Eq.\,(\ref{e247}) from the equality in Eq.\,(\ref{e243}), one arrives at the following recursive expression for the elements of the ordered sequence $\{\h{\mathcal{G}}_{\X{01}}^{\X{(\nu)}}[\mathsf{v},G] \| \nu\in \mathds{N}\}$:
\begin{align}\label{e73}
\h{\mathcal{G}}_{\X{01}}^{\X{(1)}}[\mathsf{v},G] &= \h{G} \h{\Sigma}_{\X{01}}^{\X{(1)}}[\mathsf{v},G]\hspace{0.7pt} \h{G},\nonumber\\
\h{\mathcal{G}}_{\X{01}}^{\X{(\nu)}}[\mathsf{v},G] &= \h{G} \h{\Sigma}_{\X{01}}^{\X{(\nu)}}[\mathsf{v},G]\hspace{0.7pt} \h{G} - \h{G} \sum_{\nu'=1}^{\nu-1} \h{\Sigma}_{\X{01}}^{\X{(\nu-\nu')}}[\mathsf{v},G] \hspace{0.7pt} \h{\mathcal{G}}_{\X{01}}^{\X{(\nu')}}[\mathsf{v},G],\;\; \nu \ge 2. \hspace{0.4cm}
\end{align}
We should emphasise that the expression in Eq.\,(\ref{e245}) is \textsl{not} to be identified as the direct-space representation of the expression in Eq.\,(\ref{e72}). This fact becomes evident by realising that in contrast to $G^{\X{(\nu)}}(a,b)$, which has \textsl{no} implicit dependence on $\lambda$, $\h{\mathcal{G}}_{\X{01}}^{\X{(\nu)}}[\mathsf{v},G]$ (or $\mathcal{G}_{\X{01}}^{\X{(\nu)}}(a,b;[\mathsf{v},G]) \equiv \langle a\vert \h{\mathcal{G}}_{\X{01}}^{\X{(\nu)}}[\mathsf{v},G]\vert b\rangle$) is an implicit function of $\lambda$, this as arising from the implicit dependence of $G$ on $\lambda$. The two functions can also \textsl{not} be identified when $G^{\X{(\nu)}}(a,b)$ is determined in terms of $G$, instead of $G_{\X{0}}$ (see later, p.\,\pageref{ForAnArbitrary}).

We now posit that in the direct-space representation for arbitrary $\nu \in \mathds{N}$ the operator $\h{\Sigma}_{\X{01}}^{\X{(\nu)}}[\mathsf{v},G]$, as encountered in the perturbation series expansion in Eq.\,(\ref{e70}), can be recursively determined from the following equalities (\S\,\ref{sec1}):\,\refstepcounter{dummy}\label{NoteThat}\footnote{Note that $\delta G(1,2)/\delta G(3,4) = \delta(1-3) \delta(2-4)$, where, for $j$ denoting $\bm{r}_j,t_j,\sigma_j$, Eq.\,(\protect\ref{e8}), $\delta(i-j)$ stands for $\delta^d(\bm{r}_i-\bm{r}_j) \delta(t_i-t_j) \delta_{\sigma_i,\sigma_j}$, and for $j$ denoting $\bm{r}_j,t_j,\sigma_j,\mu_j$, Eq.\,(\protect\ref{es16b}), $\delta(i-j)$ stands for $\delta^d(\bm{r}_i-\bm{r}_j) \delta(t_i,t_j) \delta_{\sigma_i,\sigma_j} \delta_{\mu_i,\mu_j}$. One similarly has $\delta \mathscr{G}(1,2)/\delta \mathscr{G}(3,4) = \delta(1-3) \delta(2-4)$, where, since $j$ denotes $\bm{r}_j,\tau_j,\sigma_j$, Eq.\,(\protect\ref{e20}), $\delta(i-j)$ stands for $\delta^d(\bm{r}_i-\bm{r}_j) \delta_{\hbar\beta}^{\protect\X{(\varsigma)}}(\tau_i-\tau_j) \delta_{\sigma_i,\sigma_j}$, where $\delta_{\tau_0}^{\protect\X{(\varsigma)}}(\tau) \doteq \sum_{k=-\infty}^{\infty} \varsigma^k \delta(\tau + k \tau_0)$, in which $\varsigma = \pm 1$ for bosons / fermions (see Eq.\,(\protect\ref{e23}) and note that in thermal equilibrium $\mathscr{G}(i,j)$ is periodic / antiperiodic in $\tau_{ij} \equiv \tau_i - \tau_j$, with period $\hbar\beta$, for bosons / fermions, Eq.\,(\ref{e26a})). Similar considerations apply in the case of the Hubbard Hamiltonian, with the exception that the above $\delta^d(\bm{r}_i-\bm{r}_j)$ are to be replaced by $\delta_{l_i,l_j}$, Eqs\,(\protect\ref{e37}), (\protect\ref{e20a}), and (\protect\ref{e42d}). \label{notef}}
\begin{align}\label{e74}
\Sigma_{\X{01}}^{\X{(1)}}(a,b;[\mathsf{v},G]) &= \Sigma_{\X{00}}^{\X{(1)}}(a,b;[\mathsf{v},G]),\nonumber\\
\Sigma_{\X{01}}^{\X{(\nu)}}(a,b;[\mathsf{v},G]) &= \Sigma_{\X{00}}^{\X{(\nu)}}(a,b;[\mathsf{v},G]) \nonumber\\
&\hspace{-1.0cm} -\sum_{\nu'=1}^{\nu-1} \int \rd 1\rd 2\; \frac{\delta \Sigma_{\X{00}}^{\X{(\nu-\nu')}}(a,b;[\mathsf{v},G])}{\delta G(1,2)}\hspace{1.0pt} \mathcal{G}_{\X{01}}^{\X{(\nu')}}(1,2;[\mathsf{v},G]),\;\; \nu \ge 2,\hspace{1.2cm}
\end{align}
where $\int \rd j$ stands for the same set of mathematical operations as discussed in \S\,\ref{s21}.\footnote{See also footnote \raisebox{-1.0ex}{\normalsize{\protect\footref{noteb}}} on p.\,\protect\pageref{ThusForInstance}.} The second expression in Eq.\,(\ref{e74}) is simplified by using the identification in Eq.\,(\ref{e8}) in the $T=0$ case, and those in Eqs\,(\ref{e20}) and (\ref{es16b}) in the $T > 0$ case,\footnote{In the case of the Hubbard Hamiltonian, using the identifications in respectively Eqs\,(\protect\ref{e37}), (\protect\ref{e20a}), and (\protect\ref{e42d}).} and the fact that functions in this expression are diagonal in the spin space.\footnote{The same arguments as in pp.\,\protect\pageref{AsIsCommon} and \protect\pageref{WithPsi} apply here.} The expression in Eq.\,(\ref{e74}) is further simplified in the case of uniform GSs and uniform ensemble of states, where functions of $(i,j)$ depend on $\bm{r}_i - \bm{r}_j$.\footnote{For the Hubbard Hamiltonian, they depend on $\bm{R}_i - \bm{R}_j$, where $\{\bm{R}_i \| i\}$ are the underlying lattice vectors. Only for a uniform lattice in $d=1$ can the lattice points be numbered in such a way that in uniform GSs and uniform ensemble of states functions of $(i,j)$ depend on $i-j$.}

Note that the self-energy functions on the RHSs of the equalities in Eq.\,(\ref{e74}) consist of the elements of the sequence $\{\Sigma_{\X{00}}^{\X{(\nu)}}(a,b;[\mathsf{v},G])\| \nu\}$. The dependence of $\h{\Sigma}_{\X{01}}^{\X{(\nu)}}[\mathsf{v},G]$ on the elements of $\{\h{\Sigma}_{\X{01}}^{\X{(\nu')}}[\mathsf{v},G] \| \nu' =1,\dots,\nu-1\}$ is implicit, through that of the elements of the sequence $\{ \h{\mathcal{G}}_{\X{01}}^{\X{(\nu')}}[\mathsf{v},G]\| \nu' =1,\dots,\nu-1\}$ on these, Eq.\,(\ref{e73}). Since the summation variable $\nu'$ on the RHS of Eq.\,(\ref{e74}) takes the values from the set $\{1,\dots,\nu-1\}$, from Eq.\,(\ref{e73}) one observes that calculation of the required elements of the set of operators $\{\h{\mathcal{G}}_{\X{01}}^{\X{(\nu)}}\| \nu\}$ is demanding of the calculation of $\h{\Sigma}_{\X{01}}^{\X{(\nu')}}[\mathsf{v},G]$ for $\nu' = 1,\dots, \nu-1$. Hence, indeed the equalities in Eqs\,(\ref{e73}) and (\ref{e74}) constitute a \textsl{recursive} formalism for the calculation of the elements of the sequence $\{\h{\Sigma}_{\X{01}}^{\X{(\nu)}}[\mathsf{v},G]\| \nu\}$.

The validity of the formalism introduced above is established as follows. The first of the two equalities in Eq.\,(\ref{e74}) is trivially valid on account of the fact that at first order proper (or 1PI) self-energy diagrams are $G$-skeleton (or 2PI). Regarding the second equality in Eq.\,(\ref{e74}), the first term on the RHS clearly accounts for the contributions of all $\nu$th-order proper self-energy diagrams in terms of $(\mathsf{v},G)$. Hence, the validity of the equality at hand rests on the second term accounting for the contributions of \textsl{all} $\nu$th-order \textsl{non-skeleton} self-energy diagrams. That this is indeed the case can be ascertained on the basis of the following observations:
\begin{quote}
\item[(i)] since $(\nu-\nu') + \nu' = \nu$, the product of $\h{\Sigma}_{\X{00}}^{\X{(\nu-\nu')}}[\mathsf{v},G]$ with $\h{\mathcal{G}}_{\X{01}}^{\X{(\nu')}}[\mathsf{v},G]$ amounts to a $\nu$th-order self-energy contribution in the interaction potential $\mathsf{v}$;
\item[(ii)] the second term on the RHS of Eq.\,(\ref{e74}) sequentially replaces all lines representing $G$ in each diagram associated with $\Sigma_{\X{00}}^{\X{(\nu-\nu')}}(a,b;[\mathsf{v},G])$ by $\mathcal{G}_{\X{01}}^{\X{(\nu')}}[\mathsf{v},G]$;
\item[(iii)] with reference to Eq.\,(\ref{e72}), $\lambda^{\nu} \h{\mathcal{G}}_{\X{01}}^{\X{(\nu)}}[\mathsf{v},G] \equiv \h{\mathcal{G}}_{\X{01}}^{\X{(\nu)}}[v,G]$ amounts to the total contribution of the $\nu$th-order Green-function diagrams contributing to $\h{G} - \h{G}_{\X{0}}$;
\item[(iv)] in view of the equalities in Eq.\,(\ref{e74}), one clearly observes that substitution of $\mathcal{G}_{\X{01}}^{\X{(\nu)}}[\mathsf{v},G]$ for a $G$ in any self-energy diagram results in a \textsl{non-$G$-skeleton} proper self-energy diagram;
\item[(v)] by inspecting the diagrammatic representations of the expressions in Eqs\,(\ref{e73}) and (\ref{e74}), one can convince oneself that the recursive calculation of the self-energy contribution $\Sigma_{\X{01}}^{\X{(\nu)}}(a,b;[\mathsf{v},G])$ on the basis of these expressions results in the complete set of $G$-skeleton self-energy diagrams for $\Sigma_{\X{01}}^{\X{(\nu)}}(a,b;[\mathsf{v},G])$, $\forall \nu\in\mathds{N}$.
\end{quote}
\vspace{0.2cm}
\noindent
We note that we have explicitly verified Eq.(\ref{e74}) (and thus Eq.\,(\ref{e76}) below) for $\nu$ up to and including $4$. In this connection, the number $N_{\X{01}}^{\X{(\nu)}}$ of diagrams contributing to $\Sigma_{\X{01}}^{\X{(\nu)}}(a,b)$ are \cite{BF16a}:\footnote{These values can be deduced from the expression in Eq.\,(17) of Ref.\,\protect\citen{MM06}. In this connection, note that the latter expression yields $1$, instead of $2$, for $N_{\X{01}}^{\X{(1)}}$. This is because this expression does not count the Hartree contribution to the self-energy. Values of $N_{\X{01}}^{\X{(\nu)}}$ for $\nu \ge 2$ are not affected by this, since for $\nu\ge 2$ a self-energy diagram containing a Hartree self-energy (or tadpole) insertion cannot be $G$-skeleton. We point out that these numbers can be determined, in principle for arbitrary values of $\nu \in \mathds{N}$, with the aid of the program \texttt{Snu} that we provide in appendix \protect\ref{sad}, p.\,\protect\pageref{Snu}.} $N_{\X{01}}^{\X{(1)}} = 2$, $N_{\X{01}}^{\X{(2)}} = 2$, $N_{\X{01}}^{\X{(3)}} = 10$, and $N_{\X{01}}^{\X{(4)}} = 82$.

For \refstepcounter{dummy}\label{ForAnArbitrary}an arbitrary $\nu$, the operator $\h{\Sigma}_{\X{00}}^{\X{(\nu)}}[\mathsf{v},G]$ is determined along the lines described in \S\,\ref{s23}, with $G$ taking the place of $G_{\X{0}}$. All $G_{\X{0}}$ in the expression in Eq.\,(\ref{e259b}), including those in the expression for $A_{r,s}^{\X{(2\nu-1)}}$, are to be replaced by $G$. The underlying sequence $\{G^{\X{(\nu)}}(a,b) \| \nu\}$, Eqs\,(\ref{e248a}), (\ref{e60c}), and (\ref{e60d}), are \textsl{not} to be identified with $\{\mathcal{G}_{\X{01}}^{\X{(\nu)}}(a,b)\| \nu\}$ (see Eq.\,(\ref{e72}) and the remark following Eq.\,(\ref{e73})), since despite $G^{\X{(\nu)}}(a,b)$ being evaluated in terms of $G$, it incorporates contributions arising from non-$G$-skeleton self-energy insertions. In other words, with $\{G^{\X{(\nu)}}(a,b)\| \nu\}$  evaluated in terms of $G$, the series in Eq.\,(\ref{e245}) no longer describes $G(a,b)$.

Although it is possible to express the functional derivative in Eq.\,(\ref{e74}) in closed form, from the perspective of computational efficiency the relevant expression offers no practical advantage. Instead, it is advantageous to make use of the formal definition of the functional derivative\,\footnote{See Appendix I, p.\,51, in Ref.\,\protect\citen{KKN60}, and Appendix A, p.\,403, in Ref.\,\protect\citen{ED11}.} and write the expression in Eq.\,(\ref{e74}) in a form convenient for numerical treatment. One has\,\footnote{It may be preferable in practice to determine the expression in Eq.\,(\protect\ref{e76}), and similar expressions, according to $\protect\h{\Sigma}_{\protect\X{01}}^{\protect\X{(\nu)}}[\mathsf{v},G] =
\left. \partial\Big(\epsilon\hspace{0.4pt}\protect\h{\Sigma}_{\protect\X{00}}^{\protect\X{(\nu)}}[\mathsf{v},G] - \sum_{\nu'=1}^{\nu-1}  \protect\h{\Sigma}_{\protect\X{00}}^{\protect\X{(\nu-\nu')}}[\mathsf{v},G+\epsilon\hspace{0.8pt}
\mathcal{G}_{\protect\X{01}}^{\protect\X{(\nu')}}]\Big)/\partial\hspace{0.4pt}\epsilon\right|_{\epsilon=0}$, $\nu \ge 2$.}
\begin{equation}\label{e76}
\h{\Sigma}_{\X{01}}^{\X{(\nu)}}[\mathsf{v},G] = \h{\Sigma}_{\X{00}}^{\X{(\nu)}}[\mathsf{v},G] -
\left.\frac{\partial}{\partial\epsilon} \sum_{\nu'=1}^{\nu-1}  \h{\Sigma}_{\X{00}}^{\X{(\nu-\nu')}}[\mathsf{v},G+\epsilon\hspace{0.8pt}\mathcal{G}_{\X{01}}^{\X{(\nu')}}]
\right|_{\epsilon=0},\;\; \nu \ge 2.
\end{equation}
For the evaluation of the derivative with respect to $\epsilon$, in practice one may employ\,\footnote{See \S\,5.7, p.\,180, in Ref.\,\protect\citen{PTVF97}, and \S\S\,5.2.24 and 5.3.21, pp.\,880 and 883, in Ref.\,\citen{AS72}. For the optimal value of $\epsilon$ appropriate to the first (second) expression in Eq.\,(\protect\ref{e77}), see Eq.\,(5.7.5) (Eq.\,(5.7.8)) on p.\,181 (p.\,182) of Ref.\,\protect\citen{PTVF97}.}
\begin{equation}\label{e77}
\frac{\partial}{\partial\epsilon} f(\epsilon) = \frac{1}{\epsilon}\big(f(\epsilon) - f(0)\big) + O(\epsilon) = \frac{1}{\epsilon}\big(f(\epsilon/2) - f(-\epsilon/2)\big) + O(\epsilon^2),
\end{equation}
or higher-order Lagrange's formula [\S\,25.3, p.\,882, in Ref.\,\citen{AS72}].

For illustration, since $\h{\Sigma}_{\X{00}}^{\X{(1)}}[\mathsf{v},G_{\X{0}}]$ is a \textsl{linear} functional of $G_{\X{0}}$, from Eq.\,(\ref{e76}) one immediately obtains that
\begin{equation}\label{e77a}
\h{\Sigma}_{\X{01}}^{\X{(2)}}[\mathsf{v},G] = \h{\Sigma}_{\X{00}}^{\X{(2)}}[\mathsf{v},G] -
\h{\Sigma}_{\X{00}}^{\X{(1)}}[\mathsf{v},\mathcal{G}_{\X{01}}^{\X{(1)}}],
\end{equation}
where $\mathcal{G}_{\X{01}}^{\X{(1)}}$ is given in Eq.\,(\ref{e73}). Using the diagrams representing the self-energies on the RHS of Eq.\,(\ref{e77a}), and those representing $\mathcal{G}_{\X{01}}^{\X{(1)}}[\mathsf{v},G]$, one immediately verifies the validity of this equality, that the second term on the RHS removes the four non-$G$-skeleton contributions associated with the first term, leaving two second-order self-energy diagrams that are indeed $G$-skeleton.\footnote{See Eq.\,(\protect\ref{e60g}) and the following remarks.}\footnote{The program \texttt{Snu}, appendix \protect\ref{sad}, p.\,\protect\pageref{Snu}, amongst others calculates the number of these diagrams.}\footnote{See Fig.\,\protect\ref{f10}, p.\,\protect\pageref{CompleteSet}.}

\refstepcounter{dummyX}
\subsection{The self-energy operator
\texorpdfstring{$\h{\Sigma}_{\protect\X{10}}[W,G_{\protect\X{0}}]$}{}}
\phantomsection
\label{s25}
With $W$ denoting the screened two-body interaction potential, \S\,\ref{sec3}, the formalism of diagram-free perturbation series expansion of the self-energy in terms of $(W,G_{\X{0}})$ to be presented in this section amounts to a straightforward generalisation of the formalism introduced in \S\,\ref{s24}. In this connection and in the light of the relevant relationship in Eq.\,(\ref{e4}) and the subsequent remark, we note that the diagrams associated with $\h{\Sigma}_{\X{10}}^{\X{(\nu)}}$ form a subset of the diagrams associated with $\h{\Sigma}_{\X{00}}^{\X{(\nu)}}$, $\forall\nu$. The sought-after functional $\h{\Sigma}_{\X{10}}^{\X{(\nu)}}[W,G_{\X{0}}]$ is thus obtained by removing from $\h{\Sigma}_{\X{00}}^{\X{(\nu)}}[W,G_{\X{0}}]$ the total contribution of the $\nu$th-order diagrams consisting of polarisation insertions.\footnote{Diagrammatically, a \textsl{polarisation insertion} is a part of a diagram that can be detached from it by cutting two lines representing the two-body interaction potential.}

We begin with the equation describing the single-particle operator $\h{W}$ associated with the screened two-particle interaction potential $W(a,b)$ in terms of the single-particle operator $\h{v}$ associated with the bare two-body interaction potential $v(a,b)$, and the operator $\h{P}$ associated with the \textsl{proper} polarisation function $P(a,b)$:\,\refstepcounter{dummy}\label{ForOrientation}\footnote{For orientation, in Ref.\,\protect\citen{FW03}, p.\,153, $P$ is denoted by $\Pi^{\star}$. For completeness, diagrammatically, a (connected) polarisation diagram is \textsl{proper} when it does not become disconnected on cutting a single line representing a two-body interaction potential (p.\,110 in Ref.\,\protect\citen{FW03}). In this context, one may introduce the notations of \textsl{one-interaction irreducible} (1II) and \textsl{two-interaction irreducible} (2II), in analogy with respectively \textsl{one-particle irreducible} (1PI) and \textsl{two-particle irreducible} (2PI); thus, a \textsl{proper} polarisation diagram is 1II. This analogy is apparent in the designations $G$-skeleton and $W$-skeleton, with the former (latter) referring to connected diagrams that do not contain any self-energy (polarisation) insertion. \label{notea}}\footnote{See appendix A in Ref.\,\protect\citen{BF16a} for some relevant details.}
\begin{equation}\label{e78}
\h{W} = \h{v} + \h{v} \h{P} \h{W}.
\end{equation}
We shall discuss some relevant aspects of the functions $W(a,b)$ and $P(a,b)$ later in this section. For now it is important to note that \textsl{structurally} the above equation is identical to the Dyson equation \cite{FW03} $\h{G} = \h{G}_{\X{0}} + \h{G}_{\X{0}} \h{\Sigma} \h{G}$, with $\h{W}$, $\h{v}$, and $\h{P}$ substituting for respectively $\h{G}$, $\h{G}_{\X{0}}$, and $\h{\Sigma}$ \cite{FEDvH91}. By the same reasoning as presented at the outset of \S\,\ref{s22}, we shall use the symbols $W$ and $P$ for both instances of $T=0$ and $T>0$, although in a different context we would use instead the symbols $\mathscr{W}$ and $\mathscr{P}$ ($\mathsf{W}$ and $\mathsf{P}$),\footnote{The $\mathsf{W}$ here is not to be confused with that in Eq.\,(\protect\ref{e80}) below.} within the Matsubara (TFD) formalism, in analogy with the symbols $\mathscr{G}$ and $\mathscr{S}$ ($\mathsf{G}$ and $\Upsigma$) denoting the one-particle Green function, Eq.\,(\ref{e18}) (Eq.\,(\ref{es7})),\footnote{\emph{Cf.} Eq.\,(\protect\ref{e11a}).} and the self-energy corresponding to thermal ensemble of states.

Similar to the self-energy, the polarisation function $P$ can be expanded in perturbation series in terms of $(v,G_{\X{0}})$, $(v,G)$, $(W,G_{\X{0}})$, and $(W,G)$. For what follows it is relevant explicitly to distinguish between these perturbation series expansions. This we do by adopting a similar notational convention as that for the self-energy, \S\,\ref{sec1}. In dealing with the \textsl{complete} relevant perturbation series expansions, one formally has (\emph{cf.} Eq.\,(\ref{e3}))
\begin{equation}\label{e79}
P = P_{\X{00}}[v,G_{\X{0}}] = P_{\X{01}}[v,G] = P_{\X{10}}[W,G_{\X{0}}] = P_{\X{11}}[W,G].
\end{equation}
Since in this section we explicitly deal with $\h{\Sigma}_{\X{10}}[W,G_{\X{0}}]$, the relevant operator to consider here is $\h{P}_{\X{10}}[W,G_{\X{0}}]$. For what follows, in analogy with the function $\mathsf{v}(i,j)$ in Eq.\,(\ref{e260a}), we introduce the function $\mathsf{W}(i,j)$, defined according to
\begin{equation}\label{e80}
W(i,j) \doteq \lambda\hspace{0.6pt} \mathsf{W}(i,j).
\end{equation}
As in the case of $v$, here the dimensionless coupling constant $\lambda$ serves as a convenient book-keeping device. In the light of the equation in Eq.\,(\ref{e78}), evidently the $\lambda$ in Eq.\,(\ref{e80}) cannot be the same coupling constant as in Eq.\,(\ref{e260a}).

\begin{figure}[t!]
\centerline{
\includegraphics*[angle=0, width=0.95\textwidth]{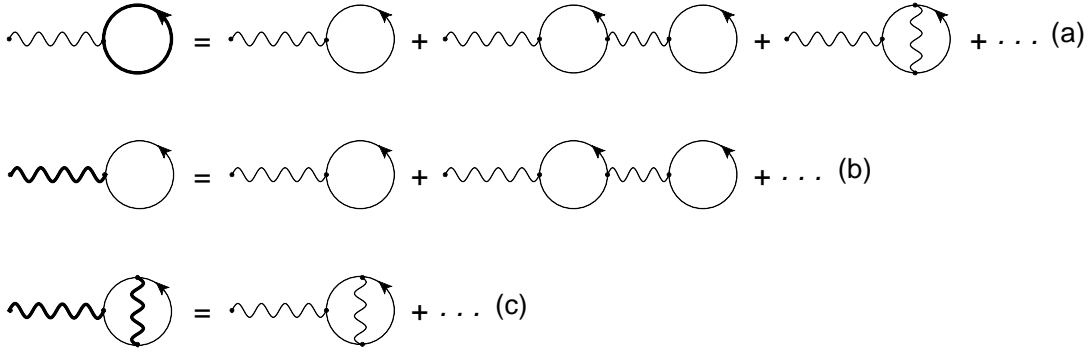}}
\caption{(a) The \textsl{exact} Hartree self-energy $\h{\Sigma}^{\textsc{h}} \equiv \h{\Sigma}^{\textsc{h}}[v,G] \equiv \h{\Sigma}_{\protect\X{01}}^{\textsc{h}}[v,G]$ (Eq.\,(\protect\ref{e301a}) below), in terms of the bare two-body interaction potential $v$ (thin wavy line) and the exact one-particle Green function $G$ (bold solid line), expanded in terms of $v$ and the non-interacting one-particle Green function $G_{\protect\X{0}}$ (thin solid line). (b) Expansion of the Hartree self-energy in $\h{\Sigma}_{\protect\X{10}}^{\protect\X{(1)}}[W,G_{\protect\X{0}}]$ (see the first equality in Eq.\,(\protect\ref{e86}) below) in terms of $v$ and $G_{\protect\X{0}}$, where $W$ is the two-body screened interaction potential (bold wavy line). (c) Expansion of a contribution to $\h{\Sigma}_{\protect\X{10}}^{\protect\X{(2)}}[W,G_{\protect\X{0}}]$ in terms of $v$ and $G_{\protect\X{0}}$. In (a), (b), and (c) only diagrams for contributions up to and including the second order in $v$ are shown. The expansions in (a), (b), and (c) shed light on three noteworthy facts. Firstly, the systematic expansion of the self-energy operator $\h{\Sigma}_{\protect\X{10}}[W,G_{\protect\X{0}}]$ must indeed include the Hartree self-energy in terms of the \textsl{screened} two-body interaction potential. Secondly, a systematic expansion of the Hartree contribution to $\h{\Sigma}_{\protect\X{10}}^{\protect\X{(1)}}[W,G_{\protect\X{0}}]$ is missing terms in $\h{\Sigma}^{\textsc{h}}$ that are taken account of by contributions in $\h{\Sigma}_{\protect\X{10}}^{\protect\X{(\nu)}}[W,G_{\protect\X{0}}]$ with $\nu \ge 2$ (here, the last diagram on the RHS of (a) that is missing on the RHS of (b) coincides with the first diagram on the RHS of (c)). Thirdly, taking into account the contribution of a Hartree self-energy diagram in terms of both $W$ \textsl{and} $G$ (as one might be inclined to do in calculating the self-energy functional $\h{\Sigma}_{\protect{11}}[W,G]$, \S\,\protect\ref{s26}) gives rise to an over-counting of certain self-energy contributions. We point out that since for spin-independent two-body potentials (such as the Coulomb potential) $\h{\Sigma}^{\textsc{h}}[v,G]$ is determined by the \textsl{total} GS number density $n(\bm{r}) = \sum_{\sigma} n_{\sigma}(\bm{r})$, for these potentials $\h{\Sigma}^{\textsc{h}}[v,G] \equiv \h{\Sigma}^{\textsc{h}}[v,G_{\protect\X{0}}]$ for any $G_{\protect\X{0}}$ that yields the exact $n(\bm{r})$. This is the case for the $G_{\protect\X{0}}$ corresponding to the Hohenberg-Kohn-Sham Hamiltonian \protect\cite{BF9799}.}
\label{f12x}
\end{figure}

For the perturbation series expansion of the self-energy operator in terms of $(W,G_{\X{0}})$, one has (\emph{cf.} Eq.\,(\ref{e70}) and see Fig.\,\ref{f12x})
\begin{equation}\label{e81}
\h{\Sigma}_{\X{10}}[W,G_{\X{0}}] = \sum_{\nu=1}^{n} \lambda^{\nu} \h{\Sigma}_{\X{10}}^{\X{(\nu)}}[\mathsf{W},G_{\X{0}}] + O(\lambda^{n+1}),
\end{equation}
where $\lambda^{\nu} \h{\Sigma}_{\X{10}}^{\X{(\nu)}}[\mathsf{W},G_{\X{0}}] \equiv \h{\Sigma}_{\X{10}}^{\X{(\nu)}}[W,G_{\X{0}}]$ (note the $\mathsf{W}$ and $W$, Eq.\,(\ref{e80})) denotes the total contribution of all $\nu$th-order proper (or 1PI) self-energy diagrams (including both $G$-skeleton and non-$G$-skeleton diagrams) that are $W$-skeleton, that is they do not contain polarisation insertions. Similarly, for the perturbation series expansion of $\h{P}_{\X{10}}[W,G_{\X{0}}]$ one has
\begin{equation}\label{e82}
\h{P}_{\X{10}}[W,G_{\X{0}}] = \sum_{\nu=0}^{n-1} \lambda^{\nu} \h{P}_{\X{10}}^{\X{(\nu)}}[\mathsf{W},G_{\X{0}}] + O(\lambda^{n}),
\end{equation}
where $\lambda^{\nu} \h{P}_{\X{10}}^{\X{(\nu)}}[\mathsf{W},G_{\X{0}}] \equiv \h{P}_{\X{10}}^{\X{(\nu)}}[W,G_{\X{0}}]$ (note the $\mathsf{W}$ and $W$, Eq.\,(\ref{e80})).

From the expression in Eq.\,(\ref{e82}) and that in Eq.\,(\ref{e78}), one obtains (\emph{cf.} Eq.\,(\ref{e71}))
\begin{equation}\label{e83}
\h{v} = \h{W} \big(\h{I} + \sum_{\nu=1}^{n} \lambda^{\nu} \h{P}_{\X{10}}^{\X{(\nu-1)}}[\mathsf{W},G_{\X{0}}]\hspace{0.5pt}  \h{\mathsf{W}}\big)^{-1} + O(\lambda^{n+2}).
\end{equation}
One thus arrives at the exact expression (\emph{cf.} Eq.\,(\ref{e72}))
\begin{equation}\label{e84}
\h{v} = \h{W} - \sum_{\nu=1}^{n} \lambda^{\nu}
\h{\mathcal{W}}_{\X{10}}^{\X{(\nu)}}[\mathsf{W},G_{\X{0}}] + O(\lambda^{n+2}),
\end{equation}
where $\{\h{\mathcal{W}}_{\X{10}}^{\X{(\nu)}}\| \nu\}$ is recursively determined from the following equalities (\emph{cf.} Eq.\,(\ref{e73})):
\begin{align}\label{e85}
\h{\mathcal{W}}_{\X{10}}^{\X{(1)}}[\mathsf{W},G_{\X{0}}] &= \h{\mathsf{W}} \h{P}_{\X{10}}^{\X{(0)}}[\mathsf{W},G_{\X{0}}]\hspace{0.7pt} \h{\mathsf{W}},\nonumber\\
\h{\mathcal{W}}_{\X{10}}^{\X{(\nu)}}[\mathsf{W},G_{\X{0}}] &= \h{\mathsf{W}} \h{P}_{\X{10}}^{\X{(\nu-1)}}[\mathsf{W},G_{\X{0}}]\hspace{0.7pt} \h{\mathsf{W}} - \h{\mathsf{W}} \sum_{\nu'=1}^{\nu-1} \h{P}_{\X{10}}^{\X{(\nu-\nu'-1)}}[\mathsf{W},G_{\X{0}}] \hspace{0.7pt} \h{\mathcal{W}}_{\X{10}}^{\X{(\nu')}}[\mathsf{W},G_{\X{0}}],\;\; \nu \ge 2.\nonumber\\
\end{align}
One observes that for $\h{W} \equiv \lambda \h{\mathsf{W}}$ given, calculation of $\h{\mathcal{W}}_{\X{10}}^{\X{(\nu)}}$, $\nu \in \mathds{N}$, requires knowledge of $\h{P}_{\X{10}}^{\X{(0)}}$ in the case of $\nu=1$, and of the sequence $\{\h{P}_{\X{10}}^{\X{(\nu')}}\| \nu' =0,\dots,\nu-1\}$ in the case of $\nu \ge 2$. As we shall see in \S\,\ref{s33}, $\h{P}_{\X{10}}^{\X{(0)}}$ and $\h{P}_{\X{10}}^{\X{(1)}}$ are calculated directly from respectively $\h{P}_{\X{00}}^{\X{(0)}}$ and $\h{P}_{\X{00}}^{\X{(1)}}$, Eq.\,(\ref{e221}) below, and calculation of $\h{P}_{\X{10}}^{\X{(\nu)}}$ for $\nu \ge 2$ is demanding of the knowledge of the sequence $\{\h{\mathcal{W}}_{\X{10}}^{\X{(\nu')}}\| \nu'=1,\dots,\nu-1\}$. It follows therefore that knowledge of the sequence $\{\h{P}_{\X{00}}^{\X{(\nu)}} \| \nu \in \mathds{N}_0\}$, determined recursively on the basis of the equalities in Eq.\,(\ref{e217}) below, suffices for the recursive determination of the sequence $\{\h{\mathcal{W}}_{\X{10}}^{\X{(\nu)}}\| \nu\in \mathds{N}\}$.\footnote{$\mathds{N}_0 \equiv \mathds{N} \cup \{0\}$. See appendix \protect\ref{sae}.}

In the light of the above discussions, one can convince oneself that in the direct-space representation the sought-after sequence $\{\h{\Sigma}_{\X{10}}^{\X{(\nu)}}[\mathsf{W},G_{\X{0}}]\| \nu\in \mathds{N}\}$ is recursively determined from the following equalities (\emph{cf.} Eq.\,(\ref{e74})):
\begin{align}\label{e86}
\Sigma_{\X{10}}^{\X{(1)}}(a,b;[\mathsf{W},G_{\X{0}}]) &= \Sigma_{\X{00}}^{\X{(1)}}(a,b;[\mathsf{W},G_{\X{0}}]),\nonumber\\
\Sigma_{\X{10}}^{\X{(\nu)}}(a,b;[\mathsf{W},G_{\X{0}}]) &= \Sigma_{\X{00}}^{\X{(\nu)}}(a,b;[\mathsf{W},G_{\X{0}}])
\nonumber\\ &\hspace{-2.0cm} -\sum_{\nu'=1}^{\nu-1} \int \rd 1\rd 2\; \frac{\delta \Sigma_{\X{00}}^{\X{(\nu-\nu')}}(a,b;[\mathsf{W},G_{\X{0}}])}{\delta \mathsf{W}(1,2)}\hspace{1.0pt} \mathcal{W}_{\X{10}}^{\X{(\nu')}}(1,2;[\mathsf{W},G_{\X{0}}]), \;\; \nu \ge 2.\hspace{1.2cm}
\end{align}
In analogy with the expression in Eq.\,(\ref{e76}), one has\,\footnote{See Appendix I, p.\,51, in Ref.\,\protect\citen{KKN60}, and Appendix A, p.\,403, in Ref.\,\protect\citen{ED11}.}
\begin{equation}\label{e87}
\h{\Sigma}_{\X{10}}^{\X{(\nu)}}[\mathsf{W},G_{\X{0}}] = \h{\Sigma}_{\X{00}}^{\X{(\nu)}}[\mathsf{W},G_{\X{0}}] -
\left.\frac{\partial}{\partial\epsilon} \sum_{\nu'=1}^{\nu-1} \h{\Sigma}_{\X{00}}^{\X{(\nu-\nu')}}[\mathsf{W}+
\epsilon\hspace{0.8pt}\mathcal{W}_{\X{10}}^{\X{(\nu')}},G_{\X{0}}]
\right|_{\epsilon=0},\;\; \nu \ge 2.
\end{equation}
Similar expressions as in Eq.\,(\ref{e77}) may be applied here.

For illustration, since $\h{\Sigma}_{\X{00}}^{\X{(1)}}[\mathsf{v},G_{\X{0}}]$ is a \textsl{linear} functional of $\mathsf{v}$, from Eq.\,(\ref{e87}) one immediately obtains that
\begin{equation}\label{e87a}
\h{\Sigma}_{\X{10}}^{\X{(2)}}[\mathsf{W},G_{\X{0}}] = \h{\Sigma}_{\X{00}}^{\X{(2)}}[\mathsf{W},G_{\X{0}}] -
\h{\Sigma}_{\X{00}}^{\X{(1)}}[\mathcal{W}_{\X{10}}^{\X{(1)}},G_{\X{0}}],
\end{equation}
where $\mathcal{W}_{\X{10}}^{\X{(1)}}$ is given in Eq.\,(\ref{e85}). Using the diagrams representing the self-energies on the RHS of Eq.\,(\ref{e87a}), and that representing $\mathcal{W}_{\X{10}}^{\X{(1)}}[\mathsf{W},G_{\X{0}}]$, one immediately verifies the validity of this equation, that the second term on the RHS removes the two non-$W$-skeleton contributions associated with the first term, appropriately leaving the four second-order $W$-skeleton proper self-energy diagrams to which the LHS corresponds.\footnote{See Eq.\,(\protect\ref{e60g}), and recall that the proper (\emph{i.e.} 1PI) self-energy operator $\protect\h{\Sigma}_{\protect\X{00}}^{\protect\X{(2)}}[\mathsf{v},G_{\protect\X{0}}]$ is described in terms of six diagrams.}

\refstepcounter{dummyX}
\subsection{The self-energy operator \texorpdfstring{$\h{\Sigma}_{\protect\X{11}}[W,G]$}{}}
\phantomsection
\label{s26}
Calculation of the self-energy operator $\h{\Sigma}_{\X{11}}[W,G]$ can be accomplished in two mathematically different but equivalent ways. In one, one relies on the self-energy operator $\h{\Sigma}_{\X{01}}[v,G]$, \S\,\ref{s24}, and in the other on the self-energy operator $\h{\Sigma}_{\X{10}}[W,G_{\X{0}}]$, \S\,\ref{s25}. In the former case, one recursively replaces $v$ by $W$, along the lines of \S\,\ref{s25}, and in the latter one recursively replaces $G_{\X{0}}$ by $G$, along the lines of \S\,\ref{s24}.

In this section we focus on the calculation of $\h{\Sigma}_{\X{11}}[W,G]$ on the basis of the self-energy operator $\h{\Sigma}_{\X{01}}[v,G]$, \S\,\ref{s24}. In doing so, it proves convenient to consider the operator
\begin{equation}\label{e300}
\h{\Sigma}' \doteq \h{\Sigma} -\h{\Sigma}^{\textsc{h}},
\end{equation}
where $\h{\Sigma}^{\textsc{h}}$ is the \textsl{exact} Hartree contribution to the self-energy operator, Fig.\,\ref{f12x}. Since we rely on the self-energy functional $\h{\Sigma}_{\protect\X{01}}[v,G]$ as the basis for the calculation of the self-energy functional $\h{\Sigma}_{\protect\X{11}}[W,G]$, we do not introduce different functional forms for $\h{\Sigma}^{\textsc{h}}$ and make the following identification:\,\footnote{See Fig.\,\protect\ref{f12x} as well as Eq.\,(\protect\ref{e250b}) below.}
\begin{equation}\label{e301a}
\h{\Sigma}^{\textsc{h}} \equiv \h{\Sigma}^{\textsc{h}}[v,G].
\end{equation}
Assuming the functional $\h{\Sigma}_{\protect\X{11}}'[W,G]$ to have been calculated for the exact $G$ and to infinite order in the coupling constant of the screened interaction potential $W$, Eq.\,(\ref{e80}), one has\,\footnote{Consult Eqs\,(10.59) and (10.60), p.\,195, in Ref.\,\protect\citen{RDM92}, as well as Eq.\,(10.25), p.\,290, in Ref.\,\protect\citen{SvL13}.}
\begin{equation}\label{e302a}
\h{\Sigma} \equiv \h{\Sigma}_{11}[W,G] \equiv \h{\Sigma}^{\textsc{h}}[v,G] + \h{\Sigma}_{\protect\X{11}}'[W,G].
\end{equation}
Diagrammatically, $\h{\Sigma}^{\textsc{h}}[v,G]$ corresponds to the first-order \textsl{tadpole} self-energy diagram. Since higher-order proper (or 1PI) self-energy diagrams containing a tadpole subdiagram cannot be $G$-skeleton (or 2PI), the diagrammatic perturbation expansion of the self-energy (\emph{cf.} Eq.\,(\ref{e300}))
\begin{equation}\label{e303a}
\h{\Sigma}_{\X{01}}'[v,G] \doteq \h{\Sigma}_{\X{01}}[v,G] -\h{\Sigma}^{\textsc{h}}[v,G]
\end{equation}
is free from tadpole subdiagrams. Similarly, the diagrammatic perturbation series expansion of $\h{\Sigma}_{\protect\X{11}}'[W,G] \equiv \h{\Sigma}_{11}[W,G]- \h{\Sigma}^{\textsc{h}}[v,G]$, Eq.\,(\ref{e302a}), is free from tadpole subdiagrams. Further, since $\h{\Sigma}^{\textsc{h}}[v,G]$ is first order in the coupling constant of $v$, with $\p{\h{\Sigma}}_{\hspace{-3.5pt}\X{11}}^{\X{(\nu)}}[W,G]$ denoting the $\nu$th-order perturbational contribution to $\h{\Sigma}_{\X{11}}'[W,G]$ in the coupling constant of $W$, with reference to Eq.\,(\ref{e302a}), one has\,\footnote{$\h{\Sigma}_{\X{11}}^{\X{(1)}}[W,G]$ constitutes Hedin's \protect\cite{LH65} $GW$ approximation of the self-energy operator.}
\begin{align}\label{e304x}
\h{\Sigma}_{\X{11}}^{\X{(1)}}[W,G] &= \h{\Sigma}^{\textsc{h}}[v,G] + \p{\h{\Sigma}}_{\hspace{-3.5pt}\X{11}}^{\X{(1)}}[W,G],\nonumber\\
\h{\Sigma}_{\X{11}}^{\X{(\nu)}}[W,G] &\equiv
\p{\h{\Sigma}}_{\hspace{-3.5pt}\X{11}}^{\X{(\nu)}}[W,G]\;\;
\text{for}\;\; \nu \ge 2.
\end{align}
With reference to Eq.\,(\ref{e303a}), clearly
\begin{equation}\label{e305d}
\p{\h{\Sigma}}_{\hspace{-3.5pt}\X{11}}^{\X{(1)}}[W,G] = \p{\h{\Sigma}}_{\hspace{-3.5pt}\X{01}}^{\X{(1)}}[W,G].
\end{equation}
Thus, following the first equality in Eq.\,(\ref{e304x}),
\begin{equation}\label{e305a}
\h{\Sigma}_{\X{11}}^{\X{(1)}}[W,G] = \h{\Sigma}^{\textsc{h}}[v,G] + \p{\h{\Sigma}}_{\hspace{-3.5pt}\X{01}}^{\X{(1)}}[W,G].
\end{equation}
Therefore, for the perturbational calculation of $\h{\Sigma}_{\X{11}}[W,G]$ it remains to consider the sequence $\{\p{\h{\Sigma}}_{\hspace{-3.5pt}\X{11}}^{\X{(\nu)}}[W,G]\| \nu \ge 2\}$ (note the second equality in Eq.\,(\ref{e304x})). Writing
\begin{equation}\label{e305b}
\h{\Sigma}_{\X{11}}'[W,G] = \sum_{\nu=1}^{n} \lambda^{\nu} \p{\h{\Sigma}}_{\hspace{-3.5pt}\X{11}}^{\X{(\nu)}}[\mathsf{W},G] + O(\lambda^{n+1}),
\end{equation}
following Eqs\,(\ref{e304x}) and (\ref{e305a}), one has
\begin{equation}\label{e305c}
\h{\Sigma}_{\X{11}}[W,G] = \h{\Sigma}^{\textsc{h}}[v,G] + \p{\h{\Sigma}}_{\hspace{-3.5pt}\X{01}}^{\X{(1)}}[W,G] +
\sum_{\nu=2}^{n} \lambda^{\nu} \h{\Sigma}_{\X{11}}^{\X{(\nu)}}[\mathsf{W},G] + O(\lambda^{n+1})\;\; \text{for}\;\; n \ge 2.\;\;\;
\end{equation}
As in other similar cases, $\lambda^{\nu} \p{\h{\Sigma}}_{\hspace{-3.5pt}\X{11}}^{\X{(\nu)}}[\mathsf{W},G] \equiv \p{\h{\Sigma}}_{\hspace{-3.5pt}\X{11}}^{\X{(\nu)}}[W,G]$ and $\lambda^{\nu}\h{\Sigma}_{\X{11}}^{\X{(\nu)}}[\mathsf{W},G] \equiv \h{\Sigma}_{\X{11}}^{\X{(\nu)}}[W,G]$ (note the $\mathsf{W}$ and $W$).

For the calculation of $\h{\Sigma}_{11}^{\X{(\nu)}}[W,G]$ for $\nu\ge 2$ we employ the procedure of \S\,\ref{s25}. In doing so, the sequence $\{\h{\mathcal{W}}_{\X{10}}^{\X{(\nu)}}[W,G_{\X{0}}]\| \nu\in\mathds{N}\}$, Eq.\,(\ref{e85}), is to be replaced by the sequence $\{\h{\mathcal{W}}_{\X{11}}^{\X{(\nu)}}[W,G]\| \nu\in\mathds{N}\}$. The sequence $\{\h{\mathcal{W}}_{\X{11}}^{\X{(\nu)}}\| \nu\in\mathds{N}\}$ can be obtained from $\{\h{\mathcal{W}}_{\X{10}}^{\X{(\nu)}}\| \nu\in\mathds{N}\}$, Eq.\,(\ref{e85}), on the basis of the recursive approach of \S\,\ref{s24}. One thus arrives at the following expressions (\emph{cf.} Eq.\,(\ref{e74})):
\begin{align}\label{e306}
\mathcal{W}_{\X{11}}^{\X{(1)}}(a,b;[W,G]) &=  \mathcal{W}_{\X{10}}^{\X{(1)}}(a,b;[W,G]), \nonumber\\
\mathcal{W}_{\X{11}}^{\X{(\nu)}}(a,b;[W,G]) &= \mathcal{W}_{\X{10}}^{\X{(\nu)}}(a,b;[W,G])\nonumber\\
&\hspace{-2.0cm}-\sum_{\nu'=1}^{\nu-1}\int \rd 1\rd 2\; \frac{\delta \mathcal{W}_{\X{10}}^{\X{(\nu-\nu')}}(a,b;[W,G])}{\delta G(1,2)}\hspace{1.0pt} \mathcal{G}_{\X{11}}^{\X{(\nu')}}(1,2;[\mathsf{W},G]),\;\; \nu \ge 2,\hspace{1.2cm}
\end{align}
where the sequence $\{\h{\mathcal{G}}_{\X{11}}^{\X{(\nu)}}\| \nu\in\mathds{N}\}$ is recursively determined from the following equalities (\emph{cf.} Eq.\,(\ref{e73})):
\begin{align}\label{e306a}
\h{\mathcal{G}}_{\X{11}}^{\X{(1)}}[\mathsf{W},G] &= \h{G} \h{\Sigma}_{\X{11}}^{\X{(1)}}[\mathsf{W},G]\hspace{0.7pt} \h{G},\nonumber\\
\h{\mathcal{G}}_{\X{11}}^{\X{(\nu)}}[\mathsf{W},G] &= \h{G} \h{\Sigma}_{\X{11}}^{\X{(\nu)}}[\mathsf{W},G]\hspace{0.7pt} \h{G} - \h{G} \sum_{\nu'=1}^{\nu-1} \h{\Sigma}_{\X{11}}^{\X{(\nu-\nu')}}[\mathsf{W},G] \hspace{0.7pt} \h{\mathcal{G}}_{\X{11}}^{\X{(\nu')}}[\mathsf{W},G],\;\; \nu \ge 2.
\end{align}
In practice, the second equality in Eq.\,(\ref{e306}) is to be calculated on the basis of the expression
\begin{equation}\label{e307}
\h{\mathcal{W}}_{\X{11}}^{\X{(\nu)}}[\mathsf{W},G] = \h{\mathcal{W}}_{\X{10}}^{\X{(\nu)}}[\mathsf{W},G] - \left.\frac{\partial}{\partial\epsilon} \sum_{\nu'=1}^{\nu-1} \h{\mathcal{W}}_{\X{10}}^{\X{(\nu-\nu')}}[\mathsf{W},G+
\epsilon\hspace{0.8pt}\mathcal{G}_{\X{11}}^{\X{(\nu')}}]
\right|_{\epsilon=0},\;\; \nu \ge 2,
\end{equation}
where the derivative with respect to $\epsilon$ may be determined on the basis of the expressions in Eq.\,(\ref{e77}).

With the ordered sequence $\{\h{\mathcal{W}}_{\X{11}}^{\X{(\nu)}}[\mathsf{W},G] \| \nu\in\mathds{N}\}$ at hand, the sought-after ordered sequence of perturbational self-energy contributions $\{\p{\h{\Sigma}}_{\hspace{-3.5pt}\X{11}}^{\X{(\nu)}}[\mathsf{W},G]\| \nu\in\mathds{N}\}$ is determined recursively from the following equalities (\emph{cf.} Eq.\,(\ref{e86})):
\begin{align}\label{e308}
\p{\Sigma}_{\hspace{-3.5pt}\X{11}}^{\X{(1)}}(a,b;[\mathsf{W},G]) &= \p{\Sigma}_{\hspace{-3.5pt}\X{01}}^{\X{(1)}}(a,b;[\mathsf{W},G]),\nonumber\\
\p{\Sigma}_{\hspace{-3.5pt}\X{11}}^{\X{(\nu)}}(a,b;[\mathsf{W},G]) &= \p{\Sigma}_{\hspace{-3.5pt}\X{01}}^{\X{(\nu)}}(a,b;[\mathsf{W},G]) \nonumber\\
&\hspace{-2.0cm} -\sum_{\nu'=1}^{\nu-1} \int \rd 1\rd 2\; \frac{\delta \p{\Sigma}_{\hspace{-3.5pt}\X{01}}^{\X{(\nu-\nu')}}(a,b;[\mathsf{W},G])}{\delta \mathsf{W}(1,2)}\hspace{1.0pt} \mathcal{W}_{\X{11}}^{\X{(\nu')}}(1,2;[\mathsf{W},G]),\;\; \nu \ge 2,\hspace{1.2cm}
\end{align}
where the first equality is a reproduction of the equality in Eq.\,(\ref{e305d}). In practice, the second equality in Eq.\,(\ref{e308}) is to be replaced by (\emph{cf.} Eq.\,(\ref{e87}))
\begin{equation}\label{e309}
\p{\h{\Sigma}}_{\hspace{-3.5pt}\X{11}}^{\X{(\nu)}}[\mathsf{W},G] =
\p{\h{\Sigma}}_{\hspace{-3.5pt}\X{01}}^{\X{(\nu)}}[\mathsf{W},G] - \left.
\frac{\partial}{\partial\epsilon} \sum_{\nu'=1}^{\nu-1} \p{\h{\Sigma}}_{\hspace{-3.5pt}\X{01}}^{\X{(\nu-\nu')}}[\mathsf{W}
+\epsilon\hspace{0.8pt}\mathcal{W}_{\X{11}}^{\X{(\nu')}},G]\right|_{\epsilon=0},\;\; \nu \ge 2,
\end{equation}
where the derivative with respect to $\epsilon$ may be determined on the basis of the expressions in Eq.\,(\ref{e77}).

For illustration, since $\p{\h{\Sigma}}_{\X{01}}^{\X{(1)}}[\mathsf{v},G]$ is a \textsl{linear} functional of $\mathsf{v}$, from Eq.\,(\ref{e309}) one immediately obtains that
\begin{equation}\label{e309b}
\p{\h{\Sigma}}_{\hspace{-3.5pt}\X{11}}^{\X{(2)}}[\mathsf{W},G] =
\p{\h{\Sigma}}_{\hspace{-3.5pt}\X{01}}^{(\X{2})}[\mathsf{W},G] - \p{\h{\Sigma}}_{\hspace{-3.5pt}\X{01}}^{\X{(1)}}[\mathcal{W}_{\X{11}}^{\X{(1)}},G],
\end{equation}
where $\mathcal{W}_{\X{11}}^{\X{(1)}}$ is given in Eq.\,(\ref{e306}), in which $\mathcal{W}_{\X{10}}^{\X{(1)}}$ is given in Eq.\,(\ref{e85}). Using the diagrams representing the self-energies on the RHS of Eq.\,(\ref{e309b}), and that representing $\mathcal{W}_{\X{11}}^{\X{(1)}}[\mathsf{W},G]$, one immediately verifies the validity of this equation, that the second term on the RHS removes the non-$W$-skeleton contribution associated with the first term,\footnote{This term is described in terms of two 2PI self-energy diagrams.} leaving the LHS to be described in terms of a single $G$- and $W$-skeleton self-energy diagram.

From the expressions in Eq.\,(\ref{e306a}), one observes that the sequence $\{\p{\h{\Sigma}}_{\hspace{-3.5pt}\X{11}}^{\X{(\nu)}}\| \nu\}$ of the self-energy operators to be calculated in turn determines the sequence $\{\h{\mathcal{W}}_{\X{11}}^{\X{(\nu)}}\| \nu\}$, Eq.\,(\ref{e306}), required for the calculation of $\{\p{\h{\Sigma}}_{\hspace{-3.5pt}\X{11}}^{\X{(\nu)}}\| \nu\}$, Eq.\,(\ref{e308}). This aspect does not affect the recursive nature of the calculations, since $\p{\h{\Sigma}}_{\hspace{-3.5pt}\X{11}}^{\X{(1)}}[W,G]$ can be calculated directly and for $\nu \ge 2$ calculation of $\p{\h{\Sigma}}_{\hspace{-3.5pt}\X{11}}^{\X{(\nu)}}[W,G]$ requires calculation of $\{\h{\mathcal{W}}_{\X{11}}^{\X{(\nu')}}\| \nu' = 1,\dots,\nu-1\}$, necessitating the knowledge of the sequence $\{\h{\Sigma}_{\X{11}}^{\X{(\nu')}}\| \nu' = 1,\dots,\nu-2\}$ in the case of $\nu\ge 3$; for $\nu=2$, the operator $\h{\mathcal{W}}_{\X{11}}^{\X{(2)}}$ is directly calculated from $\h{\mathcal{W}}_{\X{10}}^{\X{(2)}}$, Eq.\,(\ref{e306}).

\refstepcounter{dummyX}
\section{The dynamical screened interaction potential \texorpdfstring{$W$}{}}
\phantomsection
\label{sec3}
Calculation of the self-energy operators $\h{\Sigma}_{\X{10}}[W,G_{\X{0}}]$ and $\h{\Sigma}_{\X{11}}[W,G]$ necessitates calculation of the screened two-body interaction function $W$ \cite{JH57}. Diagrammatic expansions of this function are well-known \cite{}\cite{JH57,MYS95,FW03,NO98,HBKF04,JMZ69,PN89}. In this section we describe a diagram-free formalism for the calculation of this function.\footnote{Appendix A of Ref.\,\protect\citen{BF16a} is devoted to some details relevant to the considerations of this section. In the mentioned appendix only systems of fermions are considered.}

\refstepcounter{dummyX}
\subsection{Preliminaries}
\phantomsection
\label{s30}
In anticipation of what follows, we begin by introducing two-particle Green functions corresponding to $T=0$ and $T > 0$. To this end, for clarity we first recapitulate the following conventions, introduced earlier in this paper:\,\footnote{The considerations of this section apply equally to the Hubbard model, \S\S\,\protect\ref{s224}, \protect\ref{s227}, for which the $\bm{r}_j$ and $\bm{r}_j'$ in Eq.\,(\protect\ref{e90b}) are to be replaced by respectively $l_j$ and $l_j'$.}
\begin{eqnarray}\label{e90b}
&j \rightleftharpoons \bm{r}_jt_j\sigma_j,\;\; &j' \rightleftharpoons \bm{r}_j't_j'\sigma_j',
\hspace{1.4cm}\text{($T=0$ formalism)}\nonumber\\
&j \rightleftharpoons \bm{r}_j\tau_j\sigma_j,\;\; &j' \rightleftharpoons \bm{r}_j'\tau_j'\sigma_j',
\hspace{1.4cm}\text{(Matsubara formalism)}\nonumber\\
&\hspace{0.4cm}j \rightleftharpoons \bm{r}_jt_j\sigma_j\mu_j,\;\; &j' \rightleftharpoons \bm{r}_j't_j'\sigma_j'\mu_j',
\hspace{1.0cm}\text{(TFD formalism)}
\end{eqnarray}
where $j \in \mathds{N}$. Within the indicated formalisms, for the two-particle Green function one has:\,\footnote{For $G_{\protect\X{2}}$ see p.\,116 in Ref.\,\protect\citen{FW03} (\emph{cf.} Eq.\,(\protect\ref{e11a})), and for $\mathscr{G}_{\protect\X{2}}$, p.\,253 in Ref.\,\protect\citen{FW03} (\emph{cf.} Eq.\,(\protect\ref{e18})), as well as Eqs\,(3.1) and (3.2) in Ref.\,\protect\citen{MS59}, p.\,1347. We have defined $\mathsf{G}_{\protect\X{2}}$ in analogy with $G_{\protect\X{2}}$.}
\begin{align}\label{e90}
G_{\X{2}}(1,2;1',2') &\doteq (-\ii\hspace{1.0pt})^2 \langle\Psi_{N;0}\vert \mathcal{T} \big\{\h{\psi}_{\textsc{h}}^{\phantom{\dag}}(1)\h{\psi}_{\textsc{h}}^{\phantom{\dag}}(2) \h{\psi}_{\textsc{h}}^{\dag}(2')\h{\psi}_{\textsc{h}}^{\dag}(1')\big\}\vert\Psi_{N;0}\rangle,\;\;\text{($T=0$ formalism)}
\nonumber\\ \\
\label{e91}
\mathscr{G}_{\X{2}}(1,2;1',2') &\doteq (-1)^2 \Tr\!\big[\h{\varrho}\hspace{0.8pt}\mathcal{T}_{\X{\uptau}}\big\{
\h{\psi}_{\textsc{k}}^{\phantom{\dag}}(1) \h{\psi}_{\textsc{k}}^{\phantom{\dag}}(2)
\h{\psi}_{\textsc{k}}^{\dag}(2')\h{\psi}_{\textsc{k}}^{\dag}(1')\big\}\big],\;\;\text{(Matsubara formalism)}
\nonumber\\ \\
\label{e90a}
\mathsf{G}_{\X{2}}(1,2;1',2') &\doteq (-\ii\hspace{1.0pt})^2 \langle 0(\beta)\vert
\mathcal{T}_{\hspace{-1.2pt}{}_{\mathscr{C}}}\hspace{-1.0pt}\big\{\psi_{\textsc{h}}(1) \psi_{\textsc{h}}(2)\b{\psi}_{\textsc{h}}(2')
\b{\psi}_{\textsc{h}}(1')\big\}\vert 0(\beta)\rangle.\;\;\;\text{(TFD formalism)}\nonumber \\
\end{align}

With
\begin{align}\label{e92}
&\h{n}_{\textsc{h}}(j) \doteq \h{\psi}_{\textsc{h}}^{\dag}(j) \h{\psi}_{\textsc{h}}^{\phantom{\dag}}(j),
\;\;\;\;\text{($T=0$ formalism)}\\
\label{e93}
&\h{n}_{\textsc{k}}(j) \doteq \h{\psi}_{\textsc{k}}^{\dag}(j) \h{\psi}_{\textsc{k}}^{\phantom{\dag}}(j),
\;\;\;\;\text{(Matsubara formalism)}\\
\label{e92a}
&\h{\mathsf{n}}_{\textsc{h}}(j) \doteq \b{\psi}_{\textsc{h}}(j) \psi_{\textsc{h}}(j)\hspace{3.0pt}\phantom{,}
\;\;\;\;\text{(TFD formalism)}
\end{align}
denoting the number-density operators, for the density-density correlation functions $D(1,2)$, $\mathscr{D}(1,2)$, and $\mathsf{D}(1,2)$ one has [pp.\,151 and 301 in Ref.\,\citen{FW03}]\,\footnote{We have used the identity $\langle (\wh{A} - \langle\wh{A}\rangle) (\wh{B} - \langle\wh{B}\rangle)\rangle \equiv \langle \wh{A} \wh{B}\rangle - \langle\wh{A}\rangle \langle\wh{B}\rangle$.}
\begin{align}\label{e94}
D(1,2) &\doteq -\ii \big(\langle\Psi_{N;0}\vert \mathcal{T}\big\{\h{n}_{\textsc{h}}(1)\hspace{0.6pt}\h{n}_{\textsc{h}}(2) \big\}\vert\Psi_{N;0}\rangle
-\langle\Psi_{N;0}\vert \h{n}_{\textsc{h}}(1)\vert\Psi_{N;0}\rangle\hspace{0.6pt}\langle\Psi_{N;0}\vert \h{n}_{\textsc{h}}(2)  \vert\Psi_{N;0}\rangle \big),\nonumber\\
& &\hspace{7.6cm} \text{($T=0$ formalism)}\\
\label{e95}
\mathscr{D}(1,2) &\doteq -\big(\Tr\!\big[\h{\varrho}\hspace{0.8pt} \mathcal{T}_{\X{\uptau}}\big\{\h{n}_{\textsc{k}}(1)  \hspace{0.6pt}\h{n}_{\textsc{k}}(2)\big\}]
-\Tr\!\big[\h{\varrho}\hspace{0.8pt} \h{n}_{\textsc{k}}(1)\big]
\Tr\!\big[\h{\varrho}\hspace{0.8pt} \h{n}_{\textsc{k}}(2)\big]\big), \nonumber\\
& &\hspace{6.75cm}\text{(Matsubara formalism)}\\
\label{e94a}
\mathsf{D}(1,2) &\doteq -\ii \big(\langle 0(\beta)\vert \mathcal{T}_{\hspace{-1.2pt}{}_{\mathscr{C}}}\hspace{-1.0pt}\big\{\h{\mathsf{n}}_{\textsc{h}}(1)
\hspace{0.6pt}\h{\mathsf{n}}_{\textsc{h}}(2) \big\}\vert 0(\beta)\rangle
-\langle 0(\beta)\vert \h{\mathsf{n}}_{\textsc{h}}(1)\vert 0(\beta)\rangle\hspace{0.6pt}\langle 0(\beta)\vert \h{\mathsf{n}}_{\textsc{h}}(2)  \vert 0(\beta)\rangle\big).\nonumber\\
& &\hspace{7.75cm}\text{(TFD formalism)}
\end{align}
Although for bosons / fermions these functions are more concisely expressed in terms of the relevant density-fluctuation operators\,\footnote{Note that since for $\wh{H}$ independent of time $G(i,j)$ ($\mathscr{G}(i,j)$) is a function of $t_i-t_j$ ($\tau_i-\tau_j$), the right-most expression in Eq.\,(\protect\ref{e96}) ((\protect\ref{e97})) makes explicit that $\h{n}_{\textsc{h}}'(j) -\h{n}_{\textsc{h}}(j)$ ($\h{n}_{\textsc{k}}'(j) -\h{n}_{\textsc{k}}(j))$ is independent of $t_j$ ($\tau_j$). Similarly as regards $\h{\mathsf{n}}_{\textsc{h}}'(j) -\h{\mathsf{n}}_{\textsc{h}}(j)$, even though $\mathsf{G}(i,j)$ is a function of $t_i-t_j$ only if this difference is understood as signifying the difference in the path lengths of $t_i$ and $t_j$ along $\mathscr{C}$ as measured from some fixed point on this path, such as $t_{\mathrm{i}}$, Eq.\,(\protect\ref{es4}).}
\begin{align}\label{e96}
\h{n}_{\textsc{h}}'(j) &\doteq \h{n}_{\textsc{h}}(j) -\langle\Psi_{N;0}\vert \h{n}_{\textsc{h}}(j)\vert\Psi_{N;0}\rangle \equiv \h{n}_{\textsc{h}}(j) \mp \ii G(j,j^+),\\
\label{e97}
\h{n}_{\textsc{k}}'(j) &\doteq \h{n}_{\textsc{k}}(j) -\Tr\!\big[\h{\varrho}\hspace{0.8pt} \h{n}_{\textsc{k}}(j)\big] \equiv \h{n}_{\textsc{k}}(j) \pm \mathscr{G}(j,j^+),\\
\label{e96a}
\h{\mathsf{n}}_{\textsc{h}}'(j) &\doteq \h{\mathsf{n}}_{\textsc{h}}(j) -\langle 0(\beta)\vert \h{\mathsf{n}}_{\textsc{h}}(j)\vert 0(\beta)\rangle \equiv \h{\mathsf{n}}_{\textsc{h}}(j) \mp \ii \mathsf{G}(j,j^+),
\end{align}
the expressions in Eqs\,(\ref{e94}), (\ref{e95}), and (\ref{e94a}) have the advantage that they can be directly written in terms of the relevant one- and two-particle Green functions. One has
\begin{align}\label{e98}
D(1,2) &= +\ii \big(G_{\X{2}}(1,2;1^+,2^+) - G(1,1^+) G(2,2^+)\big),\\
\label{e99}
\mathscr{D}(1,2) &= -\big(\mathscr{G}_{\X{2}}(1,2;1^+,2^+) - \mathscr{G}(1,1^+) \mathscr{G}(2,2^+)\big),\\
\label{e99a}
\mathsf{D}(1,2) &= +\ii \big(\mathsf{G}_{\X{2}}(1,2;1^+,2^+) - \mathsf{G}(1,1^+) \mathsf{G}(2,2^+)\big),
\end{align}
where $j^+$ is defined in Eqs\,(\ref{e260b}), (\ref{e23a}), and (\ref{es16e}). The usefulness of these expressions will become apparent shortly.

For the \textsl{improper} polarisation function\,\footnote{
A polarisation function is \textsl{improper} when its diagrammatic representation is \textsl{improper}, \emph{i.e.} it is not 1II (see footnote \raisebox{-1.0ex}{\normalsize{\protect\footref{notea}}} on p.\,\protect\pageref{ForOrientation}).} one has [pp.\,153 and 302 in Ref.\,\citen{FW03}]\,\footnote{Here we are adopting a notational convention which is contrary to that in Ref.\,\protect\citen{FW03}, where $\star$ signifies a \textsl{proper} correlation function. See an earlier relevant remark in footnote on p.\,\protect\pageref{FNStar}.}
\begin{equation}\label{e100}
P^{\star}(1,2) \doteq \frac{1}{\hbar}\hspace{0.6pt} D(1,2),\;\;
\mathscr{P}^{\star}(1,2) \doteq \frac{1}{\hbar}\hspace{0.6pt} \mathscr{D}(1,2),\;\;
\mathsf{P}^{\star}(1,2) \doteq \frac{1}{\hbar}\hspace{0.6pt} \mathsf{D}(1,2).
\end{equation}
With these expressions at hand, \emph{from now onwards the symbol $P^{\star}$ will represent also the functions $\mathscr{P}^{\star}$ and $\mathsf{P}^{\star}$}, similar to the single-particle operator $\h{P}$ in Eq.\,(\ref{e78}), which represents the single-particle operator corresponding to the \textsl{proper} polarisation function specific to both $T=0$ and $T>0$.\refstepcounter{dummy}\label{TheFunctionDenoted}\footnote{The function $P^{\star}$ is in the condensed-matter physics literature often denoted by $\chi$.} For the single-particle operator $\h{W}$ associated with the screened two-body interaction potential $W$, one has [pp.\,154 and 302 in Ref.\,\citen{FW03}]
\begin{equation}\label{e101}
\h{W} = \h{v} + \h{v} \h{P}^{\star} \h{v}.
\end{equation}
From this equality and that in Eq.\,(\ref{e78}), one obtains\,\footnote{For $\protect\h{I}$, see footnote \raisebox{-1.0ex}{\normalsize{\protect\footref{noteb}}} on p.\,\protect\pageref{ThusForInstance}. See also appendix \protect\ref{sae}.}
\begin{equation}\label{e102}
\h{P} = \h{P}^{\star} \big(\h{I} + \h{v} \h{P}^{\star}\big)^{-1}
\equiv \big(\h{I} + \h{P}^{\star} \h{v}\big)^{-1} \h{P}^{\star}.
\end{equation}
Equivalently
\begin{equation}\label{e103}
\h{P}^{\star} = \h{P} \big(\h{I} -\h{v} \h{P}\big)^{-1} \equiv \big(\h{I} -\h{P} \h{v}\big)^{-1} \h{P},
\end{equation}
where
\begin{equation}\label{e104}
\h{I} - \h{v} \h{P} \equiv \h{\upepsilon}
\end{equation}
is the single-particle operator corresponding to the dielectric response function $\upepsilon(a,b)$.

The above considerations show the way in which the \textsl{proper} polarisation function $P(a,b)$, which takes a prominent place in the considerations of \S\S\,\ref{s25} and \ref{s26}, can be calculated from the knowledge of the interacting one- and two-particle Green functions. Calculation of the interacting one-particle Green functions $G$, $\mathscr{G}$, and $\mathsf{G}$ is the subject of \S\,\ref{s22}.\footnote{Making use of the Dyson equation, these one-particle Green functions can also be calculated on the basis of the relevant self-energy operators to be perturbationally calculated with the aid of the recursive formalisms described in  \S\S\,\protect\ref{s23}, \protect\ref{s24}, \protect\ref{s25}, and \protect\ref{s26}.} In the following we describe a formalism through which the two-particle Green functions $G_{\X{2}}$, $\mathscr{G}_{\X{2}}$, and $\mathsf{G}_{\X{2}}$ are calculated along the lines of \S\,\ref{s22}.

\refstepcounter{dummyX}
\subsection{Technicalities}
\phantomsection
\label{s31}
Some technical details regarding the two-body screened interaction potential $W$ are in place. As will become evident, the focus in this section is on the \textsl{improper} polarisation operator $\h{P}^{\star}$, instead of the \textsl{proper} polarisation operator $\h{P}$ which is central to the considerations of \S\S\,\ref{s25} and \ref{s26}. The reason for this is two-fold. Firstly, considerations based on $\h{P}^{\star}$ are more transparent than those based on $\h{P}$, and, secondly, some observations with regard to $\h{P}^{\star}$ are directly established (that for instance $P^{\star}(1,2)$ is a function of $t_1-t_2$)\,\footnote{As regards $\mathscr{P}^{\star}(1,2)$, it is a function of $\tau_1-\tau_2$, and as regards $\mathsf{P}^{\star}(1,2)$, it is a function of $t_1-t_2$ on the understanding that $t_1$ and $t_2$ are measured along $\mathscr{C}$, Eq.\,(\protect\ref{es3}), from for instance $t_{\mathrm{i}}$.} on account of the direct relationship between $\h{P}^{\star}$ and the one- and two-particle Green functions, Eqs\,(\ref{e98}), (\ref{e99}), (\ref{e99a}), and (\ref{e100}). The latter properties are subsequently immediately shown to apply to $\h{P}$ on account of the equalities in Eq.\,(\ref{e102}).

With $W(i,j) \equiv \langle i\vert \h{W}\vert j\rangle$ (and similarly for $v(i,j)$ and $P^{\star}(i,j)$),\footnote{See footnote \raisebox{-1.0ex}{\normalsize{\protect\footref{noteb}}} on \protect\pageref{ThusForInstance}.} the equation in Eq.\,(\ref{e101}) takes the form
\begin{equation}\label{e105}
W(a,b) = v(a,b) + \int \rd 1 \rd 2\; v(a,1) P^{\star}(1,2) v(2,b).
\end{equation}
Adopting a similar notation as in Eq.\,(\ref{e11a}), and with the bare interaction potential $v(i,j)$ as specified in Eq.\,(\ref{e10}), the equality in Eq.\,(\ref{e105}) can be expressed as\,\footnote{For $T > 0$, the relevant $v(i,j)$ is that specified in either Eq.\,(\protect\ref{e22}) or Eq.\,(\protect\ref{es16c}), depending on whether one employs respectively the Matsubara formalism or the TFD one. Accordingly, for $T > 0$, within the Matsubara framework $\tau$ and $\tau'$ replace the $t$ and $t'$ in Eq.\,(\protect\ref{e106}).}
\begin{align}\label{e106}
&\hspace{-0.8cm} W_{\sigma,\sigma'}(\bm{r}t,\bm{r}'t') = u_{\sigma,\sigma'}(\bm{r},\bm{r}')\hspace{0.6pt} \delta(t-t')\nonumber\\
&\hspace{0.6cm} +\sum_{\sigma_1,\sigma_2} \int \textrm{d}^d r_1\textrm{d}^d r_2\; u_{\sigma,\sigma_1}(\bm{r},\bm{r}_1) P_{\sigma_1,\sigma_2}^{\star}(\bm{r}_1t,\bm{r}_2t') u_{\sigma_2,\sigma'}(\bm{r}_2,\bm{r}').
\end{align}
For $\wh{H}$ independent of time, $P_{\sigma_1,\sigma_2}^{\star}(\bm{r}_1t,\bm{r}_2t')$ is a function of $t-t'$, and therefore so is $W_{\sigma,\sigma'}(\bm{r}t,\bm{r}'t')$. Thus, from the equalities in Eq.\,(\ref{e102}) it follows that $P_{\sigma_1,\sigma_2}(\bm{r}_1t,\bm{r}_2t')$ is also a function of $t-t'$. In the specific case where $u_{\sigma,\sigma'}$ is spin-independent, that is where $u_{\sigma,\sigma'}\equiv u$ (as in the case of the Coulomb interaction potential\,\footnote{\emph{Cf.} Eq.\,(3.19), p.\,25, in Ref.\,\protect\citen{FW03}.} $u_{\textsc{c}}$ underlying \emph{ab initio} electronic-structures calculations), the screened interaction potential is similarly independent of spin, denoted by $W$. In such case, one has
\begin{equation}\label{e108}
W(\bm{r}t,\bm{r}'t') = u(\bm{r},\bm{r}')\hspace{0.6pt} \delta(t-t') +\int \textrm{d}^d r_1\textrm{d}^d r_2\; u(\bm{r},\bm{r}_1) P^{\star}(\bm{r}_1t,\bm{r}_2t') u(\bm{r}_2,\bm{r}'),
\end{equation}
where
\begin{equation}\label{e109}
P^{\star}(\bm{r}t,\bm{r}'t') \doteq \sum_{\sigma,\sigma'} P_{\sigma,\sigma'}^{\star}(\bm{r}t,\bm{r}'t').
\end{equation}

Considering the Hubbard Hamiltonian for spin-$\tfrac{1}{2}$ particles, \S\S\,\ref{s224}, \ref{s227}, and appendix \ref{sab}, the expression in Eq.\,(\ref{e105}) takes the form
\begin{align}\label{e110}
W_{\sigma,\sigma'}^{l,l'}(t-t') &= U (1-\delta_{\sigma,\sigma'}) \delta_{l,l'}\hspace{0.6pt} \delta(t-t')
\nonumber\\
&+ U^2 \sum_{\sigma_1,\sigma_2} (1-\delta_{\sigma,\sigma_1})(1-\delta_{\sigma_2,\sigma'}) P_{\sigma_1,\sigma_2}^{\star\hspace{0.4pt}l,l'}(t-t')\nonumber\\
&\equiv  U (1-\delta_{\sigma,\sigma'}) \delta_{l,l'}\hspace{0.6pt} \delta(t-t') + U^2 P_{\b{\sigma},\b{\sigma}'}^{\star\hspace{0.4pt}l,l'}(t-t'),
\end{align}
where we have used the expressions in Eqs\,(\ref{e38}) and (\ref{e39}), and adopted the notation introduced in appendix \ref{sab} (\emph{cf.} Eq.\,(\ref{eb3b}) herein). In the last equality, $\b{\sigma} = \downarrow$ ($\uparrow$) for $\sigma = \uparrow$ ($\downarrow$). At the zeroth-order perturbation expansion of $P_{\sigma,\sigma'}^{\star\hspace{0.4pt}\X{l,l'}}(t-t')$, one has\,\footnote{See Eqs\,(\protect\ref{e100}), (\protect\ref{e201}), and (\protect\ref{e98x}). The expressions in the latter equations establish that more generally $P_{\sigma,\sigma'}^{\star\hspace{0.4pt}\protect\X{(0)}} = P_{\sigma}^{\star\hspace{0.4pt}\protect\X{(0)}}\hspace{0.4pt}\delta_{\sigma,\sigma'}$. Using diagrams, one can convince oneself that $P_{\sigma,\sigma'}^{\protect\X{(1)}} = P_{\sigma}^{\protect\X{(1)}}\hspace{0.4pt} \delta_{\sigma,\sigma'}$. The latter equality does \textsl{not} apply to $P_{\sigma,\sigma'}^{\star\hspace{0.4pt}\protect\X{(1)}}$ because of the \textsl{improper} polarisation diagram associated with $\protect\h{P}^{\protect\X{(0)}}\protect\h{v}\protect\h{P}^{\protect\X{(0)}}$. Note that $\protect\h{P}^{\protect\X{(0)}} \equiv \protect\h{P}^{\star\hspace{0.4pt}\protect\X{(0)}}$, so that $P_{\sigma,\sigma'}^{\protect\X{(0)}} = P_{\sigma}^{\protect\X{(0)}}\hspace{0.4pt}\delta_{\sigma,\sigma'}$.}
\begin{equation}\label{e106b}
P_{\sigma,\sigma'}^{\star\hspace{0.4pt}\X{(0)}\hspace{0.4pt}\X{l,l'}}(t-t') = P_{\sigma}^{\star\hspace{0.4pt}\X{(0)}\hspace{0.4pt}\X{l,l'}}(t-t')\hspace{0.6pt} \delta_{\sigma,\sigma'},
\end{equation}
from which and Eq.\,(\ref{e110}) one obtains
\begin{equation}\label{e111}
W_{\sigma,\sigma'}^{l,l'}(t-t') = U (1-\delta_{\sigma,\sigma'}) \delta_{l,l'}\hspace{0.6pt} \delta(t-t') + U^2 P_{\b{\sigma}}^{\star\hspace{0.4pt}\X{(0)}\hspace{0.4pt}\X{l,l'}}(t-t')\hspace{0.6pt} \delta_{\sigma,\sigma'} + O(U^3),
\end{equation}
making explicit that, in the case of spin-$\tfrac{1}{2}$ particles described by the Hubbard Hamiltonian, to second order (inclusive) in the on-site interaction energy the interaction between particles with anti-parallel spin is unscreened.

\refstepcounter{dummyX}
\subsection{The two-particle Green function
\texorpdfstring{$G_{\protect\X{2}}$}{} and its perturbation expansion in terms of \texorpdfstring{$v$}{} and \texorpdfstring{$G_{\protect\X{0}}$}{}}
\phantomsection
\label{s32}
The diagram-free perturbation series expansion of the two-particle Green function $G_{\X{2}}$ in terms of $(v,G_{\X{0}}$) is very similar to that of the one-particle Green function $G$ in terms of $(v,G_{\X{0}})$, considered in \S\S\,\ref{s212} and \ref{s22}. This is owing to the fact that for $G_{\X{2}}$ one has (\emph{cf.} Eq.\,(\ref{e242})) [Eq.\,(5.34), p.\,139, in Ref.\,\citen{SvL13}]
\begin{equation}\label{e200}
G_{\X{2}}(a,b;c,d) = \frac{G_{\X{2;0}}(a,b;c,d) + \sum_{\nu=1}^{\infty} \lambda^{\nu} N_{\nu}(a,b;c,d)}{1 + \sum_{\nu=1}^{\infty} \lambda^{\nu} D_{\nu}},
\end{equation}
where $c$ and $d$ are similar variables as $a$ and $b$, Eqs\,(\ref{e12}), (\ref{e24}), (\ref{es16a}), (\ref{e36}), (\ref{e24a}), and (\ref{e42c}), and $D_{\nu}$ is the same constant as defined in Eq.\,(\ref{e261}) in which the function $A_{2\nu}$ is defined in Eq.\,(\ref{e260}). The function $G_{\X{2;0}}$ is the non-interacting two-particle Green function, for which one has [Eq.\,(5.27), p.\,135, in Ref.\,\citen{SvL13}]\,\footnote{With reference to our earlier remarks, the expression in Eq.\,(\protect\ref{e201}) also applies by considering $G_{\protect\X{2}}$ and $G_{\protect\X{0}}$ to denote  respectively $\mathscr{G}_{\protect\X{2}}$ and $\mathscr{G}_{\protect\X{0}}$ in applying the Matsubara formalism, and $\mathsf{G}_{\protect\X{2}}$ and $\mathsf{G}_{\protect\X{0}}$ in applying the TFD formalism. For the case of the Matsubara formalism, compare with Eq.\,(25.1), p.\,241, in Ref.\,\protect\citen{FW03}. With reference to footnote on p.\,\protect\pageref{SeeAlsoRef}, we note that since the $\mathscr{G}_{\protect\X{0}}$ in the present work is $-\protect\ii\hspace{0.8pt}$ times its counterpart in Ref.\,\protect\citen{SvL13}, the products of two non-interacting Green functions on the RHS of Eq.\,(\protect\ref{e201}) implicitly take account of a required minus sign.}
\begin{equation}\label{e201}
G_{\X{2;0}}(a,b;c,d) =
\begin{vmatrix}
G_{\X{0}}(a,c) & G_{\X{0}}(a,d) \\
G_{\X{0}}(b,c) & G_{\X{0}}(b,d)
\end{vmatrix}_{\pm}
\equiv G_{\X{0}}(a,c) G_{\X{0}}(b,d) \pm G_{\X{0}}(a,d) G_{\X{0}}(b,c),
\end{equation}
where $+/-$ denotes permanent / determinant, corresponding to systems of bosons / fermions. The expression on the RHS of Eq.\,(\ref{e201}) is referred to as the `Hartree-Fock approximation' of $G_{\X{2}}(a,b;c,d)$ \cite{BK61}.\footnote{See Eq.\,(20), p.\,290, as well as footnote 5, p.\,288, of Ref.\,\protect\citen{BK61}. Although $G_{\X{2}}$ is not explicitly defined in this reference, the authors closely follow the conventions of their Ref.\,4, our Ref.\,\protect\citen{MS59}.} For the function $N_{\nu}(a,b;c,d)$ one has [Eq.\,(5.34), p.\,139, in Ref.\,\citen{SvL13}]
\begin{align}\label{e202}
&\hspace{0.0cm}N_{\nu}(a,b;c,d) \doteq \frac{1}{\nu !} \Big(\frac{\ii}{2\hbar}\Big)^{\nu} \int \prod_{j=1}^{2\nu} \rd j\; \mathsf{v}(1,2) \dots \mathsf{v}(2\nu-1,2\nu) \nonumber\\
&\hspace{4.0cm}\times A_{2\nu+2}^{2\textsc{b}}(a,b;c,d;1,2,\dots,2\nu-1,2\nu),
\end{align}
where
\begin{align}\label{e203}
&\hspace{-0.8cm} A_{2\nu+2}^{2\textsc{b}}(a,b;c,d;1,2,\dots,2\nu-1,2\nu)\nonumber\\
&\hspace{0.0cm} \doteq
\begin{vmatrix}
G_{\X{0}}(a,c) & G_{\X{0}}(a,d) & G_{\X{0}}(a,1^+) & G_{\X{0}}(a,2^+) & \dots & G_{\X{0}}(a,2\nu^+) \\
G_{\X{0}}(b,c) & G_{\X{0}}(b,d) & G_{\X{0}}(b,1^+) & G_{\X{0}}(b,2^+) & \dots & G_{\X{0}}(b,2\nu^+) \\
G_{\X{0}}(1,c) & G_{\X{0}}(1,d) & G_{\X{0}}(1,1^+) & G_{\X{0}}(1,2^+) & \dots & G_{\X{0}}(1,2\nu^+) \\
G_{\X{0}}(2,c) & G_{\X{0}}(2,d) & G_{\X{0}}(2,1^+) & G_{\X{0}}(2,2^+) & \dots & G_{\X{0}}(2,2\nu^+) \\
\vdots & \vdots & \vdots & \vdots & \ddots & \vdots \\
G_{\X{0}}(2\nu,c) & G_{\X{0}}(2\nu,d) & G_{\X{0}}(2\nu,1^+) & G_{\X{0}}(2\nu,2^+) & \dots & G_{\X{0}}(2\nu,2\nu^+)
\end{vmatrix}_{\pm}
\end{align}
is a \textsl{double-bordered} permanent / determinant associated with the permanent / determinant in Eq.\,(\ref{e260}) [\S\,3.7.2, p.\,49, in Ref.\,\citen{VD99}]. The function $N_{\nu}$ in Eq.\,(\ref{e200}) (and Eq.\,(\ref{e202})) is not to be confused with the function $N_{\nu}$ in Eq.\,(\ref{e242}) (and Eq.\,(\ref{e257})). The two functions are easily distinguished by the number of their arguments, four in the case of the $N_{\nu}$ in Eqs\,(\ref{e200}), and two in that case of the $N_{\nu}$ in Eq.\,(\ref{e242}).

In the light of the expression in Eq.\,(\ref{e201}), for bosons / fermions one has\,\footnote{With reference to footnote on p.\,\protect\pageref{TheFunctionDenoted}, for the non-interacting counterpart of $\chi(1,2)$ corresponding to bosons / fermions (at $T=0$) one thus has $\chi_{\protect\X{0}}(1,2) = \pm\protect\ii\hspace{0.2pt}\hbar^{-1}\hspace{0.4pt}  G_{\protect\X{0}}(1,2^+)\hspace{0.4pt} G_{\protect\X{0}}(2,1^+)$, which is a well-known result, generally referred to as the random-phase approximation (RPA) \protect\cite{JMZ69,PN89}, or the \textsl{bubble} approximation, of $\chi(1,2)$  (one can safely write $\chi_{\protect\X{0}}(1,2) = \pm\protect\ii\hspace{0.2pt}\hbar^{-1}\hspace{0.4pt} G_{\protect\X{0}}(1,2)\hspace{0.4pt} G_{\protect\X{0}}(2,1)$, suppressing the superscript $+$ of $2$ and $1$). The minus sign in the case of fermions is associated with $\chi_{\protect\X{0}}(1,2)$ being diagrammatically represented by a \textsl{closed} fermion loop.} (\emph{cf.} Eqs\,(\ref{e98}), (\ref{e99}), and (\ref{e99a}))
\begin{align}\label{e98x}
&D_{\X{0}}(1,2) \doteq +\ii \big(G_{\X{2;0}}(1,2;1^+,2^+) - G_{\X{0}}(1,1^+) G_{\X{0}}(2,2^+)\big)
\nonumber\\
&\hspace{1.4cm} \equiv \pm\ii\hspace{0,8pt} G_{\X{0}}(1,2^+) G_{\X{0}}(2,1^+),\\
\label{e99x}
&\mathscr{D}_{\X{0}}(1,2) \doteq -\big(\mathscr{G}_{\X{2;0}}(1,2;1^+,2^+) - \mathscr{G}_{\X{0}}(1,1^+) \mathscr{G}_{\X{0}}(2,2^+)\big) \nonumber\\
&\hspace{1.4cm} \equiv \mp \mathscr{G}_{\X{0}}(1,2^+) \mathscr{G}_{\X{0}}(2,1^+),\\
\label{e99ax}
&\mathsf{D}_{\X{0}}(1,2) \doteq +\ii \big(\mathsf{G}_{\X{2;0}}(1,2;1^+,2^+) - \mathsf{G}_{\X{0}}(1,1^+) \mathsf{G}_{\X{0}}(2,2^+)\big)\nonumber\\
&\hspace{1.4cm} \equiv \pm\ii\hspace{0,8pt} \mathsf{G}_{\X{0}}(1,2^+) \mathsf{G}_{\X{0}}(2,1^+).
\end{align}

With reference to the considerations of \S\,\ref{s221}, from the expression in Eq.\,(\ref{e200}) one obtains the following perturbation series expansion for $G_{\X{2}}$ (\emph{cf.} Eq.\,(\ref{e245})):
\begin{equation}\label{e203a}
G_{\X{2}}(a,b;c,d) = G_{\X{2;0}}(a,b;c,d) + \sum_{\nu=1}^{n} \lambda^{\nu}
G_{\X{2}}^{\X{(\nu)}}(a,b;c,d) + O(\lambda^{n+1}),
\end{equation}
where $G_{\X{2}}^{\X{(\nu)}}(a,b;c,d)$, $\nu \in \mathds{N}$, denotes the $\nu$th-order perturbational contribution to the two-particle Green function $G_{\X{2}}(a,b;c,d)$. The ordered sequence $\{G_{\X{2}}^{\X{(\nu)}}\| \nu\in\mathds{N}\}$ is calculated recursively in the basis of the following equalities (\emph{cf.} Eq.\,(\ref{e248})):
\begin{align}\label{e203b}
G_{\X{2}}^{\X{(1)}}(a,b;c,d) &= N_1(a,b;c,d) - F_1\hspace{0.6pt} G_{\X{2;0}}(a,b;c,d),\nonumber\\
G_{\X{2}}^{\X{(\nu)}}(a,b;c,d) &= N_{\nu}(a,b;c,d) - F_{\nu}\hspace{0.6pt} G_{\X{2;0}}(a,b;c,d) - \sum_{\nu'=1}^{\nu-1} F_{\nu-\nu'}\hspace{0.6pt} N_{\nu'}(a,b;c,d),\;\; \nu \ge 2. \nonumber\\
\end{align}
At the time of writing these lines, for the case at hand we are able to deduce the equivalent of the recursive relations in Eq.\,(\ref{e248a}) \textsl{only} for systems of fermions. The reason for this will be clarified below.

Expansion of a double-bordered \textsl{determinant} of a specific form has been described in \S\,3.7.2, p.\,49, of Ref.\,\citen{VD99}. This specific form relates to the $2\times 2$ zero matrix on the south-east corner of the matrix considered. For this reason, below we explicitly deduce the expansion relevant for the considerations of this section. For doing so, we make use of the algebra of \textsl{compound matrices} [\S\,0.8.1, p.\,21, in Ref.\,\citen{HJ13}]. Since we make use of the Sylvester identity [\S\,0.8.6, p.\,27, in Ref.\,\citen{HJ13}], which is applicable only to determinants, \emph{the following considerations are specific to fermion systems}.

We proceed by denoting the matrix of which the function $A_{2\nu+2}^{2\textsc{b}}$ in Eq.\,(\ref{e203}) is the permanent / determinant by $\mathbb{A}_{2\nu+2}^{2\textsc{b}}$, and for conciseness by $\mathbb{A}$. Introducing the index set
\begin{equation}\label{e204}
\bm{\alpha} \doteq \{3,4,\dots,2\nu+2\},
\end{equation}
one has\,\footnote{Expressing the set $\{1,2,\dots,2\nu+2\}$ as $\bm{\alpha} \cup \{1,2\}$ amounts to a \textsl{partitioning} of the former set. For partitioned sets and matrices, consult \S\,0.7, p.\,16, in Ref.\,\protect\citen{HJ13}.}
\begin{equation}\label{e205}
\mathbb{A} \equiv \mathbb{A}(\bm{\alpha} \cup \{1,2\},\bm{\alpha} \cup \{1,2\}) \equiv \mathbb{A}(\bm{\alpha} \cup \{1,2\}).
\end{equation}
One further has
\begin{equation}\label{e206}
\mathbb{A}(\bm{\alpha},\bm{\alpha}) \equiv \mathbb{A}(\bm{\alpha}) = \mathbb{A}_{2\nu},
\end{equation}
where $\mathbb{A}_{2\nu}$ is the matrix whose permanent / determinant $A_{2\nu}$ is presented in Eq.\,(\ref{e260}). Let now the $2\times 2$ matrix $\mathbb{B}$ be defined as
\begin{equation}\label{e207}
\mathbb{B} =
\begin{pmatrix}
b_{1,1} & b_{1,2} \\
b_{2,1} & b_{2,2}
\end{pmatrix},
\end{equation}
where
\begin{equation}\label{e208}
b_{i,j} \doteq \left| \mathbb{A}(\bm{\alpha} \cup \{i\},\bm{\alpha} \cup \{j\}) \right|_{-},
\end{equation}
in which $\vert\dots\vert_{-} \equiv \det(\dots)$. By the Sylvester identity [\S\,0.8.6, p.\,27, in Ref.\,\citen{HJ13}], one has
\begin{equation}\label{e209}
\left|\mathbb{A}\right|_{-} =
\frac{\left| \mathbb{B}\right|_{-}}{\left|\mathbb{A}(\bm{\alpha})\right|_{-}} \equiv
\frac{\left| \mathbb{B}\right|_{-}}{A_{2\nu}}.
\end{equation}
One can convince oneself that the equality does \textsl{not} apply on replacing the $\vert\dots\vert_{-}$ in Eq.\,(\ref{e209}) by $\vert\dots\vert_{+}$.

The matrix elements $\{b_{i,j}\| i,j = 1,2\}$ are single-bordered determinants \cite{VD99}, to be determined along the same line as those leading to the equality in Eq.\,(\ref{e259}). One obtains\,\footnote{We point out that on changing the minus signs between the two terms on the RHSs of the expressions in Eq.\,(\protect\ref{e210}), one obtains the relevant expressions for \textsl{bosons}, assuming that in such case $A_{2\nu}$ and $A_{r,s}^{\protect\X{(2\nu-1)}}$ stand for permanents (\emph{cf.} Eqs\,(\protect\ref{e258}), (\protect\ref{e260}) and (\protect\ref{e259})). Thus, the treatment here is limited to systems of fermions only because the Sylvester identity, as employed in Eq.\,(\protect\ref{e209}) for determinants, has no analogue for permanents.}
\begin{align}\label{e210}
b_{1,1} &= A_{2\nu}\hspace{0.6pt} G_{\X{0}}(a,c) - \sum_{r,s=1}^{2\nu}
A_{r,s}^{\X{(2\nu-1)}}\hspace{0.6pt} G_{\X{0}}(a,s^+) G_{\X{0}}(r,c), \nonumber\\
b_{1,2} &= A_{2\nu}\hspace{0.6pt} G_{\X{0}}(a,d) - \sum_{r,s=1}^{2\nu}
A_{r,s}^{\X{(2\nu-1)}}\hspace{0.6pt} G_{\X{0}}(a,s^+) G_{\X{0}}(r,d), \nonumber\\
b_{2,1} &= A_{2\nu}\hspace{0.6pt} G_{\X{0}}(b,c) - \sum_{r,s=1}^{2\nu}
A_{r,s}^{\X{(2\nu-1)}}\hspace{0.6pt} G_{\X{0}}(b,s^+) G_{\X{0}}(r,c), \nonumber\\
b_{2,2} &= A_{2\nu}\hspace{0.6pt} G_{\X{0}}(b,d) - \sum_{r,s=1}^{2\nu}
A_{r,s}^{\X{(2\nu-1)}}\hspace{0.6pt} G_{\X{0}}(b,s^+) G_{\X{0}}(r,d).
\end{align}
For $A_{2\nu+2}^{2\textsc{b}} \equiv \vert\mathbb{A}\vert_{-}$ one thus obtains (\emph{cf.} Eq.\,(\ref{e259}))
\begin{align}\label{e211}
&\hspace{-1.0cm} A_{2\nu+2}^{2\textsc{b}} = A_{2\nu}\hspace{0.6pt} G_{\X{2;0}}(a,b;c,d)\nonumber\\
&\hspace{0.1cm} +\sum_{r,s=1}^{2\nu} A_{r,s}^{\X{(2\nu-1)}} \left\{G_{\X{0}}(b,c) G_{\X{0}}(a,s^+) G_{\X{0}}(r,d) +G_{\X{0}}(a,d) G_{\X{0}}(b,s^+) G_{\X{0}}(r,c)\right.\nonumber\\
&\hspace{2.3cm}\left.  -G_{\X{0}}(a,c) G_{\X{0}}(b,s^+) G_{\X{0}}(r,d) - G_{\X{0}}(b,d) G_{\X{0}}(a,s^+) G_{\X{0}}(r,c)\right\}\nonumber\\
&\hspace{0.1cm} +\sum_{r,s,r',s'=1}^{2\nu}\frac{A_{r,s}^{\X{(2\nu-1)}}\hspace{0.6pt} A_{r',s'}^{\X{(2\nu-1)}}}{A_{2\nu}}\hspace{1.2pt} \left\{G_{\X{0}}(a,s^+) G_{\X{0}}(r,c) G_{\X{0}}(b,{s'}^+) G_{\X{0}}(r',d)\right.\nonumber\\
&\hspace{4.3cm}\left. - G_{\X{0}}(a,s^+) G_{\X{0}}(r,d) G_{\X{0}}(b,{s'}^+) G_{\X{0}}(r',c)\right\}.
\end{align}
In this way, for \textsl{fermions} one arrives at (\emph{cf.} Eq.\,(\ref{e259a}))
\begin{equation}\label{e212}
N_{\nu}(a,b;c,d) = D_{\nu}\hspace{0.6pt} G_{\X{2;0}}(a,b;c,d) + M_{\nu}(a,b;c,d),
\end{equation}
where $M_{\nu}(a,b;c,d)$ is in an obvious manner determined on the basis of the expressions in Eqs\,(\ref{e202}) and (\ref{e211}).

\refstepcounter{dummyX}
\subsection{The proper polarisation operators
\texorpdfstring{$\h{P}_{\protect\X{00}}[v,G_{\protect\X{0}}]$}{},  \texorpdfstring{$\h{P}_{\protect\X{01}}[v,G]$}{},
\texorpdfstring{$\h{P}_{\protect\X{10}}[W,G_{\protect\X{0}}]$}{}, and
\texorpdfstring{$\h{P}_{\protect\X{11}}[W,G]$}{} }
\phantomsection
\label{s33}
The formalism of the previous section, \S\,\ref{s32}, enables one to calculate the two-particle Green function $G_{\X{2}}$ in terms of $(v,G_{\X{0}})$ to in principle arbitrary order in the coupling constant $\lambda$ of the bare two-body interaction potential $v$. In the light of the formalism of calculating the one-particle Green function $G$ in terms of $(v,G_{\X{0}})$, described in \S\,\ref{s221}, and following the equalities in Eqs\,(\ref{e98}), (\ref{e99}), (\ref{e99a}), and (\ref{e100}), one is therefore in a position to calculate the \textsl{improper} polarisation function $P^{\star}$ in terms of $(v,G_{\X{0}})$ to in principle arbitrary order in $\lambda$. Although the \textsl{proper} polarisation function $P$ as encountered in \S\S\,\ref{s25} and \ref{s26} can be obtained from $P^{\star}$ on the basis of the equalities in Eq.\,(\ref{e102}), for the considerations of this section it proves necessary to bypass these equalities and instead calculate $P$ by relying on the following equality that follows from the equation in Eq.\,(\ref{e78}):
\begin{equation}\label{e213}
\h{P} = \h{v}^{-1} - \h{W}^{-1}.
\end{equation}
The reason for this approach can be surmised from the equality in Eq.\,(\ref{e60}) on which the formalism of \S\,\ref{s23} is founded.

For the inverse operator $\h{W}^{-1}$, from the equation in Eq.\,(\ref{e101}) one obtains
\begin{equation}\label{e214}
\h{W}^{-1} = \big(\h{I} + \h{P}^{\star} \h{v}\big)^{-1} \h{v}^{-1}.
\end{equation}
On the basis of the perturbation series expansion for $\h{P}_{\X{00}}^{\star}[v,G_{\X{0}}]$ (\emph{cf.} Eq.\,(\ref{e82})),
\begin{equation}\label{e215}
\h{P}_{\X{00}}^{\star}[v,G_{\X{0}}] = \sum_{\nu=0}^{n} \lambda^{\nu}
\h{P}_{\X{00}}^{\star \X{(\nu)}}[\mathsf{v},G_{\X{0}}] + O(\lambda^{n+1}),
\end{equation}
where $\h{\mathsf{v}}$ is related to $\h{v}$ through the relationship in Eq.\,(\ref{e260a}), from the equality in Eq.\,(\ref{e214}), one obtains (\emph{cf.} Eq.\,(\ref{e83}))
\begin{equation}\label{e216}
\h{W}^{-1} = \big(\h{I} + \sum_{\nu=1}^{n+1} \lambda^{\nu} \h{P}_{\X{00}}^{\star \X{(\nu-1)}}[\mathsf{v},G_{\X{0}}] \h{\mathsf{v}}\big)^{-1} \h{v}^{-1} + O(\lambda^{n+1}).
\end{equation}
Writing (\emph{cf.} Eq.\,(\ref{e215}))
\begin{equation}\label{e218}
\h{P}_{\X{00}}[v,G_{\X{0}}] = \sum_{\nu=0}^{n} \lambda^{\nu} \h{P}_{\X{00}}^{\X{(\nu)}}[\mathsf{v},G_{\X{0}}] + O(\lambda^{n+1}),
\end{equation}
on account of the equality in equality in Eq.\,(\ref{e213}), one arrives at the following recursive expression (\emph{cf.} Eqs\,(\ref{e60c}) and (\ref{e60f})):
\begin{align}\label{e217}
\h{P}_{\X{00}}^{\X{(0)}}[\mathsf{v},G_{\X{0}}] &=
\h{P}_{\X{00}}^{\star \X{(0)}}[\mathsf{v},G_{\X{0}}], \nonumber\\
\h{P}_{\X{00}}^{\X{(\nu)}}[\mathsf{v},G_{\X{0}}] &= \h{P}_{\X{00}}^{\star \X{(\nu)}}[\mathsf{v},G_{\X{0}}] -\sum_{\nu'=0}^{\nu-1} \h{P}_{\X{00}}^{\star \X{(\nu-\nu'-1)}}[\mathsf{v},G_{\X{0}}] \h{\mathsf{v}} \h{P}_{\X{00}}^{\X{(\nu')}}[\mathsf{v},G_{\X{0}}], \;\; \nu \ge 1.
\end{align}
The sequences $\{\h{P}_{\X{00}}^{\star\hspace{0.5pt}\X{(\nu)}}\| \nu\in \mathds{N}_0\}$, $\{\h{\mathscr{P}}_{\X{00}}^{\star\hspace{0.5pt}\X{(\nu)}}\| \nu\in\mathds{N}_0\}$, and $\{\h{\mathsf{P}}_{\X{00}}^{\star\hspace{0.5pt}\X{(\nu)}}\| \nu\in\mathds{N}_0\}$ are determined on the basis of the following expressions (see Eqs\,(\ref{e98}), (\ref{e99}), (\ref{e99a}), and (\ref{e100})):
\begin{align}\label{e218a}
&P_{\X{00}}^{\star\hspace{0.5pt} \X{(\nu)}}(1,2) = +\frac{\ii}{\hbar}
\big(G_{\X{2}}^{\X{(\nu)}}(1,2;1^+,2^+) -
\sum_{\nu'=0}^{\nu} G^{\X{(\nu-\nu')}}(1,1^+) G^{\X{(\nu')}}(2,2^+)\big),\\
\label{e218b}
&\mathscr{P}_{\X{00}}^{\star\hspace{0.5pt} (\nu)}(1,2) = -\frac{1}{\hbar} \big(\mathscr{G}_{\X{2}}^{\X{(\nu)}}(1,2;1^+,2^+) - \sum_{\nu'=0}^{\nu} \mathscr{G}^{\X{(\nu-\nu')}}(1,1^+) \mathscr{G}^{\X{(\nu')}}(2,2^+)\big),\\
\label{e216c}
&\mathsf{P}_{\X{00}}^{\star\hspace{0.5pt} \X{(\nu)}}(1,2) = +\frac{\ii}{\hbar}
\big(\mathsf{G}_{\X{2}}^{\X{(\nu)}}(1,2;1^+,2^+) -
\sum_{\nu'=0}^{\nu} \mathsf{G}^{\X{(\nu-\nu')}}(1,1^+) \mathsf{G}^{\X{(\nu')}}(2,2^+)\big).
\end{align}
The sequences $\{G^{\X{(\nu)}}\| \nu\}$, $\{G_{\X{2}}^{\X{(\nu)}}\| \nu\}$, $\{\mathscr{G}^{\X{(\nu)}}\| \nu\}$, $\{\mathscr{G}_{\X{2}}^{\X{(\nu)}}\| \nu\}$, and $\{\mathsf{G}^{\X{(\nu)}}\| \nu\}$, $\{\mathsf{G}_{\X{2}}^{\X{(\nu)}}\| \nu\}$ are determined on the basis of the recursive expressions in Eqs\,(\ref{e248}) and (\ref{e203b}). Note that $G^{\X{(0)}}(a,b) \equiv G_{\X{0}}(a,b)$ and $G_{\X{2}}^{\X{(0)}}(a,b;c,d) \equiv G_{\X{2;0}}(a,b;c,d)$, where the function $G_{\X{2;0}}$ is determined in terms of $G_{\X{0}}$ according to the expression in Eq.\,(\ref{e201}). With reference to the first equality in Eq.\,(\ref{e217}), following Eq.\,(\ref{e100}) one clearly has
\begin{equation}\label{e216d}
\h{P}_{\X{00}}^{\star \X{(0)}}(1,2) \equiv
\h{P}_{\X{00}}^{\star \X{(0)}}(1,2;[\mathsf{v},G_{\X{0}}]) = \frac{1}{\hbar} D_{\X{0}}(1,2),
\end{equation}
where $D_{\X{0}}(1,2)$ is the function presented in Eq.\,(\ref{e98x}). Similar expressions apply to $\mathscr{P}_{\X{00}}^{\star \X{(0)}}(1,2)$ and $\mathsf{P}_{\X{00}}^{\star \X{(0)}}(1,2)$ with the $D_{\X{0}}(1,2)$ on the RHS replaced by respectively $\mathscr{D}_{\X{0}}(1,2)$ and $\mathsf{D}_{\X{0}}(1,2)$, Eqs\,(\ref{e99x}) and (\ref{e99ax}).

The perturbation series expansion in Eq.\,(\ref{e218}) forms the foundation on which we construct the perturbation series expansions for $\h{P}_{\X{01}}[v,G]$, $\h{P}_{\X{10}}[W,G_{\X{0}}]$, and $\h{P}_{\X{11}}[W,G]$ along the lines of \S\S\,\ref{s24}, \ref{s25}, and \ref{s26}. Clearly, the underlying approaches constitute formalisms in which $G$ and $W$, or solely $G$, as the case may be, are to be determined self-consistently. In this connection, following Eq.\,(\ref{e213}), we introduce the functional $\h{W}_{\X{\varsigma\varsigma'}}$, defined according to
\begin{equation}\label{e219a}
\h{W}_{\X{\varsigma\varsigma'}} \doteq \big(\h{v}^{-1} - \h{P}_{\X{\varsigma\varsigma'}}\big)^{-1},\;\; \X{\varsigma}, \X{\varsigma'} \in \{\X{0},\X{1}\}.
\end{equation}
For consistency, $\h{W}_{\X{10}}[W,G_{\X{0}}]$ and $\h{W}_{\X{11}}[W,G]$ are to be employed in the calculation of respectively $\h{\Sigma}_{\X{10}}[W,G_{\X{0}}]$ and $\h{\Sigma}_{\X{11}}[W,G]$.

\refstepcounter{dummyX}
\subsubsection{The sequence \texorpdfstring{$\{\h{P}_{\protect\X{01}}^{\protect\X{(\nu)}}\| \nu\}$}{}}
\phantomsection
\label{s331}
From the considerations of \S\,\ref{s24}, one obtains (\emph{cf.} Eq.\,(\ref{e74}))
\begin{align}\label{e219}
P_{\X{01}}^{\X{(0)}}(a,b;[\mathsf{v},G]) &= P_{\X{00}}^{\X{(0)}}(a,b;[\mathsf{v},G]), \nonumber\\
P_{\X{01}}^{\X{(\nu)}}(a,b;[\mathsf{v},G]) &= P_{\X{00}}^{\X{(\nu)}}(a,b;[\mathsf{v},G])\nonumber\\
&\hspace{-2.0cm}
-\sum_{\nu'=1}^{\nu} \int \rd 1 \rd 2\; \frac{\delta P^{\X{(\nu-\nu')}}(a,b;[\mathsf{v},G])}{\delta G(1,2)}\hspace{1.0pt} \mathcal{G}_{\X{01}}^{\X{(\nu')}}(1,2;[\mathsf{v},G]),\;\; \nu \ge 1,
\end{align}
where the sequence $\{\h{\mathcal{G}}_{\X{01}}^{\X{(\nu)}} \| \nu\}$, with $\mathcal{G}_{\X{01}}^{\X{(\nu)}}(1,2) \equiv \langle 1\vert \h{\mathcal{G}}_{\X{01}}^{\X{(\nu)}}\vert 2\rangle$,\footnote{See footnote \raisebox{-1.0ex}{\normalsize{\protect\footref{noteb}}} on p.\,\protect\pageref{ThusForInstance}.} is determined recursively on the basis of the expressions in Eq.\,(\ref{e73}). Similarly to the case of $\h{\Sigma}_{\X{01}}^{\X{(\nu)}}[\mathsf{v},G]$, in practice for $\nu \ge 1$ one is to calculate $\h{P}_{\X{01}}^{\X{(\nu)}}[\mathsf{v},G]$ with the aid of the following or a mathematically equivalent equality (\emph{cf.} Eqs\,(\ref{e76}) and (\ref{e77})):
\begin{equation}\label{e220}
\h{P}_{\X{01}}^{\X{(\nu)}}[\mathsf{v},G] = \h{P}_{\X{00}}^{\X{(\nu)}}[\mathsf{v},G] -
\left. \frac{\partial}{\partial\epsilon} \sum_{\nu'=1}^{\nu} \h{P}_{\X{00}}^{\X{(\nu-\nu')}}[\mathsf{v},G+\epsilon\hspace{0.8pt}\mathcal{G}_{\X{01}}^{\X{(\nu')}}]
\right|_{\epsilon=0},\;\; \nu \ge 1.
\end{equation}

For illustration, since $\h{P}_{\X{00}}^{\X{(0)}}[\mathsf{v},G_{\X{0}}]$ is a \textsl{quadratic} functional of $G_{\X{0}}$,\footnote{See Eqs\,(\protect\ref{e98x}), (\protect\ref{e217}), and (\protect\ref{e216d}). Clearly, $\h{P}_{\protect\X{00}}^{\protect\X{(0)}}[\mathsf{v},G_{\protect\X{0}}]$ does \textsl{not} explicitly depend on $\mathsf{v}$.} the expression corresponding to $\h{P}_{\X{01}}^{\X{(1)}}[\mathsf{v},G]$ as deduced from Eq.\,(\ref{e220}) is not as simple as that corresponding to for instance $\h{\Sigma}_{\X{01}}^{\X{(2)}}[\mathsf{v},G]$ in Eq.\,(\ref{e77a}). One instead has
\begin{equation}\label{e220a}
\h{P}_{\X{01}}^{\X{(1)}}[\mathsf{v},G] = \h{P}_{\X{00}}^{\X{(1)}}[\mathsf{v},G] -
\left. \frac{\partial}{\partial\epsilon} \h{P}_{\X{00}}^{\X{(0)}}[\mathsf{v},G+\epsilon\hspace{0.8pt}\mathcal{G}_{\X{01}}^{\X{(1)}}]
\right|_{\epsilon=0},
\end{equation}
where $\mathcal{G}_{\X{01}}^{\X{(1)}}$ is given in Eq.\,(\ref{e73}). Discarding the zeroth- and the second-order terms in $\epsilon$ in the diagrammatic representation of $\h{P}_{\X{00}}^{\X{(0)}}[\mathsf{v},G+\epsilon\hspace{0.8pt}\mathcal{G}_{\X{01}}^{\X{(1)}}]$, one readily verifies that the remaining four diagrams indeed remove the four non-$G$-skeleton proper polarisation diagrams in the diagrammatic representation of $\h{P}_{\X{00}}^{\X{(1)}}[\mathsf{v},G]$, leading to the resulting $\h{P}_{\X{01}}^{\X{(1)}}[\mathsf{v},G]$ being indeed correctly represented by a single $G$-skeleton proper polarisation diagram.

\refstepcounter{dummyX}
\subsubsection{The sequence \texorpdfstring{$\{\h{P}_{\protect\X{10}}^{\protect\X{(\nu)}}\| \nu\}$}{}}
\phantomsection
\label{s332}
Without going into details, we suffice to mention that on the basis of the considerations in \S\,\ref{s25}, one obtains (\emph{cf.} Eq.\,(\ref{e86}))
\begin{align}\label{e221}
P_{\X{10}}^{\X{(\nu)}}(a,b;[\mathsf{W},G_{\X{0}}]) &= P_{\X{00}}^{\X{(\nu)}}(a,b;[\mathsf{W},G_{\X{0}}]),\;\; \nu=0,1,\nonumber\\
P_{\X{10}}^{\X{(\nu)}}(a,b;[\mathsf{W},G_{\X{0}}]) &= P_{\X{00}}^{\X{(\nu)}}(a,b;[\mathsf{W},G_{\X{0}}])
\nonumber\\ &\hspace{-2.0cm} -\sum_{\nu'=1}^{\nu-1} \int \rd 1\rd 2\; \frac{\delta P_{\X{00}}^{\X{(\nu-\nu')}}(a,b;[\mathsf{W},G_{\X{0}}])}{\delta \mathsf{W}(1,2)}\hspace{1.0pt} \mathcal{W}_{\X{10}}^{\X{(\nu')}}(1,2;[\mathsf{\mathcal{W}},G_{\X{0}}]), \;\; \nu \ge 2,\hspace{1.0cm}
\end{align}
where the sequence $\{\h{\mathcal{W}}_{\X{10}}^{\X{(\nu)}} \| \nu\}$ is recursively determined on the basis of the equalities in Eq.\,(\ref{e85}). In this connection, we recall that while calculation of the latter sequence in turn requires knowledge of the sequence $\{\h{P}_{\X{10}}^{\X{(\nu)}} \| \nu\}$, both sequences can be determined strictly recursively.\footnote{See the remarks following Eq.\,(\protect\ref{e85}).}

Regarding the first equality in Eq.\,(\ref{e221}), we note that the case corresponding to $\nu=0$ reflects the fact that $\h{P}_{\X{\varsigma\p{\varsigma}}}^{\X{(0)}}$ is explicitly independent of the interaction potential, and the case corresponding to $\nu=1$ the fact that a polarisation diagram of order less than $2$ in the interaction potential cannot contain a polarisation insertion. As regards the second equality in Eq.\,(\ref{e221}), in practical calculations this equality may be first expressed equivalently as (\emph{cf.} Eq.\,(\ref{e87}))
\begin{equation}\label{e222}
\h{P}_{\X{10}}^{\X{(\nu)}}[\mathsf{W},G_{\X{0}}] = \h{P}_{\X{00}}^{\X{(\nu)}}[\mathsf{W},G_{\X{0}}] -
\left.\frac{\partial}{\partial\epsilon}\sum_{\nu'=1}^{\nu-1} \h{P}_{\X{00}}^{\X{(\nu-\nu')}}[\mathsf{W}+\epsilon\hspace{0.8pt}\mathcal{W}_{\X{10}}^{\X{(\nu')}},G_{\X{0}}]
\right|_{\epsilon=0},\;\; \nu \ge 2,
\end{equation}
and subsequently dealt with numerically in an appropriate way (\emph{cf.} Eqs\,(\ref{e76}) and (\ref{e77})).

For illustration, since $\h{P}_{\X{00}}^{\X{(1)}}[\mathsf{v},G_{\X{0}}]$ is a \textsl{linear} functional of $\mathsf{v}$, from Eq.\,(\ref{e222}) one immediately obtains that (\emph{cf.} Eq.\,(\ref{e87a}))
\begin{equation}\label{e222a}
\h{P}_{\X{10}}^{\X{(2)}}[\mathsf{W},G_{\X{0}}] = \h{P}_{\X{00}}^{\X{(2)}}[\mathsf{W},G_{\X{0}}] - \h{P}_{\X{00}}^{\X{(1)}}[\mathcal{W}_{\X{10}}^{\X{(1)}},G_{\X{0}}],
\end{equation}
where $\mathcal{W}_{\X{10}}^{\X{(1)}}$ in given in Eq.\,(\ref{e85}). Using diagrams, one easily verifies that the second term on the RHS of Eq.\,(\ref{e222a}) corresponds to five second-order non-$W$-skeleton proper polarisation diagrams, removing the contributions of those corresponding to the first term on the RHS, thus leaving the contributions of twenty-six proper $W$-skeleton polarisation diagrams as constituting the total second-order contributions to $\h{P}_{\X{10}}^{\X{(2)}}[\mathsf{W},G_{\X{0}}]$.

\refstepcounter{dummyX}
\subsubsection{The sequence \texorpdfstring{$\{\h{P}_{\protect\X{11}}^{\protect\X{(\nu)}}\| \nu\}$}{}}
\phantomsection
\label{s333}
As in the case of $\h{\Sigma}_{\X{11}}[W,G]$, \S\,\ref{s26}, the perturbation series expansion of the operator $\h{P}_{\X{11}}[W,G]$ can be determined along two mathematically different but equivalent ways: by a systematic substitution of $W$ for $v$ in the sequence of the perturbational terms corresponding to the operator $\h{P}_{\X{01}}[v,G]$, and by a systematic substitution of $G$ for $G_{\X{0}}$ in the perturbational terms corresponding to the operator $\h{P}_{\X{10}}[W,G_{\X{0}}]$. Expressing $\h{P}_{\X{11}}[W,G]$ as
\begin{equation}\label{e223}
\h{P}_{\X{11}}[W,G] = \sum_{\nu=0}^{n} \lambda^{\nu} \h{P}_{\X{11}}^{\X{(\nu)}}[\mathsf{W},G] + O(\lambda^{n+1}),
\end{equation}
on account of the considerations of \S\,\ref{s26}, one has (\emph{cf.} Eq.\,(\ref{e308}))
\begin{align}\label{e224}
P_{\X{11}}^{\X{(\nu)}}(a,b;[\mathsf{W},G]) &= P_{\X{01}}^{\X{(\nu)}}(a,b;[\mathsf{W},G]),\;\; \nu = 0,1,\nonumber\\
P_{\X{11}}^{\X{(\nu)}}(a,b;[\mathsf{W},G]) &= P_{\X{01}}^{\X{(\nu)}}(a,b;[\mathsf{W},G]) \nonumber\\
&\hspace{-2.0cm} -\sum_{\nu'=1}^{\nu-1} \int \rd 1\rd 2\; \frac{\delta P_{\X{01}}^{\X{(\nu-\nu')}}(a,b;[\mathsf{W},G])}{\delta \mathsf{W}(1,2)}\hspace{1.0pt} \mathcal{W}_{\X{11}}^{\X{(\nu')}}(1,2;[\mathsf{W},G]),\;\; \nu \ge 2,\hspace{1.2cm}
\end{align}
where the sequence $\{\h{\mathcal{W}}_{\X{11}}^{\X{(\nu)}}\| \nu\}$ is recursively obtained from the equalities in Eq.\,(\ref{e306}). In practical applications, the second equality in Eq.\,(\ref{e224}) may be first expressed equivalently as (\emph{cf.} Eq.\,(\ref{e309}))
\begin{equation}\label{e225}
\h{P}_{\X{11}}^{\X{(\nu)}}[\mathsf{W},G] =
\h{P}_{\X{01}}^{\X{(\nu)}}[\mathsf{W},G] - \left.\frac{\partial}{\partial\epsilon} \sum_{\nu'=1}^{\nu-1} \h{P}_{\X{01}}^{\X{(\nu-\nu')}}[\mathsf{W}+\epsilon\hspace{0.8pt}\mathcal{W}_{\X{11}}^{\X{(\nu')}},G]
\right|_{\epsilon=0},\;\; \nu \ge 2,
\end{equation}
and subsequently dealt with numerically in an appropriate way (\emph{cf.} Eqs\,(\ref{e76}) and (\ref{e77})).

Alternatively, on account of the considerations in \S\S\,\ref{s24} and \ref{s26}, one has (\emph{cf.} Eqs\,(\ref{e74}) and (\ref{e306}))
\begin{align}\label{e226}
P_{\X{11}}^{\X{(0)}}(a,b;[W,G]) &=  P_{\X{10}}^{\X{(0)}}(a,b;[W,G]), \nonumber\\
P_{\X{11}}^{\X{(\nu)}}(a,b;[W,G]) &= P_{\X{10}}^{\X{(\nu)}}(a,b;[W,G])\nonumber\\
&\hspace{-2.0cm}-\sum_{\nu'=1}^{\nu}\int \rd 1\rd 2\; \frac{\delta P_{\X{10}}^{\X{(\nu-\nu')}}(a,b;[W,G])}{\delta G(1,2)}\hspace{1.0pt} \mathcal{G}_{\X{11}}^{\X{(\nu')}}(1,2;[\mathsf{W},G]),\;\; \nu \ge 1,
\end{align}
where the sequence $\{\h{\mathcal{G}}_{\X{11}}^{\X{(\nu)}}\| \nu\}$ is recursively determined from the equalities in Eq.\,(\ref{e306a}). In practical applications, the second equality in Eq.\,(\ref{e226}) may be first written equivalently as (\emph{cf.} Eq.\,(\ref{e307}))
\begin{equation}\label{e227}
\h{P}_{\X{11}}^{\X{(\nu)}}[\mathsf{W},G] = \h{P}_{\X{10}}^{\X{(\nu)}}[\mathsf{W},G] - \left. \frac{\partial}{\partial\epsilon} \sum_{\nu'=1}^{\nu} \h{P}_{\X{10}}^{\X{(\nu-\nu')}}[\mathsf{W},G+\epsilon\hspace{0.8pt}\mathcal{G}_{\X{11}}^{\X{(\nu')}}]
\right|_{\epsilon=0},\;\; \nu \ge 1,
\end{equation}
and subsequently dealt with numerically in an appropriate way (\emph{cf.} Eqs\,(\ref{e76}) and (\ref{e77})).

For illustration, on the basis of the same consideration as leading to Eq.\,(\ref{e220a}), from the expression in Eq.\,(\ref{e227}) one obtains that\,\footnote{Clearly, $\h{P}_{\protect\X{10}}^{\protect\X{(0)}}[\mathsf{W},G]$ only implicitly depends on $\mathsf{W}$, through the dependence of $G$ on $\mathsf{W}$.}
\begin{equation}\label{e227a}
\h{P}_{\X{11}}^{\X{(1)}}[\mathsf{W},G] = \h{P}_{\X{10}}^{\X{(1)}}[\mathsf{W},G] - \left. \frac{\partial}{\partial\epsilon} \h{P}_{\X{10}}^{\X{(0)}}[\mathsf{W},G+\epsilon\hspace{0.8pt}\mathcal{G}_{\X{11}}^{\X{(1)}}]
\right|_{\epsilon=0},
\end{equation}
where $\mathcal{G}_{\X{11}}^{\X{(1)}}$ is given in Eq.\,(\ref{e306a}). One easily verifies that the first term on the RHS of Eq.\,(\ref{e227a}) corresponds to five first-order proper polarisation diagrams, and the second term to four non-$G$-skeleton proper polarisation diagrams, resulting in the LHS appropriately to correspond to a single $W$- and $G$-skeleton first-order polarization diagram.

We note that, following the equalities in Eq.\,(\ref{e226}), calculation of $\h{P}_{\X{11}}^{\X{(\nu)}}$ is demanding of the prior calculation of $\{\h{\mathcal{G}}_{\X{11}}^{\X{(\nu')}} \| \nu'=1,\dots,\nu\}$. Following Eq.\,(\ref{e306a}), the calculation of the latter sequence requires calculation of $\{\p{\h{\Sigma}}_{\X{11}}^{\X{(\nu')}}\| \nu' = 1,\dots,\nu\}$. In turn, following Eq.\,(\ref{e308}), calculation of the elements of the latter set is dependent on the prior calculation of the sequence $\{\h{\mathcal{W}}_{\X{11}}^{\X{(\nu')}}\| \nu' =1,\dots, \nu-1\}$ in the case of $\nu \ge 2$. With reference to Eq.\,(\ref{e306}), according to which calculation of $\h{\mathcal{W}}_{\X{11}}^{\X{(\nu)}}$ for $\nu \ge 2$ is dependent on the knowledge of $\{\h{\mathcal{G}}_{\X{11}}^{\X{(\nu')}} \| \nu' =1,\dots,\nu-1\}$, one thus observes that indeed the elements of the sequence $\{\h{\mathcal{G}}_{\X{11}}^{\X{(\nu)}} \| \nu\}$, and thus of $\{\h{P}_{\X{11}}^{\X{(\nu)}}\| \nu\}$, can be recursively calculated on the basis of the equalities in Eq.\,(\ref{e226}).

\refstepcounter{dummyX}
\section{Summary and concluding remarks}
\phantomsection
\label{s4}
In this paper we have introduced a set of fully self-consistent \textsl{diagram-free} perturbational schemes for the calculation of the one- and two-particle Green functions, $G$ and $G_{\X{2}}$ respectively, the self-energy operator $\Sigma$, the polarisation function $P$, the dielectric response function $\upepsilon$, and the screened two-body interaction potential $W$, all corresponding to GSs and equilibrium thermal ensemble of states. In these schemes, the perturbational contributions to the relevant functions are determined recursively. The schemes are deduced within the framework of the weak-coupling perturbation series expansions of $G$ and $G_{\X{2}}$, in terms of the bare two-body interaction potential $v$ and the non-interacting one-particle Green function $G_{\X{0}}$, which are founded on the Wick decomposition theorem, appendix \ref{saa}. Despite the weak-coupling foundation on which the perturbational series expansions of $\Sigma$, $P$, $\upepsilon$ and $W$ in terms of $(v,G)$, $(W,G_{\X{0}})$, and $(W,G)$ have been based, their applicability is \textsl{not} limited to weakly-correlated GSs and thermal ensemble of states. In a forthcoming publication \cite{BF16a}\footnote{\emph{Note added to \textsf{arXiv:1912.00474v2}}: To be published simultaneously with the present v2-paper.} we present the details of a rigorous formalism for the \textsl{self-consistent} calculation of in particular self-energy operator as a functional of specifically $G$.

The considerations of the present paper have been directly related to the \textsl{normal} states of interacting systems of fermions and bosons, both for $T=0$ (zero temperature) and $T > 0$ (\S\S\,\ref{s222x} and \ref{sac01}). The generalisation of the schemes introduced in this paper for dealing with the cases of superconductive and superfluid states (or phases) of these systems will be presented in a separate publication \cite{BF16b}.

In dealing with equilibrium thermal ensemble of states, we have explicitly considered the \textsl{imaginary-time} formalism of Matsubara \cite{TM55,ESF59,MS59,FW03,AGD75}, \S\,\ref{s222x}, and the \textsl{real-time} formalism of TFD \cite{UMT82,HU95}, \S\S\,\ref{sac01}, \ref{s226}. The real-time nature of the TFD formalism enables one directly to calculate the \textsl{dynamical} correlation functions, thus bypassing the need for the analytic continuation of these functions as required within the imaginary-time formalism of Matsubara.

At least for sufficiently large orders of the perturbation theory, in practice the integrals over the internal space-`time' variables and the sums over the internal spin indices (as well as the indices corresponding to the two-component fields $\uppsi$ and $\b{\uppsi}$ within the framework of the TFD, Eq.\,(\ref{es5})) underlying the relevant expressions\,\footnote{For instance, those in Eqs\,(\protect\ref{e257}) and (\protect\ref{e261}).} are to be evaluated by means of a Monte-Carlo sampling method \cite{Note1}.\footnote{As regards the application of the Monte Carlo sampling methods in the framework of the formalisms introduced in this paper, consult the overview in \S\,\protect\ref{sec1c}.} Since the arithmetic complexity of the calculation of an arbitrary $\mathpzc{n}$-determinant scales at most\,\footnote{For some relevant remarks relating to the algorithm of Strassen, see footnote \raisebox{-1.0ex}{\normalsize{\protect\footref{notec}}} on p.\,\protect\pageref{ByAtMost}.} like $\mathpzc{n}^3$ (using for instance the standard Gaussian-elimination method \cite{GvL13}), for fermions the perturbation series expansions of this paper bypass the $\mathpzc{n} \times \mathpzc{n}!$ arithmetic complexity associated with expanding $\mathpzc{n}$-determinants and establishing the contributions corresponding to connected Green-function diagrams, and, insofar as the self-energy is concerned, those corresponding to the proper (or 1PI) and $G$-skeleton (or 2PI) self-energy diagrams.\footnote{Here $\mathpzc{n} \simeq 2\nu$, where $\nu$ is the order of the perturbation series expansion, varying between $1$ and some maximum finite value $n$ in any practical calculation. For some relevant details, consult appendices \protect\ref{sac} and \protect\ref{sad}.} As regards bosons, although the formalisms of this paper similarly bypass the $\mathpzc{n} \times \mathpzc{n}!$ arithmetic complexity inherent to diagrammatic expansions, they cannot avoid the exponential arithmetic complexity associated with the calculation of permanents; according to the algorithm of Ryser \cite{HJR63,HM78,BR91}, for a general $\mathpzc{n}$-permanent this complexity scales like $\mathpzc{n} \times 2^\mathpzc{n}$. We note that the computational complexity of the calculation of permanents is an NP-hard problem \cite{LGV79,MA06}.

For spin-$\mathsf{s}$ particles, whether fermions or bosons, with $\mathsf{s} \not=0$, a non-exhaustive sampling of the internal spin indices of the particles\,\footnote{The relevant sums over the internal spin indices are implicit in the integrals with respect $j$, $j=1,2,\dots,2\nu$, in the expressions in for instance Eqs\,(\protect\ref{e257}) and (\protect\ref{e261}) (see Eqs\,(\protect\ref{e11}) and (\protect\ref{e23})). As is evident from the expressions in Eqs\,(\protect\ref{es18}) and (\protect\ref{es19}), within the framework of the TFD one encounters additional sums over $2^{\nu}$ terms for the calculation of the contributions corresponding to the $\nu$th order of the perturbation theory.} at any given order of the perturbation theory is equivalent to discarding contributions of some specific diagrams at that order. For spin-independent interaction potentials, at the $\nu$th order of the perturbation theory the arithmetic complexity of the calculations corresponding to summations over \textsl{all} internal spin indices amounts to $(2\mathsf{s}+1)^{2\nu}$, for both fermions and bosons. For Hubbard-like models of spin-$\tfrac{1}{2}$ fermions, where the bare interaction potential is on-site and operative only between particles with opposite spin indices, \S\S\,\ref{s224}, \ref{s227}, and appendix \ref{sab}, for this arithmetic complexity one has $2^{\nu}$, equal to the \textsl{square root} of  $(2\mathsf{s}+1)^{2\nu}$ for $\mathsf{s}=\tfrac{1}{2}$. Contrasting $\mathpzc{n} \times 2^{\mathpzc{n}}$ for $\mathpzc{n} \simeq 2\nu$ with $(2\mathsf{s}+1)^{2\nu}$, one observes that for spin-$\mathsf{s}$ bosons, with $\mathsf{s} > \tfrac{1}{2}$, the arithmetic complexity of the full summation over the internal spin indices at the $\nu$th-order of the perturbation theory overwhelms that of the calculation of the required permanents as $\nu\to\infty$.\footnote{Note that the smallest integer $\mathsf{s}$ satisfying the condition $\mathsf{s} > \tfrac{1}{2}$ is $\mathsf{s} =1$.}

On account of the \textsl{recursive} nature of the perturbational schemes that we have introduced in this paper, in the Monte Carlo sampling of the underlying functions\,\footnote{Such as the sequence of functions $\{M_{\nu}(a,b) \| \nu\}$, Eqs\,(\protect\ref{e259b}) and (\protect\ref{e248a}).} the variable $\nu$, the order of the perturbation expansion, cannot be treated as a stochastic variable. In the light of the infinite summations that are implicit in the calculations of, for instance, the self-energy functionals $\Sigma_{\X{01}}[v,G]$, $\Sigma_{\X{10}}[W,G_{\X{0}}]$, and $\Sigma_{\X{11}}[W,G]$, \S\S\,\ref{s24}, \ref{s25}, and \ref{s26}, we believe that this does not impose any practical limitation on the use of these schemes.

Lastly, in appendices \ref{sac} and \ref{sad} we explicitly show how the conventional many-body perturbation expansions can be reformulated in terms of the cycle decompositions of the elements of the symmetric group $S_{\mathpzc{n}}$, with $\mathpzc{n} = 2\nu$ specific to the $\nu$th-order of the perturbation theory. In these, as well as in appendix \ref{sab}, we present a number of programs, written in the programming language of Mathematica$^{\X{\circledR}}$, for performing the perturbation series expansions of the one-particle Green function and the self-energy operator (both $\h{\Sigma}_{\X{00}}[v,G_{\X{0}}]$ and $\h{\Sigma}_{\X{01}}[v,G]$) to in principle an arbitrary order of the perturbation theory. With some minor modifications, these programs can be transformed into ones for performing the perturbation series expansions of the polarisation function.

\refstepcounter{dummyX}
\section{Acknowledgement}
\phantomsection
\label{s5}
We have drawn the Feynman diagrams for this publication with the aid of the program \textsf{JaxoDraw}.\footnote{\href{http://jaxodraw.sourceforge.net/}{\textsf{JaxoDraw:\,\textsl{Feynman Diagrams with Java}}}.}

\begin{appendix}

\refstepcounter{dummyX}
\section{On the Wick theorem}
\phantomsection
\label{saa}
%\addtocontents{toc}{\protect\setcounter{tocdepth}{0}}
%%\addtocontents{lof}{\protect\setcounter{tocdepth}{0}}

This appendix is a brief summary of a comprehensive pedagogical review \cite{BF19} of the extant works on a variety of Wick operator identities and decompositions that have been published since the original publications by Houriet and Kind \cite{HK49} and Wick \cite{GCW50}, in respectively 1949 and 1950, up to the present time.

\subsection{Statement of the theorem}
\phantomsection
\label{saa1}
The Wick theorem in its original form \cite{HK49,GCW50} is an operator identity, relating a time-ordered product of a set of $n \ge 2$ creation and annihilation operators in the interaction picture to a sum over $n$-products, in the case of $n$ \textsl{even} complemented by a sum over products of $n/2$ contractions, each such product being referred to as a fully contracted term.\footnote{As we show in \S\,\protect\ref{saa2c}, the number of fully-contracted terms is equal to $(n-1)!!$.}\refstepcounter{dummy}\label{ForCompleteness}\footnote{For completeness, we note that contractions $\{\h{\mathcal{C}}_{i,j}\| i,j\}$ (assumed to be $c$-numbers, that is $\protect\h{\mathcal{C}}_{i,j} = \mathcal{C}_{i,j}\hspace{0.6pt}\protect\h{1}$, $\forall i,j$, where $\protect\h{1}$ denotes the unit operator in the Fock space -- see the \textsl{third} general remark in \S\,\ref{saa2b}) are \textsl{not} subject to the process of normal ordering. This fact is explicit in the Wick operator identity corresponding to the product of two (field) operators, namely $\mathcal{T}(\h{\varphi}(i)\h{\varphi}(j)) \equiv \mathcal{N}(\h{\varphi}(i)\h{\varphi}(j)) + \protect\h{\mathcal{C}}_{i,j}$ (\emph{cf.} Eq.\,(\protect\ref{ea1r}) below), deduced by rearranging the terms in the definition of the contraction $\protect\h{\mathcal{C}}_{i,j}$, Eq.\,(\ref{ea1db}) below. This observation is relevant, in that in contrast to what may be perceived at first glance (and suggested in some texts), one has $\mathcal{N}(\h{1}) = \h{0}$, \emph{i.e.} the normal ordering of a $c$-number is identically vanishing. The equality $\mathcal{N}(\h{1}) = \h{1}$ is \textsl{erroneous} \protect\cite{BF19}, although at places one may for convenience \textsl{define} $\mathcal{N}(\h{1})$ as being equal to $\protect\h{1}$. \label{noted}} Each $n$-product in the former sum consists of an $(n-2p)$-product of \textsl{normal-ordered} operators and a $p$-product of $p$ contractions\,\footnote{\textsl{Contraction}, Eq.\,(\protect\ref{ea1db}), is also known as \textsl{pairing} and \textsl{covariance}.} of the remaining $2p$ operators, where $p= 0,1,\dots, m$, with $m = n/2-1$ ($m= (n-1)/2$) in the case of $n$ \textsl{even} (\textsl{odd}). For a given $p$, $0\le p \le m$, the normal-ordered $(n-2p)$-products corresponding to all possible $p$ contractions are encountered in the above-mentioned sum.

\subsection{General remarks}
\phantomsection
\label{saa2b}
The following general remarks are in order.

First, the time-ordering operation referred to above is generally understood as being the \textsl{chronological} time ordering operation $\mathcal{T}$ whose application to a product of operators effects a permutation of their original positions in such a way that the time indices of the permuted operators decrease monotonically from left to right, multiplying the ordered product by the signature of the permutation ($\pm 1$ for even / odd permutations) in the case of fermion operators. Naturally, time ordering is not defined when the time indices of at least two operators in a product to be time-ordered are equal. In such case, the desired order needs to be enforced explicitly.\footnote{If for instance the desired ordering in the case of $\mathcal{T}(\h{\varphi}(i) \h{\varphi}(j))$, with $t_i = t_j$, is $\h{\varphi}(j) \h{\varphi}(i)$, this ordering is achieved by effecting $t_j \rightharpoonup t_j^+ \equiv t_j + 0^+$.} To this end, one can adopt such convention as retaining the relative pre-time-ordering orders of the relevant operators, or letting the relative orders of these coincide with their relative orders as determined by normal ordering.

Second, the operators in a product to be time ordered do not need to be time dependent, as in the context of the Wick operator identity `time' merely refers to a parameter, or index, attached to operators for the purpose of bookkeeping.\footnote{For instance, in Ref.\,\protect\citen{BR86}, Ch.\,4, the time-ordering operation $\mathcal{T}$ is initially defined for the products of the operators $\{\wh{A}_i\| i\}$ and $\{\wh{B}_i \| i\}$, ordering the products of these operators in accordance with the values of their indices. Thus, for instance, $\mathcal{T}(\wh{B}_2 \wh{A}_1 \wh{B}_3) = \wh{B}_3 \wh{B}_2 \wh{A}_1$.} In other words, for the process of the time-ordering of operators in an operator product the \textsl{dynamics} of these operators is irrelevant. It should therefore not come as a surprise that the Wick operator identity also applies to ordinary products of operators,\footnote{For the same reason that the Wick operator identity as considered in this appendix applies to canonical operators in the interaction picture, the latter operators are to be canonical operators in the Schr\"{o}dinger picture.}\footnote{The Wick operator identity for ordinary products of operators is explicitly considered in appendix 4\hspace{0.6pt}A.I, p.\,413, of Ref.\,\protect\citen{MYS95}.} provided that the relative orders of the operators to be normal ordered are the same as those in the original product (this restriction equally applies to the definition of the relevant contractions). This assertion is immediately appreciated by viewing the operators in the original product as being already appropriately time-ordered.

Third, in the context of the Wick theorem, the contractions of the operators in a product to be time ordered are to be $c$-numbers. This implies that the operators in the product are to be canonical; more generally, for boson / fermion operators, the commutations / anti-commutations of these operators are to be equal to $c$-numbers, resulting in the contractions to be $c$-cumbers. Consequently, in dealing with time-dependent (field) operators, they are to be in the interaction picture; in dealing with time-independent (field) operators, for which the order of the operators in the relevant product is taken as representing the chronological time order (indicated in the previous paragraph), the (field) operators are to be in the Schr\"{o}dinger picture.\footnote{Or the \textsl{interaction picture} in the case the operators in the relevant product are time-dependent, however their time arguments are equal. In this case also a time ordering of the operators is to be imposed on the basis of some prescription.}

Fourth, in contrast to (the chronological) time ordering $\mathcal{T}$, normal ordering $\mathcal{N}$,\footnote{We denote the normal-ordered product of the operators $\h{\varphi}_1$ and $\h{\varphi}_2$ by $\mathcal{N}(\h{\varphi}_1 \h{\varphi}_2)$. This product is referred to as $S$-product and denoted by $\hspace{0.0pt}:\hspace{-2.0pt}\h{\varphi}_1\h{\varphi}_2\hspace{-2.0pt}:\hspace{0.0pt}$ in Ref.\,\protect\citen{GCW50}.} that is placing creation operators to the left of the annihilation operators, is subject to variation, the choice depending on the application at hand. For instance, normal ordering can be with respect to the $0$-particle vacuum state $\vert 0\rangle$, in which case the interaction-picture field operators $\h{\psi}(i)$ and $\h{\psi}^{\dag}(i)$, Eq.\,(\protect\ref{e9}), are identified as respectively annihilation and creation operators for the purpose of normal ordering.\footnote{This is the case when dealing with for instance uniform systems of spinless bosons, where at zero temperature all non-interacting bosons are condensed into the single-particle state corresponding to the wave vector $\bm{k} = \bm{0}$ and the relevant field operators $\h{\uppsi}(i)$ and $\h{\uppsi}^{\dag}(i)$ are deduced from the original ones, $\h{\psi}(i)$ and $\h{\psi}^{\dag}(i)$, by suppressing the Fourier component corresponding to $\bm{k}=\bm{0}$ of $\h{\psi}(i)$ and $\h{\psi}^{\dag}(i)$ [Refs\,\protect\citen{NNB47,STB58,HP59}, and Ch.\,6, p.\,198, in Ref.\,\protect\citen{FW03}].} In this case, for boson / fermion field operators one has $\mathcal{N}(\h{\psi}(i) \h{\psi}^{\dag}(j)) = \pm \h{\psi}^{\dag}(j) \h{\psi}(i)$ (conventionally, $\mathcal{N}(\h{\psi}(i) \h{\psi}(j)) = \h{\psi}(i) \h{\psi}(j)$ and $\mathcal{N}(\h{\psi}^{\dag}(i) \h{\psi}^{\dag}(j)) = \h{\psi}^{\dag}(i) \h{\psi}^{\dag}(j)$).\footnote{This convention is also applied to more extended products of annihilation / creation operators.} On the other hand, if the normal ordering is with respect to an uncorrelated $N$-particle (ground) state $\vert\Phi_{N;0}\rangle$, with $N > 0$, barring the condensed state of bosons \cite{BF19}, annihilation and creation operators to be ordered originate from \textsl{both} $\h{\psi}(j)$ and $\h{\psi}^{\dag}(j)$; writing\,\footnote{The operator $\protect\h{\psi}_{\varsigma}^{\dag}$, $\varsigma \in \{\protect\X{-},\protect\X{+}\}$, is generally \textsl{not} the Hermitian conjugate of $\protect\h{\psi}_{\varsigma}^{\phantom{\dag}}$.}
\begin{equation}\label{ea1s}
\h{\psi}(j) = \h{\psi}_-^{\phantom{\dag}}(j) + \h{\psi}_+^{\phantom{\dag}}(j),\;\;\;
\h{\psi}^{\dag}(j) = \h{\psi}_-^{\dag}(j) + \h{\psi}_+^{\dag}(j),
\end{equation}
depending on the notational convention (which is subject to variation in the literature concerning non-relativistic quantum field theory), one has [p.\,86 in Ref.\,\protect\citen{FW03}]
\begin{equation}\label{ea1t}
\h{\psi}_+^{\phantom{\dag}}(j) \vert\Phi_{N;0}\rangle =  0,\;\;\;
\h{\psi}_-^{\dag}(j) \vert\Phi_{N;0}\rangle =  0.
\end{equation}
In such case, for boson / fermion field operators one, for instance, has
\begin{equation}\label{ea1u}
\mathcal{N}(\h{\psi}_+^{\phantom{\dag}}(i) \h{\psi}_+^{\dag}(j)) = \pm \h{\psi}_+^{\dag}(j) \h{\psi}_+^{\phantom{\dag}}(i).
\end{equation}

Normal ordering may also be based on \textsl{arbitrary} decompositions of $\h{\psi}$ and $\h{\psi}^{\dag}$ into respectively $\h{\psi}_{\pm}^{\phantom{\dag}}$ and $\h{\psi}_{\pm}^{\dag}$ (that is, decompositions in which the latter operators are not subject to conditions similar to those in Eq.\,(\ref{ea1t})), resulting, for $\h{\varphi}(k)$ representing any of the four (field) operators $\h{\psi}_{\pm}^{\phantom{\dag}}(k)$, $\h{\psi}_{\pm}^{\dag}(k)$, in the contractions\,\footnote{Assumed to be a $c$-number, that is $\h{\mathcal{C}}_{i,j} \equiv \mathcal{C}_{i,j}\hspace{0.6pt}\protect\h{1}$.}
\begin{equation}\label{ea1db}
\h{\mathcal{C}}_{i,j} \doteq \mathcal{T}(\h{\varphi}(i)\h{\varphi}(j)) - \mathcal{N}(\h{\varphi}(i)\h{\varphi}(j))
\end{equation}
that generally (not invariably) fail to satisfy the condition $\h{\mathcal{C}}_{i,j} = \pm \h{\mathcal{C}}_{j,i}$ for boson / fermion field operators. Even in this case, the Wick operator identity has been shown to hold \protect\cite{EKS98}.\footnote{Notably, in the framework of the thermo-field dynamics (TFD) the equality $\protect\h{\mathcal{C}}_{i,j} = \pm \protect\h{\mathcal{C}}_{j,i}$ does not hold \protect\cite{RB94}. As a matter of fact, within this framework and in the context of the Wick operator identity, the normal-ordered product of a set of operators is a non-trivial \textsl{function} of the operators of this set \protect\cite{RB94}.}

Fifth, the chronological time-ordering $\mathcal{T}$ and the normal-ordering $\mathcal{N}$ are two specific operations of the more general $\mathcal{A}$-ordering and $\mathcal{B}$-ordering operations.\footnote{The theorem connecting these ordering schemes is developed in appendix A4 of Ref.\,\protect\citen{JR80} under the heading `The Ordering Theorem'. In the same appendix, a generalisation of this theorem concerning the $\mathcal{A}$-ordered products of $\mathcal{B}$-ordered products is presented. In Ref.\,\protect\citen{BF19} we discuss this theorem and its generalisation in some detail and fill in some of the gaps in their developments in Ref.\,\protect\citen{JR80}. Amongst others, we show the link between this theorem and the Wick theorem as deduced on the basis of the techniques of quantum groups \protect\cite{SM95}, or Hopf algebras, using the concept of \textsl{coproduct}. Pioneering work on the Wick theorem using these techniques is due to Brouder (consult for instance Ref.\,\protect\citen{CB05}.} Very briefly, in the framework of the Wick operator identity concerning the $\mathcal{A}$- and $\mathcal{B}$-ordering operations, one considers operators of the form
\begin{equation}\label{eay1}
\h{\varphi}_i \equiv \h{\varphi}(\alpha_i,\beta_i),\;\; i \in \mathcal{I},
\end{equation}
where $\{\alpha_i\| i\}$ and $\{\beta_i\| i\}$ are sets of (compound) variables to be specified below, and, for some integer $n$,
\begin{equation}\label{eay1a}
\mathcal{I} \equiv \{1,2,\dots,n\}
\end{equation}
is the index set. Of the above operators it is expected that
\begin{equation}\label{eay1b}
\hspace{4.0pt}\h{\hspace{-4.0pt}\mathpzc{S}}_{i,j} \doteq [\h{\varphi}_i, \h{\varphi}_j]_{\mp}
\end{equation}
be a $c$-number, $\forall i, j \in\mathcal{I}$, that is $\hspace{4.0pt}\h{\hspace{-4.0pt}\mathpzc{S}}_{i,j} = \mathpzc{S}_{i,j} \h{1}$, where $\mathpzc{S}_{i,j}$ is a real or complex number. The $\mathcal{A}$- and $\mathcal{B}$-ordering operations order the product of the operators $\{\h{\varphi}_i \| i \in \mathcal{I}\}$ on the basis of the orderings of the quantities $\{\alpha_i \| i \in \mathcal{I}\}$ and $\{\beta_i \| i\in \mathcal{I}\}$, respectively. For the operations $\mathcal{A}$ and $\mathcal{B}$ to be well-defined, it is required that the latter sets be \textsl{totally ordered} \cite{PMC05}, so that in each set all elements are \textsl{comparable}.\refstepcounter{dummy}\label{TheSymbolGE}\footnote{The symbol $\ge$ (and similarly $\le$ \dash alternative notations are respectively $\succeq$ and $\preceq$) signifies a binary relation between some or all elements of a \textsl{partially ordered} set $S$. To underline this partial ordering, one writes $(S,\ge)$. For $x, y, z \in S$, this binary relation is \textsl{reflexive} ($x \ge x$), \textsl{transitive} ($x\ge y$, $y \ge z$ $\Rightarrow$ $x\ge z$), and \textsl{anti-symmetric} ($x \ge y$, $y \ge x$ $\Rightarrow$ $x=y$) [p.\,xi and \S\,3.1, p.\,51, in Ref.\,\protect\citen{PMC05}]. Partial ordering is distinguished from an \textsl{equivalence relation} $\mathpzc{R}$ by the fact that $\mathpzc{R}$ is \textsl{symmetric} ($x \mathpzc{R} y$ $\Rightarrow$ $y \mathpzc{R} x$) [p.\,xii in Ref.\,\protect\citen{PMC05}]. When any two elements of a partially-ordered set $S$ are \textsl{comparable}, that is either $x \ge y$ or $y \ge x$, $x, y \in S$, the ordering is \textsl{total} [p.\,xi in Ref.\,\protect\citen{PMC05}]. Alternatively, a partially-ordered set (poset) $S$ is \textsl{totally} (\textsl{fully} or \textsl{linearly}) ordered if for \textsl{any} $x, y\in S$ exactly one of the relations $x <y$, $x = y$, $x > y$ is true. When $(S,\ge)$ is \textsl{totally ordered}, $S$ is also referred to as an \textsl{ordered set} or a \textsl{chain} [Def. 1.2.7, p.\,13, in Ref.\,\protect\citen{AA14}]. We note that in Ref.\,\protect\citen{PMC05} \textsl{ordered} is the short for \textsl{partially ordered} [p.\,xi in Ref.\,\protect\citen{PMC05}]. In Ref.\,\protect\citen{AA14} the property `any two elements are comparable' is introduced as the additional property beyond the above-mentioned three properties defining $\ge$ as the binary relation specific to $S$ as a poset. For completeness, when $(S,\ge)$ is partially ordered, $(S,\le)$ is also partially ordered. Further, for $x, y\in S$ and $x\ge y$ ($x \le y$), $x$ is \textsl{strictly} greater (less) than $y$ if `$x \ge y$ and $x\not=y$' (`$x \le y$ and $x\not=y$') [pp.\,xi and xii in Ref.\,\protect\citen{PMC05}]. See also Chap. 20 of Ref.\,\protect\citen{RNFK11}.} In Ref.\,\citen{JR80} the latter two sets are explicitly assumed to consist of \textsl{real} parameters, however these sets can be more general, required only to be \textsl{totally ordered} \cite{BF19}.

A generalisation of the ordering theorem described above is one concerning the $\mathcal{A}$ ordering of $\mathcal{B}$-ordered products of operators, with each $\mathcal{B}$-ordered product comprised of operators associated with the same value of the parameter $\alpha \in \{\alpha_i\| i\in \mathcal{I}\}$, briefly discussed in \S\,A4-3 of Ref.\,\citen{JR80} and to be discussed in some detail in Ref.\,\citen{BF19}. For this generalisation corresponding to the case where $\mathcal{A}$ is identified with the time-ordering operation, and $\mathcal{B}$ with the normal-ordering one, the reader is referred to Ref.\,\citen{IZ80} (p.\,181 herein).

Sixth, a Wick-type operator identity relating time-ordered products of operators in the interaction picture to \textsl{ordinary} products of these operators (as opposed to their normal-ordered products) can be developed. In this identity, the \textsl{retarded}\,\footnote{See \S\,8.3, p.\,125, in Ref.\,\protect\citen{HBKF04}. For fermions, see Eq.\,(7.62), p.\,77, in Ref.\,\protect\citen{FW03}.} non-interacting one-particle Green function takes the place of its time-ordered counterpart (which is proportional to the contraction of these operators\,\footnote{See Eqs\,(8.27) and (8.29), pp.\,88 and 89, in Ref.\,\protect\citen{FW03}.}) [p.\,182 in Ref.\,\citen{IZ80}]. For boson / fermion (field) operators, this follows from the identity\,\footnote{Contrast this with the identity $\mathcal{T}(\h{\varphi}(i)\h{\varphi}(j)) \equiv \mathcal{N}(\h{\varphi}(i)\h{\varphi}(j)) + \mathcal{C}_{i,j}\hspace{0.6pt}\h{1}$ in footnote \raisebox{-1.0ex}{\normalsize{\protect\footref{noted}}} on p.\,\protect\pageref{ForCompleteness}.}
\begin{equation}\label{ea1r}
\mathcal{T}(\h{\varphi}(i) \h{\varphi}(j)) \equiv \h{\varphi}(i) \h{\varphi}(j) \pm \Theta(t_j - t_i) [\h{\varphi}(j), \h{\varphi}(i)]_{\mp},
\end{equation}
where $\h{\varphi}(k)$ stands for either $\h{\psi}(k)$ or $\h{\psi}^{\dag}(k)$.

\subsection{Combinatorics}
\phantomsection
\label{saa2c}
In the considerations of this section, we assume that each operator in the product of $n$ operators to which the Wick operator identity, \S\,\ref{saa1}, is applied, is either an annihilation or a creation operator from the perspective of the adopted normal ordering operation (specified under the fourth general remark in \S\,\ref{saa2b}). By the linearity of $\mathcal{T}$ and $\mathcal{N}$,\footnote{Since $\mathcal{N}(\h{1}) = 0$, footnote \raisebox{-1.0ex}{\normalsize{\protect\footref{noted}}}, p.\,\protect\pageref{ForCompleteness}, the normal-ordering operation is more precisely \textsl{semi}-linear.} time- and normal-ordered products of arbitrary field operators can always be expressed as a linear superposition of respectively the time- and normal-ordered products of the latter type.

The number of terms in the Wick operator identity containing $k$ contractions is equal to\,\footnote{Compare with the parenthetic remarks on the first three lines of Eq.\,(4.10.1), p.\,93, in Ref.\,\protect\citen{MYS95}.}
\begin{equation}\label{ea1a}
\mathfrak{T}(n,k) \doteq \binom{n}{2k} (2k-1)!! \equiv \frac{n!}{2^k (n-2k)! k!},
\end{equation}
where the binomial coefficient $\binom{n}{2k}$ is the number of ways in which $2k$ distinct objects can be selected from amongst $n$ distinct objects (say, vertices in a graph), without regard to order, and $(2k -1)!! \doteq 1\cdot 3\ldots\cdot (2k-1)\equiv (2k)!/(2^k k!)$ the number of ways in which $2k$ distinct objects can be matched into $k$ disjoint pairs.\refstepcounter{dummy}\label{TheNumber}\,\footnote{The number $(2k-1)!!$, which in the present context can be established by induction, coincides with the number of \textsl{perfect matchings} \protect\cite{BM82,MA06} of the complete graph $K_{2k}$ \protect\cite{FH69,NN71,BM82,MA07}}\footnote{Note that $\int_{-\infty}^{\infty} \protect\rd x\, \exp(-ax^2/2)\hspace{0.7pt} x^{2k}/\int_{-\infty}^{\infty} \protect\rd x\, \exp(-ax^2/2) = (2k-1)!!/a^k$ for all $\re[a]> 0$ (\emph{cf.} Eq.\,(\protect\ref{ea1q}) below). The generalisation of this result for moments of multivariate Gaussian distribution functions is due to Isserlis \protect\cite{LI18}.} Clearly, the expression for $\mathfrak{T}(n,k)$ on the RHS of Eq.\,(\ref{ea1a}) correctly yields $\mathfrak{T}(n,0) = 1$ and $\mathfrak{T}(n,k) = 0$ for integer values of $k$ satisfying $k > \lfloor n/2\rfloor$, where $\lfloor x\rfloor$, the floor function,\footnote{See appendix \protect\ref{sae}.} yields the greatest integer less than or equal to $x$. Further, for $n$ \textsl{even}, $\mathfrak{T}(n,n/2) = (n-1)!!$ appropriately coincides with the number of fully contracted terms.\footnote{After the completion of the present work, it came to our attention that the details of this paragraph are presented under Definition 1.35, p.\,15, of Ref.\,\protect\citen{SJ97}.}

From the expression on the RHS of Eq.\,(\ref{ea1a}), for the total number of terms $\mathfrak{T}_n$ on the RHS of the Wick operator identity corresponding to the time-ordered product of $n$ operators, that is for (see Table \ref{t1})\,\footnote{The function $\mathfrak{T}_n$ is also equal to the number of connections a telephone exchange can offer to $n$ subscribers in pairs, with no provision of conference circuit. See Problem 17, p.\,85, in Ref.\,\protect\citen{JR02}.}
\begin{equation}\label{ea1c}
\mathfrak{T}_n \doteq \sum_{k=0}^{\lfloor n/2\rfloor} \mathfrak{T}(n,k),
\end{equation}
one obtains [pp.\,85 and 86 in  Ref.\,\citen{JR02}]
\begin{equation}\label{ea1d}
\mathfrak{T}_n = \Big(\frac{-\ii}{\sqrt{2}}\Big)^n H_n(\ii/\sqrt{2}) \equiv (-\ii)^n \mathnormal{H\hspace{-1.2pt}e}_n(\ii),
\end{equation}
where $H_n(z)$ and $\mathnormal{H\hspace{-1.2pt}e}_n(z)$ are the $n$th-order Hermite polynomials [Ch.\,22, p.\,771, in Ref.\,\citen{AS72}]\,\footnote{See in particular the entries 22.5.18 and 22.5.19, p.\,778, in Ref.\,\protect\citen{AS72}.} whose significance in the context of the Wick theorem we shall briefly discuss in \S\,\ref{saa6}. The validity of the result in Eq.\,(\ref{ea1d}) is trivially verified on the basis of the exact expression [\S\,22.3.11, p.\,775, in Ref.\,\citen{AS72}]
\begin{equation}\label{ea1da}
\mathnormal{H\hspace{-1.2pt}e}_n(x) = \sum_{k=0}^{\lfloor n/2\rfloor} (-1)^k\hspace{0.6pt}
\mathfrak{T}(n,k)\hspace{0.6pt} x^{n-2k}.
\end{equation}
To leading order one has [p.\,86 in  Ref.\,\citen{JR02}]\,\footnote{On consulting the work by Chowla \emph{et al.} \protect\cite{CHM51} [Theorem 8, p.\,333, in Ref.\,\citen{CHM51}], cited by Riordan \protect\cite{JR02}, it becomes evident that $\mathfrak{T}_n$ is also `the number of solutions of $x^2 = 1$ in $S_n$, the symmetric group of degree $n$', that is the number of elements of order $2$, or \textsl{involutions}, in $S_n$, appendix \protect\ref{sac}. An element in $S_n$ is of order $k$ if its $k$th power is equal to the identity element of $S_n$.}
\begin{equation}\label{ea1e}
\mathfrak{T}_n \sim \frac{1}{\sqrt{2}} \e^{\sqrt{\textrm{n}} -1/4}\hspace{0.7pt} \Big(\frac{n}{\textrm{e}}\Big)^{n/2}\;\; \text{for}\;\; n\to\infty.
\end{equation}
We note that some contractions are identically vanishing, and, in diagrammatic expansions, some non-vanishing contractions correspond to disconnected diagrams, \S\,\ref{sec1a}.

\begin{table}[t!]
\caption{The total number of terms $\mathfrak{T}_n$, Eq.\,(\protect\ref{ea1c}), on the RHS of the Wick operator identity corresponding to a time-ordered product of $n$ operators. For $n$ even, $\mathfrak{T}(n,n/2) = (n-1)!!$ is the total number of fully-contracted terms.}
\vspace{3pt}
\label{t1}
\begin{center}
\begin{tabular}{ccccccccccccc}
\vspace{0pt}
$n$ & 0 & 1 & 2 & 3 & 4 & 5 & 6 & 7 & 8 & 9 & 10 & \ldots\\
\hline
\vspace{3pt}
$\mathfrak{T}_n$ & 1 & 1 &2 & 4 & 10 & 26 & 76 & 232 & 764 & 2620 & 9496 & \ldots\\
$(n-1)!!$              & --&--&1 &-- & 3   & -- & 15 & --   & 105 & --     & 945   & \ldots\\
\vspace{3pt}
\end{tabular}
\end{center}
\end{table}

\subsection{The Hermite polynomials and the Gaussian integrals}
\phantomsection
\label{saa6}
The connection between the Wick decomposition theorem, \S\,\ref{saa7a} below, concerning commuting fields and the Hermite polynomials $\{\mathnormal{H\hspace{-1.2pt}e}_{m}(x) \| m\in \mathds{N}_0\}$, encountered in Eqs\,(\ref{ea1d}) and (\ref{ea1da}), has been discussed in detail by Glimm and Jaffe, \cite{GJ87} Janson \cite{SJ97}, and Simon \cite{BS74}, and succinctly reviewed by Wurm and Berg \cite{WB08}. These polynomials satisfy the orthonormality relation [\S\S\,22.1.1,2, 22.2.15, pp.\,773, 774, in Ref.\,\citen{AS72}]\footnote{See also \S\,5.5, p.\,105, in Ref.\,\protect\citen{GS75}.}
\begin{equation}\label{ea1q}
\int_{-\infty}^{\infty} \rd x\; \frac{\e^{-x^2/2}}{\sqrt{2\pi}} \hspace{1.2pt} \mathnormal{H\hspace{-1.2pt}e}_{l}(x) \hspace{1.2pt} \mathnormal{H\hspace{-1.2pt}e}_{m}(x) = l!\hspace{0.8pt} \delta_{l,m},
\end{equation}
where $\e^{-x^2/2}/\sqrt{2\pi}$ is the \textsl{standard} Gaussian probability distribution function, corresponding to the mean $\upmu =0$ and the variance $\upsigma^2 =1$ [\S\,26.2.1, p.\,931, in Ref.\,\citen{AS72}].\footnote{In the context of Gaussian Hilbert spaces \protect\cite{GJ87,SJ97,BS74}, the orthogonality relationship in Eq.\,(\ref{ea1q}) is referred to as the Wiener, or chaos, decomposition; every Gaussian Hilbert space induces an orthogonal decomposition of the corresponding square integrable random variables that are measurable \protect\cite{PRH74} with respect to the $\sigma$-field generated by the Gaussian Hilbert space (see in particular Ch.\,2 in Ref.\,\protect\citen{SJ97}).}

Consider the Schr\"{o}dinger-picture \textsl{boson} operators $\{\h{b}, \h{b}^{\dag}\}$ satisfying the canonical commutation relations\,\footnote{A similar analysis for the Schr\"{o}dinger-picture fermion operators $\{\h{f},\h{f}^{\dag}\}$ is not meaningful on account of $\h{f}^n = \h{0}$ for $n \ge 2$.}
\begin{equation}\label{ea1g}
[\h{b},\h{b}^{\dag}]_- = \h{1},\;\; [\h{b},\h{b}]_- = [\h{b}^{\dag},\h{b}^{\dag}]_- = \h{0},
\end{equation}
and describing the Hamiltonian
\begin{equation}\label{ea1ga}
\hspace{6.0pt}\h{\hspace{-6.0pt}\mathpzc{H}} = \frac{1}{2} \hbar\omega\hspace{0.6pt} \big(\h{b}^{\dag} \h{b} + \h{b} \h{b}^{\dag}\big)
\end{equation}
of the quantum harmonic oscillator corresponding to frequency $\omega$. Denoting the vacuum state of $\h{b}$ by $\vert\phi_{\X{0}}\rangle$, so that
\begin{equation}\label{ea1i}
\h{b}\hspace{0.6pt} \vert\phi_{\X{0}}\rangle = 0,
\end{equation}
from the expression on the RHS of Eq.\,(\ref{ea1ga}), making use of the first commutation relation in Eq.\,(\ref{ea1g}), one obtains
\begin{equation}\label{ea1h}
\hspace{6.0pt}\h{\hspace{-6.0pt}\mathpzc{H}}\hspace{0.6pt} \vert\phi_{\X{0}}\rangle = \frac{1}{2}\hbar\omega\hspace{0.6pt} \vert\phi_{\X{0}}\rangle,
\end{equation}
so that $\vert\phi_{\X{0}}\rangle$ is the zero-particle eigenstate of $\hspace{6.0pt}\h{\hspace{-6.0pt}\mathpzc{H}}$. In the light of the notation elsewhere in this appendix, $\vert\phi_{\X{0}}\rangle$ may thus be denoted by $\vert 0\rangle$.

With reference to Eq.\,(\ref{ea1i}), the normal (or Wick \cite{GJ87,SJ97,BS74}) ordering\,\footnote{Wick ordering plays a role in the process of renormalization of renormalizable field theories, although in general this ordering scheme is not sufficient for the task. For orientation and details, consult Refs.\,\protect\citen{VR91,MS99,FKT02}.} of an $n$-product of the operators $\{\h{b},\h{b}^{\dag}\}$ with respect to $\vert\phi_{\X{0}}\rangle$ amounts to effecting a permutation in the positions of these operators, placing the creation operators to the left of the annihilation operators. Thus, for the Hermitian operator
\begin{equation}\label{ea1j}
\h{q} \doteq \h{b} + \h{b}^{\dag},
\end{equation}
one has
\begin{equation}\label{ea1k}
\h{q}^2 = \h{b}^{\dag\hspace{0.6pt}2} + \h{b}^2 + \h{b}\hspace{0.6pt} \h{b}^{\dag} + \h{b}^{\dag} \h{b} \equiv \h{b}^{\dag\hspace{0.6pt}2} + \h{b}^2 + 2 \h{b}^{\dag} \h{b} + [\h{b},\h{b}^{\dag}]_- \equiv \mathcal{N}(\h{q}^{2}) + \h{1},
\end{equation}
from which one obtains
\begin{equation}\label{ea1ka}
\mathcal{N}(\h{q}^{2}) = \h{q}^2 - \h{1}.
\end{equation}
Similarly, one obtains
\begin{equation}\label{ea1kb}
\h{q}^3 = \mathcal{N}(\h{q}^3) + 3\hspace{0.6pt}\h{q},
\end{equation}
or equivalently
\begin{equation}\label{ea1kc}
\mathcal{N}(\h{q}^3) = \h{q}^3 -  3\hspace{0.6pt}\h{q}.
\end{equation}
Along the same lines as above, one arrives at
\begin{equation}\label{ea1k1}
\h{q}^4 = \mathcal{N}(\h{q}^4) + 6 \hspace{0.6pt}\mathcal{N}(\h{q}^2) + 3\hspace{0.6pt} \h{1},
\end{equation}
which in combination of the equality in Eq.\,(\ref{ea1ka}) results in
\begin{equation}\label{ea1kd}
\mathcal{N}(\h{q}^4) = \h{q}^4 -6 \h{q}^2 + 3\hspace{0.6pt} \h{1}.
\end{equation}
In this way, on the basis of the recurrence relation [\S\,22.7.14, p.\,782, in Ref.\,\citen{AS72}]
\begin{equation}\label{ea1m}
\mathnormal{H\hspace{-1.2pt}e}_{n+1}(x) = x \hspace{1.2pt} \mathnormal{H\hspace{-1.2pt}e}_{n}(x) - n \hspace{1.2pt} \mathnormal{H\hspace{-1.2pt}e}_{n-1}(x),
\end{equation}
combined with [\S\S\,22.4.8, 22.5.18, pp.\,777, 778, in Ref.\,\citen{AS72}]
\begin{equation}\label{ea1n}
\mathnormal{H\hspace{-1.2pt}e}_{0}(x) = 1,\;\; \hspace{1.2pt} \mathnormal{H\hspace{-1.2pt}e}_{1}(x) = x,
\end{equation}
and $\h{q}^0 = \h{1}$, for a general $n$ one deduces that \cite{GJ87,WB08}
\begin{equation}\label{ea1k1a}
\mathcal{N}(\h{q}^n) = \mathnormal{H\hspace{-1.2pt}e}_{n}(\h{q}),
\end{equation}
where $\mathnormal{H\hspace{-1.2pt}e}_{n}(x)$ is the Hermite polynomial as encountered in Eqs\,(\ref{ea1d}) and (\ref{ea1da}). The relationship in Eq.\,(\ref{ea1k1a}) is equivalent to the following explicit one \cite{GJ87,WB08} based on the binomial expansion of $(\h{b}+\h{b}^{\dag})^n$ subject to the limitation that in each constituent term $\h{b}^{\dag}$ is to stand to the left of $\h{b}$:\,\footnote{\textsl{Defining} $\mathcal{N}(\h{1}) = \h{1}$ (see footnote \raisebox{-1.0ex}{\normalsize{\protect\footref{noted}}} on p.\,\protect\pageref{ForCompleteness}), the equality in Eq.\,(\protect\ref{ea1k1b}) applies on account of the linearity of the normal-ordering operation and the fact that $\{\h{b},\h{b}^{\dag}\}$ are bosonic operators.}
\begin{equation}\label{ea1k1b}
\mathcal{N}(\h{q}^n) = \sum_{l=0}^n \binom{n}{l} (\h{b}^{\dag})^{n-l} (\h{b})^l.
\end{equation}

In the light of the equality in Eq.\,(\ref{ea1k1a}), and of the recurrence relation in Eq.\,(\ref{ea1m}), one has
\begin{equation}\label{ea1mb}
\h{q}\hspace{0.6pt}\mathcal{N}(\h{q}^n) = \mathcal{N}(\h{q}^{n+1}) + n\hspace{0.6pt} \mathcal{N}(\h{q}^{n-1}),
\end{equation}
so that by induction one deduces that
\begin{equation}\label{ea1k2}
\h{q}^n = \sum_{k=0}^{\lfloor n/2\rfloor} \mathfrak{T}(n,k)\hspace{0.7pt} \mathcal{N}(\h{q}^{n-2k})
\equiv \sum_{k=0}^{\lfloor n/2\rfloor} \mathfrak{T}(n,k)\hspace{0.7pt}
\mathnormal{H\hspace{-1.2pt}e}_{n-2k}(\h{q}),
\end{equation}
where $\mathfrak{T}(n,k)$ is defined in Eq.\,(\ref{ea1a}). For simplicity of notation, in the cases of $n$ \textsl{even} the first equality in Eq.\,(\ref{ea1k2}) relies on the \textsl{definition} $\mathcal{N}(\h{1}) \doteq \h{1}$.\footnote{As we have indicated earlier (see footnote \raisebox{-1.0ex}{\normalsize{\protect\footref{noted}}} on p.\,\protect\pageref{ForCompleteness}), the correct equality is $\mathcal{N}(\h{1}) = \h{0}$.} For $\h{q}$ identified with the $c$-number $x \h{1}$, the right-most expression in Eq.\,(\ref{ea1k2}) amounts to the expansion of $x^n$ in terms of the Hermite polynomials.\footnote{Employing the expansion $x^n = \sum_{k=0}^{\lfloor n/2\rfloor} \alpha_{n,k} \mathnormal{H\hspace{-1.2pt}e}_{n-2k}(x)$, where the lower bound of the sum reflects the fact that $\mathnormal{H\hspace{-1.2pt}e}_{m}(x)$ is a polynomial of order $m$, on the basis of the orthogonality relation in Eq.\,(\protect\ref{ea1q}) and the closed forms of the integrals 7.376.2 and 7.376.3 on p.\,804 of Ref.\,\protect\citen{GR07}, noting that $\mathnormal{H\hspace{-1.2pt}e}_{m}(x) \equiv \mathnormal{H}_m(x/\sqrt 2)/2^{m/2}$ and that $\mathnormal{H}_m(x)$ is an even / odd function of $x$ for even / odd integer values of $m$, one arrives at $\alpha_{n,k} \equiv \mathfrak{T}(n,k)$ for $k \in\{0,1,\dots,\lfloor n/2\rfloor\}$. See also Table 22.12, p.\,801, in Ref.\,\protect\citen{AS72}.}

For the time-independent operator $\h{q}$, one can define
\begin{equation}\label{ea1k3}
\mathcal{T}(\h{q}^2) = \h{q}^2,
\end{equation}
so that, following Eq.\,(\ref{ea1ka}), for the contraction
\begin{equation}\label{ea1k4}
\h{\mathcal{C}} \doteq \mathcal{T}(\h{q}^2) - \mathcal{N}(\h{q}^2)
\end{equation}
one obtains (see Eq.\,(\ref{ea1ka}))
\begin{equation}\label{ea1k5}
\h{\mathcal{C}} = \h{1}.
\end{equation}
In the light of this result, and $\mathcal{T}(\h{q}^n) = \h{q}^n$, the first equality in Eq.\,(\ref{ea1k2}) amounts to the Wick operator identity for the Hermitian operator $\h{q}^n \equiv \mathcal{T}(\h{q}^n)$. The coefficient $\mathfrak{T}(n,k)$ is hereby equal to the number of $n$-products consisting of the normal-ordered products of $n-2k$ operators, and $k$ contractions of the remaining $2k$ operators (see \S\,\ref{saa1}).

In introducing the number $\mathfrak{T}(n,k)$ in \S\,\ref{saa2c}, we explicitly considered the Wick operator identity corresponding to a product of $n$ operators, with operator being either a creation or an annihilation operator. Interestingly, while the above considerations are based on the operator $\h{q}$, which is neither a creation nor an annihilation operator with respect to the underlying vacuum state $\vert\phi_{\X{0}}\rangle$, nonetheless the operator identity in Eq.\,(\ref{ea1k2}) has reproduced the combinatorial factor $\mathfrak{T}(n,k)$ corresponding to the Wick operator identity subject to the above-mentioned restriction. To shed light on this observation, consider $\mathpzc{C}(\h{b}\hspace{0.6pt}\h{b})$, $\mathpzc{C}(\h{b}^{\dag}\h{b})$, $\mathpzc{C}(\h{b}\hspace{0.6pt}\h{b}^{\dag})$, and $\mathpzc{C}(\h{b}^{\dag}\h{b}^{\dag})$ as denoting the relevant contractions, for which one has
\begin{equation}\label{ea1k6}
\mathpzc{C}(\h{b}\hspace{0.6pt}\h{b}) = \mathpzc{C}(\h{b}^{\dag}\h{b}) = \mathpzc{C}(\h{b}^{\dag}\h{b}^{\dag}) = \h{0},\;\; \mathpzc{C}(\h{b}\hspace{0.6pt}\h{b}^{\dag}) = [\h{b},\h{b}^{\dag}]_- \equiv \h{1}.
\end{equation}
One observes that from the possible four distinct contractions of $\{\h{b},\h{b}^{\dag}\}$, only one is non-vanishing,\footnote{Compare with the results in for instance Eq.\,(8.22), p.\,88, of Ref.\,\protect\citen{FW03}.} and it is further identical to the contraction $\h{\mathcal{C}}$ of $\h{q}^2 \equiv \h{q}\hspace{0.6pt}\h{q}$ in Eq.\,(\ref{ea1k5}).

\subsection{The Wick decomposition theorem}
\phantomsection
\label{saa7a}
In the previous sections of this appendix we considered the Wick \textsl{operator identity}, relating a `time'-ordered product of canonical (field) operators in the interaction picture to a superposition of terms, each consisting of products of the contractions of an even number of these operators times the normal-ordered product of the remaining operators. In many applications of theoretical and practical interest, it is not the time-ordered products of the above-mentioned operators that are directly relevant, but their \textsl{expectation values} with respect to uncorrelated many-body states,\footnote{For the generalisation of the Wick theorem concerning expectation values of the time-ordered product of canonical (field) operators with respect to arbitrary many-body states, the reader is referred to Refs\,\protect\citen{PAH90} and \protect\citen{vLS12}. See also Refs\,\protect\citen{SF69} and \protect\citen{SF71}, and appendix H, p.\,298, in Ref.\,\protect\citen{PD84}. For the generalisation of the Wick theorem for the \textsl{matrix elements} of the time-ordered products of operators with respect to uncorrelated many-body states, consult Ref.\,\protect\citen{BB69}. For a detailed exposition of the underlying theory of non-unitary Bogoliubov transformations, consult Ref.\,\protect\citen{BR86}.}\footnote{We note that the considerations in Ref.\,\protect\citen{BF02}, with regard to the asymptotic behaviour of the time-Fourier transform of the self-energy operator $\Sigma$ for fermions at large values of the absolute value of the energy $\varepsilon$ (reciprocal to time $t$), can be greatly simplified by relying on the formalisms in Refs\,\protect\citen{PAH90} and \protect\citen{vLS12}. In this connection, we note that the GS correlation function $\Gamma^{\protect\X{(n)}}$ (appendix B, p.\,1538, in Ref.\,\protect\citen{BF02}) is directly related to the interacting $n$-particle Green function $G_n$ (see Eq.\,(5.1), p.\,125, in Ref.\,\protect\citen{SvL13}), and the GS correlation function $\Gamma_{\textsc{s}}^{\protect\X{(n)}}$ (appendix C, p.\,1544, in Ref.\,\citen{BF02}) to $G_{n;\protect\X{0}}$, the non-interacting counterpart of $G_n$. The expression for $\Gamma_{\textsc{s}}^{\X{(n)}}$ in terms of a determinant of the single-particle (Slater-Fock) density matrices (appendix C is Ref.\,\protect\citen{BF02}) is a direct consequence of the conventional Wick decomposition theorem (Eq.\,(5.27), p.\,135, in Ref.\,\protect\citen{SvL13}).} or their ensemble averages, notably the averages with respect to the equilibrium ensemble of states. Considering the expectation values of the time-ordered products of canonical (field) operators in the interaction picture with respect to \textsl{normalised} uncorrelated many-body states, one can, through the application of an appropriate canonical transformation of the (field) operators, achieve that the uncorrelated state under consideration is rendered the vacuum state of the new (transformed) annihilation (field) operators.  In this connection, the decomposition of the interaction-picture canonical field operators $\h{\psi}(j)$ and $\h{\psi}^{\dag}(j)$ into respectively $\h{\psi}_{\pm}^{\phantom{\dag}}(j)$ and $\h{\psi}_{\pm}^{\dag}(j)$, Eq.\,(\ref{ea1s}), corresponds to a unitary canonical transformation of this kind.\footnote{The doubling of the field operators here corresponds to a doubling of the underlying operators in the occupation-number representation (examples of this can be seen in Eqs\,(7.34) and (37.1), pp.\,70 and 326, of Ref.\,\protect\citen{FW03}). The latter operators are comprised of the \textsl{linear} combinations of the original creation and annihilation operators, subject to the condition that (a) these linear combinations amount to a canonical linear transformation of the original operators, and (b) the underlying uncorrelated $N$-particle state is the vacuum state of all the new annihilation operators in the occupation-number representation. We shall discuss this subject in some detail in Ref.\,\protect\citen{BF19}. For now we only mention that in quantum electrodynamics the field operators $\h{\psi}_{\pm}^{\phantom{\dag}}$ and $\h{\psi}_{\pm}^{\dag}$ (more precisely, for spinor fields, $\psi_{\pm}$ and $\b{\psi}_{\pm}$, where $\b{\psi} \doteq \psi^{\dag} \upgamma_4$ is the adjoint spinor, with $\upgamma_4 \equiv \protect\ii \upgamma_0$, where $\upgamma_0$ is the anti-Hermitian Dirac matrix in the Pauli representation) corresponding to electrons (as well as the field operators corresponding to the radiation field) are described in terms of contour integrals (see Eqs\,(1.47) and (1.48) in Ref.\,\protect\citen{JS49}).} Since contractions of canonical operators in the interaction picture are $c$-numbers, application of the Wick operator identity, in which the normal ordering operation normal orders the new (field) operators, leads to the equality of the expectation value of the time-ordered operator product under consideration with a superposition of the contributions associated with the fully-contracted terms in the Wick operator identity. \emph{This equality is also generally referred to as the Wick theorem, or the Wick decomposition theorem.} The perturbation series expansions for $G(a,b)$ and $G_{\X{2}}(a,b;c,d)$, in respectively Eqs\,(\ref{e242}) and (\ref{e200}), are directly related to the Wick \textsl{decomposition} referred to here.

As regards the ensemble averages of the time-ordered products of the canonical (field) operators in the interaction picture, the Wick \textsl{decomposition} theorem applies to ensembles characterised by the density operators (or statistical operators) $\h{\rho}$ expressible as \cite{PD84}
\begin{equation}\label{ea5f}
\h{\rho} = \frac{\exp(\wh{\mathcal{A}})}{\Tr[\exp(\wh{\mathcal{A}})]},
\end{equation}
where $\wh{\mathcal{A}}$ stands for a one-particle operator\,\footnote{That is, a second-quantised operator that is quadratic in the field operators.} that satisfies $\Tr[\exp(\wh{\mathcal{A}})] < \infty$ and commutes with the total-number operator $\wh{N} = \sum_{\sigma} \wh{N}_{\sigma}$, Eq.\,(\ref{e11d}). The grand-canonical density operator $\h{\varrho}_{\X{0}}$ corresponding to the non-interacting thermodynamic Hamiltonian $\wh{\mathcal{K}}_{\X{0}}$, Eq.\,(\ref{e17a}), falls into this category of density operators, for which one has $\wh{\mathcal{A}} \equiv -\beta\hspace{0.6pt} \wh{\mathcal{K}}_{\X{0}}$.\footnote{With reference to Eq.\,(\protect\ref{e11f}), $\wh{H}_{\X{0}}$ commutes not only with $\wh{N}$, but also with $\wh{N}_{\sigma}$, $\forall\sigma$.} Notably, in Ref.\,\citen{PD84} it has been shown how on effecting an appropriate limit in the parameters specifying the single-particle operator $\wh{\mathcal{A}}$, one can achieve that the average of an operator in the ensemble of states specified by the $\h{\rho}$ in Eq.\,(\ref{ea5f}) reduces to the expectation value with respect to \textsl{any} $N$-particle eigenstate of $\wh{H}_{\X{0}}$ (or equivalently $\h{\mathcal{K}}_{\X{0}}$), including its GS $\vert\Phi_{N;0}\rangle$.

For the Wick decomposition of the ensemble averages of the time-ordered products of canonical operators in the interaction picture corresponding to ensemble of states characterised by the density operator $\h{\rho}$ in Eq.\,(\ref{ea5f}), with $\wh{\mathcal{A}}$ as specified above, introduction of the process of normal ordering is \textsl{redundant}. In this connection, the original approach by Matsubara \cite{TM55}\footnote{Matsubara's approach is based on that by J.\,L. Anderson \protect\cite{JLA54} in quantum electrodynamics.} (as opposed to the approaches by Bloch and De Dominicis \cite{BdD58}, and Gaudin \cite{MG60}) to Wick \textsl{decomposition} of the thermal ensemble average of the product of operators, \textsl{unnecessarily} relies on a normal ordering operation.\footnote{In his original publication, Matsubara \cite{TM55} erroneously concluded that the relevant Wick \textsl{decomposition} were exact only in the thermodynamic limit. This error was rectified in a subsequent publication by Thouless \protect\cite{DJT57}. For a comprehensive discussion of this problem, consult the work by Evans and Steer \protect\cite{ES96}. See also the \textsl{historical note} in Ref.\,\protect\citen{Note2}.}

In the light of the considerations regarding the TFD formalism in \S\S\,\ref{sac01} and \ref{s226}, we note that the Wick decomposition theorem within this formalism has been explicitly discussed in Refs\,\citen{PAH90} and \citen{RB94} (see also Ref.\,\citen{XT90}).

\subsection{Pfaffians and Hafnians}
\phantomsection
\label{saa7}
The Wick theorem underlying the expressions in Eqs\,(\ref{e242}) and (\ref{e200}), describing respectively the one- and two-particle Green function in terms of permanents / determinants in the case of bosons / fermions, is equivalent to the formulation of this theorem in terms of Hafnians \cite{DLRV70,ERC59}\footnote{Also written as `Haffnian', or `haffnian', in analogy with Pfaffian.} / Pfaffians \cite{NW78,MA06,TM60,VD99,ERC59}, as presented in Ref.\,\citen{IZ80}.\footnote{See pp.\,182 and 185 herein.} This equivalence can be established with the aid of the formal expansion of permanents / determinants in terms of Hafnians / Pfaffians \cite{DLRV70}. As regards the relationship between determinants and Pfaffians, one has the following two relevant theorems:
\begin{itemize}
\item[(1)] An arbitrary determinant of order $n$ can be expressed as a Pfaffian of the same order [\S\,418, p.\,396, in Ref.\,\citen{TM60}] [Theorem, p.\, 77, in Ref.\,\citen{VD99}], and
\item[(2)] An arbitrary determinant of order $2n$ can be expressed as a Pfaffian of order $n$ [\S\,417, p.\,395, in Ref.\,\citen{TM60}] [Theorem, p.\,78, in Ref.\,\citen{VD99}].\footnote{For skew-symmetric determinants of order $2n$, see the theorem on p.\,75 of Ref.\,\protect\citen{VD99}.}
\end{itemize}
For completeness, consider the array
\begin{equation}\label{ea1f}
\mathpzc{A} \doteq \{a_{i,j} \| 1 \le i < j \le 2k\}.
\end{equation}
The Hafnian \cite{DLRV70,ERC59} and Pfaffian \cite{NW78,MA06,TM60,VD99,ERC59,EHL68} of $\mathpzc{A}$ are defined as\,\footnote{See in particular Eqs\,(4.3.13), (4.3.14), and (4.3.15) on p.\,73 of Ref.\,\protect\citen{VD99}. See also Ref.\,\protect\citen{DC09}.}\footnote{It has been pointed out [p.\,v in Ref.\,\protect\citen{VD99}] that while Pfaffians are often tacitly assumed to correspond to matrices in texts on linear algebra, the correspondence is an unnecessarily restrictive one. See our later reference to $\protect\Pf(\mathbb{A})$, where $\mathbb{A}$ is the skew-symmetric $2k\times 2k$ matrix whose upper diagonal part is comprised of $\mathpzc{A}$.}
\begin{equation}\label{ea2a}
\Hf(\mathpzc{A}) \doteq \sum_{\bm{\mu}} a_{\bm{\mu}},
\end{equation}
\begin{equation}\label{ea2b}
\Pf(\mathpzc{A}) \doteq \sum_{\bm{\mu}} \sign(\bm{\mu})\hspace{0.6pt} a_{\bm{\mu}},
\end{equation}
where $\sum_{\bm{\mu}}$ denotes summation over all \textsl{matchings}
\begin{equation}\label{ea1by}
\bm{\mu} \equiv (i_1 j_1, i_2j_2, \dots, i_k j_k),\;\; 1 \le i_s < j_s \le k,\;\, 1 \le s \le k,
\end{equation}
and
\begin{equation}\label{ea1bz}
a_{\bm{\mu}} \doteq a_{i_1,j_1} a_{i_2,j_2} \dots a_{i_k,j_k}.
\end{equation}
Further, in Eq.\,(\ref{ea2b}) $\sign(\bm{\mu})$ denotes the signature\,\footnote{The signature $\sign(\bm{\mu})$ is easily determined graphically (see the figures on p.\,209 of Ref.\,\protect\citen{MA07}).} of the $2k$-permutation $\mathscr{P}_{\bm{\mu}}^{\X{(2k)}}$ (\emph{cf.} Eq.\,(\ref{e330}) below)
\begin{equation}\label{ea1b}
\mathscr{P}_{\bm{\mu}}^{\X{(2k)}} \doteq
\begin{pmatrix}
1 & 2 & 3 & 4 & \ldots & 2k-1 & 2k \\
i_1 & j_1 & i_2 & j_2 & \ldots & i_k & j_k
\end{pmatrix}\hspace{-0.6pt},
\end{equation}
of which there are $(2k-1)!!$ distinct ones \cite{DC09}.\footnote{See also p.\,209 in Ref.\,\citen{MA07}.} With [p.\,258 in Ref.\,\citen{AS72}]
\begin{equation}\label{ea1bwa}
(2k-1)!! = \frac{2^k}{\sqrt{\pi}}\hspace{0.6pt}\Gamma(k+1/2),
\end{equation}
to leading order one has [\S\,6.1.39, p.\,257, in Ref.\,\citen{AS72}]
\begin{equation}\label{ea1bw}
\frac{(2k -1)!!}{(2k)!} \sim \frac{1}{2^{k+1/2}\hspace{0.6pt}k!}\sim
\frac{1}{\sqrt{\pi\!\e}} \Big(\frac{\e/2}{k}\Big)^{k+1/2}\;\;\text{for}\;\; k\to\infty,
\end{equation}
where $\e = \ln^{-1}(1) = 2.718\,\dots\,$.

By assuming the array $\mathpzc{A}$ in Eq.\,(\ref{ea1f}) to constitute the upper diagonal part of the \textsl{skew-symmetric} $2k \times 2k$ matrix $\mathbb{A}$,\footnote{See for instance Eq.\,(4.3.25), p.\,76, in Ref.\,\protect\citen{VD99}.} one can alternatively denote $\Pf(\mathpzc{A})$ by $\Pf(\mathbb{A})$. In this light, the observation in Eq.\,(\ref{ea1bw}) is interesting in that for the $2k\times 2k$ skew-symmetric matrix $\mathbb{A}$ at hand one has $(\Pf(\mathbb{A}))^2 = \det(\mathbb{A})$.\footnote{According to a theorem by Cayley [Theorems 4.9 and 4.10, p.\,66, and the Theorem on p.\,75 of Ref.\,\protect\citen{VD99}; see also Ref.\,\protect\citen{JHH66}], for the determinant of an $n\times n$ \textsl{skew-symmetric} matrix $\mathbb{M}_n$ one has $\det(\mathbb{M}_n) = (\Pf(\mathbb{M}_n))^2$ when $n$ \textsl{even} \protect\cite{EHL68}, and $\det(\mathbb{M}_n) = 0$ when $n$ \textsl{odd}. For  an algorithmic approach towards construction of Pfaffians, the reader is referred to Ref.\,\protect\citen{DLRV70}.} In this connections, while the \textsl{explicit} evaluation of $\det(\mathbb{A})$ involves a summation over $(2k)!$ terms (the number of $2k$-permutations), that of $\Pf(\mathbb{A})$ involves a summation over $(2k-1)!!$ terms.

We note in passing that in the perturbational treatment of the Anderson impurity model by Yosida and Yamada \cite{YY70I,YY75III} and Yamada \cite{KY75II}, the authors employ the Wick decomposition theorem for fermion operators in terms of Pfaffians. In this connection, we remark that in these works the authors exploit an anti-symmetry property of the underlying non-interacting Green function that in general does not obtain.\footnote{See for instance the equalities in Eq.\,(2.10), p.\,972, of Ref.\,\protect\citen{KY75II}.}

We close this section by presenting an interesting result concerning Pfaffians. To this end, let $\{\h{C}_j \| j =1,\dots,m\}$, with $m$ an \textsl{even} integer, be a set of linear operators on an $M$-dimensional vector space, with $M < \infty$. Let further
\begin{equation}\label{ea2}
\big[\h{C}_i, \h{C}_j\big]_{+} = a_{i,j} \h{\mathscr{I}}, \;\; i\not= j,
\end{equation}
where $\{a_{i,j}\| i,j\}$ are real or complex constants, and $\h{\mathscr{I}}$ the identity operator in the said vector space, for which one has $\Tr[\h{\mathscr{I}}] = M$. With (\emph{cf.} Eq.\,(\ref{ea1f}))
\begin{equation}\label{ea2x1}
\mathpzc{A}_m \doteq\{ a_{i,j} \| 1 \le i < j \le m\},
\end{equation}
one has [Lemma 1 in Ref.\,\citen{EHL68}]
\begin{equation}\label{ea2x2}
\frac{1}{M} \Tr[\h{C}_1 \h{C}_2\dots\h{C}_m] = \frac{1}{2^{m/2}} \Pf(\mathpzc{A}_m).
\end{equation}
This result is interesting in particular because for $\{\h{f}_i^{\phantom{\dag}},\h{f}_i^{\dag} \| i = 1,2,\dots,m\}$ a set of \textsl{canonical} fermion annihilation and creation operators, the operators $\{\h{C}_i \| i=1,2, \dots,m\}$ defined according to
\begin{equation}\label{ea2x3}
\h{C}_i \doteq \h{f}_i^{\phantom{\dag}} + \frac{1}{2} \sum_{\substack{j=1 \\ j\not= i}}^m a_{i,j}\hspace{0.6pt} \h{f}_j^{\dag},\;\; i =1,2,\dots, m,
\end{equation}
where $a_{i,j} \doteq (\mathbb{A}_m)_{i,j}$, with $\mathbb{A}_m$ an $m\times m$ \textsl{symmetric} matrix, satisfy the equality in Eq.\,(\ref{ea2}) \cite{EHL68}. For the dimension $M$ of the vector space at hand, one has $M = 2^m$.\footnote{With $\h{f}_i \vert 0\rangle = 0$, $\forall i$, in the occupation-number representation for a normalised $N$-particle state of the fermions at hand one has $\vert n_1,n_2,\dots n_m\rangle = (\h{f}_1^{\dag})^{n_1} (\h{f}_2^{\dag})^{n_2}\dots (\h{f}_m^{\dag})^{n_m}\vert 0\rangle$, where $n_i \in \{0,1\}$ and $\sum_{i=1}^m n_i = N$ [\S\,1.2, p.\,4, in Ref.\,\protect\citen{NO98}]. The dimension of the Fock space corresponding to $N=0,1,\dots,m$ is thus immediately seen to be equal to $M = 2^m$.}

\subsection{Closing remarks}
\phantomsection
\label{saa9}
While the Wick theorem applies to canonical (field) operators,\footnote{Generally, for operators in the interaction picture. For specific details, see however \S\,\protect\ref{saa2b}.} it some instances it can be fruitfully employed for calculating the correlation functions of other operators. One prominent example of such application of the Wick theorem is encountered in the work by Lieb, Schultz, and Mattis \cite{LSM61} concerning antiferromagnetic linear chains of quantum spin-$\tfrac{1}{2}$ operators with nearest-neighbour interaction,\refstepcounter{dummy}\label{InRef}\,\footnote{In Ref.\,\protect\citen{LSM61} two distinct one-dimensional models with nearest-neighbour interaction have been considered: the (anisotropic) $XY$-model, and the Heisenberg-Ising model. In the latter model, the coupling between the spin-$\tfrac{1}{2}$ operators is alternately of the Heisenberg and the Ising type.} where these operators are represented in terms of canonical fermions operators.\footnote{These canonical fermion operators, $\{\h{c}_i^{\phantom{\dag}},\h{c}_i^{\dag}\| i\}$, are constructed in two steps. To highlight these, with $\{\h{S}_i^{\alpha}\| \alpha = x, y, z\}$ denoting the Cartesian components of the spin-$\tfrac{1}{2}$ operators, for the matrix representation of these operators one has $\mathbb{S}_i^{\alpha} = \tfrac{1}{2} \bbsigma^{\alpha}$, where $\{\bbsigma^{\alpha}\|\alpha = x,y,z\}$ are the $2\times 2$ Pauli matrices satisfying $[\bbsigma^{\alpha},\bbsigma^{\beta}]_- = \epsilon_{\alpha\beta\gamma}\hspace{0.6pt} \bbsigma^{\gamma}$, where $\epsilon_{\alpha\beta\gamma}$ is the Levi-Civita symbol, equal to $+1$ ($-1$) for $\alpha = x$, $\beta=y$, $\gamma = z$ and the even (odd) permutations of these, and $0$ otherwise. In the first step, one defines $\h{a}_i^{\phantom{\dag}} \doteq \h{S}_i^x -\protect\ii \h{S}_i^y \iff \h{a}_i^{\dag} \doteq \h{S}_i^x + \protect\ii \h{S}_i^y$, which, following the commutation relations of the Pauli matrices, can be shown to behave as canonical fermion (boson) operators for equal (unequal) indices. In the second step, the above-mentioned canonical fermion operators $\{\h{c}_i^{\phantom{\dag}},\h{c}_i^{\dag}\| i\}$ are obtained from $\{\h{a}_i^{\phantom{\dag}},\h{a}_i^{\dag}\| i\}$ by means of the Jordan-Wigner transformation  \protect\cite{JW28,LSM61,SML64,HAK64}. With $\h{S}_i^x = (\h{a}_i^{\dag} + \h{a}_i^{\phantom{\dag}})/2$, $\h{S}_i^y = (\h{a}_i^{\dag} - \h{a}_i^{\phantom{\dag}})/2\protect\ii$, and $\h{S}_i^z = \h{a}_i^{\dag}\h{a}_i^{\phantom{\dag}} -\tfrac{1}{2}$, expressing $\{\h{a}_i^{\phantom{\dag}},\h{a}_i^{\dag}\| i\}$ in terms of $\{\h{c}_i^{\phantom{\dag}},\h{c}_i^{\dag}\| i\}$, one thus arrives at the expression for $\h{S}_i^{\alpha}$, $\alpha = x, y, z$, in terms of the latter canonical fermion operators. For relevant details, the reader may also consult Refs\,\protect\citen{DCM88} and \protect\citen{EF13}.}\footnote{We note that correlation functions of quantum spin operators can also be similarly dealt with on the basis of the Holstein-Primakoff \cite{HP40} representation of these operators in terms of canonical boson operators (see also \S\,3.11, p.\,88, and \S\,5.11, p.\,181, in Ref.\,\protect\citen{DCM88}; compare the Bogoliubov transformation in Eq.\,(5.178), p.\,183, of the latter reference, with that in Eq.\,(A.5) of Ref.\,\protect\citen{GA59}).} Hereby, the correlation functions of the former operators are expressed in terms of those of the latter operators. Another example concerns the two-dimensional Ising model on a rectangular lattice and subject to the periodic boundary condition, dealt with by Schultz, Mattis, and Lieb \cite{SML64}. Here, the density operator $\h{\rho}$ of an $N$-row lattice is related, by means of the transfer matrix, to that of the $(N-1)$-row lattice, which is expressed in terms of the spin operators of the last row.\footnote{In Ref.\,\protect\citen{SML64}, the Cartesian components of the spin operators corresponding to the $n$th row and $m$th column of the rectangular $N \times M$ lattice are denoted by $\{\sigma_{nm}^{\alpha}\| \alpha = x, y, z\}$. For $n = N-1$, $\sigma_{nm}^{\alpha}$ is denoted by $\sigma_{m}^{\alpha}$. As in the case of the one dimensional $XY$ model, here the operator $\sigma_{m}^{\alpha}$, $\alpha = x, y, z$, is expressed in terms of canonical fermion operators. Note that here $\sigma_{m}^{\alpha}$ stands for the operator $\h{S}_m^{\alpha}$ referred to in the previous footnote.} Following the representation of $\sigma_{nm}^x$, with $n=N-1$, in terms of canonical fermion operators (deduced along the same lines as in the case of the one-dimensional $XY$ model, referred to above), making use of the Wick theorem, the GS expectation value of\,\footnote{As indicated in the previous footnote, $\sigma_{m}^{\alpha} \doteq \sigma_{nm}^{\alpha}\vert_{n=N-1}$.} $\sigma_{m\phantom{'}}^x\!\! \sigma_{m'}^x$ is determined and expressed in terms of the determinant of an $\vert m-m'\vert \times \vert m-m'\vert$ Toeplitz matrix.\footnote{See also Ref.\,\protect\citen{MPW63}.} The GS correlation function of $\sigma_{nm\phantom{'}}^x\!\! \sigma_{nm'}^x$ in the limit $N, M\to \infty$ is subsequently expressed in terms of a superposition of two correlation functions corresponding to $\sigma_{m\phantom{'}}^x\!\! \sigma_{m'}^x$ and $\sigma_{m\phantom{'}}^y\!\! \sigma_{m'}^y$, which are expressible in terms of two determinants introduced by Kaufman and Onsager \cite{KO49}.\footnote{See Eqs\,(44) and (45) in Ref.\,\protect\citen{KO49}.}
$\hfill\Box$

%\addtocontents{toc}{\protect\setcounter{tocdepth}{2}}

\refstepcounter{dummyX}
\section{The connected and disconnected Green-function diagrams}
\phantomsection
\label{sac}
As we have discussed in \S\S\,\ref{s221} and \ref{s222}, the denominator of the expression in Eq.\,(\ref{e242}) is responsible for the suppression of the contributions of disconnected diagrams in the weak-coupling perturbation series expansion of the one-particle Green function $G(a,b)$ in terms of $(v,G_{\X{0}})$.\footnote{The same applies to the two-particle Green function as expressed in Eq.\,(\protect\ref{e200}).} One therefore obtains the same perturbation series expansion for $G(a,b)$ by identifying $D_{\nu}$ with zero for all $\nu \in\mathds{N}$ and simultaneously explicitly discarding the contributions in the numerator of the expression in Eq.\,(\ref{e242}) that correspond to disconnected Green-function diagrams. In this appendix we discuss an approach whereby the latter contributions are identified and thus marked for disposal. Although this approach brings one back to the conventional diagrammatic perturbation series expansion of $G(a,b)$ in terms of $(v,G_{\X{0}})$, nonetheless, since the discussions of this appendix shed light on the computational complexity of the conventional diagrammatic approach, we believe that these discussions are not out of place in this paper, in particular because they amount to a very novel practical approach in regard to diagrammatic many-body expansions. Strictly speaking however, \emph{the discussions in this appendix are not essential to those in the main body of this paper.} In the closing part of this appendix, we present some programs, written in the programming language of Mathematica$^{\X{\circledR}}$, that implement the approach presented in this appendix.\footnote{$\copyright$ 2019, 2021 \textsf{All methods, algorithms and programs presented in this appendix, as well as elsewhere in this publication, are intellectual property of the author. Any commercial use of these without his written permission is strictly prohibited. All academic and non-commercial uses of the codes in this publication, or modifications thereof, must be appropriately cited. The same restrictions apply to the contents of the Mathematica Notebook that we publish alongside this paper.}}

On identifying $D_{\nu}$ with zero for all $\nu \in \mathds{N}$, following Eq.\,(\ref{e247}), one has $F_{\nu} = 0$ for all $\nu \in \mathds{N}$. Hence, in the light of Eq.\,(\ref{e248a}),
\begin{equation}\label{e319}
G^{\X{(\nu)}}(a,b) = \mathcal{M}_{\X{0};\nu}(a,b),\;\; \forall \nu\in \mathds{N},
\end{equation}
where (\emph{cf.} Eq.\,(\ref{e259b}))
\begin{align}\label{e320}
\mathcal{M}_{\X{0;\nu}}(a,b) &\doteq \sum_{r=1}^{2\nu}\int \rd r\;\mathfrak{M}_{\X{0;\nu}}(r,r) \hspace{0.6pt}G_{\X{0}}(a,r^+) G_{\X{0}}(r,b) \nonumber \\
&+ \sum_{\substack{r,s=1 \\ r\not= s}}^{2\nu} \int \rd r\rd s \;\mathfrak{M}_{\X{0;\nu}}(r,s) \hspace{0.6pt}G_{\X{0}}(a,s^+) G_{\X{0}}(r,b),
\end{align}
in which\,\footnote{For $r=s$ the condition $j\not= r,s$ is to be understood as denoting $j\not= r$, or $j\not= s$.}
\begin{equation}\label{e320d}
\mathfrak{M}_{\X{0;\nu}}(r,s) \doteq \pm\frac{1}{\nu !} \Big(\frac{\ii}{2\hbar}\Big)^{\nu}
\int \prod_{\substack{ j=1 \\ j\not= r,s }}^{2\nu} \rd j\; \mathsf{v}(1,2) \dots \mathsf{v}(2\nu-1,2\nu)
\hspace{0.6pt}\mathcal{A}_{\X{0};r,s}^{\X{(2\nu-1)}}(1,2,\dots,2\nu-1,2\nu).
\end{equation}
The function $\mathcal{A}_{\X{0};r,s}^{\X{(2\nu-1)}}$ is obtained from the function $A_{r,s}^{\X{(2\nu-1)}}$ in Eqs\,(\ref{e259b}) and (\ref{e254x}) by discarding the contributions associated with disconnected Green-function diagrams. Below we first describe an approach for identifying these contributions, and subsequently present the relevant practical details in the form of programs written in the Mathematica$^{\X{\circledR}}$ programming language.

In the light of the diagrammatic expansion of $G(a,b)$, one can convince oneself that the function $\mathfrak{M}_{\X{0;\nu}}(r,s)$ need not be calculated for all values of $r, s \in\{1,2,\dots,2\nu\}$, and that calculations corresponding to merely $(r,s) = (1,1)$, $(1,2)$, and $(1,3)$ suffice.\footnote{More generally, corresponding to respectively $(r,s) = (j,j)$, $\forall j \in \{1,2,\dots,2\nu\}$, $(r,s) = (2j-1,2j)$ or $(2j,2j-1)$, $\forall j\in \{1,2,\dots,\nu\}$, and $(r,s) = (2j,j')$ or $(2j-1,j')$, with $j'$ satisfying respectively $j'\not=2j-1, 2j$ and $j'\not= 2j-1, 2j$, $\forall j\in \{1,2,\dots,\nu\}$ and $\forall j'\in \{1,2,\dots,2\nu\}$.} The pair $(r,s) = (1,1)$ corresponds to the set of connected diagrams that are linked through the directed lines representing $G_{\X{0}}(a,1^+)$ and $G_{\X{0}}(1,b)$ to the external vertices $a$ and $b$ (\textsl{collectively}\,\footnote{Note that for $\nu=1$ the transition from Eq.\,(\ref{e250}) below to Eq.\,(\ref{e250b}) below involves an infinite summation, whereby $G_{\protect\X{0}}$ is transformed into $G$. Hence the use here of the qualification \textsl{collective}.} describing the local Hartree self-energy $\Sigma_{\X{01}}^{\textsc{h}}[v,G]$, Eq.\,(\ref{e250b}) below), the pair $(r,s) = (1,2)$ to the diagrams shunted by the line representing $\mathsf{v}(1,2)$ (\textsl{collectively} describing the Fock, or the \textsl{exchange}, self-energy diagram $\Sigma_{\X{01}}^{\textsc{f}}[v,G]$, Eq.\,(\ref{e250b}) below), and $(r,s) = (1,3)$ to all the remaining diagrams. In the light of these observations, a simple enumeration yields\,\footnote{Note that $\mathfrak{M}_{\X{0;\nu}}(2j-1,2j)$ is a simple multiple of $\mathsf{v}(2j-1,2j)$, $\forall j\in \{1,2,\dots,\nu\}$.}\footnote{One appropriately has $2\nu + 2\nu  + 2\nu (2\nu-2) \equiv (2\nu)^2$.}
\begin{align}\label{e320e}
\mathcal{M}_{\X{0;\nu}}(a,b) &= 2\nu \int \rd 1\;\mathfrak{M}_{\X{0;\nu}}(1,1) \hspace{0.6pt}G_{\X{0}}(a,1^+) G_{\X{0}}(1,b) \nonumber\\
&+ 2\nu \int \rd 1\rd 2 \;\mathfrak{M}_{\X{0;\nu}}(1,2) \hspace{0.6pt}G_{\X{0}}(a,2^+) G_{\X{0}}(1,b)\nonumber\\
&+ 2\nu (2\nu-2)  \int \rd 1\rd 3 \;\mathfrak{M}_{\X{0;\nu}}(1,3) \hspace{0.6pt}G_{\X{0}}(a,3^+) G_{\X{0}}(1,b).
\end{align}
\emph{We must emphasise that while the $1$ in the first expression on the RHS of Eq.\,(\ref{e320e}) can be replaced by any integer from the set $\{1,2,\dots,2\nu\}$, the $1$ and $2$ in the second expression, and the $1$ and $3$ in the third one, \textsl{cannot}.}\footnote{These limitations are necessary for avoiding \textsl{any} renumbering of the pairs $(2j-1,2j)$, $j \in \{1,2,\dots,\nu\}$, in the expression on the RHS of Eq.\,(\protect\ref{e320d}), that would render in particular the programs to be introduced later in this appendix unnecessarily complex and intransparent. Any renumbering that would amount to a permutation of the $\nu$ pairs $(1,2), \dots, (2\nu-1,2\nu)$, followed possibly by swaps inside the pairs, $(2j-1,2j) \rightharpoonup (2j,2j-1)$, is in principle harmless and therefore permitted.} The $1$ and $2$ in the second expression can however be replaced by respectively $2j-1$ and $2j$, or $2j$ and $2j-1$, $\forall j\in\{1,2,\dots \nu\}$, and the $1$ and $3$ in the third expression by respectively $j$ and $j' \in \mathfrak{S}_j^{\X{(\nu)}}$, where\,\footnote{See appendix \protect\ref{sae}.}
\begin{equation}\label{e320f}
\mathfrak{S}_j^{\X{(\nu)}} \doteq \left\{\begin{array}{ll} \{1,2,\dots,2\nu\} \backslash \{j, j+1\}, & j = \text{odd},\\ \\ \{1,2,\dots,2\nu\} \backslash \{j-1, j\}, & j = \text{even}.\end{array}\right.
\end{equation}
For clarity, for instance once $\mathfrak{M}_{\X{0;\nu}}(1,3)$ has been calculated, one can replace for instance $3$ by, say, $2$. Doing so prior to the calculation would \textsl{incorrectly} imply that the $2$ on the RHS of Eq.\,(\ref{e320d}) were to be replaced by $3$. We note that the simplified expression for $\mathcal{M}_{\X{0;\nu}}(a,b)$ in Eq.\,(\ref{e320e}) is a direct consequence of the formalism under consideration corresponding to \textsl{two-body} interaction potentials. There are similar, but not identical, simplifications possible for $n$-body interaction potentials, with $n>2$.

In anticipation of what follows, let
\begin{equation}\label{e321}
\mathscr{P}^{\X{(n)}} \equiv \big\{\mathscr{P}_l^{\X{(n)}}\| l=1,2,\dots, n!\big\}
\end{equation}
denote the permutation group on a set of $n$ elements.\footnote{Generally referred to as the \textsl{symmetric group} \protect\cite{SR12} on $n$ elements, and denoted by $S_n$.} With $P_l^{\X{(n)}}(i)$ denoting the permutation function associated with the group element $\mathscr{P}_l^{\X{(n)}}$ defined on $\{1,2,\dots,n\}$, one has
\begin{equation}\label{e322}
P_l^{\X{(n)}}(i) = j,\;\; \text{where}\;\; i, j \in \{1,2,\dots,n\},\; l \in \{1,2,\dots,n!\}.
\end{equation}
The function $P_l^{\X{(n)}}(i)$ is an automorphism of $\{1,2,\dots,n\}$. In the following, $\upsigma_l^{\X{(n)}}$ will stand for unity in the case of bosons, and for the signature of $\mathscr{P}_l^{\X{(n)}}$ in the case of fermions, with $\upsigma_l^{\X{(n)}} = \pm 1$ for even / odd permutations of the \textsl{ordered} set $\{1,2,\dots,n\}$. This specification of $\upsigma_l^{\X{(n)}}$ reflects the fact that the function $A_{r,s}^{\X{(2\nu-1)}}$ in Eq.\,(\ref{e254x}) corresponds to a permanent \cite{HM78,NW78,MA06} in the case of bosons, and a determinant \cite{TM60,VD99} in the case of fermions (\emph{cf.} Eqs\,(\ref{e260}) and (\ref{e254x})).

With\refstepcounter{dummy}\label{ForDeterminants}\,\footnote{For determinants, see p.\,4 in Ref.\,\protect\citen{VD99}. As regards permanents, here the difference with the case of determinants is restricted to the definition of $\upsigma_l^{\X{(2\nu)}}$, to be encountered in Eq.\,(\protect\ref{e324}) below, which in the case of permanents takes the value $+1$ for all $l$.}\footnote{See appendix \protect\ref{sae}.}
\begin{equation}\label{e323}
\mathcal{G}_{\X{0};r,s}(i,j) \doteq \left\{\begin{array}{cc} G_{\X{0}}(i,j), & i\not=r \wedge j \not= s,\\ \\
0, & (i=r \wedge j\not=s) \vee (i\not= r \wedge j=s),\\ \\
1, & i=r \wedge j=s,\end{array} \right.
\end{equation}
one can express the cofactor\,\footnote{See the footnote on p.\,\protect\pageref{InRefVD99} regarding `cofactor', in particular in relation to permanents.} $A_{r,s}^{\X{(2\nu-1)}}$ as encountered in the defining expression for $M_{\nu}(a,b)$, Eq.\,(\ref{e259b}), as
\begin{equation}\label{e324}
A_{r,s}^{\X{(2\nu-1)}}(1,2,\dots,2\nu) = \sum_{l=1}^{(2\nu)!} \upsigma_l^{\X{(2\nu)}} \hspace{0.8pt}\Psi_{\X{0};r,s}^{\X{(l;\nu)}}(1,2,\dots,2\nu),
\end{equation}
where
\begin{equation}\label{e325}
\Psi_{\X{0};r,s}^{\X{(l;\nu)}}(1,2,\dots,2\nu) \doteq \prod_{j=1}^{2\nu} \mathcal{G}_{\X{0};r,s}(j,P_{l}^{\X{(2\nu)}}(j)^+).
\end{equation}
For the superscripts $+$ on the RHS, see Eqs\,(\ref{e258}) and (\ref{e260}). With reference to Eq.\,(\ref{e323}), we note that for expressing a $(2\nu-1)$-permanent / -determinant as a $2\nu$-permanent / -determinant, we have identified the external vertices $a$ and $b$, thereby suppressing the indices $a$ and $b$ and equating the resulting Green-function line $G_{\X{0}}(r,s^+)$ with $1$ (compare the first and the third line on the RHS of Eq.\,(\ref{e323})). We further note that the function $\Psi_{\X{0};r,s}^{\X{(l;\nu)}}$ consists of a product of $2\nu -1$ non-interacting Green functions, leading to the integrand of the function $\mathfrak{M}_{\X{0;\nu}}(r,s)$ in Eq.\,(\ref{e320d}) consisting of a product of $2\nu+1$ non-interacting Green functions, conform the fact that a $\nu$th-order Green-function diagram consists of $2\nu+1$ lines representing Green functions.

In the light of the expression in Eq.\,(\ref{e325}), and on account of the middle entry on the RHS of Eq.\,(\ref{e323}), one notes that\,\footnote{Using De Morgan's law \protect\cite{PP13} $\neg (\mathpzc{a}\vee \mathpzc{b}) \equiv (\neg \mathpzc{a}) \wedge (\neg \mathpzc{b})$.}
\begin{equation}\label{e325e}
\mathcal{G}_{0;r,s}(j,P_l^{\X{(2\nu}}(j)^+) \not=0\;\;\text{when}\;\; (j\not=r \vee P_l^{\X{(2\nu)}}(j) = s) \wedge (j = r \vee P_l^{\X{(2\nu)}}(j) \not= s).
\end{equation}
Thus, writing the expression in Eq.\,(\ref{e325}) as\,\footnote{With reference to the third line on the RHS of Eq.\,(\protect\ref{e323}), the function $\mathcal{G}_{\protect\X{0};r,s}(r,P_{l}^{\protect\X{(2\nu)}}(r)^+)$ on the second line of the RHS of Eq.\,(\protect\ref{e325c}) is identically equal to $1$ for $l \in S_{r,s}^{\protect\X{(\nu)}}$, Eq.\,(\protect\ref{e325a}) below.}
\begin{equation}\label{e325c}
\Psi_{\X{0};r,s}^{\X{(l;\nu)}}(1,2,\dots,2\nu) = \left\{\begin{array}{ll} \!\!\!\mathcal{G}_{\X{0};r,r}(r,P_{l}^{\X{(2\nu)}}(s)^+) \prod_{\substack{ j=1 \\ j\not=r }}^{2\nu} \mathcal{G}_{\X{0};r,r}(j,P_{l}^{\X{(2\nu)}}(j)^+), &
\hspace{-2.6cm}\text{for}\;\; r=s, \\ \\
\!\!\!\mathcal{G}_{\X{0};r,s}(r,P_{l}^{\X{(2\nu)}}(r)^+) \hspace{0.8pt}\mathcal{G}_{\X{0};r,s}(s,P_{l}^{\X{(2\nu)}}(s)^+)
\prod_{\substack{ j=1 \\ j\not=r, s}}^{2\nu} \mathcal{G}_{\X{0};r,s}(j,P_{l}^{\X{(2\nu)}}(j)^+), & {}\\
{} & \hspace{-2.6cm}\text{for}\;\; r\not=s, \end{array}\right.
\end{equation}
on the basis of the observation in Eq.\,(\ref{e325e}), it follows that the equality in Eq.\,(\ref{e324}) can be equivalently expressed as
\begin{equation}\label{e325b}
A_{r,s}^{\X{(2\nu-1)}}(1,2,\dots,2\nu) = \sum_{l \in S_{r,s}^{\X{(\nu)}}} \upsigma_l^{\X{(2\nu)}} \hspace{0.8pt}\Psi_{\X{0};r,s}^{\X{(l;\nu)}}(1,2,\dots,2\nu),
\end{equation}
where (see Eq.\,(\ref{e330}) below)
\begin{equation}\label{e325a}
S_{r,s}^{\X{(\nu)}} \doteq
\big\{l \hspace{1.0pt}\big\|\hspace{1.0pt} P_l^{\X{(2\nu)}}(r) = s ,\hspace{0.8pt} l \in \{1,2,\dots,(2\nu)!\} \big\}.
\end{equation}
The simplicity of the defining expression for $S_{r,s}^{\X{(\nu)}}$ is mainly a consequence of the multiplicative nature of the expression for $\Psi_{\X{0};r,s}^{\X{(l;\nu}}$ in Eq.\,(\ref{e325}), and of the permutation $P_l^{\X{(2\nu))}}(j)$ being an automorphism of $\{1,2,\dots,2\nu\}$ for all values of $l \in \{1,2,\dots,(2\nu)!\}$. Evidently, the set $S_{r,s}^{\X{(\nu)}}$ consists of $\vert S_{r,s}^{\X{(\nu)}}\vert = (2\nu-1)!$ elements,\footnote{By fixing the image of $r$ to be $s$, we have effectively reduced a $2\nu$- to a $(2\nu-1)$-permutation.}\footnote{This number is not to be confused with the \emph{recontres} number $D_{2\nu,1}$ [pp.\,57-65 in Ref.\,\protect\citen{JR02}], which is equal to $1$ for $\nu=1$ and satisfies $D_{2\nu,1} > (2\nu-1)!$ for $\nu \ge 2$.} implying that the summation in Eq.\,(\ref{e325b}) explicitly discards $2\nu$ identically-vanishing summands of the summation on the RHS of Eq.\,(\ref{e324}). For later reference, we point out that in the light of the expressions in Eq.\,(\ref{e323}), for $l \in S_{r,s}^{\X{(\nu)}}$ the expression in Eq.\,(\ref{e325c}) can be recast as
\begin{equation}\label{e325d}
\Psi_{\X{0};r,s}^{\X{(l;\nu)}}(1,2,\dots,2\nu) =  \prod_{\substack{ j=1 \\ j\not=r }}^{2\nu} G_{\X{0}}(j,P_{l}^{\X{(2\nu)}}(j)^+),\;\; \forall l \in S_{r,s}^{\X{(\nu)}}.
\end{equation}

The sought-after expression for the function $\mathcal{A}_{\X{0};r,s}^{\X{(2\nu-1)}}$ is deduced from that in Eq.\,(\ref{e325b}) by discarding the summands on the RHS that correspond to disconnected Green-function diagrams. With $\mathcal{S}_{r,s}^{\X{(\nu)}}$ (not to be confused with $S_{r,s}^{\X{(\nu)}}$) denoting the set of values of $l$ corresponding to connected Green-function diagrams, by definition one has (\emph{cf.} Eq.\,(\ref{e320d}))
\begin{equation}\label{e327}
\mathcal{A}_{\X{0};r,s}^{\X{(2\nu-1)}}(1,2,\dots,2\nu) = \sum_{l\in \mathcal{S}_{r,s}^{\X{(\nu)}}} \upsigma_l^{\X{(2\nu)}} \hspace{0.8pt}\Psi_{\X{0};r,s}^{\X{(l;\nu)}}(1,2,\dots,2\nu),
\end{equation}
where clearly
\begin{equation}\label{e326}
\mathcal{S}_{r,s}^{\X{(\nu)}} \subset S_{r,s}^{\X{(\nu)}},\;\; \forall\nu \in \mathds{N}.
\end{equation}

Connected Green-function diagrams corresponding to given indices $r$ and $s$, and thereby to the given set $\mathcal{S}_{r,s}^{\X{(\nu)}}$, are conveniently characterised by expressing each permutation $\mathscr{P}_l^{\X{(2\nu)}}$, with $l \in S_{r,s}^{\X{(\nu)}}$, in terms of its cycles \cite{SR12}. Here, each cycle corresponds to a boson / fermion loop. Of these cycles one contains both $r$ and $s$ in the cases where $r\not=s$; in the cases where $r=s \equiv j$, $j \in \{1,2,\dots,2\nu\}$, $j$ necessarily forms a $1$-cycle. The loop to which $r$ and $s$ belong (or to which $j$ belongs in the case of $r=s \equiv j$) originates from the above-mentioned process of identifying the vertices marked by $a$ and $b$, followed by discarding these indices and equating with unity the Green-function line representing the function $G_{\X{0}}(r,s^+)$ thus brought about, Eq.\,(\ref{e323}).\footnote{See Fig.\,\protect\ref{f11} below, where the process of \textsl{identifying} the vertices $a$ and $b$, discarding the indices $a$ and $b$, and equating the resulting Green-function line $G_{\X{0}}(r,s^+)$ (represented by a broken line) with $1$, results in a \textsl{loop} of which one segment consists of a broken line and the remaining segments of solid lines.}

In view of the above remarks, let\,\refstepcounter{dummy}\label{UsingCombinatorica}\footnote{Using \textsl{Combinatorica} \protect\cite{PS03}, the cycle decomposition of a permutation is obtained by the command \texttt{ToCycles}. Further, the integers $m_l$ and $n_l^{\protect\X{(j)}}$ are both obtained with the aid of \texttt{Length}: with \texttt{Pl} $= \{P_l^{\protect\X{(2\nu)}}(1), \dots, P_l^{\protect\X{(2\nu)}}(2\nu)\}$, $m_l$ is obtained by \texttt{Length[Cl]}, and $n_l^{\protect\X{(j)}}$ by \texttt{Length[Cl][[j]]}, where \texttt{Cl} $=$ \texttt{ToCycles[Pl]}. \label{notee}}
\begin{equation}\label{e328}
\mathscr{P}_l^{\X{(2\nu)}} = \mathscr{C}_l^{\X{(1)}} \dots \mathscr{C}_l^{\X{(m_l)}}
\end{equation}
denote the cycle-decomposition \cite{SR12} of $\mathscr{P}_l^{\X{(2\nu)}}$, where $m_l$ is the number of the cycles corresponding to the $l$th $2\nu$-permutation of $\{1,2,\dots,2\nu\}$, satisfying $1 \le m_l \le 2\nu$. With $n_l^{\X{(j)}}$ denoting the length of the cycle $\mathscr{C}_l^{\X{(j)}}$, $j \in \{1,\dots,m_l\}$,\footnote{$\mathscr{C}_l^{\X{(j)}}$ is thus an $n_l^{\X{(j)}}$-cycle.} one has
\begin{equation}\label{e329}
\sum_{j=1}^{m_l} n_l^{\X{(j)}} = 2\nu,\;\;\forall l \in \{1,2,\dots, (2\nu)!\},
\end{equation}
so that in the case where $m_l = 1$, one has $n_l^{\X{(1)}} = 2\nu$, and in the case where $m_l = 2\nu$, $n_l^{\X{(j)}} = 1$ for all $j \in \{1,2,\dots,2\nu\}$.\footnote{The condition $n_l^{\protect\X{(j)}} = 1$, $\forall j$, applies only in the case of the identity permutation.} Thus $1 \le n_l^{\X{(j)}} \le 2\nu$ for all $j \in \{1,\dots,m_l\}$. To $\mathscr{C}_l^{\X{(j)}}$ one can associate the function $C_l^{\X{(j)}}(k)$, with $k$ varying over $\{1,\dots, n_l^{\X{(j)}}\}$, whereby in particular for each $l \in S_{r,s}^{\X{(\nu)}}$ one has\,\footnote{Note that the integer $r$ in the first row standing directly above the integer $s$ in the second row, is implied by the definition of $S_{r,s}^{\X{(\nu)}}$ in Eq.\,(\protect\ref{e325a}).} (\emph{cf.} Eq.\,(\ref{e325a}))
\begin{equation}\label{e330}
\mathscr{P}_l^{\X{(2\nu)}}:\;
\begin{pmatrix}
1 & 2 & \dots & r & \dots & 2\nu \\
P_l^{\X{(2\nu)}}(1) &  P_l^{\X{(2\nu)}}(2) & \dots & s & \dots & P_l^{\X{(2\nu)}}(2\nu)
\end{pmatrix}
= \prod_{j=1}^{m_l} \mathfrak{C}_l^{\X{(j)}},
\end{equation}
where
\begin{equation}\label{e331}
 \mathfrak{C}_l^{\X{(j)}} \doteq \big(C_l^{\X{(j)}}(1),\dots, C_l^{\X{(j)}}(n_l^{\X{(j)}})\big).
\end{equation}
The RHS of the equality in Eq.\,(\ref{e330}) describes the cycle decomposition of the permutation on the LHS, associated with the permutation group element $\mathscr{P}_l^{\X{(2\nu)}}$, here corresponding to some $l\in S_{r,s}^{\X{(\nu)}}$, Eq.\,(\ref{e325a}). We note in passing that the number of $n$-permutations that have precisely $k$ cycles is equal to $(-1)^{n-k}$ times the Stirling number of the first kind \cite{MA07,JR02}\footnote{See also Ch.\,24, p.\,821, in Ref.\,\protect\citen{AS72}.}\footnote{The present observation can be utilised to count the number of loops in the $\nu$th-order Feynman diagrams under discussion.} $\mathpzc{S}_{\hspace{0.3pt}n}^{\X{(k)}}$. With
\begin{equation}\label{e331a}
\mathpzc{s}_{\hspace{0.4pt}n,k} \doteq (-1)^{n-k} \mathpzc{S}_{\hspace{0.4pt}n}^{\X{(k)}},
\end{equation}
one has
\begin{equation}\label{e331b}
\sum_{k=0}^{n} \mathpzc{s}_{\hspace{0.4pt}n,k} = n!.
\end{equation}
In the case at hand, where $n =2\nu$,
\begin{equation}\label{e331c}
\sum_{l=1}^{(2\nu)!} \delta_{m^l,k} = \mathpzc{s}_{\hspace{0.4pt}\X{2\nu},k}.
\end{equation}
Multiplying both sided of this equality by $k^p$, where $p$ may in principle be integer, real or complex, and summing both sides of the resulting equality  with respect to $k$ over $\{0,1,\dots, 2\nu\}$, one obtains the following non-trivial sum-rule to be satisfied by $\{m^l \| l\}$:
\begin{equation}\label{e331d}
\upmu_{\X{\nu}}^{\X{(p)}} \doteq \sum_{l=1}^{(2\nu)!} (m^{l})^p = \sum_{k=0}^{2\nu}  \mathpzc{s}_{\hspace{0.4pt}\X{2\nu},k}\hspace{0.8pt} k^p.
\end{equation}
With reference to Eq.\,(\ref{e331b}), for $p = 0$ this sum-rule appropriately reduces to the identity $(2\nu)! \equiv (2\nu)!$.\footnote{One has $\upmu_{\X{1}}^{\X{(1)}} = 3$, $\upmu_{\X{2}}^{\X{(1)}} = 50$, $\upmu_{\X{3}}^{\X{(1)}} = 1764$, \dots, $\upmu_{\X{1}}^{\X{(2)}} = 5$, $\upmu_{\X{2}}^{\X{(2)}} = 120$, $\upmu_{\X{3}}^{\X{(2)}} = 5012$, \dots\,.}

\begin{figure}[t!]
\centerline{
\includegraphics*[angle=0, width=0.65\textwidth]{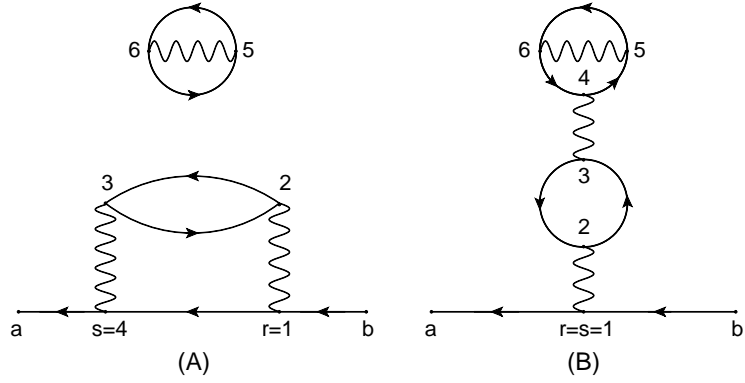}}
\caption{Two third-order Green-function diagrams corresponding to the perturbation series expansion of $G(a,b)$ in terms of $G_{\protect\X{0}}$ (solid line) and the two-particle interaction potential $\mathsf{v}$ (solid wavy line): (A) disconnected, (B) connected. The indices $r$ and $s$ are those featuring as subscripts in the cofactor $A_{r,s}^{\protect\X{(5)}}(1,2,\dots,6)$, Eqs\,(\protect\ref{e259b}) and (\protect\ref{e325b}). Note that, following the expression in Eq.\,(\protect\ref{e259b}), every interaction function is of the form $\mathsf{v}(2j-1,2j)$, $j=1, 2, 3$.}
\label{f12}
\end{figure}

For illustration, let us consider the two third-order Green-function diagrams in Fig.\,\ref{f12}. \emph{For the diagrammatic representation of $\Psi_{\X{0};r,s}^{\X{(l;\nu)}}$, as well as of $\Phi_l^{\X{(j)}}(r,s)$, Eq.\,(\ref{e335}) below, we adopt the common convention\,\footnote{See for instance Fig.\,9.7, p.\,99, in Ref.\,\protect\citen{FW03}.} according to which $G(i,j)$ is represented by a solid line connecting the vertices $i$ and $j$ and directed from $j$ to $i$.} Identifying the external vertices $a$ and $b$, followed by suppressing $a$ and $b$,\footnote{For this process, see Fig.\,\protect\ref{f11} below, p.\,\protect\pageref{f11x}.} for $\nu=3$ one has the following permutations associated with the diagrams A and B in Fig.\,\ref{f12}:\,\footnote{The cycles below are written in \textsl{canonical} order, with the largest element within each cycle appearing first. In this connection, permutations are invariant under the \textsl{cyclic} permutations of the elements in each of their cycles.}
\begin{equation}\label{e332}
\mathscr{P}_{l_1}^{\X{(6)}}:\;
\begin{pmatrix}
1 & 2 & 3 & 4 & 5 & 6 \\
4 & 3 & 2 & 1 & 6 & 5
\end{pmatrix}
= (4,1) (3,2) (6,5), \;\;\;\;\; \text{(A)}
\end{equation}
\begin{equation}\label{e333}
\mathscr{P}_{l_2}^{\X{(6)}}:\;
\begin{pmatrix}
1 & 2 & 3 & 4 & 5 & 6 \\
1 & 3 & 2 & 6 & 4 & 5
\end{pmatrix}
= (1) (3,2) (6,5,4), \;\;\;\;\; \text{(B)}
\end{equation}
where $l_1$ and $l_2$ are two integers from the set $\{1,2,\dots,6!\}$.\refstepcounter{dummy}\label{UsingTheLexicographic}\footnote{Using the lexicographic ranking system and identifying the rank of the identity permutation with $1$, one has $l_1 = 416$ and $l_2 = 29$. Using \textsl{Combinatorica} \protect\cite{PS03}, the lexicographic rank of the $n$-permutation \texttt{Pl} is obtained by applying the command \texttt{RankPermutation[Pl]+1}.} With reference to the equality in Eq.\,(\ref{e330}), one thus has\,\footnote{With reference to Eq.\,(\protect\ref{e331}), $C_{l_1}^{\protect\X{(1)}}(1) = 4$, $C_{l_1}^{\protect\X{(1)}}(2) = 1$, $C_{l_1}^{\protect\X{(2)}}(1) = 3$, $C_{l_1}^{\protect\X{(2)}}(2) = 2$, $C_{l_1}^{\protect\X{(3)}}(1) = 6$, and $C_{l_1}^{\protect\X{(3)}}(2) = 5$.}
\begin{equation}\label{e334}
\mathfrak{C}_{l_1}^{\X{(1)}} = (4,1),\; \mathfrak{C}_{l_1}^{\X{(2)}} = (3,2),\;
\mathfrak{C}_{l_1}^{\X{(3)}} = (6,5),
\end{equation}
so that $m_{l_1} = 3$, $n_{l_1}^{\X{(j)}} = 2$, for $j = 1,2,3$, Eq.\,(\ref{e329}). Similarly
\begin{equation}\label{e334a}
\mathfrak{C}_{l_2}^{\X{(1)}} = (1),\; \mathfrak{C}_{l_2}^{\X{(2)}} = (3,2),\;
\mathfrak{C}_{l_2}^{\X{(3)}} = (6,5,4),
\end{equation}
so that $m_{l_2} = 3$, $n_{l_2}^{\X{(1)}} = 1$, $n_{l_2}^{\X{(2)}} = 2$ and $n_{l_2}^{\X{(3)}} = 3$, Eq.\,(\ref{e329}).

To a given cycle $\mathfrak{C}_l^{\X{(j)}}$, Eq.\,(\ref{e330}), with $l\in S_{r,s}^{\X{(\nu)}}$, we associate the following function (\emph{cf.} Eqs\,(\ref{e323}) and (\ref{e325})):\,\footnote{We note that $\Phi_{\X{0};l}^{\protect\X{(j)}}(r,s)$ is invariant under the \textsl{cyclic} permutations of the elements of the $j$th cycle of $\mathscr{P}_l^{\X{(2\nu)}}$. Thus, for instance, one obtains the same function $\Phi_{\X{0};l_2}^{\X{(3)}}(1,1)$ by writing the cycle $\mathfrak{C}_{l_2}^{\protect\X{(3)}}$ in Eq.\,(\ref{e334a}) as $(4,6,5)$ and $(5,4,6)$.}
\begin{equation}\label{e335}
\Phi_{\X{0};l}^{\X{(j)}}(r,s) \doteq
\prod_{k=1}^{n_l^{\X{(j)}}-1} G_{\X{0}}(C_l^{\X{(j)}}(k), C_l^{\X{(j)}}(k+1)^+) \times \left\{\begin{array}{ll} 1, & r,s \in \mathfrak{C}_l^{\X{(j)}}, \\ \\
G_{\X{0}}(C_l^{\X{(j)}}(n_l^{\X{(j)}}), C_l^{\X{(j)}}(1)^+), & r,s \not\in \mathfrak{C}_l^{\X{(j)}}, \end{array} \right.
\end{equation}
where, by definition,
\begin{equation}\label{e335a}
\prod_{k=1}^{0} f_k \equiv 1.
\end{equation}
This definition is relevant for the cases where $n_l^{\X{(j)}} = 1$. With reference to Eqs\,(\ref{e330}) and (\ref{e331}), the arguments of the Green functions on the RHS of Eq.\,(\ref{e335}) are in accordance with the arguments of the function $\mathcal{G}_{r,s}^{\X{(0)}}$ on the RHS of Eq.\,(\ref{e325}).\refstepcounter{dummy}\label{ExplicitlyAssuming}\footnote{Explicitly, assuming $n_l^{\protect\X{(j)}} \ge 2$, $\forall j \in \{1,\dots,m_l\}$, for $k \in \{1,\dots,n_l^{\protect\X{(j)}}-1\}$ there exists a $j' \in \{1,2,\dots, 2\nu\}$ such that $(C_l^{\protect\X{(j)}}(k),C_l^{\protect\X{(j)}}(k+1)) \equiv (j',P_l^{\protect\X{(2\nu)}}(j'))$.} Following Eq.\,(\ref{e330}), in the light of the expressions in Eqs\,(\ref{e325}), (\ref{e325a}), and (\ref{e325d}), one has (see Fig.\,\ref{f12})
\begin{equation}\label{e336}
\Psi_{\X{0};r,s}^{\X{(l;\nu)}}(1,2,\dots,2\nu) = \prod_{j=1}^{m_{l}} \Phi_{\X{0};l}^{\X{(j)}}(r,s),\;\; l \in S_{r,s}^{\X{(\nu)}}.
\end{equation}
One readily verifies that for $\nu=3$ and $(r,s) = (1,4)$ ($(r,s) = (1,1)$) the function $\Psi_{\X{0};r,s}^{\X{(l_1;\nu)}}$ ($\Psi_{\X{0};r,s}^{\X{(l_2;\nu)}}$), constructed on the basis of the expressions in Eqs\,(\ref{e335}) and (\ref{e336}), indeed represents the contribution of diagram A (B) in Fig.\,\ref{f12} to the function $M_{\nu}(a,b)$ in Eq.\,(\ref{e259b}) through the expression for $A_{r,s}^{\X{(2\nu-1)}}$ in Eq.\,(\ref{e325b}).\footnote{Note the locations of $r$ and $s$ in the matrix representing the $\mathscr{P}_l^{\protect\X{(2\nu)}}$ in Eq.\,(\protect\ref{e330}).} Clearly, however, in the case at hand while for $(r,s) = (1,1)$, $l_2 \in S_{r,s}^{\X{(\nu)}}$, for $(r,s) = (1,4)$ one has $l_1 \not\in \mathcal{S}_{r,s}^{\X{(\nu)}}$, so that by definition the contribution of $\Psi_{\X{0;1,1}}^{\X{(l_1;\nu)}}(1,2,\dots,6)$ to $\mathcal{M}_{\X{0};\nu}(a,b)$ is suppressed through the expression for $\mathcal{A}_{\X{0};r,s}^{\X{(2\nu-1)}}$ in Eq.\,(\ref{e327}).

In the light of the above considerations, and with reference to the expression for the function $\mathcal{M}_{\X{0};\nu}(a,b)$ in Eq.\,(\ref{e320}), for a given pair $(r,s)$, $\forall r, s \in \{1,2,\dots,2\nu\}$, disconnected Green-function diagrams correspond to those values of $l$ in $S_{r,s}^{\X{(\nu)}}$, Eq.\,(\ref{e325a}), for which at least one cycle is \textsl{disconnected} from the remaining cycles on the RHS of Eq.\,(\ref{e330}). Here we define the cycle $\mathfrak{C}_l^{\X{(j)}}$, corresponding to fixed values of $l$ and $j$, Eq.\,(\ref{e331}), as being \textsl{disconnected} from the remaining cycles\,\footnote{All corresponding to the \textsl{same} value of $l$ but \textsl{different} values of $j$, assuming that $m_l > 1$.} in the cycle decomposition in Eq.\,(\ref{e330}) when for \textsl{all} $k\in \{1,\dots,n_l^{\X{(j)}}\}$
\begin{equation}\label{e336b}
\upnu(C_l^{\X{(j)}}(k)) = C_l^{\X{(j)}}(k') \;\; \text{for \textsl{some}}\;\; k' \in \{1,\dots,n_l^{\X{(j)}}\},
\end{equation}
where
\begin{equation}\label{e336a}
\upnu(n) \doteq n - (-1)^n.
\end{equation}
Conversely, the cycle $\mathfrak{C}_l^{\X{(j)}}$ is \textsl{connected} with at least another cycle $\mathfrak{C}_l^{\X{(j')}}$, $j'\not= j$ (with $j, j' \in \{1,\dots, m_l\}$), when for \textsl{some} $k\in \{1,\dots,n_l^{\X{(j)}}\}$
\begin{equation}\label{e336c}
\upnu(C_l^{\X{(j)}}(k)) \not= C_l^{\X{(j)}}(k')\;\;\text{for \textsl{all}}\;\; k' \in \{1,\dots,n_l^{\X{(j)}}\}.
\end{equation}
To visualise the idea underlying the above definition of \textsl{disconnected cycles}, one may proceed as follows. One first prepares a \textsl{primary} graph \cite{FH69}\footnote{For \textsl{permutation graphs}, see also Refs\,\protect\citen{PS03} and \protect\citen{JFM72} (pp.\,123-152 of the latter reference).} corresponding to a given pair $(r,s)$, $r, s \in\{1,2,\dots,2\nu\}$, consisting of $2\nu$ vertices, indexed by the integers $\{1,2,\dots,2\nu\}$, and $\nu$ edges \cite{FH69} connecting the pairs of vertices $(2j-1,2j)$, $j \in\{1,\dots,\nu\}$, representing the two-body interaction potentials in the expressions on the RHSs of Eqs\,(\ref{e259b}) and (\ref{e320d}). Identifying $l$ with an element of the set $S_{r,s}^{\X{(\nu)}}$ in Eq.\,(\ref{e325a}) (a process that is to be repeated for all elements of $S_{r,s}^{\X{(\nu)}}$), one completes the above primary graph by adding edges to it, with each edge representing a Green function encountered in the expression for the function $\Psi_{\X{0};r,s}^{\X{(l;\nu)}}$ in Eq.\,(\ref{e336}) (through the functions $\big\{\Phi_{\X{0};l}^{\X{(j)}}(r,s) \| j\in \{1,2,\dots,m_l\}\big\}$, Eq.\,(\ref{e335})), connecting the vertices that feature as argument of the relevant Green function. With reference to the expressions in Eqs\,(\ref{e335}) and (\ref{e336}), in this graph the $j$th relevant cycle is represented by the edges connecting the vertices $C_l^{\X{(j)}}(k)$ and $C_l^{\X{(j)}}(k+1)$, with $k \in \{1,\dots, n_l^{\X{(j)}}\}$; in the case of $r,s \not\in \mathfrak{C}_l^{\X{(j)}}$,\footnote{With reference to Eq.\,(\protect\ref{e330}), recall that $r \in \mathfrak{C}_l^{\X{(j)}}$ implies $s \in \mathfrak{C}_l^{\X{(j)}}$, and \emph{vice versa}.} the graph must also include an edge connecting the vertices $C_l^{\X{(j)}}(n_l^{\X{(j)}})$ and $C_l^{\X{(j)}}(1)$. We refer to the graph thus obtained as \textsl{secondary} graph. Since $l \in S_{r,s}^{\X{(\nu)}}$, with reference to Eqs\,(\ref{e259b}), (\ref{e325b}), and (\ref{e336}), this secondary graph represents the function (see Fig.\,\ref{f12})
\begin{equation}\label{e336d}
\mathsf{v}(1,2) \dots \mathsf{v}(2\nu -1,2\nu)\hspace{1.2pt} \Psi_{\X{0};r,s}^{\X{(l;\nu)}}(1,2,\dots,2\nu).
\end{equation}
On the basis of the secondary graph thus obtained, one immediately observes that for given values of $\nu\in \mathds{N}$, $l \in \{1,2,\dots,(2\nu)!\}$, and $r, s \in \{1,2,\dots, 2\nu\}$, indeed a secondary graph corresponds to a connected Feynman diagram contributing to $G^{\X{(\nu)}}(a,b)$, according to the expressions in Eqs\,(\ref{e319}), (\ref{e320}), and (\ref{e320d}), provided that the cycle decomposition in Eq.\,(\ref{e330}) is comprised of cycles of which none is \textsl{disconnected} from the rest according to the definition introduced above.

In view of the above observations, for given values of $\nu$, and $r, s \in \{1,2,\dots,2\nu\}$, the elements of the set $\mathcal{S}_{r,s}^{\X{(\nu)}}$ are determined by sequentially choosing the elements of the set $S_{r,s}^{\X{(\nu)}}$, Eq.\,(\ref{e325a}), determining the cycle decomposition of the relevant $2\nu$-permutation, Eq.\,(\ref{e330}), and testing the relevant cycles on connectivity according to the definition described above. With reference to the expression in Eq.\,(\ref{e320e}), we recall that one needs only to consider the sets $\mathcal{S}_{r,s}^{\X{(\nu)}}$ corresponding to the pairs $(r,s) = (1,1)$, $(1,2)$, and $(1,3)$.

In the closing part of this appendix, p.\,\pageref{GraphG}, the program \texttt{GraphG}, written in the programming language of Mathematica$^{\X{\circledR}}$, constructs a graph corresponding to the permutation \texttt{P}.\footnote{The input \texttt{P} is a \textsl{list} (in the Mathematica terminology), here of integers, enclosed by curly braces, representing the a permutation in \textsl{array notation}. Thus, for the permutations in Eqs\,(\protect\ref{e332}) and (\protect\ref{e333}) one has \texttt{P = $\{4, 3, 2, 1, 6, 5\}$} and \texttt{P = $\{1, 3, 2, 6, 4, 5\}$}, respectively.} When the latter \texttt{P} coincides with the $\mathscr{P}_l^{\X{(2\nu)}}$ corresponding to fixed values of $r$ and $s$, Eq.\,(\ref{e330}), the constructed graph represents a $\nu$th-order Feynman diagram that potentially contributes to $G^{\X{(\nu)}}(a,b)$. It contributes when this diagram is connected,\footnote{From the perspective of the graph theory \protect\cite{FH69}, the connectivity relevant here is the \textsl{weak} one.} a property that is established by \texttt{ConnG}, p.\,\pageref{ConnG}. This program calculates also the set $\mathcal{S}_{r,s}^{\X{(\nu)}}$, as well as some symmetry factors $\{\Lambda_{r,s}^{\X{(l;\nu)}}\| l\}$, to be introduced below, Eq.\,(\ref{e327a}). We note that in order to execute these and the subsequent programs presented in the closing part of this appendix, the package \protect\verb|Combinatorica| is to be loaded, using the instruction \protect\verb|Needs["Combinatorica`"]|, or \protect\verb|<< Combinatorica`|.\footnote{Reference \protect\citen{PS03} amounts to a detailed manual of this package. We point out that some functionalities of the package \texttt{Combinatorica} have been superseded by preloaded functionalities in later versions of Mathematica. Consequently, Mathematica documentations of these differ depending on the version number of the Mathematica package used.}\footnote{For some relevant details, see footnote \raisebox{-1.0ex}{\normalsize{\protect\footref{notee}}} on p.\,\protect\pageref{UsingCombinatorica}.}\footnote{We note that the following and subsequent programs have \textsl{not} been optimised, as doing so would diminish their transparency. In this connection, we note that parts of the present programs can be parallelised on computers equipped with multi-core processors through using the Mathematica instruction \texttt{Parallelize}.}

Having determined the set $\mathcal{S}_{r,s}^{\X{(\nu)}}$, corresponding to connected diagrams contributing to $G^{\X{(\nu)}}(a,b)$ upon being attached to lines representing $G_{\X{0}}(a,s^+)$ and $G_{\X{0}}(r,b)$ (implying integration with respect to $r$ and $s$), one can exploit a permutation symmetry in the diagrams of the same order\,\refstepcounter{dummy}\label{ForARelevant}\footnote{For a relevant discussion based on diagrams, see item 3 on pp.\,96 and 97 of Ref.\,\protect\citen{FW03}. The symmetry that we utilise here is however \textsl{not} identical to that in Ref.\,\protect\citen{FW03}, as well as elsewhere. This is related to the fact that here both the \textsl{internal} vertices $r$ and $s$ and the \textsl{external} vertices $a$ and $b$ are fixed, while conventionally only the \textsl{external} vertices $a$ and $b$ are fixed.}  and express the function $\mathcal{A}_{\X{0};r,s}^{\X{(2\nu-1)}}$ in Eq.\,(\ref{e327}) as (\emph{cf.} Eq.\,(\ref{e320d}))
\begin{equation}\label{e327a}
\mathcal{A}_{\X{0};r,s}^{\X{(2\nu-1)}}(1,2,\dots,2\nu) \doteq \sum_{l\in \mathcal{S}_{r,s}^{\X{(\nu)\nparallel}}} \upsigma_l^{\X{(2\nu)}}\hspace{0.8pt}\Lambda_{r,s}^{\X{(l;\nu)}} \hspace{0.8pt}\Psi_{\X{0};r,s}^{\X{(l;\nu)}}(1,2,\dots,2\nu),
\end{equation}
where $\mathcal{S}_{r,s}^{\X{(\nu)\nparallel}}$ is the set deduced from $\mathcal{S}_{r,s}^{\X{(\nu)}}$ by retaining only the representatives of the disjoint classes of permutations that are related through the $\nu$-permutations of the elements of the set $\{\mathpzc{v}_{\X{1}},\mathpzc{v}_{\X{2}},\dots,\mathpzc{v}_{\X{\nu}}\}$, where $\mathpzc{v}_{\X{j}}$ may either be $(2j-1,2j)$ or $(2j,2j-1)$,\footnote{See the programs \texttt{Perm} and \texttt{RangeX} in the closing part of this appendix, beginning on page\,\protect\pageref{Perm}.} for fixed values of $r$ and $s$. Note that since the interaction potentials $\mathsf{v}(i,j)$ and $\mathsf{v}(k,l)$ commute and are assumed to satisfy $\mathsf{v}(i,j) \equiv \mathsf{v}(j,i)$, the product $\mathsf{v}(1,2)\dots \mathsf{v}(2\nu-1,2\nu)$ is invariant under the mentioned $\nu$-permutations, whereby use of these has not necessitated use of a different expression for $\mathfrak{M}_{\X{0};\nu}(r,s)$ than that in Eq.\,(\ref{e320d}).

The constant $\Lambda_{r,s}^{\X{(l;\nu)}}$ is a symmetry factor, for which one has
\begin{equation}\label{e320c}
\sum_{l\in \mathcal{S}_{r,s}^{\X{(\nu)\nparallel}}} \Lambda_{r,s}^{\X{(l;\nu)}} =
\sum_{l\in \mathcal{S}_{r,s}^{\X{(\nu)}}} 1 \equiv \vert \mathcal{S}_{r,s}^{\X{(\nu)}}\vert,\;\; \forall r, s \in \{1,2,\dots, 2\nu\}.
\end{equation}
We have \textsl{empirically} obtained\,\footnote{One has $2^d d! \equiv (2 d)!!$. It is to be noted that $(2d)!!$ is the size of the automorphism group of the $d$-dimensional hypercube ($d$-cube) $Q_d$, consisting of $2^d$ vertices and $2^{d-1} d$ edges \protect\cite{HHW88,FH00}. See also footnote on p.\,\protect\pageref{TheNumber} regarding the significance of $(2k-1)!!$.}
\begin{equation}\label{e320bx}
\frac{\vert \mathcal{S}_{\X{1,1}}^{\X{(\nu)}}\vert}{\vert \mathcal{S}_{\X{1,1}}^{\X{(\nu)\nparallel}}\vert} =
\frac{\vert \mathcal{S}_{\X{1,2}}^{\X{(\nu)}}\vert}{\vert \mathcal{S}_{\X{1,2}}^{\X{(\nu)\nparallel}}\vert} = 2^{\nu-1} (\nu-1)!,\; \forall \nu \ge 1,\;\;\text{and}\;\;\;
\frac{\vert \mathcal{S}_{\X{1,3}}^{\X{(\nu)}}\vert}{\vert \mathcal{S}_{\X{1,3}}^{\X{(\nu)\nparallel}}\vert} = 2^{\nu-2} (\nu-2)!,\; \forall \nu \ge 2.
\end{equation}
\textsl{Assuming} $\Lambda_{r,s}^{\X{(l;\nu)}}$ to take the same value for all $l \in \mathcal{S}_{r,s}^{\X{(\nu)\nparallel}}$, an assumption that is fully supported by empirical evidence, from the equalities in Eqs\,(\ref{e320c})  and (\ref{e320bx}) one deduces that
\begin{equation}\label{e320bz}
\Lambda_{\X{1,1}}^{\X{(l;\nu)}} = \Lambda_{\X{1,2}}^{\X{(l;\nu)}} = 2^{\nu-1} (\nu-1)!,\;\forall \nu \ge 1,\;\;\text{and}\;\;\; \Lambda_{\X{1,3}}^{\X{(l;\nu)}} = 2^{\nu-2} (\nu-2)!,\;\forall \nu \ge 2.
\end{equation}
By conjecturing the equalities in Eq.\,(\ref{e320bz}) as being exact for arbitrary $\nu$ and $l\in \mathcal{S}_{r,s}^{\X{(\nu)\nparallel}}$, with $(r,s) \in \{ (1,1), (1,2), (1,3)\}$, they imply that to leading order the symmetry factors $\Lambda_{r,s}^{\X{(l;\nu)}}$ grow like $2^{\nu} \nu!$ for increasing values of $\nu$. In addition, one obtains (\emph{cf.} Eq.\,(\ref{e320e}))
\begin{align}\label{e320by}
&\hspace{-1.0cm} 2\nu\vert \mathcal{S}_{\X{1,1}}^{\X{(\nu)}}\vert + 2\nu \vert \mathcal{S}_{\X{1,2}}^{\X{(\nu)}}\vert + 2\nu (2\nu-2) \vert \mathcal{S}_{\X{1,3}}^{\X{(\nu)}}\vert \nonumber\\
&\hspace{1.0cm} = 2\nu \Lambda_{\X{1,1}}^{\X{(l;\nu)}}\vert \mathcal{S}_{\X{1,1}}^{\X{(\nu)\nparallel}}\vert + 2\nu  \Lambda_{\X{1,2}}^{\X{(l;\nu)}}\vert \mathcal{S}_{\X{1,2}}^{\X{(\nu)\nparallel}}\vert + 2\nu (2\nu-2)  \Lambda_{\X{1,3}}^{\X{(l;\nu)}}\vert \mathcal{S}_{\X{1,3}}^{\X{(\nu)\nparallel}}\vert,
\end{align}
which is supportive of the general validity of the empirical results in Eq.\,(\ref{e320bx}).

With reference to the extant diagrammatic Monte Carlo method \cite{vHKPS10,KvHGPPST10},\footnote{In \S\,4 of Ref.\,\protect\citen{vHKPS10}, under ``Connectivity and irreducibility'', p.\,101, the authors amongst others indicate that they verify the irreducibility of the diagrams by looking up in a hash table of the momenta associated with the lines in the diagrams. A Green-function line whose associated momentum is equal to the external momentum signifies an \textsl{improper} Green-function diagram.} we point out that the Monte Carlo sampling of the Green-function diagrams considered in this appendix can be achieved by means of the Monte Carlo sampling of $\nu$ (where $\nu \in \{1,\dots,n\}$ for some fixed value of $n$), and of the elements of $\mathcal{S}_{r,s}^{\X{(\nu)}}$, with $(r,s) \in \{(1,1), (1,2), (1,3)\}$, for any given value of $\nu$. In this connection, the package \textsl{Combinatorica} \cite{PS03} provides the possibility of generating random permutations.\footnote{\texttt{RandomPermutation[n]} generates a random \texttt{n}-permutation of the ordered set $\{1,2,...,\texttt{n}\}$ [\S\,2.1.3, p.\,60, in Ref.\,\protect\citen{PS03}].}

We close this appendix  by presenting a series of documented programs, written in the programming language of Mathematica. Below, \texttt{Gnu}, p.\,\pageref{Gnu}, is the main program, calculating all the relevant quantities (stored in the \textsl{list} \texttt{T}) required for the calculation of $G^{\X{(\nu)}}$ on the basis of the formalism introduced in this appendix. This program relies on \texttt{ConnG}, p.\,\pageref{ConnG}, which calculates the set $\mathcal{S}_{r,s}^{\X{(\nu)\nparallel}}$ of independent symmetry-limited permutations and the associated set of symmetry factors $\{\Lambda_{r,s}^{\X{(l;\nu)}}\| l\}$ for given values of $\nu\in \mathds{N}$, and $r, s \in \{1,2,\dots, 2\nu\}$.\footnote{Recall the empirical equalities in Eq.\,(\protect\ref{e320bz}).} The output of this program is a \textsl{list} of integers representing the relevant values of $l$ in $\mathcal{S}_{r,s}^{\X{(\nu)\nparallel}}$; each of these integers is equal to the lexicographic rank\,\footnote{See \S\,2.1.1, p.\,56, in Ref.\,\protect\citen{PS03}. The $j$th output is equal to \texttt{l = RankPermutation[j]+1}. The addition of $1$ is necessary in order for \texttt{Permutations[Range[1,2 nu]][[l]]} coinciding with the relevant $2\nu$-permutation.} of a relevant permutation of the ordered set $\{1,2,\dots,2\nu\}$. These integers are followed by the list of the relevant symmetry factors.\footnote{The last element of the output of \texttt{ConnG} is deduced by the Mathematica instruction \texttt{X = ConnG[r,s,nu]} followed by \texttt{Last[X]} (or simply \texttt{Last[ConnG[r,s,nu]]}, but we do not recommend this, since repeated calls to \texttt{ConnG} for the same triplet (\texttt{r, s, nu}) is unnecessarily wasteful of the computational resources), which is equal to $\vert \mathcal{S}_{r,s}^{\X{(\nu)\nparallel}}\vert$, the cardinal number of $\mathcal{S}_{r,s}^{\X{(\nu)\nparallel}}$. With \texttt{m = Last[X]}, the elements of this set, which are the lexicographic ranks of the relevant $2\nu$-permutations, are \texttt{X[[1;;m]]}. The symmetry weight factors are obtained through \texttt{X[[m+1;;2m]]}. With \texttt{l} $\in \{1,2,\dots,\texttt{m}\}$, \texttt{X[[l]]]} and \texttt{X[[m+l]]} coincide with respectively $l$ (the lexicographic rank of the permutation of the type in Eq.\,(\protect\ref{e330})) and $\Lambda_{r,s}^{\protect\X{(l;\nu)}}$. The integer \texttt{X[[2m+1]]} is equal to the cardinal number $\vert\mathcal{S}_{r,s}^{\X{(\nu)}}\vert$, Eq.\,(\protect\ref{e320c}). Note that \texttt{X[[2m+2]]} is equal to \texttt{m}, that is $\vert\mathcal{S}_{r,s}^{\X{(\nu)\nparallel}}\vert$. Thus, the ratio $\vert \mathcal{S}_{r,s}^{\protect\X{(\nu)}}\vert / \vert \mathcal{S}_{r,s}^{\protect\X{(\nu)\nparallel}}\vert$ is equal to \texttt{X[[2m+1]]/X[[2m+2]]}.} The program \texttt{ConnG} relies on the programs \texttt{GraphG}, p.\,\pageref{GraphG} (which in turn relies on the program \texttt{GraphGX}), and \texttt{Perm}, p.\,\pageref{Perm} (which in turn crucially relies on the program \texttt{RangeX}), all presented below. We point out that program \texttt{GraphG} in \texttt{ConnG} can be replaced by the simpler program \texttt{GraphGX}, with no consequence; we have used \texttt{GraphG} merely for logical consistency and transparency. The programs to be presented below can be directly generalised for the calculation of the \textsl{improper} polarisation function $P^{\star}$, \S\,\ref{s30}.

Before presenting the programs indicated above, we point out that a crude visualisation of the $\nu$th-order Feynman diagrams associated with the $2\nu$-permutations of the form in Eq.\,(\ref{e330}) is possible through using one of the following two Mathematica instructions, or some complex variants of these:\,\footnote{For the visualisation of these diagrams in conventional form, the package \textsl{FeynArts 3} \cite{TH01} may be employed. We point out that the two instructions presented here result in graphs in which $2$-cycles (diagrammatically, the polarisation bubbles) are represented by lines furnished with two arrows pointing in opposite directions.}\\

{\footnotesize
\begin{verbatim}
ShowGraph[Graph[P,r,s], VertexNumber -> True, VertexStyle -> Red]

GraphPlot[Graph[P,r,s], DirectedEdges -> True, VertexLabeling -> True]

\end{verbatim}}

\noindent
Here, \texttt{P} is a \textsl{list} consisting of $2\nu$ integer entries belonging to the set $\{1,2,\dots,2\nu\}$ and separated by commas, representing the relevant $2\nu$-permutation in the array notation.\footnote{In this notation, \texttt{P = $\{$4, 3, 2, 1, 6, 5$\}$} and \texttt{P = $\{$1, 3, 2, 6, 4, 5$\}$} denote the $6$-permutations in respectively Eq.\,(\protect\ref{e332}) and Eq.\,(\protect\ref{e333}).} We point out that \texttt{GraphG}, p.\,\pageref{GraphG}, produces \textsl{directed} graphs, implying that the above two instructions produce some idealisation of Feynman diagrams in which interaction lines are also directed. Without elaborating, we mention that it is possible to produce graphs in which these lines are visually distinguished from the lines representing one-particle Green functions.\\

{\footnotesize\refstepcounter{dummy}\label{Gnu}
\begin{verbatim}
(* Program `Gnu'. *)
Clear[Gnu];
Gnu[nu_] :=
 Module[(* Calculates all connected diagrams contributing to the
 one-particle Green function G at the nu-th order of the perturbation
 theory. It prints some relevant data and returns the lexicographic ranks
 of all the 2nu-permutations of (1,2,...,2nu} describing the relevant
 diagrams, along with the corresponding weights Lambda_{r,s}^{(l;nu)}.
 For the reasons specified in the paper (B. Farid, Many-body perturbation
 expansions without diagrams. I. Normal states, appendix B), the three
 pairs (r,s) = (1,1), (1,2) and (1,3) suffice. *)
 {tx, G11, G12, G13, m11, m12, m13, num, m, T},
  tx[n_] :=
   Module[{ld, t}, ld = Last[IntegerDigits[n]];
    t = Which[ld == 1, "st", ld == 2, "nd", ld == 3, "rd", ld > 3,
      "th"]; t]; G11 = ConnG[1, 1, nu]; G12 = ConnG[1, 2, nu];
  m11 = Last[G11]; m12 = Last[G12];
  If[nu > 1, (G13 = ConnG[1, 3, nu]; m13 = Last[G13]), (m13 = 0)];
  m = m11 + m12 + m13;
  (* For the exact values of m to be printed below, consult e.g. Eq. (3.34)
  and the 2nd column from left of Table I of the paper by Cvitanović et al.
  (Phys. Rev. D 18, 1939 (1978)). In the latter publication, `order' k
  coincides with our 2nu so that for nu = 1, 2, 3, ... the output value of
  m must be equal to respectively 2, 10, 74, ... . *)
  Print["The total number m of the ", nu, tx[nu],
   "-order diagrams contributing to G(a,b): ", m];
  Print["The total number m = ", m, " is the sum of m11: ", m11,
   ", m12: ", m12, ", and m13: ", m13];
  T = Table[{G11[[j]], G11[[m11 + j]]}, {j, 1, m11}];
  T = Append[T, Table[{G12[[j]], G12[[m12 + j]]}, {j, 1, m12}]];
  If[nu > 1, (T =
     Append[T, Table[{G13[[j]], G13[[m13 + j]]}, {j, 1, m13}]])];
  T = Flatten[T]; num = {m11, m12, m13, m};
  T = Flatten[Append[T, num]]; T]

\end{verbatim}}

{\footnotesize\refstepcounter{dummy}\label{ConnG}
\begin{verbatim}

(* Program `ConnG'. *)
Clear[ConnG];
ConnG[r_, s_, nu_] :=
 Module[(* First determines all 2nu-permutations of {1,2,...,2nu}
 corresponding to all connected diagrams contributing to G^{(nu)}(a,b)
 on being linked to the external vertices a and b by means of G_0(a,s^+)
 and G_0(r,b). The integers r and s, which may or may not be equal,
 must be elements of {1,2,...,2nu}. All the above permutations satisfy
 P(r) = s. Subsequently subjects the relevant components of these
 permutations to all 2nu-permutations appropriately determined by
 Perm (which crucially relies on RangeX) and selects the representatives
 of the disjoint classes of the former 2nu-permutations that are related by
 the latter 2nu-permutations. Generally, hereby the factor 1/nu! in the
 relevant expression for G^{(nu)}(a,b) is partially compensated, through
 the symmetry factors Lambda_{r,s}^{(l;nu)} (below collected in the
 list W). The last element of the output list T is the number m of independent
 diagrams; the first m elements are the lexicographic ranks of the
 independent permutations of {1,2,...,2nu}, and the following m elements
 the relevant symmetry factors {Lambda_{r,s}^{(l;nu)} | l}. Thus, T[[j]]
 and T[[m+j]], with j in {1,2,...,m}, correspond to each other. The one but
 last element of T is equal to the number of connected nu-th-order diagrams
 connected to G_0(a,s^+) and G_0(r,b) WITHOUT symmetry reduction,
 characterised by Lambda_{r,s}^{(l;nu)} = 1 for all l. *)
 {j, l, li, ls, k, n, rank, U, P, gr, QX, Q, T, W},
  U = Range[1, 2 nu]; k = 0;
  Do[(*l*) P = Permutations[U][[l]];
   If[P[[r]] == s, (gr = GraphG[P, r, s];
     If[ConnectedQ[gr, Weak], (k = k + 1;
       rank[k] = RankPermutation[P] + 1)])], {l, 1, (2 nu)!}];
  T = Table[rank[j], {j, 1, k}]; W = Table[1, {j, 1, k}]; m = k;
  Do[(*j*)  If[j <= m, P = Permutations[U][[T[[j]]]], Break[]];
   Do[(*l*)
    Do[(*li*) QX = Perm[P, l, li, r, s]; Q = QX[[1]];
     If[ QX[[2]], (If[
        Q != P, (n = RankPermutation[Q] + 1;
         Do[(*i*) If[i <=  m,
           If[T[[i]] == n , (T = Delete[T, i]; W = Delete[W, m];
             m = m - 1; W[[j]] = W[[j]] + 1; Break[])]], {i, 1,
           k}])])], {li, 0, 2^nu - 1}], {l, 1, nu!}], {j, 1, k}];
  Do[T = Append[T, W[[j]]], {j, 1, m}]; T = Append[T, k];
  T = Append[T, m]; T]

\end{verbatim}}

{\footnotesize\refstepcounter{dummy}\label{GraphG}
\begin{verbatim}

(* Programs `GraphG' and `GraphGX'. *)
Clear[GraphG];
GraphG[P_, r_, s_] :=
 Module[(* Returns the graph corresponding to the contribution to
 G^{(nu})(a,b) described by the 2nu-permutations P of {1,2,...,2nu},
 satisfying P(r) = s. The integers r and s correspond to the vertices
 in G_0(a,s^+) and G_0(r,b), the latter connecting the graph with the
 external vertices a and b. *)
  {i, j, Q, gr, ex}, Q = ToCycles[P]; gr = GraphGX[P]; ex = {s, r};
  Do[(*j*) T = Table[(*i*) P[[Q[[j, i]]]], {i, 1, Length[Q[[j]]]}];
   If[MemberQ[T, r] && MemberQ[T, s], gr = DeleteEdge[gr, ex]], {j,
    1, Length[Q]}]; gr]

GraphGX[P_] :=
 Module[(* Returns the graph corresponding to the one-particle
 Green function associated with the 2nu-permutation P of
 {1,2,...,2nu}. Includes the directed edge {s,r} (from s to r)
 representing the G_0(r,s^+) (note the change in the positions of
 s and r) that is to be identified with 1 in the case of r and s
 belonging to the same cycle of P. This task is carried out by
 GraphG. Note that P is to satisfy P(r) = s. *)
 {tnu, i, j, Q, gr},
  tnu = Length[P] (*=2nu*); Q = ToCycles[P];
  gr = MakeGraph[
    Range[1, tnu], (Mod[#2, 2] == 0 && #2 - 1 == #1) &];
  Do[T = Table[{Q[[j, i]], P[[Q[[j, i]]]]}, {i, 1, Length[Q[[j]]]}];
   gr = AddEdges[gr, T], {j, 1, Length[Q]}]; gr]

\end{verbatim}}

{\footnotesize\refstepcounter{dummy}\label{Perm}
\begin{verbatim}
(* Programs `Perm' and `RangeX'. *)
Clear[Perm];
Perm[Pin_, l_, li_, r_, s_] :=
 Module[(* Returns Pout, the 2nu-permutation of {1,2,...,2nu}
 deduced from Pin by subjecting nu entries of this to a nu-permutation
 specified by the inputs l and li (with l in {1,2,...,nu!} and li in
 {0,1, ..., 2^nu -1}), and the remaining nu entries to the
 nu-permutation attendant to the former permutation. Variation
 of li over the entire relevant set is necessary on account of the
 algorithmic design of ConnG. It also returns a flag, which is False
 if the diagram represented by Pout amounts to a non-topological
 transformation of the diagram represented by Pin (even in the case
 of flag = False, Pout satisfies P(r) = s). *)
 {nu, i, j, k, A, T, flag, Pout}, nu = Length[Pin]/2;
  A = ToCycles[Pin]; Q = Flatten[Permutations[RangeX[li, nu]][[l]]];
  T = Table[
    Table[Q[[A[[j, i]]]], {i, 1, Length[A[[j]]]} ], {j, 1,
     Length[A]}]; Pout = FromCycles[T];
  flag = If[Q[[r]] != r || Q[[s]] != s, False, True];
  T = {Pout, flag}; T]

Clear[RangeX];
RangeX[li_, nu_] :=
 Module[(* Returns a permutation of nu pairs, each of which is of the
 form (2j-1,2j) or (2j,2j-1), depending on the value of li. The integer li
 belongs to the set {0,1,...,2^nu-1}. All pairs are of the form (2j-1,2j)
 in the specific case of li=0, and of the form (2j,2j-1) in the specific
 case of li=2^nu-1. *) {j, k, A, R},
  A = IntegerDigits[li, 2] + 1; k = Length[A];
  If[k < nu, Do[A = Prepend[A, 1], {j, k + 1, nu}]];
  R = Table[Permutations[{2 j - 1, 2 j}][[A[[j]]]], {j, 1, nu}]; R]

\end{verbatim}}
$\hfill\Box$

\refstepcounter{dummyX}
\section{The diagrammatic perturbation expansion of \texorpdfstring{$\Sigma$}{} in terms of
\texorpdfstring{$v$}{} and \texorpdfstring{$G$}{}}
\phantomsection
\label{sad}
In appendix \ref{sac} we have presented an approach whereby the conventional diagrammatic perturbation series expansion of the interacting Green function $G(a,b)$ in terms of connected Green-function diagrams and $(v,G_{\X{0}})$ is deduced from the rigorous weak-coupling perturbational expression in Eq.\,(\ref{e242}). Building on the work in appendix \ref{sac}, in this appendix we present an approach whereby the diagrammatic perturbation series expansion of the self-energy operator $\h{\Sigma}$ in terms of $G$-skeleton (or 2PI) self-energy diagrams and $(v,G)$, that is $\h{\Sigma}_{\X{01}}[v,G]$, \S\,\ref{s24}, is obtained. As in the case of appendix \ref{sac}, \emph{the discussions in this appendix are not essential to the discussions in the main body of this paper.} In the closing part of the main part of this appendix, we present some programs, written in the programming language of Mathematica$^{\X{\circledR}}$, that implement the approach to be introduced below. For completeness, in \S\,\ref{sad1} we present some additional programs relevant to the calculation of the self-energy in terms of $(v, G_{\X{0}})$, where $G_{\X{0}}$ may be both the one-particle Green function corresponding to the truly non-interacting Hamiltonian and to the non-interacting Hamiltonian that further takes account of the exact Hartree self-energy.\footnote{\S\,\protect\ref{sad1} is an addition to the text of \href{https://arxiv.org/abs/1912.00474}{\textsf{arXiv:\,1912.0074v1}}. The programs presented in \S\,\protect\ref{sad1} find application in appendix D of Ref.\,\protect\citen{BF16a}.}

We begin with the operator equation in Eq.\,(\ref{e60e}), which provides the link between the $\nu$th-order perturbational contributions to the one-particle Green function and those of the \textsl{improper} self-energy. In the light of the equalities in Eqs\,(\ref{e319}) and (\ref{e320e}), one immediately obtains
\begin{align}\label{e250}
\Sigma_{\X{00}}^{\star\X{(\nu)}}(a,b;[\mathsf{v},G_{\X{0}}]) &= 2\nu \left.\mathfrak{M}_{\X{0;\nu}}(1,1;[G_{\X{0}}])\right\arrowvert_{\substack{1=b}} \delta(a,b)\nonumber\\
&+  2\nu \left.\mathfrak{M}_{\X{0;\nu}}(1,2;[G_{\X{0}}])\right\arrowvert_{\substack{1=b \\ 2 = a}}
+ 2\nu (2\nu-2) \left.\mathfrak{M}_{\X{0;\nu}}(1,3;[G_{\X{0}}])\right\vert_{\substack{1=b \\ 3 = a}},
\end{align}
where we have denoted $\mathfrak{M}_{\X{0;\nu}}(r,s)$ as $\mathfrak{M}_{\X{0;\nu}}(r,s;[G_{\X{0}}])$ so as to make explicit its functional dependence on the non-interacting Green function $G_{\X{0}}$. Further\,\footnote{See footnote \raisebox{-1.0ex}{\normalsize{\protect\footref{notef}}} on p.\,\protect\pageref{NoteThat}.}
\begin{align}\label{e250a}
\delta(a,b) \doteq \left\{\begin{array}{ll}
\delta^d(\bm{r}-\bm{r}') \delta(t-t') \delta_{\sigma,\sigma'}, & \text{($T=0$ formalism)}\\ \\
\delta^d(\bm{r}-\bm{r}') \delta_{\hbar\beta}^{\X{(\varsigma)}}(\tau-\tau') \delta_{\sigma,\sigma'}, & \text{(Matsubara formalism)}\\ \\
\delta^d(\bm{r}-\bm{r}') \delta(t,t') \delta_{\sigma,\sigma'}\delta_{\mu,\mu'}. & \text{(TFD formalism)} \end{array}
\right.
\end{align}
For the Hubbard model (in general, lattice models), the $\delta^d(\bm{r}-\bm{r}')$ on the RHS of the above equality is to be replaced by $\delta_{l,l'}$, Eqs\,(\ref{e36}), (\ref{e24a}), and (\ref{e42c}). As regards $\delta(t,t')$, see the remark following Eq.\,(\ref{es6}). For the reasons specified following Eq.\,(\ref{e320e}), \emph{the substitutions $1=b$, $2=a$, and $3=a$ on the RHS of Eq.\,(\ref{e250}) are to be effected only \textsl{after} having evaluated $\mathfrak{M}_{\X{0;\nu}}(1,1;[G_{\X{0}}])$, $\mathfrak{M}_{\X{0;\nu}}(1,2;[G_{\X{0}}])$, and $\mathfrak{M}_{\X{0;\nu}}(1,3;[G_{\X{0}}])$.}

In the light of the diagrammatic expansion of $\Sigma_{\X{01}}(a,b;[v,G])$, and on account of the first equalities in Eqs\,(\ref{e60f}) and (\ref{e74}), from Eq.\,(\ref{e250}) one immediately infers that
\begin{align}\label{e250b}
\Sigma_{\X{01}}^{\X{(1)}}(a,b;[\mathsf{v},G]) &= 2 \left.\mathfrak{M}_{\X{0;1}}(1,1;[G])\right\arrowvert_{\substack{1=b}} \delta(a,b) + 2 \left.\mathfrak{M}_{\X{0;1}}(1,2;[G])\right\arrowvert_{\substack{1=b \\ 2 = a}}\nonumber\\
&\equiv \Sigma^{\textsc{h}}(a,b;[\mathsf{v},G]) + \Sigma^{\textsc{f}}(a,b;[\mathsf{v},G]) \nonumber\\
&\equiv \Sigma^{\textsc{hf}}(a,b;[\mathsf{v},G]),
\end{align}
where the exact Hartree self-energy $\Sigma^{\textsc{h}}(a,b;[v,G])$ has been introduced in Eq.\,(\ref{e301a}). The prefactors $2$ following the first equality in Eq.\,(\ref{e250b}) are compensated by the $2$ in the prefactor $(\tfrac{\ii}{2\hbar})^{\nu}\vert_{\nu=1}$ on the RHS of Eq.\,(\ref{e320d}). Note that the argument $[G]$ of the functions $\mathfrak{M}_{\X{0;1}}(1,1;[G])$ and $\mathfrak{M}_{\X{0;1}}(1,2;[G])$ conveys the fact that the relevant function $\mathcal{A}_{\X{0};r,s}^{\X{(2\nu-1)}}$ on the RHS of  Eq.\,(\ref{e320d}) is to be evaluated in terms of $G$, instead of $G_{\X{0}}$.

Defining (\emph{cf.} Eq.\,(\ref{e300}))
\begin{equation}\label{e250c}
\Sigma_{\X{01}}''(a,b;[v,G]) \doteq \Sigma_{\X{01}}(a,b;[v,G]) - \Sigma^{\textsc{hf}}(a,b;[v,G]),
\end{equation}
from Eq.\,(\ref{e250}) one immediately obtains that\,\footnote{By definition ${\Sigma_{\X{01}}''}^{\hspace{-1.2pt}\X{(1)}}(a,b;[v,G])\equiv 0$. }
\begin{equation}\label{e250d}
{\Sigma_{\X{01}}''}^{\hspace{-1.2pt}\X{(\nu)}}(a,b;[v,G]) = 2\nu (2\nu-2) \left.\wb{\mathfrak{M}}_{\X{\nu}}(1,3)\right\vert_{\substack{1=b \\ 3 = a}},\;\; \forall\nu \ge 2,
\end{equation}
where (\emph{cf.} Eq.\,(\ref{e320d}))
\begin{equation}\label{e250e}
\wb{\mathfrak{M}}_{\X{\nu}}(r,s) \doteq \pm\frac{1}{\nu !} \Big(\frac{\ii}{2\hbar}\Big)^{\nu}
\int \prod_{\substack{ j=1 \\ j\not= r,s }}^{2\nu} \rd j\; \mathsf{v}(1,2) \dots \mathsf{v}(2\nu-1,2\nu)
\hspace{0.6pt}\b{\mathcal{A}}_{r,s}^{\X{(2\nu-1)}}(1,2,\dots,2\nu-1,2\nu),
\end{equation}
in which (\emph{cf.} Eq.\,(\ref{e327a}))
\begin{equation}\label{e250f}
\b{\mathcal{A}}_{r,s}^{\X{(2\nu-1)}}(1,2,\dots,2\nu) \doteq \sum_{l\in \b{\mathcal{S}}_{r,s}^{\X{(\nu)\nparallel}}} \upsigma_l^{\X{(2\nu)}}\hspace{0.8pt}\Lambda_{r,s}^{\X{(l;\nu)}} \hspace{0.8pt}\Psi_{r,s}^{\X{(l;\nu)}}(1,2,\dots,2\nu).
\end{equation}
The set $\b{\mathcal{S}}_{r,s}^{\X{(\nu)\nparallel}}$ is deuced from $\mathcal{S}_{r,s}^{\X{(\nu)\nparallel}}$ by discarding all values of $l$ in the latter set for which the diagram associated with the $2\nu$-permutation in Eq.\,(\ref{e330}) is not $G$-skeleton (or 2PI). Since a diagram that is not 1PI is necessarily not a 2PI one, in the process of deducing $\b{\mathcal{S}}_{r,s}^{\X{(\nu)\nparallel}}$ from $\mathcal{S}_{r,s}^{\X{(\nu)\nparallel}}$ one automatically discards all self-energy diagrams that are not 1PI, that is those that are not \textsl{proper}.

The function $\Psi_{r,s}^{\X{(l;\nu)}}$ on the RHS of Eq.\,(\ref{e250f}) is defined as (\emph{cf.} Eq.\,(\ref{e336}))
\begin{equation}\label{e403}
\Psi_{r,s}^{\X{(l;\nu)}}(1,2,\dots,2\nu) \doteq \prod_{j=1}^{m_l} \Phi_l^{\X{(j)}}(r,s),
\end{equation}
where (\emph{cf.} Eq.\,(\ref{e335}))
\begin{equation}\label{e335b}
\Phi_{l}^{\X{(j)}}(r,s) \doteq
\prod_{k=1}^{n_l^{\X{(j)}}-1} G(C_l^{\X{(j)}}(k), C_l^{\X{(j)}}(k+1)^+) \times \left\{\begin{array}{ll} 1, & r,s \in \mathfrak{C}_l^{\X{(j)}}, \\ \\
G(C_l^{\X{(j)}}(n_l^{\X{(j)}}), C_l^{\X{(j)}}(1)^+), & r,s \not\in \mathfrak{C}_l^{\X{(j)}}. \end{array} \right.
\end{equation}
\refstepcounter{dummy}\label{SinceA1Cycle}In this connection, since in the cases of $r\not=s$ a $1$-cycle in the cycle decomposition of the permutation in Eq.\,(\ref{e330}) to which the expression in Eq.\,(\ref{e335}) corresponds\,\footnote{Here with the $G_{\X{0}}$ herein replaced by $G$, in view of the definition of $\Psi_{r,s}^{\X{(l;\nu)}}$.} invariably represents a particle loop associated with a tadpole diagram representing a Hartree self-energy insertion, and self-energy diagrams of order $\nu \ge 2$ with this insertion \textsl{cannot} be 2PI, in assembling the set $\b{\mathcal{S}}_{r,s}^{\X{(\nu)\nparallel}}$ corresponding to $r\not= s$ and $\nu\ge 2$ from the elements of the set $\mathcal{S}_{r,s}^{\X{(\nu)\nparallel}}$, an $l \in \mathcal{S}_{r,s}^{\X{(\nu)\nparallel}}$ for which the cycle decomposition of the associated $2\nu$-permutation, Eqs\,(\ref{e328}) and (\ref{e330}), contains a $1$-cycle, can be immediately discarded.\footnote{The condition $r\not=s$ excludes the cases where $(j)$, with $j\equiv r=s$, is a $1$-cycle. Recall that in such cases by the convention $G(r,s^+) \equiv 1$ the loop at issue does \textsl{not} constitute a Hartree self-energy.}

\begin{figure}[t!]
\centerline{
\includegraphics*[angle=0, width=0.32\textwidth]{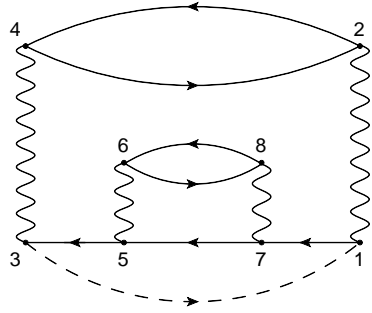}}
\caption{A fourth-order non-skeleton self-energy diagram described by the function $\Psi_{1,3}^{\protect\X{(l;4)}}$, Eq.\,(\protect\ref{e403}), with the integer $l \in \mathcal{S}_{1,3}^{\protect\X{(4)\nparallel}}$ taking such value that the corresponding $8$-permutation coincides with that in Eq.\,(\protect\ref{e253a}). Here directed solid line represents $G$ (instead of $G_{\protect\X{0}}$ as in Fig.\,\protect\ref{f12}). On account of the cycles of this permutation, one has: $\Psi_{1,3}^{\protect\X{(l;4)}}(1,2,\dots,8) = G(7,1^+) G(3,5^+) G(5,7^+) G(4,2^+) G(2,4^+) G(8,6^+) G(6,8^+)$, where, following the prescription in Eq.\,(\protect\ref{e323}), we have identified the Green function $G(1,3^+)$ (depicted by broken line) with $1$. For a general pair $(r,s)$, the Green function $G(r,s^+)$ follows from identifying the external vertices $a$ and $b$ in the Green functions $G_{\protect\X{0}}(r,b)$ and $G_{\protect\X{0}}(a,s^+)$ on the RHS of Eq.\,(\protect\ref{e259b}). Conform the fact that a $\nu$th-order self-energy diagram consists of $2\nu-1$ Green-function lines, the above function $\Psi_{1,3}^{\protect\X{(l;4)}}$ consists of a product of $7$ Green functions. Note that the numbers attached to the vertices linked by an interaction line (wavy solid line) form pairs of the form $(2j-1,2j)$, $j\in\{1,2,3,4\}$, in conformity with the expressions in Eqs\,(\protect\ref{e320d}) and (\protect\ref{e250e}).}
\label{f11}
\refstepcounter{dummy}\label{f11x}
\end{figure}

For illustration, let us consider the case of $\nu=4$ and the following $8$-permutation of $\{1,2,\dots,8\}$ that corresponds to the pair $(r,s) = (1,3)$, Eq.\,(\ref{e330}):\,\footnote{With reference to the relevant footnote on p.\,\pageref{UsingTheLexicographic}, here $l = 11801$.}
\begin{equation}\label{e253a}
\mathscr{P}_l^{\X{(8)}}:\;
\begin{pmatrix}
\bm{1} & 2 & 3 & 4 & 5 & 6 & 7 & 8 \\
\bm{3} & 4 & 5 & 2 & 7 & 8 & 1 & 6
\end{pmatrix}
= (7,\bm{1},\bm{3},5) (4,2) (8,6).
\end{equation}
The diagram corresponding to this permutation is depicted in Fig.\,\ref{f11}. Upon suppressing the broken line representing the function $G(r,s^+) \equiv 1$, the resulting diagram is seen to represent a \textsl{proper} (\emph{i.e.} 1PI) self-energy diagram, which however is \textsl{not} $G$-skeleton (\emph{i.e.} not 2PI). Therefore, the corresponding $l \in \mathcal{S}_{\X{1,3}}^{\X{(4)\nparallel}}$ is not an element of $\b{\mathcal{S}}_{\X{1,3}}^{\X{(4)\nparallel}}$.

Construction of the set $\b{\mathcal{S}}_{r,s}^{\X{(\nu)\nparallel}}$ from the set $\mathcal{S}_{r,s}^{\X{(\nu)\nparallel}}$ is straightforward: for each $l \in \mathcal{S}_{r,s}^{\X{(\nu)\nparallel}}$ one sequentially removes two Green-function lines from the self-energy diagram corresponding to the relevant $2\nu$-permutation $\mathscr{P}_l^{\X{(2\nu)}}$ (such as the diagram in Fig.\,\ref{f11}, which corresponds to the $8$-permutation in Eq.\,(\ref{e253a})) and tests the resulting diagram for connectedness.\footnote{As in the case of connected graphs corresponding to one-particle Green function considered in appendix \protect\ref{sac}, here also the relevant connectivity is the \textsl{weak} one.}\footnote{The function $\Psi_{r,s}^{\protect\X{(l;\nu)}}$ consisting of a product of $(2\nu-1)$ Green functions, for $\nu\ge 2$ at most $\binom{2\nu-1}{2} \equiv (\nu-1) (2\nu-1) = \mathcal{O}(\nu^2)$ iterations are required to establish whether the diagram corresponding to $l \in \mathcal{S}_{r,s}^{\protect\X{(\nu)\nparallel}}$ is $G$-skeleton, \emph{i.e.} 2PI.}

We close this appendix by presenting the programs, written in the programming language of Mathematica$^{\X{\circledR}}$, that in conjunction with those presented in appendix \ref{sac} determine the elements of the sets $\{\b{\mathcal{S}}_{\X{r,s}}^{\X{(\nu)\nparallel}}\| r,s \}$ and $\{\Lambda_{r,s}^{\X{(l;\nu)}} \| l, r,s\}$ as encountered on the RHS of Eq.\,(\ref{e250f}).\footnote{The calculated symmetry factors are to be contrasted with values deduced from the equalities in Eq.\,(\protect\ref{e320bz}). } The program \texttt{Snu}, p.\,\pageref{Snu}, is similar to the program \texttt{Gnu}, p.\,\pageref{Gnu}, in appendix \ref{sac}, however concerns $G$-skeleton self-energy diagrams. This program relies on the program \texttt{SkeletonS}, which in turn relies on the program \texttt{SkeletonG}, both of which are presented below, p.\,\pageref{SkeletonS}. These programs can be directly generalised for the determination of the diagrams associated with the \textsl{proper} polarisation function $P$ from those associated with the \textsl{improper} one $P^{\star}$, \S\,\ref{s30}.\\

{\footnotesize\refstepcounter{dummy}\label{Snu}
\begin{verbatim}
(* Program `Snu'. *)
Clear[Snu];
Snu[nu_] :=
 Module[(* Returns the permutations and the associated weights
 Lambda_{r,s}^{(l;nu)} corresponding to G-skeleton self-energy
 diagrams associated with the three relevant pairs (r,s) = (1,1),
 (1,2), (1,3). It also prints some relevant details. *)
 {tx, S11, S12, S13, m11, m12, m13, num, m, T},
  tx[n_] :=
   Module[{ld, t}, ld = Last[IntegerDigits[n]];
    t = Which[ld == 1, "st", ld == 2, "nd", ld == 3, "rd", ld > 3,
      "th"]; t]; S11 = SkeletonS[1, 1, nu]; S12 = SkeletonS[1, 2, nu];
   m11 = Last[S11]; m12 = Last[S12];
  If[nu > 1, (S13 = SkeletonS[1, 3, nu]; m13 = Last[S13]), (m13 = 0)];
  m = m11 + m12 + m13; (* For the exact values of m printed below,
  consult e.g. Eq. (17) of the paper by Molinari and Manini (Eur. Phys.
  J. B 51, 331 (2006)). Thus, for nu = 1, 2, 3, 4, ... the m below must be
  equal to respectively 2, 2, 10, 82, ... (by leaving out the Hartree, or
  the tadpole, diagram, the m for nu = 1 would be 1). For nu > 1, m11 and
  m12 must be equal to 0. To save computation time in the cases of nu > 1,
  it is advisable to comment out the instructions below that concern S11,
  m11, and S12, m12. *)
  Print["The total number m of the ", nu, tx[nu],
   "-order G-skeleton diagrams contributing to \[CapitalSigma](a,b): ", m];
   Print["The total number m = ", m, " is the sum of m11: ", m11,
   ", m12: ", m12, ", and m13: ", m13];
  T = Table[{S11[[j]], S11[[m11 + j]]}, {j, 1, m11}];
  T = Append[T, Table[{S12[[j]], S12[[m12 + j]]}, {j, 1, m12}]];
  If[nu > 1, (T =
     Append[T, Table[{S13[[j]], S13[[m13 + j]]}, {j, 1, m13}]])];
  T = Flatten[T]; num = {m11, m12, m13, m};
  T = Flatten[Append[T, num]]; T]

\end{verbatim}}

{\footnotesize\refstepcounter{dummy}\label{SkeletonS}
\begin{verbatim}
(* Programs `SkeletonS' and `SkeletonG'. *)
Clear[SkeletonS];
SkeletonS[r_, s_, nu_] :=
 Module[(* By considering the nu-th-order connected Green-function
 diagrams that through the Green functions G_0(a,s^+) and G_0(r,b) are
 linked to the external vertices a and b, selects out the
 \[CapitalSigma]^{(nu)}(s,r) that are G-skeleton. Returns the
 corresponding 2nu-permutations of {1,2,...,2nu} (their lexicographic
 ranks) and the associated weights Lambda_{r,s}^{(l;nu)}. *)
 {m, n, v, R, U, T, P, rS, wS},
  R = Range[1, 2 nu]; U = ConnG[r, s, nu]; m = Last[U]; n = 0;
  Do[(*j*) P = Permutations[R][[U[[j]]]]; v = SkeletonG[P, r, s][[2]];
    If[v, (n = n + 1; rS[n] = U[[j]]; wS[n] = U[[m + j]])], {j, 1, m}];
    T = Table[rS[j], {j, 1, n}];
  T = Flatten[Append[T, Table[wS[j], {j, 1, n}]]]; T = Append[T, n]; T]


Clear[SkeletonG];
SkeletonG[P_, r_, s_] :=
 Module[(* Returns True if the self-energy diagram contributing to
 \[CapitalSigma](s,r) is G-skeleton, False otherwise. *)
 {i, j, k, l, ex, e1, e2, Q, v, gr, grx},
  Q = ToCycles[P]; gr = GraphG[P, r, s]; ex = {s, r}; v = True;
  Do[(*k*) Do[(*l*) e2 = {Q[[k, l]], P[[Q[[k, l]]]]};
    If[e2 !=
      ex, (Do[(*i*)
       Do[(*j*) e1 = {Q[[i, j]], P[[Q[[i, j]]]]};
        If[e1 != ex,
         If[e1 != e2, (grx = DeleteEdges[gr, {e1, e2}];
           v = ConnectedQ[grx, Weak];
           If[v == False, Goto[end]])]], {j, 1, Length[Q[[i]]]}], {i,
        1, Length[Q]}])], {l, 1, Length[Q[[k]]]}], {k, 1, Length[Q]}];
   Label[end]; T = {gr, v}; T]

\end{verbatim}}

\refstepcounter{dummyX}
\subsection{The diagrammatic perturbation expansion of \texorpdfstring{$\Sigma$}{} in terms of
\texorpdfstring{$v$}{} and \texorpdfstring{$G_{\protect\X{0}}$}{}}
\phantomsection
\label{sad1}
In this brief section\,\footnote{This subsection is an addition to the text of \href{https://arxiv.org/abs/1912.00474}{\textsf{arXiv:\,1912.0074v1}}. The programs presented here are included in the notebook \texttt{PermutationsAndDiagramsV2.nb} (\texttt{PermutationsAndDiagrams.nb} has been published alongside \textsf{arXiv:\,1912.0074v1}, which is also available at \href{https://www.notebookarchive.org/permutationsanddiagrams-nb--2020-02-b2t1tpc/}{\textsf{Wolfram Notebook Archive}}).} we present programs, written in the programming language of Mathematica$^{\X{\circledR}}$, concerning the perturbational calculation of $\Sigma_{\X{00}}(s,r;[v,G_{\X{0}}])$. The underlying self-energy diagrams are therefore 1PI. The program \texttt{Snu1PI} below, p.\,\pageref{Snu1PI}, is the counterpart of the program \texttt{Snu} above, p.\,\pageref{Snu}. The program \texttt{Snu1PI} relies on two external sub-programs \texttt{S1PI} and \texttt{Snux1PI}, p.\,\pageref{S1PI}. We also include a program, named \texttt{Snu1PInoTP}, p.\,\pageref{Snu1PInoTP}, that discards from the output of \texttt{Snu1PI} the permutations (and the associated weight factors) corresponding to 1PI diagrams with tadpole self-energy insertion(s). The $2\nu$-permutations as calculated by \texttt{Snu1PInoTP} correspond to the $\nu$th-order self-energy diagrams contributing to $\Sigma_{\X{00}}(s,r;[v,G_{\X{0}}])$ wherein $G_{\X{0}} \equiv \{G_{\X{0};\sigma}\| \sigma\}$ is the set of non-interacting one-particle Green functions that take account of the \textsl{exact} Hartree self-energy $\Sigma^{\textsc{h}}$. The program \texttt{Snu1PInoTP} relies on two external sub-programs \texttt{CountL} and \texttt{Tadpole}, p.\,\pageref{CountL}.

\begin{figure}[t!]
\centerline{
\includegraphics*[angle=0, width=0.6\textwidth]{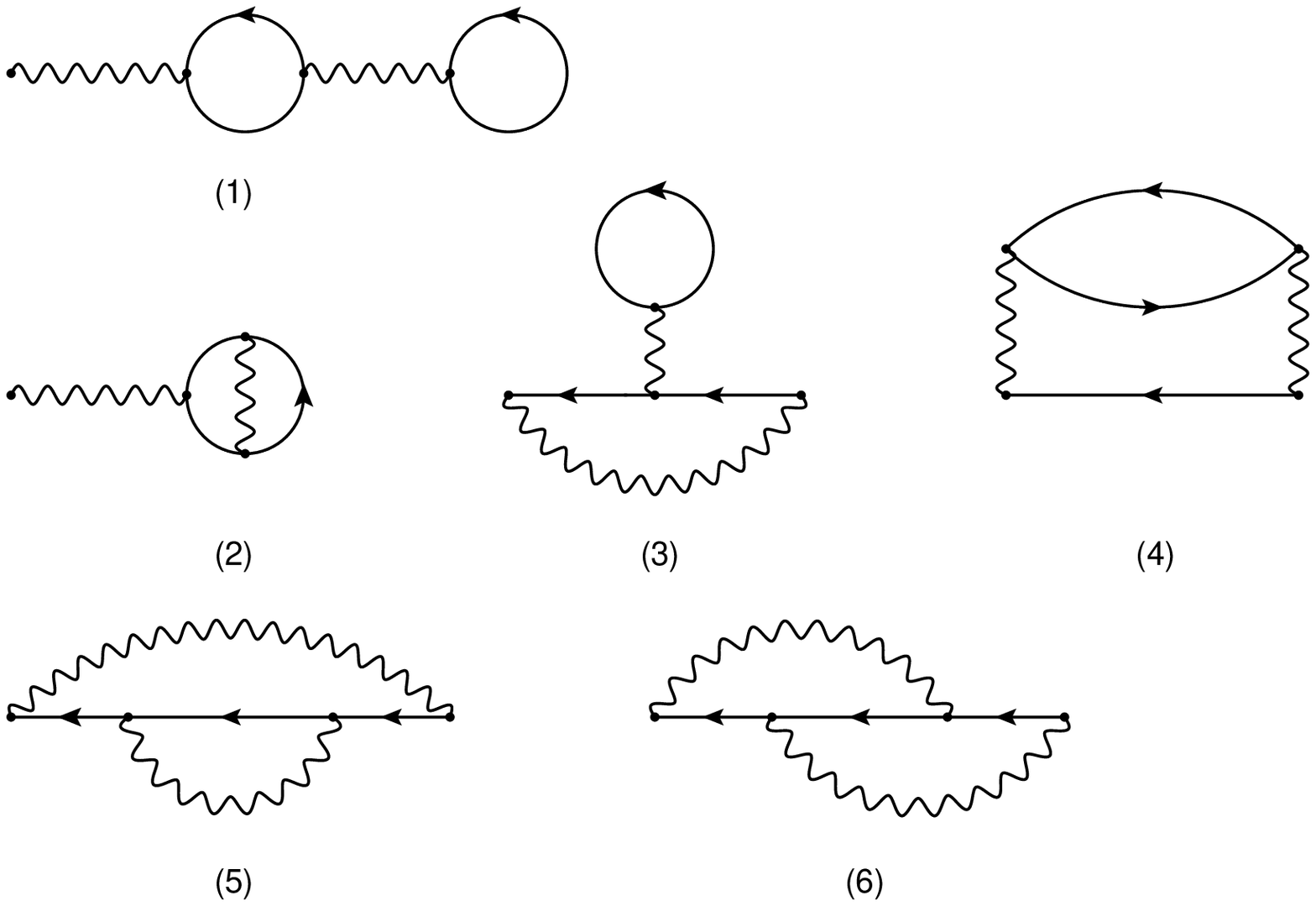}}
\caption{The \protect\refstepcounter{dummy}\label{CompleteSet}complete set of $2$nd-order proper self-energy diagrams ordered in accordance with their decreasing number of loops. Program \texttt{Snu1PI[nu]}, p.\,\protect\pageref{Snu1PI}, generates these six diagrams for \texttt{nu = 2}. Using the output of this program, that is using \texttt{T}, where \texttt{T = Snu1PI[nu]}, program \texttt{Snu1PInoTP[T,nu]}, p.\,\protect\pageref{Snu1PInoTP}, discards the diagrams (1), (2), and (3) for containing the  one-corner (diagrams (1) and (3)) and three-corner (diagram (2)) tadpole diagrams. Note that only diagrams (4) and (6) are skeleton.}
\label{f10}
\end{figure}

For illustration, in Fig.\,\ref{f10} we present the complete set of $2$nd-order proper self-energy diagrams. In the caption of this figure we discuss the link between these diagrams and the programs \texttt{Snu1PI} and \texttt{Snu1PInoTP}. The following are the $4$-permutations associated with the diagrams in Fig.\,\ref{f10}, along with the corresponding $(r,s)$:\,\footnote{From left to right in Eq.\,(\protect\ref{eac1}): the number ($j$) associated with the diagram in Fig.\,\protect\ref{f10}, the corresponding $4$-permutation, and the relevant pair $(r,s)$ in $\protect\t{\Sigma}^{\protect\X{(2.j)}}(s,r)$, where $j = 1,2,\dots, 6$.}
\begin{align}\label{eac1}
(1):\;&
\begin{pmatrix}
1 & 2 & 3 & 4 \\
1 & 3 & 2 & 4
\end{pmatrix},\;\;\;
(1,1),\nonumber\\
(2):\;&
\begin{pmatrix}
1 & 2 & 3 & 4 \\
1 & 3 & 4 & 2
\end{pmatrix},\;\;\;
(1,1),\nonumber\\
(3):\;&
\begin{pmatrix}
1 & 2 & 3 & 4 \\
2 & 3 & 1 & 4
\end{pmatrix},\;\;\;
(1,2),\nonumber\\
(4):\;&
\begin{pmatrix}
1 & 2 & 3 & 4 \\
3 & 4 & 1 & 2
\end{pmatrix},\;\;\;
(1,3),\nonumber\\
(5):\;&
\begin{pmatrix}
1 & 2 & 3 & 4 \\
2 & 3 & 4 & 1
\end{pmatrix},\;\;\;
(1,2),\nonumber\\
(6):\;&
\begin{pmatrix}
1 & 2 & 3 & 4 \\
3 & 4 & 2 & 1
\end{pmatrix},\;\;\;
(1,3).
\end{align}

\noindent
We recall that the variation in the values of $r$ and $s$ in Eq.\,(\ref{eac1}) is necessitated by the convention according to which the arguments of the two-body interaction potentials are to be of the form $(2j-1,2j)$, $j \in \{1,2,\dots,\nu\}$ (see \emph{e.g.} Eqs\,(\ref{e257}) and (\ref{e261})).

{\footnotesize\refstepcounter{dummy}\label{Snu1PI}
\begin{verbatim}

(* Program `Snu1PI'. *)
Clear[Snu1PI];
Snu1PI[nu_] :=
 Module[(* Returns the permutations and the associated weights Lambda_{r,s}^{(l;nu)}
 corresponding to one-particle irreducible (1PI), or proper, self-energy diagrams
 associated with the three relevant pairs (r,s) = (1,1), (1,2), (1,3). It also prints
 some relevant details. Based in <Snu>. *)
 {tx, S11, S12, S13, m11, m12, m13, num, m, T},
  tx[n_] :=
   Module[{ld, t}, ld = Last[IntegerDigits[n]];
    t = Which[ld == 1, "st", ld == 2, "nd", ld == 3, "rd", ld > 3,
      "th"]; t]; S11 = Snux1PI[1, 1, nu]; S12 = Snux1PI[1, 2, nu];
  m11 = Last[S11]; m12 = Last[S12];
  If[nu > 1, (S13 = Snux1PI[1, 3, nu]; m13 = Last[S13]), (m13 = 0)];
  m = m11 + m12 + m13; (* For nu = 1, 2, 3, 4, ... the m = m_nu below must be
  equal to m_1 = 2, m_2 = 6, m_3 = 42, m_4 = 414, ... . *)
  Print["The total number m of the ", nu, tx[nu], "-order 1PI diagrams contributing
  to \[CapitalSigma](a,b): ", m];
  Print["The total number m = ", m, " is the sum of m11: ", m11, ", m12: ", m12, ",
  and m13: ", m13];
  T = Table[{S11[[j]], S11[[m11 + j]]}, {j, 1, m11}];
  T = Append[T, Table[{S12[[j]], S12[[m12 + j]]}, {j, 1, m12}]];
  If[nu > 1, (T =
     Append[T, Table[{S13[[j]], S13[[m13 + j]]}, {j, 1, m13}]])];
  T = Flatten[T]; num = {m11, m12, m13, m};
  T = Flatten[Append[T, num]]; T]

\end{verbatim}}

{\footnotesize\refstepcounter{dummy}\label{S1PI}
\begin{verbatim}
(* Programs `S1PI' and `Snux1PI'. *)
Clear[S1PI];
S1PI[P_, r_, s_] :=
 Module[(* Returns True if the self-energy diagram contributing to \[CapitalSigma](s,r)
 is one-particle irreducible (1PI), False otherwise. Based on <SkeletonG>. *)
 {i, j, k, l, nu, ex, e1, Q, v, gr, grx},
  nu = Length[P]/2; Q = ToCycles[P]; gr = GraphG[P, r, s];
  ex = {s, r}; v = True;
  Do[(*i*) Do[(*j*) e1 = {Q[[i, j]], P[[Q[[i, j]]]]};
    If[e1 !=
      ex, (grx =
       If[nu == 1, DeleteEdges[gr, {e1}],
        DeleteEdges[gr, {e1, {r, s}}]]; v = ConnectedQ[grx, Weak];
      If[v == False, Goto[end]])], {j, 1, Length[Q[[i]]]}], {i, 1, Length[Q]}];
      Label[end]; T = {gr, v}; T]

Clear[Snux1PI];
Snux1PI[r_, s_, nu_] :=
 Module[(* By considering the nu-th-order connected Green-function diagrams
 that through the Green functions G_0(a,s^+) and G_0(r,b) are linked to the
 external vertices a and b, selects out the \[CapitalSigma]^{(nu)}(s,r) that
 are one-particle irreducible (1PI). Returns the corresponding 2nu-permutations
 of {1,2,...,2nu} (their lexicographic ranks) and the associated weights
Lambda_{r,s}^{(l;nu)}. Based on <SkeletonS>. *)
{m, n, v, R, U, T, P, rS, wS}, R = Range[1, 2 nu];
  U = ConnG[r, s, nu]; m = Last[U]; n = 0;
  Do[(*j*)P = Permutations[R][[U[[j]]]]; v = S1PI[P, r, s][[2]];
   If[v, (n = n + 1; rS[n] = U[[j]]; wS[n] = U[[m + j]])], {j, 1, m}];
   T = Table[rS[j], {j, 1, n}];
  T = Flatten[Append[T, Table[wS[j], {j, 1, n}]]]; T = Append[T, n]; T]

\end{verbatim}}

{\footnotesize\refstepcounter{dummy}\label{Snu1PInoTP}
\begin{verbatim}
(* Program `Snu1PInoTP'. *)
Clear[Snu1PInoTP];
Snu1PInoTP[T_, nu_] :=
 Module[(* Using the output of <Snu1PI> (that is, with T obtained by
 T = Snu1PI[nu]), returns a similar output as <Snu1PI> however with
 the permutations corresponding to the 1PI self-energy diagrams
 containing tadpole insertions discarded. The number m = m_nu of the
 1PI self-energy diagrams of order nu without tadpole insertions as
 calculated by this program must coincide with the exact numbers
 m_1 = 1, m_2 = 3, m_3 = 20, m_4 = 189, ... [see Eqs (11) and (12)
 in L. G. Molinari, Phys. Rev. B 71, 113102 (2005)]. *)
 {tx, m, n, i, j, k, l, l1, TY, TC, U, v},
  tx[n_] := Module[{ld, t}, ld = Last[IntegerDigits[n]];
    t = Which[ld == 1, "st", ld == 2, "nd", ld == 3, "rd", ld > 3,
      "th"]; t]; n = Last[T]; TX = Range[1, 2 nu]; U = {}; m = 0;
  Do[(*j*) v = Tadpole[Permutations[TX][[T[[2 j - 1]]]]];
   If[v, (U = Flatten[Append[U, {T[[2 j - 1]], T[[2 j]]}]];
     m = m + 1), 98], {j, 1, n}]; U = Flatten[Append[U, m]];
  Print["The total number m of the ", nu, tx[nu],
   "-order 1PI diagrams without tadpole insertions contributing
   to \[CapitalSigma](a,b): ", m]; U];

\end{verbatim}}

{\footnotesize\refstepcounter{dummy}\label{CountL}
\begin{verbatim}
(* Programs `CountL' and `Tadpole'. *)
Clear[CountL];
CountL[T_] :=
 Module[(* Given a cycle [T] of a 2nu-permutation representing a
 self-energy diagram (whose vertices are indexed in such a way
 that the vertices of a two-body interaction potential are of the
 form (2j-1,2j) or (2j,2j-1)) returns False if the part of the
 self-energy diagram represented by T is linked to the remaining
 part of the diagram by a single interaction line. *)
 {l, m, i, k, , k1, v},
  l = Length[T]; m = 0;
  Do[(*i*) k = T[[i]]; k1 = If[Mod[k, 2] == 0, k - 1, k + 1];
   v = MemberQ[T, k1]; m = m + If[v, 0, 1], {i, 1, l}];
  v = If[m == 1, False, True]; v];

Clear[Tadpole];
Tadpole[P_] :=
 Module[(* Given a nu-th-order self-energy diagram as represented by
 the 2nu-permutation P, returns False if the diagram has at least one
 tadpole insertion. *)
 {i, TC, v}, TC = ToCycles[P]; v = True;
  Do[(*i*)v = v && CountL[TC[[i]]], {i, 1, Length[TC]}]; v];

\end{verbatim}}
$\hfill\Box$

\refstepcounter{dummyX}
\section{The Hubbard Hamiltonian for spin-\texorpdfstring{$\tfrac{1}{2}$}{} fermions}
\phantomsection
\label{sab}
In this appendix we deduce a simplified expression for the function $M_{\nu}(a,b)$, Eq.\,(\ref{e259b}), as suited for the calculation of the interacting one-particle Green function $G[U,G_{\X{0}}]$ and the self-energies $\Sigma_{\X{00}}[U,G_{\X{0}}]$ and $\Sigma_{\X{01}}[U,G]$ corresponding to the Hubbard Hamiltonian for spin-$\tfrac{1}{2}$ fermions in $d$ space dimensions, \S\S\,\ref{s224}, \ref{s227}. Calculation of these functions is also demanding of the calculation of the elements of the sequence $\{D_{\nu} \| \nu\}$ (Eq.\,(\ref{e46}) and Eq.\,(\ref{eb3}) below), which is immediate and requires no special treatment. The considerations of this appendix are directly applicable to both the $T=0$ formalism and the Matsubara formalism for $T > 0$.\footnote{For applying the Matsubara formalism, below $t_j$ is to be replaced by $\tau_j$, $\forall j$, with $\int \textrm{d} \tau_j$ understood as denoting $\int_{0}^{\hbar\beta} \textrm{d}\tau_j$ (\emph{cf.} Eqs\,(\protect\ref{e39}) and (\protect\ref{e42})).} Extension of these considerations to encompass calculations within the framework of the TFD, \S\S\,\ref{sac01}, \ref{s226}, is straightforward.\footnote{See the relevant remarks in the penultimate paragraph of \S\,\protect\ref{s227}, p.\,\protect\pageref{InAppendixD}.}

The details to be presented in this appendix are applicable to arbitrary spatial dimensions $d$, with $d = \infty$,\footnote{For taking the limit of $d\to\infty$, the hopping matrix elements $\{T_{l,l'} \| l,l'\}$ in the Hubbard Hamiltonian, Eq.\,(\protect\ref{e27}), are to be appropriately scaled, like $T_{l.l'}/\sqrt{\mathpzc{Z}}$ (see for instance \S\,II.C, p.\,20, in Ref.\,\protect\citen{GKKR96}). Since however we only deal with the non-interacting Green function $G_{\X{0}}$ in abstract form, we do not need explicitly to deal with this scaling in this appendix.} or infinite coordination number $\mathpzc{Z}= 2 d$ on the $d$-cubic lattice, corresponding to the framework of the dynamical mean-field theory (DMFT) \cite{EMH89b,MV89,EMH89a,GKKR96,PF03}. This framework takes account of the spatial fluctuations at a mean-field level, however of the temporal quantum fluctuations in principle exactly \cite{GKKR96}. In the following, we first consider the general case, \S\,\ref{sab1}, and subsequently the case of the atomic limit,\footnote{See Eq.\,(32) \emph{et seq.} in Ref.\,\protect\citen{JH63}.} for which the single-particle energy dispersion $\varepsilon_{\bm{k}}$ in Eq.\,(\ref{e29}) is independent of $\bm{k}$, whereby $T_{l,l'} \propto \delta_{l,l'}$ and thus $G_{\X{0};\sigma}^{l,l'}(t-t') \propto \delta_{l,l'}$, Eq.\,(\ref{eb13}) below. At half-filling, the exact $G$ corresponding to this limit is well-known \cite{JH63}.\footnote{Away from half-filling, the site-occupation numbers corresponding to different spin species are to be explicitly calculated as functions of the on-site interaction energy $U$.} The expansion of this function in powers of $U$ coinciding with the weak-coupling perturbation series expansion of $G[U,G_{\X{0}}]$, the considerations in \S\,\ref{sab2} result in an infinite sequence of exact sum-rules that may be fruitfully utilised for establishing the accuracy of the computational methods to be employed in the calculations away from the atomic limit. We note in passing that when truncated at any finite order, the mentioned perturbation expansion in the atomic limit is pathological \cite{BF16a}. A similar simplified approach as in \S\,\ref{sab2} applies to a formalism centred on the calculation of the \textsl{self-energy} within the framework of the DMFT, where $\Sigma_{\sigma}^{l,l'}(t-t') \propto \delta_{l,l'}$ (\emph{cf.} Eq.\,(\ref{eb13}) below) \cite{EMH89b,MV89,EMH89a,GKKR96,PF03}.

In the closing part of \S\,\ref{sab1}, we present some programs, written in the programming language of Mathematica$^{\X{\circledR}}$, that implement a significant function, $A_{\alpha_r,\alpha_s}^{\X{(2\nu-1)}}$, encountered in this appendix. The program \texttt{A2num1}, to be presented below, p.\,\pageref{A2num1}, can be used to verify the validity of the identities that are central to this appendix. To facilitate this task, we also present three sets of instructions, in the programming language of Mathematica, for performing the relevant verifications, p.\,\pageref{Set1}.

\refstepcounter{dummyX}
\subsection{The general case}
\phantomsection
\label{sab1}
We begin by introducing a convenient set $\{\alpha_j\| j\}$ of compound variables for $j \in \{1,2,\dots,2\nu\}$, defined according to
\begin{equation}\label{eb1}
\alpha_j \rightleftharpoons \left\{ \begin{array}{ll} l_{\theta_j} t_{\theta_j} \sigma_{\theta_j}, &
j = \text{odd}, \\ \\ l_{\theta_j} t_{\theta_j} \b{\sigma}_{\theta_j}, & j = \text{even}, \end{array} \right.
\end{equation}
where\,\footnote{$\theta_{2k-1} = \theta_{2k} = k$, $\forall k \in \mathds{N}$. See appendix \protect\ref{sae}.} (\emph{cf.} Eq.\,(\ref{es20a}))
\begin{equation}\label{es20}
\theta_j \doteq \Big\lfloor \frac{j+1}{2} \Big\rfloor.
\end{equation}
One thus has
\begin{equation}\label{eb1b}
\alpha_1 \rightleftharpoons l_1  t_1\sigma_1,\; \alpha_2 \rightleftharpoons l_1 t_1\b{\sigma}_1,\; \alpha_3 \rightleftharpoons l_2 t_2\sigma_2,\; \alpha_4 \rightleftharpoons l_2 t_2\b{\sigma}_2,\; \dots\;.
\end{equation}
In terms of the compound variables $\{\alpha_j \| j\}$, the expressions in Eqs\,(\ref{e45}) and (\ref{e46}) can be equivalently written as
\begin{align}\label{eb2}
&\hspace{-1.0cm}N_{\nu}(a,b) = \frac{1}{\nu!} \Big(\frac{\ii U}{2\hbar}\Big)^{\nu} \sum_{\sigma_1,\dots, \sigma_{\nu}}\; \sum_{l_1,\dots,l_{\nu} =1}^{N_{\textsc{s}}} \int \prod_{j=1}^{\nu} \rd t_{j}\; \nonumber\\
&\hspace{4.0cm}\times A_{2\nu+1}^{\textsc{b}}(a,b;\alpha_1,\alpha_2,\dots,\alpha_{2\nu-1},\alpha_{2\nu}),
\end{align}
\begin{equation}\label{eb3}
D_{\nu} = \frac{1}{\nu!} \Big(\frac{\ii U}{2\hbar}\Big)^{\nu} \sum_{\sigma_1,\dots, \sigma_{\nu}}\; \sum_{l_1,\dots,l_{\nu} =1}^{N_{\textsc{s}}} \int \prod_{j=1}^{\nu} \rd t_{j}\; A_{2\nu}(\alpha_1,\alpha_2,\dots,\alpha_{2\nu-1},\alpha_{2\nu}).
\end{equation}

With reference to the specifications in Eq.\,(\ref{e36}), for the function $M_{\nu}(a,b)$, Eq.\,(\ref{e259b}), we introduce the alternative notation
\begin{equation}\label{eb3b}
M_{\nu}(a,b) \equiv M_{\nu;\sigma,\sigma'}^{l,l'}(t-t').
\end{equation}
For the cases where the non-interacting Green function $G_{\X{0}}(\alpha_i,\alpha_j)$ is diagonal in the spin space,\footnote{See the discussions centred on Eq.\,(\protect\ref{e11a1}).}\footnote{Following Eq.\,(\protect\ref{e37}), in this appendix $G_{\sigma_i,\sigma_j}^{l_i,l_j}(t_i-t_j) \equiv G(i,j)$. Similarly as regards $G_{\protect\X{0}}(i,j)$.} that is for (see Eq.\,(\ref{eb1}))
\begin{equation}\label{e47}
G_{\X{0}}(\alpha_i,\alpha_j) = \left\{\begin{array}{ll}
G_{\X{0};\sigma_{\theta_i}}^{l_{\theta_i},l_{\theta_j}}(t_{\theta_i}-t_{\theta_j})\hspace{0.6pt} \delta_{\sigma_{\theta_i},\sigma_{\theta_j}}, & i = \text{odd},\, j = \text{odd}, \\ \\
G_{\X{0};\sigma_{\theta_i}}^{l_{\theta_i},l_{\theta_j}}(t_{\theta_i}-t_{\theta_j})\hspace{0.6pt} \delta_{\sigma_{\theta_i},\b{\sigma}_{\theta_j}}, & i = \text{odd},\, j = \text{even}, \\ \\
G_{\X{0};\b{\sigma}_{\theta_i}}^{l_{\theta_i},l_{\theta_j}}(t_{\theta_i}-t_{\theta_j})\hspace{0.6pt} \delta_{\b{\sigma}_{\theta_i},\sigma_{\theta_j}}, & i = \text{even},\, j = \text{odd}, \\ \\
G_{\X{0};\b{\sigma}_{\theta_i}}^{l_{\theta_i},l_{\theta_j}}(t_{\theta_i}-t_{\theta_j})\hspace{0.6pt} \delta_{\b{\sigma}_{\theta_i},\b{\sigma}_{\theta_j}}, & i = \text{even},\, j = \text{even}, \end{array} \right.
\end{equation}
from the defining expression for $M_{\nu}(a,b)$, Eq.\,(\ref{e259b}), one obtains\,\footnote{As regards to the minus sign directly following the equality sign, note that here we are explicitly dealing with fermions.}
\begin{align}\label{eb4}
&\hspace{-1.5cm}M_{\nu;\sigma,\sigma'}^{l,l'}(t-t') = -\frac{1}{\nu !} \Big(\frac{\ii U}{2\hbar}\Big)^{\nu} \sum_{r,s=1}^{\nu}
\sum_{\sigma_1,\dots, \sigma_{\nu}} \sum_{l_1,\dots,l_{\nu} =1}^{N_{\textsc{s}}} \int \prod_{j=1}^{\nu} \rd t_{j}\; \nonumber\\
&\hspace{1.6cm} \times \Big\{ A_{\alpha_{2r-1},\alpha_{2s-1}}^{\X{(2\nu-1)}}(\alpha_1,\alpha_2,\dots,\alpha_{2\nu-1},\alpha_{2\nu}) \hspace{0.7pt} \hspace{0.7pt} \delta_{\sigma,\sigma_{s}} \delta_{\sigma',\sigma_{r}} \nonumber\\
&\hspace{1.76cm} +A_{\alpha_{2r},\alpha_{2s-1}}^{\X{(2\nu-1)}}(\alpha_1,\alpha_2,\dots,\alpha_{2\nu-1},\alpha_{2\nu})\hspace{0.7pt} \hspace{0.7pt} \delta_{\sigma,\sigma_{s}} \delta_{\sigma',\b{\sigma}_{r}} \nonumber\\
&\hspace{1.76cm} +A_{\alpha_{2r-1},\alpha_{2s}}^{\X{(2\nu-1)}}(\alpha_1,\alpha_2,\dots,\alpha_{2\nu-1},\alpha_{2\nu})\hspace{0.7pt} \hspace{0.7pt} \delta_{\sigma,\b{\sigma}_{s}} \delta_{\sigma',\sigma_{r}} \nonumber\\
&\hspace{1.76cm} +A_{\alpha_{2r},\alpha_{2s}}^{\X{(2\nu-1)}}(\alpha_1,\alpha_2,\dots,\alpha_{2\nu-1},\alpha_{2\nu})\hspace{0.7pt} \hspace{0.7pt} \delta_{\sigma,\b{\sigma}_{s}} \delta_{\sigma',\b{\sigma}_{r}}\Big\} \nonumber\\
&\hspace{5.6cm} \times G_{\X{0};\sigma}^{l,l_{s}}(t-t_{s}^+) G_{\X{0};\sigma'}^{l_{r},l'}(t_{r}-t'),
\end{align}
where (\emph{cf.} Eq.\,(\ref{e254x}))
\begin{equation}\label{eb5}
A_{\alpha_r,\alpha_s}^{\X{(2\nu-1)}} = \frac{\partial A_{2\nu}}{\partial G_{\X{0}}(\alpha_r,\alpha_s^+)},
\end{equation}
in which $\alpha_{s}^+ \doteq l_{\theta_s}^{\phantom{+}} t_{\theta_s}^+ \sigma_{\theta_s}^{\phantom{+}}$ for $s$ \textsl{odd}, and  $\alpha_{s}^+ \doteq l_{\theta_s}^{\phantom{+}} t_{\theta_s}^+ \b{\sigma}_{\theta_s}^{\phantom{+}}$ for $s$ \textsl{even}.

To simplify the expression in Eq.\,(\ref{eb4}), let
\begin{equation}\label{eb7a}
\sideset{}{{}^{\X{(r)}}}{\sum}_{\sigma_1,\dots,\sigma_{\nu}}
\end{equation}
denote the $(\nu-1)$-fold summation with respect to $\{\sigma_1$, \dots, $\sigma_{\nu}\}\backslash \{\sigma_r\}$.\footnote{See appendix \protect\ref{sae}.} Similarly, for $r \not= s$ let
\begin{equation}\label{eb7b}
\sideset{}{{}^{\X{(r,s)}}}{\sum}_{\sigma_1,\dots,\sigma_{\nu}}
\end{equation}
denote the $(\nu-2)$-fold summation with respect to $\{\sigma_1$, \dots, $\sigma_{\nu}\}\backslash \{\sigma_{r}, \sigma_{s}\}$. For $\nu =1$, the sum in Eq.\,(\ref{eb7a}) is to be identified with \textsl{unity}, and that in Eq.\,(\ref{eb7b}) with \textsl{zero}. For $\nu=2$, the sum in Eq.\,(\ref{eb7b}) is to be identified with \textsl{unity}. Following these specifications, the expression in Eq.\,(\ref{eb4}) can be written as
\begin{align}\label{eb4a}
&\hspace{-1.5cm}M_{\nu;\sigma,\sigma'}^{l,l'}(t-t') = -\frac{1}{\nu !} \Big(\frac{\ii U}{2\hbar}\Big)^{\nu} \sum_{r=1}^{\nu}\hspace{0.2cm}
\sideset{}{{}^{\X{(r)}}}{\sum}_{\sigma_1,\dots, \sigma_{\nu}} \sum_{l_1,\dots,l_{\nu} =1}^{N_{\textsc{s}}} \int \prod_{j=1}^{\nu} \rd t_{j}\; \nonumber\\
&\hspace{1.6cm} \times \Big\{ \left. A_{\alpha_{2r-1},\alpha_{2r-1}}^{\X{(2\nu-1)}}(\alpha_1,\alpha_2,\dots,\alpha_{2\nu-1},\alpha_{2\nu})
\right|_{\alpha_{2r-1} = l_r t_r\sigma} \hspace{0.7pt} \delta_{\sigma,\sigma'} \nonumber\\
&\hspace{1.76cm} + \left. A_{\alpha_{2r},\alpha_{2r-1}}^{\X{(2\nu-1)}}(\alpha_1,\alpha_2,\dots,\alpha_{2\nu-1},\alpha_{2\nu})\right|_{\substack{ \hspace{0.35cm}\alpha_{2r}= l_r t_r\b{\sigma} \\ \alpha_{2r-1}= l_r t_r\sigma}}\hspace{0.7pt} \delta_{\sigma,\b{\sigma}'} \nonumber\\
&\hspace{1.76cm} + \left. A_{\alpha_{2r-1},\alpha_{2r}}^{\X{(2\nu-1)}}(\alpha_1,\alpha_2,\dots,\alpha_{2\nu-1},\alpha_{2\nu})\right|_{\substack{ \alpha_{2r-1}=  l_r t_r\b{\sigma} \\ \hspace{0.35cm}\alpha_{2r}=  l_r t_r\sigma}} \hspace{0.7pt} \delta_{\b{\sigma},\sigma'} \nonumber\\
&\hspace{1.76cm} +\left. A_{\alpha_{2r},\alpha_{2r}}^{\X{(2\nu-1)}}(\alpha_1,\alpha_2,\dots,\alpha_{2\nu-1},\alpha_{2\nu})\right|_{\alpha_{2r}=  l_r t_r\sigma}\hspace{0.7pt} \delta_{\sigma,\sigma'} \Big\} \nonumber\\
&\hspace{5.2cm} \times G_{\X{0};\sigma}^{l,l_{r}}(t-t_{r}^+) G_{\X{0};\sigma'}^{l_{r},l'}(t_{r}-t')\nonumber\\
&\hspace{1.2cm} -\frac{1}{\nu !} \Big(\frac{\ii U}{2\hbar}\Big)^{\nu} \sum_{r=1}^{\nu} \sum_{\substack{s=1 \\ s \not= r}}^{\nu} \hspace{0.2cm}
\sideset{}{{}^{\X{(r,s)}}}{\sum}_{\sigma_1,\dots, \sigma_{\nu}} \sum_{l_1,\dots,l_{\nu} =1}^{N_{\textsc{s}}} \int \prod_{j=1}^{\nu} \rd t_{j}\; \nonumber\\
&\hspace{1.6cm} \times \Big\{ \left. A_{\alpha_{2r-1},\alpha_{2s-1}}^{\X{(2\nu-1)}}(\alpha_1,\alpha_2,\dots,\alpha_{2\nu-1},\alpha_{2\nu})
 \right|_{\substack{\alpha_{2r-1} = l_r t_r\sigma' \\ \alpha_{2s-1} = l_s t_s\sigma}}\nonumber\\
&\hspace{1.76cm} + \left. A_{\alpha_{2r},\alpha_{2s-1}}^{\X{(2\nu-1)}}(\alpha_1,\alpha_2,\dots,\alpha_{2\nu-1},\alpha_{2\nu})\right|_{\substack{ \hspace{0.35cm}\alpha_{2r}= l_r t_r\sigma' \\ \alpha_{2s-1}= l_s t_s\sigma}}
 \nonumber\\
&\hspace{1.76cm} + \left. A_{\alpha_{2r-1},\alpha_{2s}}^{\X{(2\nu-1)}}(\alpha_1,\alpha_2,\dots,\alpha_{2\nu-1},\alpha_{2\nu}) \right|_{\substack{ \alpha_{2r-1}=  l_r t_r\sigma' \\ \hspace{0.35cm}\alpha_{2s}=  l_s t_s\sigma}} \nonumber\\
&\hspace{1.76cm} +\left. A_{\alpha_{2r},\alpha_{2s}}^{\X{(2\nu-1)}}(\alpha_1,\alpha_2,\dots,\alpha_{2\nu-1},\alpha_{2\nu})
\right|_{\substack{ \alpha_{2r}=  l_r t_r\sigma' \\ \alpha_{2s}=  l_s t_s\sigma}} \Big\} \nonumber\\
&\hspace{5.8cm} \times G_{\X{0};\sigma}^{l,l_{s}}(t-t_{s}^+) G_{\X{0};\sigma'}^{l_{r},l'}(t_{r}-t').
\end{align}
This expression simplifies considerably in the light of the following observations. Firstly, the two terms on the $3$rd and $4$th lines (that is, those multiplying respectively $\delta_{\sigma,\b{\sigma}'}$ and $\delta_{\b{\sigma},\sigma'}$) are identically vanishing on account of the identities in Eq.\,(\ref{eb7}) below. Secondly, the two terms on the $2$nd and $5$th lines are identically equal, that is
\begin{equation}\label{eb4d}
\left. A_{\alpha_{2r-1},\alpha_{2r-1}}^{\X{(2\nu-1)}}\right|_{\alpha_{2r-1} = l_r t_r\sigma}
\equiv \left. A_{\alpha_{2r},\alpha_{2r}}^{\X{(2\nu-1)}}\right|_{\alpha_{2r}=  l_r t_r\sigma}.
\end{equation}
Thirdly, the terms on the $8$th, $9$th, $10$th and $11$th lines of the expression in Eq.\,(\ref{eb4a}) are identically equal, that is for $r\not= s$
\begin{align}\label{eb4e}
&\hspace{0.0cm}
\left. A_{\alpha_{2r-1},\alpha_{2s-1}}^{\X{(2\nu-1)}}\right|_{\substack{\alpha_{2r-1} = l_r t_r\sigma' \\ \alpha_{2s-1} = l_s t_s\sigma}} \equiv \left.A_{\alpha_{2r},\alpha_{2s-1}}^{\X{(2\nu-1)}}\right|_{\substack{ \hspace{0.35cm}\alpha_{2r}= l_r t_r\sigma' \\ \alpha_{2s-1}= l_s t_s\sigma}} \nonumber\\
&\hspace{-0.5cm}
\equiv \left. A_{\alpha_{2r-1},\alpha_{2s}}^{\X{(2\nu-1)}}\right|_{\substack{ \alpha_{2r-1}=  l_r t_r\sigma' \\ \hspace{0.35cm}\alpha_{2s}=  l_s t_s\sigma}} \equiv
\left. A_{\alpha_{2r},\alpha_{2s}}^{\X{(2\nu-1)}}
\right|_{\substack{ \alpha_{2r}=  l_r t_r\sigma' \\ \alpha_{2s}=  l_s t_s\sigma}}.
\end{align}
Fourthly, the latter four functions are proportional to $\delta_{\sigma,\sigma'}$. On the basis of these observations, for the function $M_{\nu;\sigma}^{\X{l,l'}}(t-t')$ defined according to
\begin{equation}\label{eb3c}
M_{\nu;\sigma,\sigma'}^{l,l'}(t-t') = M_{\nu;\sigma}^{l,l'}(t-t')\hspace{0.7pt} \delta_{\sigma,\sigma'},
\end{equation}
one obtains
\begin{align}\label{eb4b}
&\hspace{-1.05cm}M_{\nu;\sigma}^{l,l'}(t-t') = -\frac{2}{\nu !} \Big(\frac{\ii U}{2\hbar}\Big)^{\nu} \sum_{r=1}^{\nu} \,\sum_{l_1,\dots,l_{\nu} =1}^{N_{\textsc{s}}} \int \prod_{j=1}^{\nu} \rd t_{j}\; G_{\X{0};\sigma}^{l,l_{r}}(t-t_{r}^+) G_{\X{0};\sigma}^{l_{r},l'}(t_{r}-t') \nonumber\\
&\hspace{1.6cm}\times \hspace{0.2cm}
\sideset{}{{}^{\X{(r)}}}{\sum}_{\sigma_1,\dots, \sigma_{\nu}} \left. A_{\alpha_{2r-1},\alpha_{2r-1}}^{\X{(2\nu-1)}}(\alpha_1,\alpha_2,\dots,\alpha_{2\nu-1},\alpha_{2\nu})
\right|_{\alpha_{2r-1} = l_r t_r\sigma} \nonumber\\
&\hspace{1.2cm} -\frac{4}{\nu !} \Big(\frac{\ii U}{2\hbar}\Big)^{\nu} \sum_{r=1}^{\nu} \sum_{\substack{s=1 \\ s \not= r}}^{\nu}
\,\sum_{l_1,\dots,l_{\nu} =1}^{N_{\textsc{s}}} \int \prod_{j=1}^{\nu} \rd t_{j}\; G_{\X{0};\sigma}^{l,l_{s}}(t-t_{s}^+) G_{\X{0};\sigma}^{l_{r},l'}(t_{r}-t') \nonumber\\
&\hspace{1.6cm} \times \hspace{0.2cm}
\sideset{}{{}^{\X{(r,s)}}}{\sum}_{\sigma_1,\dots, \sigma_{\nu}} \left. A_{\alpha_{2r-1},\alpha_{2s-1}}^{\X{(2\nu-1)}}(\alpha_1,\alpha_2,\dots,\alpha_{2\nu-1},\alpha_{2\nu})
 \right|_{\substack{\alpha_{2r-1} = l_r t_r\sigma \\ \alpha_{2s-1} = l_s t_s\sigma}}.
\end{align}

The results in Eqs\,(\ref{eb4d}) and (\ref{eb4e}) are directly related to the way in which the compound variables $\{\alpha_j\| j\}$ are defined, Eq.\,(\ref{eb1}) (see also Eq.\,(\ref{e47})).\footnote{These results can be explicitly established on the basis of the expression in Eq.\,(\protect\ref{e325b}) in conjunction with that in Eq.\,(\protect\ref{e325d}) (\emph{cf.} Eq.\,(\protect\ref{eb11}) below). For instance, to establish the identity in Eq.\,(\protect\ref{eb4d}), one defines an integer-valued mapping $f$ that satisfies $f(2r-1)= 2r$ and maps the ordered set $\{1,2,\dots,2\nu\}\backslash \{2r-1\}$ onto the ordered set $\{1,2,\dots,2\nu\}\backslash \{2r\}$. In this way, the summation with respect to the elements of the set $S_{2r,2r}^{\protect\X{(\nu)}}$ underlying the expression for $A_{\alpha_{2r},\alpha_{2r}}^{\protect\X{(2\nu-1)}}$ can be expressed as one with respect to the elements of the set $S_{2r-1,2r-1}^{\protect\X{(\nu)}}$. On rearranging the relevant product, Eq.\,(\protect\ref{e325d}), through introducing the variable $j' \equiv f(j)$, whereby $j = f^{\protect\X{-1}}(j')$, making use of the associations in Eq.\,(\protect\ref{eb1b}),  one arrives at the identity in Eq.\,(\protect\ref{eb4d}). The identities in Eq.\,(\protect\ref{eb4e}) are explicitly established similarly. We point out that this approach is rendered considerably more transparent by considering the set $S_{r,s}^{\protect\X{(\nu)}}$ as consisting of the relevant $2\nu$-permutations, rather than of their indices. In this way, the sum $\sum_{l\in S_{r,s}^{\protect\X{(2\nu)}}}$ in Eq.\,(\ref{e325b}) is expressed as $\sum_{P\in S_{r,s}^{\protect\X{(2\nu)}}}$.} According to this definition, the elements $G_{\X{0}}(\alpha_i,\alpha_{2j-1}^+)$ and $G_{\X{0}}(\alpha_i,\alpha_{2j}^+)$, $\forall i \in \{1,2,\dots,2\nu\}$ and $\forall j \in \{1,2,\dots,\nu\}$, of the $2\nu \times 2\nu$ matrix $\mathbb{A}_{2\nu}$ differ only in the different spin indices associated with $\alpha_{2j-1}$ and $\alpha_{2j}$: $\sigma_j$ with the former, and $\b{\sigma}_j$ with the latter. Similarly as regards the matrix elements $G_{\X{0}}(\alpha_{2i-1},\alpha_{j}^+)$ and $G_{\X{0}}(\alpha_{2i},\alpha_{j}^+)$,  $\forall i \in \{1,2,\dots,\nu\}$, $\forall j \in \{1,2,\dots,2\nu\}$. Considering in this light the cofactors in Eq.\,(\ref{eb4d}), one immediately realises that, neglecting the substitutions, the cofactor on the left is a function of $\b{\sigma}_r$ and that the cofactor on the right is the same function however of $\sigma_r$ (compare with the first two equalities in Eq.\,(\ref{ebx2}) below).\footnote{Also compare $A_{\alpha_1,\alpha_1}^{\protect\X{(3)}}$ with $A_{\alpha_2,\alpha_2}^{\protect\X{(3)}}$, and $A_{\alpha_3,\alpha_3}^{\protect\X{(3)}}$ with $A_{\alpha_4,\alpha_4}^{\protect\X{(3)}}$ in respectively Eqs\,(\protect\ref{eby1}), (\protect\ref{eby6}), (\protect\ref{eby11}), and (\protect\ref{eby16}) below.} The identity follows on account of the substitutions: the one on the left substitutes $\sigma$ for $\b{\sigma}_r$, and that on the right $\sigma$ for $\sigma_r$. With some slight modifications, similar arguments apply to the identities in Eq.\,(\ref{eb4e}). In this connection, we note that the first two and the last two cofactors in Eq.\,(\ref{eb4e}) correspond to two matrices whose two columns are interchanged. The minus signs arising from this are compensated by the fact that cofactors are \textsl{signed} minors. The sign associated with the first and fourth cofactors in Eq.\,(\ref{eb4e}) is $+$ (since $2r-1+2s-1$ and $2r+2s$ are \textsl{even}), and that with the second and third cofactors $-$ (since $2r+2s-1$ is \textsl{odd}). For the case of $\nu=2$, in regard to the identity in Eq.\,(\ref{eb4d}) (for $r=2$), compare the expressions in Eqs\,(\ref{eby11}) and (\ref{eby16}) below, and in regard to the identities in Eq.\,(\ref{eb4e}) (for $r=1$ and $s=2$), compare the expressions in Eqs\,(\ref{eby3}), (\ref{eby7}), (\ref{eby4}), and (\ref{eby8}) below.

It remains to establish that
\begin{equation}\label{eb7}
A_{\alpha_{2r-1},\alpha_{2r}}^{\X{(2\nu-1)}} \equiv A_{\alpha_{2r},\alpha_{2r-1}}^{\X{(2\nu-1)}}\equiv 0,\;\;\; \forall r \in \{1,2,\dots,\nu\}.
\end{equation}
We emphasize that while in these identities the indices of $\alpha$ are odd-even and even-odd, the even index is to be the greater of the two indices. Thus while $A_{\alpha_1,\alpha_2}^{\X{(2\nu-1)}} \equiv A_{\alpha_2,\alpha_1}^{\X{(2\nu-1)}} \equiv 0$, neither of the two functions $A_{\alpha_2,\alpha_3}^{\X{(2\nu-1)}}$ and $A_{\alpha_3,\alpha_2}^{\X{(2\nu-1)}}$ is necessarily identically vanishing. We establish the validity of the identities in Eq.\,(\ref{eb7}) by expressing $A_{\alpha_{2r-1},\alpha_{2r}}^{\X{(2\nu-1)}}$ and $A_{\alpha_{2r},\alpha_{2r-1}}^{\X{(2\nu-1)}}$ in explicit form. Focussing on  $A_{\alpha_{2r-1},\alpha_{2r}}^{\X{(2\nu-1)}}$, following Eqs\,(\ref{e325b}), (\ref{e325a}), and (\ref{e325d}), one has
\begin{equation}\label{eb11}
A_{\alpha_{2r-1},\alpha_{2r}}^{\X{(2\nu-1)}}\hspace{-2.0pt} = \hspace{-3.8pt}\sum_{l \in S_{2r-1,2r}^{\X{(\nu)}}} \hspace{-3.8pt}\upsigma_l^{\X{(2\nu)}} G_{\X{0}}(\alpha_{2r},\alpha_{P_l^{\X{(2\nu)}}(2r)}^+)\hspace{-2.0pt} \prod_{\substack{j=1 \\ j\not= r}}^{\nu}\hspace{-2.0pt} G_{\X{0}}(\alpha_{2j-1},\alpha_{P_l^{\X{(2\nu)}}(2j-1)}^+)\hspace{0.4pt} G_{\X{0}}(\alpha_{2j},\alpha_{P_l^{\X{(2\nu)}}(2j)}^+).
\end{equation}
With reference to Eq.\,(\ref{e47}), in the light of Eqs\,(\ref{eb1}) and (\ref{eb1b}) one observes that in order for the summand of the summation with respect to $l$, with $l \in S_{2r-1,2r}^{\X{(\nu)}}$, not be identically vanishing, it is \textsl{necessary} that the following conditions be \textsl{simultaneously} satisfied:\,\footnote{These inequalities are based on the following considerations. For $P_l^{\protect\X{(2\nu)}}(2j-1) = 2j$ for \textsl{some} $j$ inside the set indicated, the function $G_{\protect\X{0}}(\alpha_{2j-1},\alpha_{P_l^{\protect\X{(2\nu)}}(2j-1)}^+)$ on the RHS of Eq.\,(\protect\ref{eb11}) will be proportional to $\delta_{\sigma_j,\protect\b{\sigma}_j}$, which is identically vanishing. By the same reasoning, $G_{\protect\X{0}}(\alpha_{2j},\alpha_{P_l^{\protect\X{(2\nu)}}(2j)}^+)$ is identically vanishing for some $j$ inside the indicated set satisfying $P_l^{\protect\X{(2\nu)}}(2j) = 2j-1$.}
\begin{align}\label{eb12}
&\hspace{-1.0cm}P_l^{\X{(2\nu)}}(2j-1) \not= 2j, \;\;\forall j \in \{1,2,\dots,\nu\}\backslash \{r\}, \nonumber\\
&\hspace{-1.0cm}P_l^{\X{(2\nu)}}(2j) \not= 2j-1, \;\;\forall j \in \{1,2,\dots,\nu\}.
\end{align}
The combination of these inequalities, which constitute $2\nu-1$ conditions, with the equality $P_l^{\X{(2\nu)}}(2r-1) = 2r$, satisfied for all $l \in  S_{2r-1,2r}^{\X{(\nu)}}$,\footnote{See Eq.\,(\protect\ref{e325a}).} gives rise to $A_{\alpha_{2r-1},\alpha_{2r}}^{\X{(2\nu-1)}} \equiv 0$. A similar reasoning establishes the validity of $A_{\alpha_{2r},\alpha_{2r-1}}^{\X{(2\nu-1)}} \equiv 0$. For clarity, the cycle decomposition of the $2\nu$-permutation $\mathscr{P}_l^{\X{(2\nu)}}$, with $l\in S_{2r-1,2r}^{\X{(\nu)}}$ (with $l\in S_{2r,2r-1}^{\X{(\nu)}}$ in considering $A_{\alpha_{2r},\alpha_{2r-1}}^{\X{(2\nu-1)}}$), satisfying the inequalities in Eq.\,(\ref{eb12}) invariably includes a cycle of length greater than $2$, involving both $2r-1$ and $2r$, for which the corresponding product of Green functions is identically vanishing in consequence of the conservation of the spin.\footnote{See the discussions centred on Eq.\,(\protect\ref{e11a1}).} For illustration, consider the case of $\nu= 4$ and $r=2$, for which one has the following generic $8$-permutation whose $l$ belongs to the set $S_{3,4}^{\X{(8)}}$:
\begin{equation}\label{eb12a}
\mathscr{P}_l^{\X{(8)}}:\;
\begin{pmatrix}
1 & 2 & \bm{3} & 4 & 5 & 6 & 7 & 8 \\
\X{\neq}\,2 & \X{\neq}\,1 & \bm{4} & \X{\neq}\,3 & \X{\neq}\,6 & \X{\neq}\,5 & \X{\neq}\,8 & \X{\neq}\,7
\end{pmatrix},\;\;\;  l \in S_{3,4}^{\X{(8)}}.
\end{equation}
One observes that an $8$-permutation satisfying the specified conditions cannot contain the $2$-cycle $(4,3)$ in its cycle decomposition. For instance, for the specific $8$-permutation\,\footnote{Clearly, $l_1 \in S_{3,4}^{\protect\X{(8)}}$.}
 \begin{equation}\label{eb12b}
\mathscr{P}_{l_1}^{\X{(8)}}:\;
\begin{pmatrix}
1 & 2 & \bm{3} & 4 & 5 & 6 & 7 & 8 \\
1 & 3 & \bm{4} & 2 & 8 & 7 & 6 & 5
\end{pmatrix}
= (1) (\bm{4},2,\bm{3}) (8,5) (7,6),
\end{equation}
$3$ and $4$ are members of a $3$-cycle. With reference to Eqs\,(\ref{e335}) and (\ref{e336}), the contribution of the cycle $(4,2,3)$ to $A_{\alpha_3,\alpha_4}^{\X{(7)}}$ amounts to $G_{\X{0}}(\alpha_4,\alpha_2^+) G_{\X{0}}(\alpha_2,\alpha_3^+)$,\footnote{On account of the third line on the RHS of Eq.\,(\protect\ref{e323}), $G_{\protect\X{0}}(\alpha_3,\alpha_4^+)$ is identified with $1$ (see also the RHS of Eq.\,(\protect\ref{e335})).} which, following the specifications in Eq.\,(\ref{e47}), is proportional to $\delta_{\b{\sigma}_2,\b{\sigma}_1} \delta_{\b{\sigma}_1,\sigma_2} \equiv 0$. For the specific case of $\nu=2$, see Eqs\,(\ref{eby2}), (\ref{eby5}), (\ref{eby12}), and (\ref{eby15}) below.

We note that by introducing the compound indices $\{\gamma_j \| j\}$, where
\begin{equation}\label{eb19}
\gamma_j \rightleftharpoons \alpha_{2j-1}, \alpha_{2j},\;\, j\in \{1,2,\dots,\nu\},
\end{equation}
for convenience one may identify the $2\nu \times 2\nu$ matrix $\mathbb{A}_{2\nu}$ with the $\nu \times\nu$ matrix $\mathbbmss{A}_{\nu}$ of $2\times 2$ matrices, that is
\begin{equation}\label{eb29}
\mathbb{A}_{2\nu}(\alpha_1,\alpha_2,\dots,\alpha_{2\nu}) \equiv \mathbbmss{A}_{\nu}(\gamma_1,\dots,\gamma_{\nu})
\doteq
\begin{pmatrix}
(\mathbbmss{A}_{\nu})_{\gamma_1;\gamma_1} &  (\mathbbmss{A}_{\nu})_{\gamma_1;\gamma_2} &\dots &  (\mathbbmss{A}_{\nu})_{\gamma_1;\gamma_{\nu}} \\
(\mathbbmss{A}_{\nu})_{\gamma_2;\gamma_1} &  (\mathbbmss{A}_{\nu})_{\gamma_2;\gamma_2} &\dots &  (\mathbbmss{A}_{\nu})_{\gamma_2;\gamma_{\nu}} \\
\vdots & \vdots & \ddots & \vdots \\
(\mathbbmss{A}_{\nu})_{\gamma_{\nu};\gamma_1} &  (\mathbbmss{A}_{\nu})_{\gamma_{\nu};\gamma_2} &\dots &  (\mathbbmss{A}_{\nu})_{\gamma_{\nu};\gamma_{\nu}}
\end{pmatrix},
\end{equation}
where
\begin{equation}\label{eb30}
(\mathbbmss{A}_{\nu})_{\gamma_i;\gamma_j} \equiv (\mathbbmss{A}_{\nu})_{\alpha_{2i-1},\alpha_{2i};\alpha_{2j-1},\alpha_{2j}} \doteq
\begin{pmatrix}
G_{\X{0}}(\alpha_{2i-1},\alpha_{2j-1}^+) & G_{\X{0}}(\alpha_{2i-1},\alpha_{2j}^+) \\ \\
G_{\X{0}}(\alpha_{2i},\alpha_{2j-1}^+) & G_{\X{0}}(\alpha_{2i},\alpha_{2j}^+)
\end{pmatrix}.
\end{equation}
With $\mathsf{A}_{\nu} \doteq \det(\mathbbmss{A}_{\nu})$, one naturally has $\mathsf{A}_{\nu} = A_{2\nu} \equiv \det(\mathbb{A}_{2\nu})$. For the Green function $G_{\X{0}}$ in Eq.\,(\ref{e47}), one thus has
\begin{align}\label{eb31}
(\mathbbmss{A}_{\nu})_{\gamma_i;\gamma_j} &=
\begin{pmatrix}
G_{0;\sigma_i}^{l_i,l_j}(t_i-t_j^+)\hspace{0.6pt} \delta_{\sigma_i,\sigma_j} & G_{0;\sigma_i}^{l_i,l_j}(t_i-t_j^+)\hspace{0.6pt} \delta_{\sigma_i,\b{\sigma}_j} \\ \\
G_{0;\b{\sigma}_i}^{l_i,l_j}(t_i-t_j^+)\hspace{0.6pt} \delta_{\b{\sigma}_i,\sigma_j} & G_{0;\b{\sigma}_i}^{l_i,l_j}(t_i-t_j^+)\hspace{0.6pt} \delta_{\b{\sigma}_i,\b{\sigma}_j}
\end{pmatrix} \nonumber\\
&\equiv
\begin{pmatrix}
G_{0;\sigma_i}^{l_i,l_j}(t_i-t_j^+) & 0 \\ \\
0 & G_{0;\b{\sigma}_i}^{l_i,l_j}(t_i-t_j^+)
\end{pmatrix} \hspace{0.0pt} \delta_{\sigma_i,\sigma_j} \nonumber\\
&+
\begin{pmatrix}
0 & G_{0;\sigma_i}^{l_i,l_j}(t_i-t_j^+) \\ \\
G_{0;\b{\sigma}_i}^{l_i,l_j}(t_i-t_j^+) & 0
\end{pmatrix} \hspace{0.0pt} \delta_{\sigma_i,\b{\sigma}_j}.
\end{align}
In a spin-unpolarised case, where $G_{0;\sigma}^{l,l'}(t-t') \equiv G_{0;\b{\sigma}}^{l,l'}(t-t')$, the expression in Eq.\,(\ref{eb31}) can be written in terms of the $2\times 2$ unit matrix $\bbsigma^{\X{0}}$ and the $2\times 2$ Pauli matrix $\bbsigma^{\X{x}}$, p.\,\pageref{InRef},\footnote{$(\bbsigma^{\protect\X{x}})_{1,1}  = (\bbsigma^{\protect\X{x}})_{2,2} = 0$, $(\bbsigma^{\protect\X{x}})_{1,2}  = (\bbsigma^{\protect\X{x}})_{2,1} = 1$.} for which one has $\bbsigma^{\X{x}} \bbsigma^{\X{x}} = \bbsigma^{\X{0}}$. The result in Eq.\,(\ref{eb31}) exposes the degree of sparsity \cite{SP84} of the matrix $\mathbbmss{A}_{\nu}$. We point out that since \textsl{at most} one of the two terms on the RHS of Eq.\,(\ref{eb31}) is non-vanishing, multiplications of two arbitrary $2\times 2$ blocks of $\mathbbmss{A}_{\nu}$ involves $2$ scalar multiplications in the general case, and $1$ scalar multiplication in the spin-unpolarised case. This is to be contrasted with the $8$ scalar multiplications required for multiplying two general $2\times 2$ matrices ($7$ multiplications by employing the approach by Strassen \cite{VS69}). For some relevant theoretical considerations regarding \textsl{partitioned matrices}, the reader is referred to the book by Horn and Johnson \cite{HJ13} [see \emph{e.g.} \S\,0.7, p.\,16, herein].

For illustration, below we consider the cases of $\nu=1$ and $\nu=2$. Where on account of the relevant Kronecker delta on the RHS of Eq.\,(\ref{e47}) the function $G_{\X{0}}(\alpha_i,\alpha_j)$ is identically vanishing, we shall replace this function by $0$. Similarly, we shall suppress the Kronecker deltas that are identically equal to $1$.

For $\nu=1$, from the expressions in Eqs\,(\ref{e260}) and (\ref{e47}) one has\,\footnote{Here $0^{\protect\X{-}} \equiv t_i^{\phantom{\protect\X{+}}} - t_i^{\protect\X{+}}$, $\forall i \in\{1,2,\dots,\nu\}$.} (\emph{cf.} Eqs\,(\ref{eb29}) and (\ref{eb31}))
\begin{equation}\label{ebx1}
A_2(\alpha_1,\alpha_2) =
\begin{vmatrix}
\X{G_{\X{0};\sigma_1}^{l_1,l_1}(0^-)} & \X{0} \\ \\
\X{0} & \X{G_{\X{0};\b{\sigma}_1}^{l_1,l_1}(0^-)}
\end{vmatrix}
\equiv G_{\X{0};\sigma_1}^{l_1,l_1}(0^-)\hspace{0.7pt} G_{\X{0};\b{\sigma}_1}^{l_1,l_1}(0^-),
\end{equation}
leading to, following the equality in Eq.\,(\ref{eb5}),
\begin{equation}\label{ebx2}
A_{\alpha_1,\alpha_1}^{\X{(1)}} = G_{\X{0};\b{\sigma}_1}^{l_1,l_1}(0^-),\;\; A_{\alpha_2,\alpha_2}^{\X{(1)}} = G_{\X{0};\sigma_1}^{l_1,l_1}(0^-),\;\; A_{\alpha_1,\alpha_2}^{\X{(1)}} \equiv A_{\alpha_2,\alpha_1}^{\X{(1)}} \equiv 0.
\end{equation}
Note that the last two identities are in conformity with those in Eq.\,(\ref{eb7}). Further, since $\alpha_1 \rightleftharpoons l_1 t_1\sigma_1$ and $\alpha_2 \rightleftharpoons l_1 t_1\b{\sigma}_1$, Eq.\,(\ref{eb1b}), it follows that (note the $\b{\sigma}$ on the RHS)
\begin{equation}\label{ebx3}
\left. A_{\alpha_1,\alpha_1}^{\X{(1)}}\right|_{\alpha_1 = l_1 t_1\sigma} \equiv \left. A_{\alpha_2,\alpha_2}^{\X{(1)}}\right|_{\alpha_2 = l_1 t_1\sigma} \equiv  G_{\X{0};\b{\sigma}}^{l_1,l_1}(0^-),
\end{equation}
in conformity with the identity in Eq.\,(\ref{eb4d}).

For $\nu=2$, from the expressions in Eqs\,(\ref{e260}) and (\ref{e47}) one has (\emph{cf.} Eqs\,(\ref{eb29}) and (\ref{eb31}))
\begin{align}\label{e48}
&\hspace{0.2cm}A_{4}(\alpha_1,\alpha_2,\alpha_3,\alpha_4) \nonumber\\
&\hspace{1.0cm} =
\begin{vmatrix}
\X{G_{\X{0};\sigma_1}^{l_1,l_1}(0^-)} & \X{0} & \X{G_{\X{0};\sigma_1}^{l_1,l_2}(t_1-t_2^+) \delta_{\sigma_1,\sigma_2}} &  \X{G_{\X{0};\sigma_1}^{l_1,l_2}(t_1-t_2^+) \delta_{\sigma_1,\b{\sigma}_2}} \\ \\
\X{0} & \X{G_{\X{0};\b{\sigma}_1}^{l_1,l_1}(0^-)} & \X{G_{\X{0};\b{\sigma}_1}^{l_1,l_2}(t_1-t_2^+) \delta_{\b{\sigma}_1,\sigma_2}} & \X{G_{\X{0};\b{\sigma}_1}^{l_1,l_2}(t_1-t_2^+) \delta_{\b{\sigma}_1,\b{\sigma}_2}} \\ \\
\X{G_{\X{0};\sigma_2}^{l_2,l_1}(t_2-t_1^+) \delta_{\sigma_1,\sigma_2}} & \X{G_{\X{0};\sigma_2}^{l_2,l_1}(t_2-t_1^+) \delta_{\b{\sigma}_1,\sigma_2}} & \X{G_{\X{0};\sigma_2}^{l_2,l_2}(0^-)} & \X{0} \\ \\
\X{G_{\X{0};\b{\sigma}_2}^{l_2,l_1}(t_2-t_1^+) \delta_{\sigma_1,\b{\sigma}_2}} & \X{G_{\X{0};\b{\sigma}_2}^{l_2,l_1}(t_2-t_1^+) \delta_{\b{\sigma}_1,\b{\sigma}_2}} & \X{0} & \X{G_{\X{0};\b{\sigma}_2}^{l_2,l_2}(0^-)}
\end{vmatrix}_{-}, \nonumber\\
\end{align}
leading to (\emph{cf.} Eq.\,(\ref{eb1b}))
\begin{align}\label{e49}
&\hspace{-0.1cm} A_{4}(l_1 t_1\!\uparrow,l_1 t_1\!\downarrow,l_2 t_2\!\uparrow,l_2 t_2\!\downarrow) =
\begin{vmatrix}
\X{G_{\X{0};\uparrow}^{l_1,l_1}(0^-)} & \X{0} & \X{G_{\X{0};\uparrow}^{l_1,l_2}(t_1-t_2^+)} &  \X{0} \\ \\
\X{0} & \X{G_{\X{0};\downarrow}^{l_1,l_1}(0^-)} & \X{0} & \X{G_{\X{0};\downarrow}^{l_1,l_2}(t_1-t_2^+)} \\ \\
\X{G_{\X{0};\uparrow}^{l_2,l_1}(t_2-t_1^+)} & \X{0} & \X{G_{\X{0};\uparrow}^{l_2,l_2}(0^-)} & \X{0} \\ \\
\X{0} & \X{G_{\X{0};\downarrow}^{l_2,l_1}(t_2-t_1^+)} & \X{0} & \X{G_{\X{0};\downarrow}^{l_2,l_2}(0^-)}
\end{vmatrix}_{-} \nonumber\\
&\hspace{2.4cm} =
\begin{vmatrix}
\X{G_{\X{0};\uparrow}^{l_1,l_1}(0^-)} & \X{G_{\X{0};\uparrow}^{l_1,l_2}(t_1-t_2^+)} \\ \\
\X{G_{\X{0};\uparrow}^{l_2,l_1}(t_2-t_1^+)} &  \X{G_{\X{0};\uparrow}^{l_2,l_2}(0^-)}
\end{vmatrix}_{-} \cdot
\begin{vmatrix}
\X{G_{\X{0};\downarrow}^{l_1,l_1}(0^-)} & \X{G_{\X{0};\downarrow}^{l_1,l_2}(t_1-t_2^+)} \\ \\
\X{G_{\X{0};\downarrow}^{l_2,l_1}(t_2-t_1^+)} &  \X{G_{\X{0};\downarrow}^{l_2,l_2}(0^-)}
\end{vmatrix}_{-},
\end{align}
\begin{align}\label{e50}
&\hspace{-0.1cm} A_{4}(l_1 t_1\!\uparrow,l_1 t_1\!\downarrow,l_2 t_2\!\downarrow,l_2 t_2\uparrow) =
\begin{vmatrix}
\X{G_{\X{0};\uparrow}^{l_1,l_1}(0^-)} & \X{0} & \X{0} &  \X{G_{\X{0};\uparrow}^{l_1,l_2}(t_1-t_2^+)} \\ \\
\X{0} & \X{G_{\X{0};\downarrow}^{l_1,l_1}(0^-)} & \X{G_{\X{0};\downarrow}^{l_1,l_2}(t_1-t_2^+)} & \X{0} \\ \\
\X{0} & \X{G_{\X{0};\downarrow}^{l_2,l_1}(t_2-t_1^+)} & \X{G_{\X{0};\downarrow}^{l_2,l_2}(0^-)} & \X{0} \\ \\
\X{G_{\X{0};\uparrow}^{l_2,l_1}(t_2-t_1^+)} & \X{0} & \X{0} & \X{G_{\X{0};\uparrow}^{l_2,l_2}(0^-)}
\end{vmatrix}_{-} \nonumber\\
&\hspace{2.4cm} =
\begin{vmatrix}
\X{G_{\X{0};\uparrow}^{l_1,l_1}(0^-)} & \X{G_{\X{0};\uparrow}^{l_1,l_2}(t_1-t_2^+)} \\ \\
\X{G_{\X{0};\uparrow}^{l_2,l_1}(t_2-t_1^+)} &  \X{G_{\X{0};\uparrow}^{l_2,l_2}(0^-)}
\end{vmatrix}_{-} \cdot
\begin{vmatrix}
\X{G_{\X{0};\downarrow}^{l_1,l_1}(0^-)} & \X{G_{\X{0};\downarrow}^{l_1,l_2}(t_1-t_2^+)} \\ \\
\X{G_{\X{0};\downarrow}^{l_2,l_1}(t_2-t_1^+)} &  \X{G_{\X{0};\downarrow}^{l_2,l_2}(0^-)}
\end{vmatrix}_{-}.
\end{align}
The second equalities in Eqs\,(\ref{e49}) and (\ref{e50}) follow from the fact that by \textsl{even} permutations of rows and columns of the relevant $4$-matrices, these can be transformed into block-diagonal forms. Regarding the functions $A_{4}$ corresponding to $\sigma_1,\sigma_2 = \downarrow, \uparrow$ and $\sigma_1, \sigma_2 = \downarrow, \downarrow$, the former is obtained from the expression on the RHS of Eq.\,(\ref{e50}) and the latter from that on the RHS of Eq.\,(\ref{e49}) on replacing $\uparrow$ by $\downarrow$, and \emph{vice versa}. One thus obtains\,\footnote{Note that the expression in Eq.\,(\protect\ref{e51}) is directly relevant to the calculation of $D_2$, Eq.\,(\protect\ref{eb3}).}
\begin{align}\label{e51}
\sum_{\sigma_1,\sigma_2} A_{4}(l_1 t_1\sigma_1,l_1 t_1\b{\sigma}_1,l_2 t_2\sigma_2,l_2 t_2\b{\sigma}_2) = &4
\begin{vmatrix}
\X{G_{\X{0};\uparrow}^{l_1,l_1}(0^-)} & \X{G_{\X{0};\uparrow}^{l_1,l_2}(t_1-t_2^+)} \\ \\
\X{G_{\X{0};\uparrow}^{l_2,l_1}(t_2-t_1^+)} &  \X{G_{\X{0};\uparrow}^{l_2,l_2}(0^-)}
\end{vmatrix}_{-} \nonumber\\
&\times
\begin{vmatrix}
\X{G_{\X{0};\downarrow}^{l_1,l_1}(0^-)} & \X{G_{\X{0};\downarrow}^{l_1,l_2}(t_1-t_2^+)} \\ \\
\X{G_{\X{0};\downarrow}^{l_2,l_1}(t_2-t_1^+)} &  \X{G_{\X{0};\downarrow}^{l_2,l_2}(0^-)}
\end{vmatrix}_{-}.
\end{align}

For the $16$ cofactors of the $4$-determinant $A_4$ in Eq.\,(\ref{e48}), one trivially obtains:
\begin{align}\label{eby1}
&\hspace{-0.4cm}A_{\alpha_1,\alpha_1}^{\X{(3)}}(\alpha_1,\alpha_2,\alpha_3,\alpha_4) \nonumber\\
&\hspace{0.8cm} = +
\begin{vmatrix}
\X{G_{\X{0};\b{\sigma}_1}^{l_1,l_1}(0^-)} & \X{G_{\X{0};\b{\sigma}_1}^{l_1,l_2}(t_1-t_2^+) \delta_{\b{\sigma}_1,\sigma_2}} & \X{G_{\X{0};\b{\sigma}_1}^{l_1,l_2}(t_1-t_2^+) \delta_{\b{\sigma}_1,\b{\sigma}_2}} \\ \\
\X{G_{\X{0};\sigma_2}^{l_2,l_1}(t_2-t_1^+) \delta_{\b{\sigma}_1,\sigma_2}} & \X{G_{\X{0};\sigma_2}^{l_2,l_2}(0^-)} & \X{0} \\ \\
\X{G_{\X{0};\b{\sigma}_2}^{l_2,l_1}(t_2-t_1^+) \delta_{\b{\sigma}_1,\b{\sigma}_2}} & \X{0} & \X{G_{\X{0};\b{\sigma}_2}^{l_2,l_2}(0^-)}
\end{vmatrix}_{-},
\end{align}
\begin{align}\label{eby2}
&\hspace{-0.4cm}A_{\alpha_1,\alpha_2}^{\X{(3)}}(\alpha_1,\alpha_2,\alpha_3,\alpha_4) \nonumber\\
&\hspace{0.4cm} = -
\begin{vmatrix}
\X{0} & \X{G_{\X{0};\b{\sigma}_1}^{l_1,l_2}(t_1-t_2^+) \delta_{\b{\sigma}_1,\sigma_2}} & \X{G_{\X{0};\b{\sigma}_1}^{l_1,l_2}(t_1-t_2^+) \delta_{\b{\sigma}_1,\b{\sigma}_2}} \\ \\
\X{G_{\X{0};\sigma_2}^{l_2,l_1}(t_2-t_1^+) \delta_{\sigma_1,\sigma_2}} & \X{G_{\X{0};\sigma_2}^{l_2,l_2}(0^-)} & \X{0} \\ \\
\X{G_{\X{0};\b{\sigma}_2}^{l_2,l_1}(t_2-t_1^+) \delta_{\sigma_1,\b{\sigma}_2}} & \X{0} & \X{G_{\X{0};\b{\sigma}_2}^{l_2,l_2}(0^-)}
\end{vmatrix}_{-} \equiv 0,
\end{align}
\begin{align}\label{eby3}
&\hspace{-0.4cm}A_{\alpha_1,\alpha_3}^{\X{(3)}}(\alpha_1,\alpha_2,\alpha_3,\alpha_4) \nonumber\\
&\hspace{1.0cm} = +
\begin{vmatrix}
\X{0} & \X{G_{\X{0};\b{\sigma}_1}^{l_1,l_1}(0^-)} & \X{G_{\X{0};\b{\sigma}_1}^{l_1,l_2}(t_1-t_2^+) \delta_{\b{\sigma}_1,\b{\sigma}_2}} \\ \\
\X{G_{\X{0};\sigma_2}^{l_2,l_1}(t_2-t_1^+) \delta_{\sigma_1,\sigma_2}} & \X{G_{\X{0};\sigma_2}^{l_2,l_1}(t_2-t_1^+) \delta_{\b{\sigma}_1,\sigma_2}} & \X{0} \\ \\
\X{G_{\X{0};\b{\sigma}_2}^{l_2,l_1}(t_2-t_1^+) \delta_{\sigma_1,\b{\sigma}_2}} & \X{G_{\X{0};\b{\sigma}_2}^{l_2,l_1}(t_2-t_1^+) \delta_{\b{\sigma}_1,\b{\sigma}_2}} & \X{G_{\X{0};\b{\sigma}_2}^{l_2,l_2}(0^-)}
\end{vmatrix}_{-},
\end{align}
\begin{align}\label{eby4}
&\hspace{-0.4cm}A_{\alpha_1,\alpha_4}^{\X{(3)}}(\alpha_1,\alpha_2,\alpha_3,\alpha_4) \nonumber\\
&\hspace{1.0cm} = -
\begin{vmatrix}
\X{0} & \X{G_{\X{0};\b{\sigma}_1}^{l_1,l_1}(0^-)} & \X{G_{\X{0};\b{\sigma}_1}^{l_1,l_2}(t_1-t_2^+) \delta_{\b{\sigma}_1,\sigma_2}} \\ \\
\X{G_{\X{0};\sigma_2}^{l_2,l_1}(t_2-t_1^+) \delta_{\sigma_1,\sigma_2}} & \X{G_{\X{0};\sigma_2}^{l_2,l_1}(t_2-t_1^+) \delta_{\b{\sigma}_1,\sigma_2}} & \X{G_{\X{0};\sigma_2}^{l_2,l_2}(0^-)} \\ \\
\X{G_{\X{0};\b{\sigma}_2}^{l_2,l_1}(t_2-t_1^+) \delta_{\sigma_1,\b{\sigma}_2}} & \X{G_{\X{0};\b{\sigma}_2}^{l_2,l_1}(t_2-t_1^+) \delta_{\b{\sigma}_1,\b{\sigma}_2}} & \X{0}
\end{vmatrix}_{-},
\end{align}
\begin{align}\label{eby5}
&\hspace{-0.4cm}A_{\alpha_2,\alpha_1}^{\X{(3)}}(\alpha_1,\alpha_2,\alpha_3,\alpha_4) \nonumber\\
&\hspace{0.4cm} = -
\begin{vmatrix}
\X{0} & \X{G_{\X{0};\sigma_1}^{l_1,l_2}(t_1-t_2^+) \delta_{\sigma_1,\sigma_2}} &  \X{G_{\X{0};\sigma_1}^{l_1,l_2}(t_1-t_2^+) \delta_{\sigma_1,\b{\sigma}_2}} \\ \\
\X{G_{\X{0};\sigma_2}^{l_2,l_1}(t_2-t_1^+) \delta_{\b{\sigma}_1,\sigma_2}} & \X{G_{\X{0};\sigma_2}^{l_2,l_2}(0^-)} & \X{0} \\ \\
\X{G_{\X{0};\b{\sigma}_2}^{l_2,l_1}(t_2-t_1^+) \delta_{\b{\sigma}_1,\b{\sigma}_2}} & \X{0} & \X{G_{\X{0};\b{\sigma}_2}^{l_2,l_2}(0^-)}
\end{vmatrix}_{-} \equiv 0,
\end{align}
\begin{align}\label{eby6}
&\hspace{-0.4cm}A_{\alpha_2,\alpha_2}^{\X{(3)}}(\alpha_1,\alpha_2,\alpha_3,\alpha_4) \nonumber\\
&\hspace{0.8cm} = +
\begin{vmatrix}
\X{G_{\X{0};\sigma_1}^{l_1,l_1}(0^-)} & \X{G_{\X{0};\sigma_1}^{l_1,l_2}(t_1-t_2^+) \delta_{\sigma_1,\sigma_2}} &  \X{G_{\X{0};\sigma_1}^{l_1,l_2}(t_1-t_2^+) \delta_{\sigma_1,\b{\sigma}_2}} \\ \\
\X{G_{\X{0};\sigma_2}^{l_2,l_1}(t_2-t_1^+) \delta_{\sigma_1,\sigma_2}} & \X{G_{\X{0};\sigma_2}^{l_2,l_2}(0^-)} & \X{0} \\ \\
\X{G_{\X{0};\b{\sigma}_2}^{l_2,l_1}(t_2-t_1^+) \delta_{\sigma_1,\b{\sigma}_2}} & \X{0} & \X{G_{\X{0};\b{\sigma}_2}^{l_2,l_2}(0^-)}
\end{vmatrix}_{-},
\end{align}
\begin{align}\label{eby7}
&\hspace{-0.4cm}A_{\alpha_2,\alpha_3}^{\X{(3)}}(\alpha_1,\alpha_2,\alpha_3,\alpha_4) \nonumber\\
&\hspace{0.8cm} = -
\begin{vmatrix}
\X{G_{\X{0};\sigma_1}^{l_1,l_1}(0^-)} & \X{0} & \X{G_{\X{0};\sigma_1}^{l_1,l_2}(t_1-t_2^+) \delta_{\sigma_1,\b{\sigma}_2}} \\ \\
\X{G_{\X{0};\sigma_2}^{l_2,l_1}(t_2-t_1^+) \delta_{\sigma_1,\sigma_2}} & \X{G_{\X{0};\sigma_2}^{l_2,l_1}(t_2-t_1^+) \delta_{\b{\sigma}_1,\sigma_2}} & \X{0} \\ \\
\X{G_{\X{0};\b{\sigma}_2}^{l_2,l_1}(t_2-t_1^+) \delta_{\sigma_1,\b{\sigma}_2}} & \X{G_{\X{0};\b{\sigma}_2}^{l_2,l_1}(t_2-t_1^+) \delta_{\b{\sigma}_1,\b{\sigma}_2}} & \X{G_{\X{0};\b{\sigma}_2}^{l_2,l_2}(0^-)}
\end{vmatrix}_{-},
\end{align}
\begin{align}\label{eby8}
&\hspace{-0.4cm}A_{\alpha_2,\alpha_4}^{\X{(3)}}(\alpha_1,\alpha_2,\alpha_3,\alpha_4) \nonumber\\
&\hspace{1.0cm} = +
\begin{vmatrix}
\X{G_{\X{0};\sigma_1}^{l_1,l_1}(0^-)} & \X{0} & \X{G_{\X{0};\sigma_1}^{l_1,l_2}(t_1-t_2^+) \delta_{\sigma_1,\sigma_2}} \\ \\
\X{G_{\X{0};\sigma_2}^{l_2,l_1}(t_2-t_1^+) \delta_{\sigma_1,\sigma_2}} & \X{G_{\X{0};\sigma_2}^{l_2,l_1}(t_2-t_1^+) \delta_{\b{\sigma}_1,\sigma_2}} & \X{G_{\X{0};\sigma_2}^{l_2,l_2}(0^-)} \\ \\
\X{G_{\X{0};\b{\sigma}_2}^{l_2,l_1}(t_2-t_1^+) \delta_{\sigma_1,\b{\sigma}_2}} & \X{G_{\X{0};\b{\sigma}_2}^{l_2,l_1}(t_2-t_1^+) \delta_{\b{\sigma}_1,\b{\sigma}_2}} & \X{0}
\end{vmatrix}_{-},
\end{align}
\begin{align}\label{eby9}
&\hspace{-0.4cm}A_{\alpha_3,\alpha_1}^{\X{(3)}}(\alpha_1,\alpha_2,\alpha_3,\alpha_4) \nonumber\\
&\hspace{1.0cm} = +
\begin{vmatrix}
\X{0} & \X{G_{\X{0};\sigma_1}^{l_1,l_2}(t_1-t_2^+) \delta_{\sigma_1,\sigma_2}} &  \X{G_{\X{0};\sigma_1}^{l_1,l_2}(t_1-t_2^+) \delta_{\sigma_1,\b{\sigma}_2}} \\ \\
\X{G_{\X{0};\b{\sigma}_1}^{l_1,l_1}(0^-)} & \X{G_{\X{0};\b{\sigma}_1}^{l_1,l_2}(t_1-t_2^+) \delta_{\b{\sigma}_1,\sigma_2}} & \X{G_{\X{0};\b{\sigma}_1}^{l_1,l_2}(t_1-t_2^+) \delta_{\b{\sigma}_1,\b{\sigma}_2}} \\ \\
\X{G_{\X{0};\b{\sigma}_2}^{l_2,l_1}(t_2-t_1^+) \delta_{\b{\sigma}_1,\b{\sigma}_2}} & \X{0} & \X{G_{\X{0};\b{\sigma}_2}^{l_2,l_2}(0^-)}
\end{vmatrix}_{-},
\end{align}
\begin{align}\label{eby10}
&\hspace{-0.4cm}A_{\alpha_3,\alpha_2}^{\X{(3)}}(\alpha_1,\alpha_2,\alpha_3,\alpha_4) \nonumber\\
&\hspace{1.0cm} = -
\begin{vmatrix}
\X{G_{\X{0};\sigma_1}^{l_1,l_1}(0^-)} & \X{G_{\X{0};\sigma_1}^{l_1,l_2}(t_1-t_2^+) \delta_{\sigma_1,\sigma_2}} &  \X{G_{\X{0};\sigma_1}^{l_1,l_2}(t_1-t_2^+) \delta_{\sigma_1,\b{\sigma}_2}} \\ \\
\X{0} & \X{G_{\X{0};\b{\sigma}_1}^{l_1,l_2}(t_1-t_2^+) \delta_{\b{\sigma}_1,\sigma_2}} & \X{G_{\X{0};\b{\sigma}_1}^{l_1,l_2}(t_1-t_2^+) \delta_{\b{\sigma}_1,\b{\sigma}_2}} \\ \\
\X{G_{\X{0};\b{\sigma}_2}^{l_2,l_1}(t_2-t_1^+) \delta_{\sigma_1,\b{\sigma}_2}} & \X{0} & \X{G_{\X{0};\b{\sigma}_2}^{l_2,l_2}(0^-)}
\end{vmatrix}_{-},
\end{align}
\begin{align}\label{eby11}
&\hspace{-0.4cm}A_{\alpha_3,\alpha_3}^{\X{(3)}}(\alpha_1,\alpha_2,\alpha_3,\alpha_4) \nonumber\\
&\hspace{1.0cm} = +
\begin{vmatrix}
\X{G_{\X{0};\sigma_1}^{l_1,l_1}(0^-)} & \X{0} & \X{G_{\X{0};\sigma_1}^{l_1,l_2}(t_1-t_2^+) \delta_{\sigma_1,\b{\sigma}_2}} \\ \\
\X{0} & \X{G_{\X{0};\b{\sigma}_1}^{l_1,l_1}(0^-)} & \X{G_{\X{0};\b{\sigma}_1}^{l_1,l_2}(t_1-t_2^+) \delta_{\b{\sigma}_1,\b{\sigma}_2}} \\ \\
\X{G_{\X{0};\b{\sigma}_2}^{l_2,l_1}(t_2-t_1^+) \delta_{\sigma_1,\b{\sigma}_2}} & \X{G_{\X{0};\b{\sigma}_2}^{l_2,l_1}(t_2-t_1^+) \delta_{\b{\sigma}_1,\b{\sigma}_2}} & \X{G_{\X{0};\b{\sigma}_2}^{l_2,l_2}(0^-)}
\end{vmatrix}_{-},
\end{align}
\begin{align}\label{eby12}
&\hspace{-0.4cm}A_{\alpha_3,\alpha_4}^{\X{(3)}}(\alpha_1,\alpha_2,\alpha_3,\alpha_4) \nonumber\\
&\hspace{0.4cm} = -
\begin{vmatrix}
\X{G_{\X{0};\sigma_1}^{l_1,l_1}(0^-)} & \X{0} & \X{G_{\X{0};\sigma_1}^{l_1,l_2}(t_1-t_2^+) \delta_{\sigma_1,\sigma_2}} \\ \\
\X{0} & \X{G_{\X{0};\b{\sigma}_1}^{l_1,l_1}(0^-)} & \X{G_{\X{0};\b{\sigma}_1}^{l_1,l_2}(t_1-t_2^+) \delta_{\b{\sigma}_1,\sigma_2}} \\ \\
\X{G_{\X{0};\b{\sigma}_2}^{l_2,l_1}(t_2-t_1^+) \delta_{\sigma_1,\b{\sigma}_2}} & \X{G_{\X{0};\b{\sigma}_2}^{l_2,l_1}(t_2-t_1^+) \delta_{\b{\sigma}_1,\b{\sigma}_2}} & \X{0}
\end{vmatrix}_{-} \equiv 0,
\end{align}
\begin{align}\label{eby13}
&\hspace{-0.4cm}A_{\alpha_4,\alpha_1}^{\X{(3)}}(\alpha_1,\alpha_2,\alpha_3,\alpha_4) \nonumber\\
&\hspace{1.0cm} = -
\begin{vmatrix}
\X{0} & \X{G_{\X{0};\sigma_1}^{l_1,l_2}(t_1-t_2^+) \delta_{\sigma_1,\sigma_2}} &  \X{G_{\X{0};\sigma_1}^{l_1,l_2}(t_1-t_2^+) \delta_{\sigma_1,\b{\sigma}_2}} \\ \\
\X{G_{\X{0};\b{\sigma}_1}^{l_1,l_1}(0^-)} & \X{G_{\X{0};\b{\sigma}_1}^{l_1,l_2}(t_1-t_2^+) \delta_{\b{\sigma}_1,\sigma_2}} & \X{G_{\X{0};\b{\sigma}_1}^{l_1,l_2}(t_1-t_2^+) \delta_{\b{\sigma}_1,\b{\sigma}_2}} \\ \\
\X{G_{\X{0};\sigma_2}^{l_2,l_1}(t_2-t_1^+) \delta_{\b{\sigma}_1,\sigma_2}} & \X{G_{\X{0};\sigma_2}^{l_2,l_2}(0^-)} & \X{0}
\end{vmatrix}_{-},
\end{align}
\begin{align}\label{eby14}
&\hspace{-0.4cm}A_{\alpha_4,\alpha_2}^{\X{(3)}}(\alpha_1,\alpha_2,\alpha_3,\alpha_4) \nonumber\\
&\hspace{1.0cm} = +
\begin{vmatrix}
\X{G_{\X{0};\sigma_1}^{l_1,l_1}(0^-)} & \X{G_{\X{0};\sigma_1}^{l_1,l_2}(t_1-t_2^+) \delta_{\sigma_1,\sigma_2}} &  \X{G_{\X{0};\sigma_1}^{l_1,l_2}(t_1-t_2^+) \delta_{\sigma_1,\b{\sigma}_2}} \\ \\
\X{0} & \X{G_{\X{0};\b{\sigma}_1}^{l_1,l_2}(t_1-t_2^+) \delta_{\b{\sigma}_1,\sigma_2}} & \X{G_{\X{0};\b{\sigma}_1}^{l_1,l_2}(t_1-t_2^+) \delta_{\b{\sigma}_1,\b{\sigma}_2}} \\ \\
\X{G_{\X{0};\sigma_2}^{l_2,l_1}(t_2-t_1^+) \delta_{\sigma_1,\sigma_2}} & \X{G_{\X{0};\sigma_2}^{l_2,l_2}(0^-)} & \X{0}
\end{vmatrix}_{-},
\end{align}
\begin{align}\label{eby15}
&\hspace{-0.4cm}A_{\alpha_4,\alpha_3}^{\X{(3)}}(\alpha_1,\alpha_2,\alpha_3,\alpha_4) \nonumber\\
&\hspace{0.4cm} = -
\begin{vmatrix}
\X{G_{\X{0};\sigma_1}^{l_1,l_1}(0^-)} & \X{0} & \X{G_{\X{0};\sigma_1}^{l_1,l_2}(t_1-t_2^+) \delta_{\sigma_1,\b{\sigma}_2}} \\ \\
\X{0} & \X{G_{\X{0};\b{\sigma}_1}^{l_1,l_1}(0^-)} & \X{G_{\X{0};\b{\sigma}_1}^{l_1,l_2}(t_1-t_2^+) \delta_{\b{\sigma}_1,\b{\sigma}_2}} \\ \\
\X{G_{\X{0};\sigma_2}^{l_2,l_1}(t_2-t_1^+) \delta_{\sigma_1,\sigma_2}} & \X{G_{\X{0};\sigma_2}^{l_2,l_1}(t_2-t_1^+) \delta_{\b{\sigma}_1,\sigma_2}} & \X{0}
\end{vmatrix}_{-} \equiv 0,
\end{align}
\begin{align}\label{eby16}
&\hspace{-0.4cm}A_{\alpha_4,\alpha_4}^{\X{(3)}}(\alpha_1,\alpha_2,\alpha_3,\alpha_4) \nonumber\\
&\hspace{1.0cm} = +
\begin{vmatrix}
\X{G_{\X{0};\sigma_1}^{l_1,l_1}(0^-)} & \X{0} & \X{G_{\X{0};\sigma_1}^{l_1,l_2}(t_1-t_2^+) \delta_{\sigma_1,\sigma_2}} \\ \\
\X{0} & \X{G_{\X{0};\b{\sigma}_1}^{l_1,l_1}(0^-)} & \X{G_{\X{0};\b{\sigma}_1}^{l_1,l_2}(t_1-t_2^+) \delta_{\b{\sigma}_1,\sigma_2}} \\ \\
\X{G_{\X{0};\sigma_2}^{l_2,l_1}(t_2-t_1^+) \delta_{\sigma_1,\sigma_2}} & \X{G_{\X{0};\sigma_2}^{l_2,l_1}(t_2-t_1^+) \delta_{\b{\sigma}_1,\sigma_2}} & \X{G_{\X{0};\sigma_2}^{l_2,l_2}(0^-)}
\end{vmatrix}_{-}.
\end{align}
One observes that conform the identities in Eq.\,(\ref{eb7}), the cofactors $A_{\alpha_1,\alpha_2}^{\X{(3)}}$, $A_{\alpha_2,\alpha_1}^{\X{(3)}}$, $A_{\alpha_3,\alpha_4}^{\X{(3)}}$ and $A_{\alpha_4,\alpha_3}^{\X{(3)}}$ are indeed identically vanishing, Eqs\,(\ref{eby2}), (\ref{eby5}), (\ref{eby12}), and (\ref{eby15}). Further, one verifies that the explicit expressions in Eqs\,(\ref{eby11}) and (\ref{eby16}) are in conformity with the exact result in Eq.\,(\ref{eb4d}) (for $r=2$), and those in Eqs\,(\ref{eby3}), (\ref{eby7}), (\ref{eby4}), and (\ref{eby8}) are in conformity with the exact results in Eq.\,(\ref{eb4e}) (for $r=1$ and $s=2$).

As in appendices \ref{sac} and \ref{sad}, we close this section by presenting a number of programs written in the programming language of Mathematica$^{\X{\circledR}}$. Program \texttt{A2num1}, p.\,\pageref{A2num1}, calculates the function $A_{\alpha_r,\alpha_s}^{\X{(2\nu-1)}}(\alpha_1,\alpha_2,\dots,\alpha_{2\nu})$ as encountered in Eq.\,(\ref{eb4}) and subsequent equations.\footnote{In \texttt{A2num1}, for given value of \texttt{nu} $\equiv \nu$ the input variable \texttt{ir} (\texttt{is}) may take the values $2r$ and $2r -1$ ($2s$ and $2s-1$), where $r, s \in \{1,2,\dots,\nu\}$.} The spin configuration\,\footnote{See Eq.\,(\protect\ref{eb1b}).}
\begin{equation}\label{eby16a}
\{\sigma_1,\sigma_2,\dots,\sigma_{2\nu}\} \equiv \{\sigma_1,\b{\sigma}_1,\dots,\sigma_{\nu},\b{\sigma}_{\nu}\}
\end{equation}
is generated by the program \texttt{Sgen}, p.\,\pageref{Sgen}. Any spin configuration generated by \texttt{Sgen} can be visualised in the arrow notation by means of the program \texttt{Arr}, p.\,\pageref{Arr}. Program \texttt{A2num1} can be used for instance to verify the identities in Eq.\,(\ref{eb7}). To verify the identities in Eqs\,(\ref{eb4d}) and (\ref{eb4e}), the underlying relevant spin configurations are to be constrained, on account of the conditions on both sides of each identity sign. Program \texttt{SgenPP}, p.\,\pageref{SgenPP}, enforces the relevant constraints. The last three sets of Mathematica instructions, presented below, p.\,\pageref{Set1}, can be used to verify the identities in respectively Eqs\,(\ref{eb4d}), (\ref{eb4e}), and (\ref{eb7}).

We note that program \texttt{A2num1} generates the analytic expression for $A_{\alpha_r,\alpha_s}^{\X{(2\nu-1)}}$, from which, on the basis of the considerations in appendices \ref{sac} and \ref{sad}, one can straightforwardly deduce the analytic expressions of $G^{\X{(\nu)}}[v,G_{\X{0}}]$ and $\Sigma_{\X{01}}^{\X{(\nu)}}[v,G]$ for in principle arbitrary values of $\nu \in \mathds{N}$. In practice, however, the factorial increase in the required computation time limits the upper bound of $\nu$ to be used in practice. In Ref.\,\citen{BF16a} we present a formalism, and the relevant programs written in the programming language of Mathematica, with the aid of which the diagrams contributing to $\Sigma_{\X{01}}^{\X{(\nu)}}[v,G]$ can be sorted into a set of disjoint classes of \textsl{algebraically} (as distinct from \textsl{topologically}) identical diagrams. The method to be presented in Ref.\,\citen{BF16a} is purely combinatorial and therefore the relevant numerical computations do \textsl{not} involve any floating-point operations.
\\

{\footnotesize\refstepcounter{dummy}\label{A2num1}
\begin{verbatim}
(* Program `A2num1'. *)

Clear[A2num1];
A2num1[ir_, is_, nu_, S_] :=
 Module[(* Returns
 A_{alpha_ir,alpha_is}^{(2nu-1)}(alpha_1,alpha_2,....,alpha_{2nu}) in
 symbolic form for given values of ir, is, nu, and the spin configuration
 S = {sigma_1, sigma_2, ...,sigma_{2nu}}. The integers ir and is are
 elements of {1,2,...,2nu}. With i = l_i t_i, and j = l_j t_j, in the output
 it is assumed that the second argument j in G_{sigma_i}(i,j) represents j^+,
 signifying l_j t_j+0^+. Further, the symbol G_{sigma_i}(i,j) generally represents
 the non-interacting one-particle Green function. Here G_{sigma_i}(i,j) is
 defined on the basis of the equality G_{sigma_i,sigma_j}(i,j) =
 G_{sigma_i}(i,j) delta_{sigma_i,sigma_j}. *)
 {g, thetj, Gx, j, l, sum, sumx, R, P},
  g[ix_, jx_, Tx_] := If[Tx[[ix]] == Tx[[jx]], 1, 0];
  thetj[jx_] := Floor[(jx + 1)/2];
  Gx[nx_, lx_] :=
   If[lx == 0,
    Table["G\[DownArrow]"[ix, jx], {ix, 1, nx}, {jx, 1, nx}],
    Table["G\[UpArrow]"[ix, jx], {ix, 1, nx}, {jx, 1, nx}]];
  R = Range[1, 2 nu]; sum = 0;
  Do[(*l*) P = Permutations[R][[l]];
   If[P[[ir]] ==
     is, (sumx =
      Signature[P] Product[
        If[j != ir, (g[j, P[[j]], S] Gx[nu, S[[j]]][[thetj[j],
            thetj[P[[j]]]]]), 1], {j, 1, 2 nu}]), (sumx = 0)];
   sum = sum + sumx, {l, 1, (2 nu)!}]; sum]

\end{verbatim}}

{\footnotesize\refstepcounter{dummy}\label{Arr}
\begin{verbatim}
(* Program `Arr'. *)

Clear[Arr];
Arr[S_] :=
 Module[(*Returns a spin configuration S generated by Sgen or SgenPP in
 arrow notation. *)
 {l, li, y, S1}, l = Length[S];
  Do[y[li] = If[S[[li]] == 0, "\!\(\*
StyleBox[\"\[DownArrow]\",\nFontColor->RGBColor[1, 0, 0]]\)", "\!\(\*
StyleBox[\"\[UpArrow]\",\nFontColor->RGBColor[1, 0, 0]]\)"], {li, 1,
    l}]; S1 = Table[y[li], {li, 1, l}]; S1]

\end{verbatim}}

{\footnotesize\refstepcounter{dummy}\label{Sgen}
\begin{verbatim}
(* Program `Sgen'. *)

Clear[Sgen];
Sgen[nu_, li_] :=
 Module[(*Returns the li-th spin configuration
 S = {sigma_1,sigma_2,...,sigma_{2nu}}, out of the possible 2^{nu}
 distinct spin configurations, for 2nu spin-1/2 particles, satisfying
 sigma_{2j} = 1-sigma_{2j-1}. Here sigma_j = 0 represents spin-down,
 and sigma_j = 1 spin-up. *)
 {k, j, A, B, S}, A = IntegerDigits[li, 2]; k = Length[A];
  If[k < nu, Do[A = Prepend[A, 0], {j, k + 1, nu}]];
  B = Table[(1 - A[[j]]), {j, 1, nu}]; S = Riffle[A, B]; S]

\end{verbatim}}

{\footnotesize\refstepcounter{dummy}\label{SgenPP}
\begin{verbatim}
(* Program `SgenPP'. *)

Clear[SgenPP];
SgenPP[ir_, is_, sigir_, sigis_, S_] :=
 Module[(*Given the spin configuration
 S = {sigma_1,sigma_2,...,sigma_{2nu}}, where sigma_j = 0 stands for
 spin-down, and sigma_j = 1 for spin-up, and where in the present application
 the indices satisfy sigma_{2j} = 1-sigma_{2j-1}, replaces sigma_{ir} by
 sigir and sigma_{is} by sigis. Subsequently adjusts the relevant neighbouring
 spins in such a way that the indices in the resulting spin configuration,
 to be returned, satisfy sigma_{2j} = 1-sigma_{2j-1}. *)
 {Spp}, Spp = S;
  If[Mod[ir, 2] ==
    0, (Spp =
     ReplacePart[Spp, {ir - 1 -> 1 - sigir, ir -> sigir}]), (Spp =
     ReplacePart[Spp, {ir -> sigir, ir + 1 -> 1 - sigir}])];
  If[Mod[is, 2] ==
    0, (Spp =
     ReplacePart[Spp, {is - 1 -> 1 - sigis, is -> sigis}]), (Spp =
     ReplacePart[Spp, {is -> sigis, is + 1 -> 1 - sigis}])]; Spp]

\end{verbatim}}

The following are three sets of instructions for testing the validity of the identities in respectively Eqs\,(\ref{eb4d}), (\ref{eb4e}), and (\ref{eb7}):\,\footnote{The integer \texttt{nu} ($\equiv \nu$) can be changed from its present value $3$ in all three sets. In \texttt{Set 2}, the present equalities concerning \texttt{ir1}, \texttt{is1}, \texttt{ir2}, and \texttt{is2} are suited for testing the first identity in Eq.\,(\protect\ref{eb4e}). For testing the remaining identities, these equalities are to be changed accordingly. For instance, one can maintain the present equalities concerning \texttt{ir1} and \texttt{is1}, and only appropriately change the equalities concerning \texttt{ir2} and \texttt{is2}. In this way, the equivalence is tested of the last three functions with the first one in Eq.\,(\protect\ref{eb4e}). This is sufficient for the purpose on account of the \textsl{transitive} property of the binary relation $\equiv$.}\\

{\footnotesize\refstepcounter{dummy}\label{Set1}
\begin{verbatim}
(* Program `Set 1'. *)

nu = 3; Do[(*sigir*) sigis = sigir; Do[(*r*)
   ir1 = 2 r - 1; is1 = 2 r - 1; ir2 = 2 r; is2 = 2 r;
  Print["ir1, is1: ", ir1, ", ", is1, ";  ir2, is2: ", ir2, ", ",
   is2]; Do[(*li*) S = Sgen[nu, li];
   Spp1 = SgenPP[ir1, is1, sigir, sigis, S];
   Spp2 = SgenPP[ir2, is2, sigir, sigis, S]; Print["\!\(\*
StyleBox[\"li\",\nFontColor->RGBColor[0, 0.67, 0]]\): ", li,
    ".  \!\(\*SubscriptBox[\(\[Sigma]\), \(ir\)]\):", Arr[{sigir}],
    ", \!\(\*SubscriptBox[\(\[Sigma]\), \(is\)]\):", Arr[{sigis}] ,
    ",   S: ", Arr[S], ", Spp1: ", Arr[Spp1], ", Spp2: ", Arr[Spp2]];
   A1 = A2num1[ir1, is1, nu, Spp1]; Print["\!\(\*
StyleBox[\"A1\",\nFontColor->RGBColor[1, 0.5, 0]]\): ", A1];
   A2 = A2num1[ir2, is2, nu, Spp2] ; Print["\!\(\*
StyleBox[\"A2\",\nFontColor->RGBColor[1, 0.5, 0]]\): ", A2];
   Print["\!\(\*
StyleBox[\"\[RightArrow]\",\nFontColor->RGBColor[1, 0, 0]]\) \!\(\*
StyleBox[\"A1\",\nFontColor->RGBColor[1, 0, 0]]\)\!\(\*
StyleBox[\"-\",\nFontColor->RGBColor[1, 0, 0]]\)\!\(\*
StyleBox[\"A2\",\nFontColor->RGBColor[1, 0, 0]]\): ",
    Simplify[A1 - A2]], {li, 0, 2^nu - 1}], {r, 1, nu}], {sigir, 0,
  1}]

\end{verbatim}}

{\footnotesize
\begin{verbatim}
(* Program `Set 2'. *)

nu = 3; Do[(*sigir*)
 Do[(*sigis*) Do[(*r*) Do[(*s*)If[r == s, Goto[end]];
     ir1 = 2 r - 1; is1 = 2 s - 1; ir2 = 2 r; is2 = 2 s - 1;
    Print["ir1, is1: ", ir1, ", ", is1, ";  ir2, is2: ", ir2, ", ",
     is2]; Do[(*li*) S = Sgen[nu, li];
     Spp1 = SgenPP[ir1, is1, sigir, sigis, S];
     A1 = A2num1[ir1, is1, nu, Spp1] ; Print["\!\(\*
StyleBox[\"A1\",\nFontColor->RGBColor[1, 0.5, 0]]\): ", A1];
     Spp2 = SgenPP[ir2, is2, sigir, sigis, S]; Print["\!\(\*
StyleBox[\"li\",\nFontColor->RGBColor[0, 0.67, 0]]\): ", li,
      ".  \!\(\*SubscriptBox[\(\[Sigma]\), \(ir\)]\):", Arr[{sigir}],
      ", \!\(\*SubscriptBox[\(\[Sigma]\), \(is\)]\):", Arr[{sigis}] ,
      ",   Spp1: ", Arr[Spp1], ",  Spp2: ", Arr[Spp2]];
     A2 = A2num1[ir2, is2, nu, Spp2] ; Print["\!\(\*
StyleBox[\"A2\",\nFontColor->RGBColor[1, 0.5, 0]]\): ", A2];
     Print["\!\(\*
StyleBox[\"\[RightArrow]\",\nFontColor->RGBColor[1, 0, 0]]\) \!\(\*
StyleBox[\"A1\",\nFontColor->RGBColor[1, 0, 0]]\)\!\(\*
StyleBox[\"-\",\nFontColor->RGBColor[1, 0, 0]]\)\!\(\*
StyleBox[\"A2\",\nFontColor->RGBColor[1, 0, 0]]\): ",
      Simplify[A1 - A2]], {li, 0, 2^nu - 1}];
    Label[end], {s, 1, nu}], {r, 1, nu}], {sigis, 0, 1}], {sigir, 0,
  1}]

\end{verbatim}}

{\footnotesize
\begin{verbatim}
(* Program `Set 3'. *)

nu = 3; Do[(*sigir*) Do[(*sigis*) Do[(*r*)
    ir1 = 2 r - 1; is1 = 2 r; ir2 = 2 r - 1; is2 = 2 r;
   Print["ir1, is1: ", ir1, ", ", is1, ";  ir2, is2: ", ir2, ", ",
    is2]; Do[(*li*) S = Sgen[nu, li]; Print["\!\(\*
StyleBox[\"li\",\nFontColor->RGBColor[0, 0.67, 0]]\): ", li,
     ".  \!\(\*SubscriptBox[\(\[Sigma]\), \(ir\)]\):", Arr[{sigir}],
     ", \!\(\*SubscriptBox[\(\[Sigma]\), \(is\)]\):", Arr[{sigis}] ,
     ",   S: ", Arr[S]]; A1 = A2num1[ir1, is1, nu, S]; Print["\!\(\*
StyleBox[\"A1\",\nFontColor->RGBColor[1, 0.5, 0]]\): ", A1];
    A2 = A2num1[ir2, is2, nu, S] ; Print["\!\(\*
StyleBox[\"A2\",\nFontColor->RGBColor[1, 0.5, 0]]\): ", A2];
    Print["\!\(\*
StyleBox[\"\[RightArrow]\",\nFontColor->RGBColor[1, 0, 0]]\) \!\(\*
StyleBox[\"A1\",\nFontColor->RGBColor[1, 0, 0]]\)\!\(\*
StyleBox[\"-\",\nFontColor->RGBColor[1, 0, 0]]\)\!\(\*
StyleBox[\"A2\",\nFontColor->RGBColor[1, 0, 0]]\): ",
     Simplify[A1 - A2]], {li, 0, 2^nu - 1}], {r, 1, nu}], {sigis, 0,
   1}], {sigir, 0, 1}]

\end{verbatim}}

\refstepcounter{dummyX}
\subsection{The atomic limit}
\phantomsection
\label{sab2}
The formalism discussed in the previous subsection is greatly simplified in the atomic limit, at which for the function $G_{\X{0};\sigma}^{l,l'}(t-t')$, Eq.\,(\ref{e47}), one has\,\footnote{The result in Eq.\,(\ref{eb13}) is a consequence of the fact that for uniform GSs / ensemble of states the only source of the $\bm{k}$ dependence of the spatial Fourier transform of the non-interacting Green function $G_{\protect\X{0}}$ is the non-interacting energy dispersions $\varepsilon_{\bm{k}}$ in Eq.\,(\protect\ref{e29}); with $\varepsilon_{\bm{k}}$ independent of $\bm{k}$, the equality in Eq.\,(\ref{eb13}) follows immediately. See also \S\,4.2.1 in Ref.\,\protect\citen{BF16a} (in particular Eq.\,(4.45) herein).}
\begin{equation}\label{eb13}
G_{\X{0};\sigma}^{l,l'}(t-t') = G_{\X{0};\sigma}^{l}(t-t')\hspace{0.6pt} \delta_{l,l'}.
\end{equation}
Introducing\,\footnote{For $\nu=1$, the sum in Eq.\,(\protect\ref{eb14}) is to be identified with unity.} (\emph{cf.} Eq.\,(\ref{eb7a}))
\begin{equation}\label{eb14}
\sideset{}{{}^{\X{(r)}}}{\sum}_{l_1,\dots,l_{\nu}}
\end{equation}
as the $(\nu-1)$-fold summation with respect to $\{l_1,\dots, l_{\nu}\}\backslash \{l_r\}$, and, for $r\not=s$ (\emph{cf.} Eq.\,(\ref{eb7b})),\footnote{For $\nu=1$ ($\nu=2$), the sum in Eq.\,(\protect\ref{eb15}) is to be identified with zero (unity).}
\begin{equation}\label{eb15}
\sideset{}{{}^{\X{(r,s)}}}{\sum}_{l_1,\dots,l_{\nu}}
\end{equation}
as the $(\nu-2)$-fold summation with respect to $\{l_1$, \dots, $l_{\nu}\}\backslash \{l_{r}, l_{s}\}$, from the expression in Eq.\,(\ref{eb4b}) one arrives at
\begin{equation}\label{eb16}
M_{\nu;\sigma}^{l,l'}(t-t') = M_{\nu;\sigma}^{l}(t-t')\hspace{0.7pt} \delta_{l,l'},
\end{equation}
where
\begin{align}\label{eb17}
&\hspace{-0.4cm}M_{\nu;\sigma}^{l}(t-t') = -\frac{2}{\nu !} \Big(\frac{\ii U}{2\hbar}\Big)^{\nu} \sum_{r=1}^{\nu} \int \prod_{j=1}^{\nu} \rd t_{j}\; G_{\X{0};\sigma}^{l}(t-t_{r}^+) G_{\X{0};\sigma}^{l}(t_{r}-t') \nonumber\\
&\hspace{2.1cm}\times \hspace{0.2cm}\sideset{}{{}^{\X{(r)}}}{\sum}_{l_1,\dots,l_{\nu}} \hspace{0.1cm}
\sideset{}{{}^{\X{(r)}}}{\sum}_{\sigma_1,\dots, \sigma_{\nu}} \left. \t{A}_{\alpha_{2r-1},\alpha_{2r-1}}^{\X{(2\nu-1)}}(\alpha_1,\alpha_2,\dots,\alpha_{2\nu-1},\alpha_{2\nu})
\right|_{\alpha_{2r-1} = l t_r\sigma} \nonumber\\
&\hspace{1.75cm} -\frac{4}{\nu !} \Big(\frac{\ii U}{2\hbar}\Big)^{\nu} \sum_{r=1}^{\nu} \sum_{\substack{s=1 \\ s \not= r}}^{\nu}
\int \prod_{j=1}^{\nu} \rd t_{j}\; G_{\X{0};\sigma}^{l}(t-t_{s}^+) G_{\X{0};\sigma}^{l}(t_{r}-t') \nonumber\\
&\hspace{2.1cm} \times \hspace{0.2cm}\sideset{}{{}^{\X{(r,s)}}}{\sum}_{l_1,\dots,l_{\nu}} \hspace{0.1cm}
\sideset{}{{}^{\X{(r,s)}}}{\sum}_{\sigma_1,\dots, \sigma_{\nu}} \left. \t{A}_{\alpha_{2r-1},\alpha_{2s-1}}^{\X{(2\nu-1)}}(\alpha_1,\alpha_2,\dots,\alpha_{2\nu-1},\alpha_{2\nu})
 \right|_{\substack{\alpha_{2r-1}\hspace{1.2pt} = l t_r\sigma \\ \alpha_{2s-1} = l t_s\sigma}}, \nonumber\\
\end{align}
in which $\t{A}_{\alpha_i,\alpha_j}^{\X{(2\nu-1)}}$ is the same function as $A_{\alpha_i,\alpha_j}^{\X{(2\nu-1)}}$ except that it has been determined in terms of the non-interacting Green functions satisfying the equality in Eq.\,(\ref{eb13}). Alternatively, the function $\t{A}_{\alpha_i,\alpha_j}^{\X{(2\nu-1)}}$ is deduced from $A_{\alpha_i,\alpha_j}^{\X{(2\nu-1)}}$ by identifying all off-diagonal elements of the underlying non-interacting Green functions in the $l$ space with zero. In arriving at the expressions in Eqs\,(\ref{eb16}) and (\ref{eb17}), we have made use of the relationship
\begin{equation}\label{eb18}
\hspace{0.1cm}
\sideset{}{{}^{\X{(r,s)}}}{\sum}_{\sigma_1,\dots, \sigma_{\nu}} \left.\t{A}_{\alpha_{2r-1},\alpha_{2s-1}}^{\X{(2\nu-1)}}(\alpha_1,\alpha_2,\dots,\alpha_{2\nu-1},\alpha_{2\nu})
 \right|_{\substack{\alpha_{2r-1}\hspace{1.2pt} = l' t_r\sigma \\ \alpha_{2s-1} = l t_s\sigma}} \propto \delta_{l,l'}.
\end{equation}

With reference to the considerations following Eq.\,(\ref{eb19}) above, we note that for the cases where the non-interacting Green function $G_{\X{0}}$ satisfies the equality in Eq.\,(\ref{eb13}), one has (\emph{cf.} Eq.\,(\ref{eb31}))
\begin{align}\label{eb32}
(\t{\mathbbmss{A}}_{\nu})_{\gamma_i;\gamma_j} = \Big\{&
\begin{pmatrix}
G_{\X{0};\sigma_i}^{l_i}(t_i-t_j^+) & 0 \\ \\
0 & G_{\X{0};\b{\sigma}_i}^{l_i}(t_i-t_j^+)
\end{pmatrix} \hspace{0.0pt} \delta_{\sigma_i,\sigma_j} \nonumber \\
&+
\begin{pmatrix}
0 & G_{\X{0};\sigma_i}^{l_i}(t_i-t_j^+) \\ \\
G_{\X{0};\b{\sigma}_i}^{l_i}(t_i-t_j^+) & 0
\end{pmatrix} \hspace{0.0pt} \delta_{\sigma_i,\b{\sigma}_j} \Big\} \hspace{0.7pt} \delta_{l_i.l_j}.
\end{align}
Similarly as in case of the expression in Eq.\,(\ref{eb31}), in the spin-unpolarised case the expression in Eq.\,(\ref{eb32}) can be expressed in terms of the $2\times 2$ unit matrix $\bbsigma^{\X{0}}$ and the $2\times 2$ Pauli matrix $\bbsigma^{\X{x}}$, p.\,\pageref{InRef}.\footnote{$(\bbsigma^{\protect\X{x}})_{1,1}  = (\bbsigma^{\protect\X{x}})_{2,2} = 0$, $(\bbsigma^{\protect\X{x}})_{1,2}  = (\bbsigma^{\protect\X{x}})_{2,1} = 1$.} Owing to the $\delta_{l_i,l_j}$ on the RHS of Eq.\,(\ref{eb32}), in general the matrix $\t{\mathbbmss{A}}_{\nu}$, corresponding to the atomic limit of the Hubbard Hamiltonian, is far more sparse \cite{SP84} than the matrix $\mathbbmss{A}_{\nu}$, Eqs\,(\ref{eb29}) -- (\ref{eb31}).
$\hfill\Box$
%\pagebreak\vfill

\refstepcounter{dummyX}
\section{List of acronyms and mathematical symbols (not exhaustive)}
\phantomsection
\label{sae}

\scriptsize{
\label{sax}
\begin{longtable}{ll}
DMFT & Dynamical mean-field theory \\
GS & Ground state, Ground-state \\
GSs & Ground states\\
LHS & Left-hand side\\
RHS & Right-hand side\\
TFD & Thermo-field dynamics\\
1PI & One-particle irreducible (diagrams representing $\Sigma$ are 1PI, those representing $\Sigma^{\star}$\\
{} & are not in general)\\
2PI & Two-particle irreducible ($G$-skeleton)\\ \\
Ch. & Chapter\\
p. / pp. & Page / Pages\\
Ref. & Reference \\
\S, \S\S & Section, Sections\\ \\
$\mathds{N}$ & Set of positive integers, $\{1,2,3, \dots\}$\\
$\mathds{N}_0$ & Set of non-negative integers, $\mathds{N} \cup \{0\} = \{0,1,2,\dots\}$\\
$\mathds{R}$ & Set of real numbers\\ \\
$\lfloor{\,}\rfloor$ & The floor function: $\lfloor x\rfloor =$ the greatest integer less than or equal to $x$\\
$\wedge$ & The logical \textsl{and}: $p_1 \wedge p_2$ is true only if both propositions $p_1$ and $p_2$ are true\\
$\vee$ & The logical \textsl{or}: $p_1 \vee p_2$ is true if \textsl{at least} one of the propositions $p_1$ and $p_2$ is true\\
$\neg$ & Negation, with $\neg\hspace{0.4pt} p =$ true (false) when the proposition $p =$ false (true) \\
$\forall$ & For all\\
$\subseteq$, $\subset$ & Subset, Proper subset\\
$A \backslash B$ & The subset of the set $A$ from which the elements of the set $B$ have been removed\\
$\vert A\vert$ & Number of elements of the set $A$, the cardinal number of $A$ \\
$\doteq$ & Equality by definition \\
$\rightleftharpoons$ & Association, a binary relation generally not expressible by $\equiv$ (as in $a \rightleftharpoons \bm{r}t\sigma$)\\
$\left| \mathbb{A}\right|_{\pm}$ &  Permanent / Determinant of the square matrix $\mathbb{A}$\\
$[{\,},{\,}]_{\mp}$ & Commutation / Anti-commutation: $[a,b]_{\mp} \doteq a b \mp b a$\\
$\Hf$ / $\Pf$ & Hafnian / Pfaffian \\ \\
$a, b, c, d$ & Unindexed compound variables, which may be primed. Thus, $a$ may represent $\bm{r}t\sigma$ or $\bm{r}'t'\sigma'$\\
$\beta$ & $1/(k_{\textsc{b}} T)$, where $k_{\textsc{b}}$ is the constant of Boltzmann, and $T$ the absolute temperature\\
$\mathscr{C}$ & Complex contour within the TFD formalism\\
$d$ & Generally, the dimension of the spatial space (a subspace of the Euclidean space $\mathds{R}^d$)\\
{} & into which the system of interest is confined; mostly encountered in $\mathrm{d}^dr$ to denote the\\
{} & integration measure in $\mathds{R}^d$\\
$G$ & The interacting one-particle Green function at $T=0$ (adiabatic approximation)\\
{} & Serves also as the generic symbol representing $\mathscr{G}$ and $\mathsf{G}$\\
$\h{G}$ & The single-particle Green \textsl{operator} associated with $G$\\
{} & Serves also as the generic symbol representing $\h{\mathscr{G}}$ and $\h{\mathsf{G}}$ \\
$G_{\X{0}}, G^{\X{(0)}}$ & The non-interacting counterpart of $G$\\
{} & Serve also as the generic symbols representing $\mathscr{G}_{\X{0}}$, $\mathscr{G}^{\X{(0)}}$, and $\mathsf{G}_{\X{0}}$, $\mathsf{G}^{\X{(0)}}$\\
$G^{\X{(\nu)}}$ & The total $\nu$th-order perturbational contribution to $G$ \\
{} & Serves also as the generic symbol representing $\mathscr{G}^{\X{(\nu)}}$ and $\mathsf{G}^{\X{(\nu)}}$\\
$\h{G}^{\X{(\nu)}}$ & The single-particle operator associated with $G^{\X{(\nu)}}$, $\nu \in\mathds{N}_0$\\
$\mathscr{G}$ & The interacting one-particle Green function for $T > 0$ (Matsubara formalism)\\
$\mathsf{G}$ & The interacting one-particle Green function for $T > 0$ (TFD formalism)\\
$G_{\X{2}}$ & The interacting two-particle Green function at $T=0$ (adiabatic approximation)\\
{} & Serves also as the generic symbol representing $\mathscr{G}_{\X{2}}$ and $\mathsf{G}_{\X{2}}$\\
$G_{\X{2;0}}$, $G_{\X{2}}^{\X{(0)}}$  & The non-interacting counterpart of $G_{\X{2}}$\\
{} & Serve also as the generic symbols representing $\mathscr{G}_{\X{2;0}}$, $\mathscr{G}_{\X{2}}^{\X{(0)}}$, and $\mathsf{G}_{\X{2;0}}$, $\mathsf{G}_{\X{2}}^{\X{(0)}}$\\
$\mathscr{G}_{\X{2}}$ & The interacting two-particle Green function for $T > 0$ (Matsubara formalism)\\
$\mathsf{G}_{\X{2}}$ & The interacting two-particle Green function for $T > 0$ (TFD formalism)\\
$\mathbb{G}$ & The $2\times 2$ matrix of the interacting one-particle Green functions within the TFD formalism\\
$\mathbb{G}_{\X{0}}$ & The non-interacting counterpart of $\mathbb{G}$\\
$\ii$ & $\sqrt{-1}$, the imaginary unit (not to be confused with $i$)\\
$\h{0}$ & The Fock-space zero, $0 \times \h{1}$\\
$\h{I}$, $\h{1}$ & The identity operators in respectively the single-particle Hilbert space and the Fock space\\
$i, j$ & Integers that also represent compound variables\\
{} & Thus, $j$ may represent $\bm{r}_jt_j\sigma_j$, or $\bm{r}_j\tau_j\sigma_j\mu_j$ (where $\mu_j \in \{\X{1},\X{2}\}$), \emph{etc.}\\
$j'$ & Similar to $j$, except that when representing a compound variable, the prime is \textsl{not} part\\
{} & of $j'$. In such case, one for instance has $\bm{r}_j't_j'\sigma_j'$, which is distinct from $\bm{r}_{j'}t_{j'}\sigma_{j'}$\\
$\b{j}$ & With $j \in \mathds{N}$ a \textsl{simple} variable or index, $\b{j} \equiv -j$ \\
$\mathcal{N}$ & Normal-ordering operation, with $\mathcal{N}(\dots)$ also widely denoted by $:\dots:$\\
$\h{P}$ & The single-particle operator associated with the \textsl{proper} polarisation function\\
{} & Serves also as the generic symbol representing $\mathscr{P}$ (Matsubara formalism) and \\
{} & $\mathsf{P}$ (TFD formalism)\\
$\h{P}_{\X{\varsigma\varsigma'}}$ &  The operator $\h{P}$ viewed as a functional of: $v$ and $G_{\X{0}}$ for $\X{\varsigma}=\X{0}$, $\X{\varsigma'} = \X{0}$; $v$ and $G$ for\\
{} & $\X{\varsigma}=\X{0}$, $\X{\varsigma'} = \X{1}$; $W$ and $G_{\X{0}}$ for $\X{\varsigma}=\X{1}$, $\X{\varsigma'} = \X{0}$; $W$ and $G$ for $\X{\varsigma}=\X{1}$, $\X{\varsigma'} = \X{1}$\\
$\h{P}_{\!\X{\varsigma\varsigma'}}^{\X{(\nu)}}$ & The total $\nu$th-order perturbational contribution to $\h{P}_{\X{\varsigma\varsigma'}}$\\
$\h{P}^{\star}$ & The single-particle operator associated with the \textsl{improper} polarisation function \\
{} & ($\star$ is not to be confused with $*$ for complex conjugation)\\
{} & Serves also as the generic symbol representing $\mathscr{P}^{\star}$ (Matsubara formalism) and\\
{} & $\mathsf{P}^*$ (TFD formalism)\\
$\h{P}_{\!\X{\varsigma\varsigma'}}^{\star}$ & The single-particle operator $\h{P}^{\star}$ viewed as a functional similar to $\h{P}_{\X{\varsigma\varsigma'}}$ in relation to $\h{P}$\\
$\h{P}_{\!\X{\varsigma\varsigma'}}^{\star \X{(\nu)}}$ & The total $\nu$th-order perturbational contribution to $\h{P}_{\X{\varsigma\varsigma'}}^{\star}$\\
$\h{\varrho}$ & The interacting density, or statistical, operator in the grand canonical ensemble\\
$\h{\varrho}_{\X{0}}$ & The non-interacting counterpart of $\h{\varrho}$\\
$\sigma, \sigma_j$ & Spin indices. For spin-$\tfrac{1}{2}$ particles, $\sigma, \sigma_j \in \{\uparrow,\downarrow\}$\\
$\b{\sigma}, \b{\sigma}_j$ & For spin-$\tfrac{1}{2}$ particles, indices complementary to $\sigma$ and $\sigma_j$. Thus, for $\sigma =\uparrow$, $\b{\sigma} = \downarrow$\\
$\bbsigma^{\X{0}}$ & The $2\times 2$ unit matrix\\
$\bbsigma^{\X{\alpha}}$ & With $\alpha = x, y, z$, a $2\times 2$  Pauli matrix\\
$\h{\Sigma}$ & The single-particle \textsl{proper} self-energy operator\\
{} & Serves also as the generic symbol representing $\h{\mathscr{S}}$ (Matsubara formalism), and\\
{} & $\h{\Upsigma}$ (TFD formalism) \\
$\h{\Sigma}_{\X{\varsigma\varsigma'}}$ & The single-particle operator $\h{\Sigma}$ viewed as a functional similar to $\h{P}_{\X{\varsigma\varsigma'}}$ in relation to $\h{P}$\\
$\h{\Sigma}_{\!\X{\varsigma\varsigma'}}^{\X{(\nu)}}$ & The total $\nu$th-order perturbational contribution to $\h{\Sigma}_{\X{\varsigma\varsigma'}}$. For $\X{\varsigma} = \X{0}$ ($\X{\varsigma} = \X{1}$), $\forall\X{\varsigma'}$, the\\
{} & order of the perturbation theory is that of the coupling constant of $v$ ($W$) \\
$\h{\Sigma}^{\star}$ & The \textsl{improper} self-energy operator ($\star$ is not to be confused with $*$ for complex conjugation)\\
$\h{\Sigma}_{\!\X{\varsigma\varsigma'}}^{\star}$ & Similar to $\h{\Sigma}_{\X{\varsigma\varsigma'}}$ however concerning $\h{\Sigma}^{\star}$\\
$\h{\Sigma}_{\!\X{\varsigma\varsigma'}}^{\star\X{(\nu)}}$ & The total $\nu$th-order perturbational contribution to $\h{\Sigma}_{\X{\varsigma\varsigma'}}^{\star}$\\
$t$, $t_j$ & Real times\\
$\tau$, $\tau_j$ & `Imaginary' times within the Matsubara formalism, although $\tau, \tau_j \in \mathds{R}$\\
$T$ & The absolute temperature \\
$\mathcal{T}$ & Chronological time-ordering operator ($T=0$ formalism) \\
$\mathcal{T}_{\tau}$ & Chronological time-ordering operator (Matsubara formalism) \\
$\mathcal{T}_{\hspace{-1.2pt}{}_{\mathscr{C}}}$ &  Chronological time-ordering operator along the contour $\mathscr{C}$ (TFD formalism)\\
$\Tr$ & Trace over the states in the relevant Fock space\\
$U$ & Magnitude of the on-site interaction potential energy in the Hubbard Hamiltonian $\wh{\mathcal{H}}$\\
$v$ & Two-body interaction potential, including the dimensionless coupling constant $\lambda$: $v = \lambda \mathsf{v}$\\
$\mathsf{v}$ & Two-body interaction potential stripped of the dimensionless coupling constant $\lambda$\\
$\h{v}$ & The single-particle operator associated with $v$\\
$W$ & Two-body \textsl{screened} interaction potential, including the dimensionless coupling constant $\lambda$:\\
{} & $W = \lambda \mathsf{W}$\\
$\mathsf{W}$ & Two-body screened interaction potential stripped of the dimensionless coupling constant $\lambda$\\
$\h{W}$ & The single-particle operator associated with $W$\\
$\h{W}_{\X{\!\varsigma\varsigma'}}$ & The single-particle operator $\h{W}$ viewed as a functional similar to $\h{P}_{\X{\varsigma\varsigma'}}$ in relation to $\h{P}$\\
$\h{W}_{\!\X{\varsigma\varsigma'}}^{\X{(\nu)}}$ & The total $\nu$th-order perturbational contribution to $\h{W}_{\X{\varsigma\varsigma'}}$\\
$\b{x}$ & With $x\in \mathds{R}$, $\b{x} = -x$. For instance, with $x=\tfrac{1}{2}$, $\b{x}=-\tfrac{1}{2}$\\
$\mathcal{Z}$ & The grand partition function \\
$\mathpzc{Z}$ & Coordination number, equal to $2d$ on the $d$-cubic lattice\\
$\vert\Psi_{N;0}\rangle$ & Interacting $N$-particle ground state in the Heisenberg picture ($T=0$ formalism)\\
$\vert\Phi_{N;0}\rangle$ & The non-interacting counterpart of $\vert\Psi_{N;0}\rangle$\\
$\vert 0(\beta)\rangle$ & The $T$-dependent interacting vacuum state in the Heisenberg picture (TFD formalism)\\
$\vert \X{0}(\beta)\rangle$ & The non-interacting counterpart of $\vert 0(\beta)\rangle$ (TFD formalism)\\
\end{longtable}
}
$\hfill\Box$

\end{appendix}

\refstepcounter{dummyX}
\addcontentsline{toc}{section}{References}


\begin{thebibliography}{999}
%
% 1.
\bibitem{NvK}
U. Felderhof, and H. van Beijeren, \href{http://dx.doi.org/10.1007/s10955-013-0902-x}{\emph{In Memoriam Nico van Kampen}}, \emph{J. Stat. Phys.} \textbf{154}, 656 (2014). G. 't Hooft, \href{https://www.knaw.nl/nl/actueel/publicaties/levensberichten-en-herdenkingen-2014}{\emph{Nicolaas Godfried van Kampen}}, \emph{Levensberichten en Herdenkingen 2014}, pp. 22-29 (The Royal Netherlands Academy of Sciences, Amsterdam, 2014). I. Oppenheim, \href{https://doi.org/10.1063/PT.3.2320}{\emph{Nicolaas Godfried van Kampen}}, \emph{Physics Today} \textbf{67} (3), 66 (2014).

% 2.
\bibitem{MYS95}
N.\,H. March, W.\,H. Young, and S. Sampanthar, \emph{The Many-Body Problem in Quantum Mechanics} (Dover, New York, 1995).

% 3.
\bibitem{FW03}
A.\,L. Fetter, and J.~D. Walecka, \emph{Quantum Theory of Many-Particle Systems} (Dover, New York, 2003).

% 4.
\bibitem{RM69}
R. Mills, \emph{Propagators for Many-Particle Systems: An Elementary Treatment} (Gordon and Breach, New York, 1969).

% 5.
\bibitem{NO98}
J.\,W. Negele, and H. Orland, \emph{Quantum Many-Particle Systems} (Westview Press, Boulder, Colorado, 1998).

% 6.
\bibitem{HBKF04}
H. Bruus, and K. Flensberg, \emph{Many-Body Quantum Theory in Condensed Matter Physics: An Introduction} (Oxford University Press, 2004).

% 7.
\bibitem{SvL13}
G. Stefanucci, and R. van Leeuwen, \emph{Nonequilibrium Many-Body Theory of Quantum Systems: A Modern Introduction} (Cambridge University Press, 2013).

% 8.
\bibitem{RPF51}
R.\,P. Feynman, \emph{Phys. Rev.} \textbf{84}, 108 (1951).

% 9.
\bibitem{FJD49}
F.\,J. Dyson, \emph{Phys. Rev.} \textbf{75}, 486 and 1736 (1949).

% 10.
\bibitem{MS59}
P.\,C. Martin, and J. Schwinger, \emph{Phys. Rev.} \textbf{115}, 1342 (1959).

% 11.
\bibitem{AGD75}
A.\,A. Abrikosov, L.\,P. Gorkov [Gor'kov], and I.\,E. Dzyaloshinski [Dzyaloshinskii], \emph{Methods of Quantum Field Theory in Statistical Physics}, revised English edition, translated and edited by Richard A. Silverman (Dover, New York, 1975). [See \emph{Historical note} in Ref.\,\protect\citen{Note2} below.]

% 12.
\bibitem{TM55}
T. Matsubara, \emph{Prog. Theor. Phys.} \textbf{14}, 351 (1955).

% 13.
\bibitem{ESF59}
E.\,S. Fradkin, \emph{Sov. Phys. JETP} \textbf{36}, 912 (1959).

% 14.
\bibitem{KP61}
O.\,V. Konstantinov, and V.\,I. Perel', \emph{Sov. Phys. JETP} \textbf{12}, 142 (1961).

% 15.
\bibitem{LVK65}
L.\,V. Keldysh, \emph{Sov. Phys. JETP} \textbf{20}, 1018 (1965).

% 16.
\bibitem{RAC68}
R.\,A. Craig, \emph{J. Math. Phys.} \textbf{9}, 605 (1968).

% 17.
\bibitem{RS86}
J. Rammer, and H. Smith, \emph{Rev. Mod. Phys.} \textbf{58}, 323 (1986).

% 18.
\bibitem{JR07}
J. Rammer, \emph{Quantum Filed Theory of Non-equilibrium States} (Cambridge University Press, 2007).

% 19.
\bibitem{PD84}
P. Danielewicz, \emph{Ann. Phys.} (N.\,Y.) \textbf{152}, 239 (1984).

% 20.
\bibitem{MW91}
M. Wagner, \emph{Phys. Rev.} B\,\textbf{44}, 6104 (1991).

% 21.
\bibitem{UMT82}
H. Umezawa, H. Matsumoto, and M. Tachiki, \emph{Thermo Field Dynamics and Condensed States} (North-Holland, Amsterdam, 1982).

% 22.
\bibitem{HU95}
H. Umezawa, \emph{Advanced Field Theory: Micro, Macro, and Thermal Physics} (American Institute of Physics, New York, 1995).

% 23.
\bibitem{RPF53}
R.\,P. Feynman, \emph{Phys. Rev.} \textbf{91}, 1291 (1953); \emph{loc. cit.}, 1301 (1953).

% 24.
\bibitem{GP98}
G. Parisi, \emph{Statistical Field Theory} (Perseus Books, Reading, 1998).

% 25.
\bibitem{IZ80}
C. Itzykson, and J.-B. Zuber, \emph{Quantum Field Theory} (McGraw-Hill, New York, 1980).

% 26.
\bibitem{HK49}
A. Houriet, and A. Kind, \emph{Helv. Phys. Acta} \textbf{22}, 319 (1949).

% 27.
\bibitem{GCW50}
G.\,C. Wick, \emph{Phys. Rev.} \textbf{80}, 268 (1950).

% 28.
\bibitem{BF16a}
B. Farid, \emph{On the Luttinger-Ward functional and the convergence of skeleton diagrammatic series expansion of the self-energy for Hubbard-like models}, to be published.

% 29.
\bibitem{HM78}
H. Minc, \emph{Permanents} (Addison-Wesley, London, 1978).

% 30.
\bibitem{NW78}
A. Nijenhuis, and H.\,S. Wilf, \emph{Combinatorial Algorithms for Computers and Calculators}, 2nd edition (Academic Press, New York, 1978).

% 31.
\bibitem{MA06}
M. Agrawal, \emph{Determinant versus permanent}, \href{http://www.icm2006.org/proceedings/}{Proceedings of the ICM\,2006}, Madrid, Spain, Vol.\,3, 985-997 (European Mathematical Society, 2006).

% 32.
\bibitem{TM60}
T. Muir, \emph{A Treatise on the Theory of Determinants}, revised and enlarged by W.\,H. Metzler (Dover, New York, 1960).

% 33.
\bibitem{VD99}
R. Vein, and P. Dale, \emph{Determinants and Their Applications in Mathematical Physics} (Springer, New York, 1999).

% 34.
\bibitem{DLRV70}
A. De Luca, L.\,M. Ricciardi, and R. Vasudevan, \emph{J. Math. Phys.} \textbf{11}, 530 (1970).

% 35.
\bibitem{PP13}
P. Pudl\'{a}k, \emph{Logical Foundations of Mathematics and Computational Complexity: A Gentle Introduction} (Springer, Heidelberg, 2013).

% 36.
\bibitem{CLP78}
P. Cvitanovi\'{c}, B. Lautrup, and R.\,B. Pearson, \emph{Phys. Rev.} D\,\textbf{18}, 1939 (1978).

% 37.
\bibitem{PH07}
Y. Pavlyukh, and W. H\"{u}bner, \emph{J. Math. Phys.} \textbf{48}, 052109 (2007).

% 38.
\bibitem{LGM05}
L.\,G. Molinari, \emph{Phys. Rev.} B\,\textbf{71}, 113102 (2005).

% 39.
\bibitem{MM06}
L.\,G. Molinari, and N. Manini, \emph{Eur. Phys. J.} B\,\textbf{51}, 331 (2006).

% 40.
\bibitem{AS72}
M. Abramowitz, I.\,A. Stegun, editors, \emph{Handbook of Mathematical Functions} (Dover, New York, 1972).

% 41.
\bibitem{GvL13}
G.\,H. Golub, and C.\,F. van Loan, \emph{Matrix Computations}, 4th edition (The Johns Hopkins University Press, Baltimore, 2013).

% 42.
\bibitem{VS69}
V. Strassen, \emph{Numer. Math.} \textbf{13}, 354 (1969).

% 43.
\bibitem{CW90}
D. Coppersmith, and S. Winograd, \emph{J. Symbolic Computation} \textbf{9}, 251 (1990).

% 44.
\bibitem{VVW11}
V.\,V. Williams, \emph{Breaking the Coppersmith-Winograd barrier} (\href{https://citeseerx.ist.psu.edu/viewdoc/summary?doi=10.1.1.228.9947}{CiteSeer$^{\textrm{x}}$ -- Nov. 2011}).

% 45.
\bibitem{DEK98}
D.\,E. Knuth, \emph{The Art of Computer Programming}, Vol.\,2: \emph{Seminumerical Algorithms}, 3rd edition (Addison Wesley, Reading, 1998).

% 46.
\bibitem{NJH02}
N.\,J. Higham, \emph{Accuracy and Stability of Numerical Algorithms}, 2nd edition (SIAM, Philadelphia, 2002).

% 47.
\bibitem{LGV79}
L.\,G. Valiant, \emph{Theor. Comput. Sci.} \textbf{8}, 189 (1979).

% 48.
\bibitem{BF16b}
B. Farid, \emph{Many-body perturbation expansions without diagrams. II. Superfluid and superconductive states}, to be published.

% 49.
\bibitem{MP934}
P. Monthoux, and D. Pines, \emph{Phys. Rev.} B\,\textbf{47}, 6069 (1993); \emph{loc. cit.} \textbf{49}, 4261 (1994).

% 50.
\bibitem{DP97}
D. Pines, \emph{Physica} C\,\textbf{282}-\textbf{287}, 273 (1997).

% 51.
\bibitem{DJS99}
D.\,J. Scalapino, \emph{J. Low Temp. Phys.} \textbf{117}, 179 (1999).

% 52.
\bibitem{NP10}
N. Plakida, \emph{High-Temperature Cuprate Superconductors: Experiment, Theory and Applications} (Springer, Berlin, 2010).

% 53.
\bibitem{HM02}
H. Monien, \emph{J. Low Temp. Phys.} \textbf{126}, 1123 (2002).

% 54.
\bibitem{JMZ69}
J.\,M. Ziman, \emph{Elements of Advanced Quantum Theory} (Cambridge University Press, 1969).

% 55.
\bibitem{PN89}
D. Pines, and P. Nozi\`{e}res, \emph{The Theory of Quantum Liquids}, Vol. I: \emph{Normal Fermi Liquids} (Westview Press, Boulder, Colorado, 1989).

% 56.
\bibitem{BF13}
B. Farid, \emph{Some rigorous results concerning the uniform metallic ground states of single-band
Hamiltonians in arbitrary dimensions}, \href{http://www.arxiv.org/abs/1305.2089v1}{arXiv:1305.2089}.

% 57.
\bibitem{JH57}
J. Hubbard, \emph{Proc. Roy. Soc. London}, A\,\textbf{240}, 539 (1957).

% 58.
\bibitem{LW60}
J.\,M. Luttinger, and J.\,C. Ward, \emph{Phys. Rev.} \textbf{118}, 1417 (1960).

% 59.
\bibitem{BK61}
G. Baym, and L.\,P. Kadanoff, \emph{Phys. Rev.} \textbf{124}, 287 (1961).

% 60.
\bibitem{KB62}
L.\,P. Kadanoff, and G. Baym, \emph{Quantum Statistical Mechanics} (W.\,A. Benjamin, New York, 1962).

% 61.
\bibitem{ERC53I}
E.\,R. Caianiello, \emph{Nuovo Cimento} \textbf{10}, 1634 (1953).

% 62.
\bibitem{ERC59}
E.\,R. Caianiello, \emph{Nuovo Cimento Suppl.} \textbf{14}, 177 (1959).

% 63.
\bibitem{FMPR81}
F. Fucito, E. Marinari, G. Parisi, and C. Rebbi, \emph{Nucl. Phys.} B\,\textbf{180}, 369 (1981).

% 64.
\bibitem{SS81}
D.\,J. Scalapino, and R.\,L. Sugar, \emph{Phys. Rev. Lett.} \textbf{46}, 519 (1981).

% 65.
\bibitem{BSS81}
R. Blankenbecler, D.\,J. Scalapino, and R.\,L. Sugar, \emph{Phys. Rev.} D\,\textbf{24}, 2278 (1981).

% 66.
\bibitem{HSSB82}
J.\,E. Hirsch, R.\,L. Sugar, D.\,J. Scalapino, and R. Blankenbecler, \emph{Phys. Rev.} B\,\textbf{26}, 5033 (1982).

% 67.
\bibitem{JEH83}
J.\,E. Hirsch, \emph{Phys. Rev.} B\,\textbf{28}, 4059 (1983).

% 68.
\bibitem{JEH85}
J.\,E. Hirsch, \emph{Phys. Rev.} B\,\textbf{31}, 4403 (1985).

% 69.
\bibitem{JEG85}
J.\,E. Gubernatis, \emph{Monte Carlo Simulations of Fermion Systems: The Determinant Method}, pp.\,212-221, in \emph{Monte-Carlo Methods and Applications in Neutronics, Photonics and Statistical Physics}, edited by R. Alcouffe, R. Dautray, A. Forster, G. Ledanois, and B. Mercier (Springer, Berlin, 1985).

% 70.
\bibitem{SK86}
G. Sugiyama, S.\,E. Koonin, \emph{Ann. Phys.} (N.\,Y.) \textbf{168}, 1 (1986).

% 71.
\bibitem{LG92}
E.\,Y. Loh, Jr, and J.\,E. Gubernatis, \emph{Stable Numerical Simulations of Models of Interacting Electrons in Condensed-Matter Physics}, Ch.\,4, pp.\,177-235, in \emph{Electronic Phase Transitions}, edited by W. Hanke, and Yu.\,V. Kopaev (North-Holland, Amsterdam, 1992).

% 72.
\bibitem{BPS04} % 'Bourovski' is as published.
E. Bourovski, N.\,[V.] Prokof'ev, and B. Svistunov, \emph{Phys. Rev.} B\,\textbf{70}, 193101 (2004).

% 73.
\bibitem{AE08}
F.\,F. Assaad, and H.\,G. Evertz, \emph{World-line and Determinantal Quantum Monte Carlo Methods for Spins, Phonons and Electrons}, pp.\,277-356, in \emph{Computational Many-Particle Physics}, edited by H. Fehske, R. Schneider, and A. Wei{\ss}e (Springer, Berlin, 2008).

% 74.
\bibitem{HF86}
J.\,E. Hirsch, and R.\,M. Fye, \emph{Phys. Rev. Lett.} \textbf{56}, 2521 (1986).

% 75.
\bibitem{FHS87}
R.\,M. Fye, J.\,E. Hirsch, and D.\,J. Scalapino, \emph{Phys. Rev.} B\,\textbf{35}, 4901 (1987).

% 76.
\bibitem{GKKR96}
A. Georges, G. Kotliar, W. Krauth, and M.~J. Rozenberg, \emph{Rev. Mod. Phys.} \textbf{68}, 13 (1996).

% 77.
\bibitem{KSHOPM06}
G. Kotliar, S.\,Y. Savrasov, K. Haule, V.\,S. Oudovenko, O. Parcollet, and C.\,A. Marianetti, \emph{Rev. Mod. Phys.} \textbf{78}, 865 (2006).

% 78.
\bibitem{SBS88}
R.\,T. Scalettar, N.\,E. Bickers, and D.\,J. Scalapino, \emph{Quantum Monte Carlo Studies of the Holstein Model}, pp.\,166-171, in \emph{Computer Simulation Studies in Condensed Matter Physics: Recent Developments}, edited by D.\,P. Landau, K.\,K. Mon, and H.-B. Sch\"{u}ttler (Springer, Berlin, 1988).

% 79.
\bibitem{TC90}
N. Trivedi, and D.\,M. Ceperley, \emph{Phys. Rev.} B\,\textbf{41}, 4552 (1990).

% 80.
\bibitem{PST96}
N.\,V. Prokof'ev, B.\,V. Svistunov, and I.\,S. Tupitsyn, \emph{JETP Lett.} \textbf{64}, 911 (1996).

% 81.
\bibitem{PST98}
N.\,V. Prokof'ev, B.\,V. Svistunov, and I.\,S. Tupitsyn, \emph{Phys. Lett.} A\,\textbf{238}, 253 (1998);  \emph{JETP Lett.} \textbf{87}, 310 (1998).

% 82.
\bibitem{BPS06}
M. Boninsegni, N.\,V. Prokof'ev, and B.\,V. Svistunov, \emph{Phys. Rev. Lett.} \textbf{96}, 070601 (2006); \emph{Phys. Rev.} E\,\textbf{74}, 036701 (2006).

% 83.
\bibitem{BPST06a} % 'Burovski' is as published.
E. Burovski, N.\,[V.] Prokof'ev, B.\,[V.] Svistunov, and M. Troyer, \emph{Phys. Rev. Lett.} \textbf{96}, 160402 (2006); Erratum: \emph{Phys. Rev. Lett.} \textbf{97}, 239902 (2006).

% 84.
\bibitem{BPST06b} % 'Buroski' is as published.
E. Burovski, N.\,[V.] Prokof'ev, B.\,[V.] Svistunov, and M. Troyer, \emph{New J. Phys.} \textbf{8}, 153 (2006); Erratum added on 21 December 2006: \emph{New J. Phys.} \textbf{8}, 182 (2006). [\href{https://doi.org/10.1088/1367-2630/8/8/153}{Open Access}]

% 85.
\bibitem{PTVF97}
W.\,H. Press, S.\,A. Teukolsky, W.\,T. Vetterling, and B.\,P. Flannery, \emph{Numerical Recipes in Fortran 77: The Art of Scientific Computing}, 2nd edition, reprinted with corrections (Cambridge University Press, 2001).

% 86.
\bibitem{JH59}
J. Hubbard, \emph{Phys. Rev. Lett.} \textbf{3}, 77 (1959).

% 87.
\bibitem{PWA59}
P.\,W. Anderson, \emph{Phys. Rev.} \textbf{115}, 2 (1959).

% 88.
\bibitem{ThWR62}
Th.\,W. Ruijgrok, \emph{Physica} \textbf{28}, 877 (1962).

% 89.
\bibitem{JH63}
J. Hubbard, \emph{Proc. Roy. Soc.} (London), A\hspace{0.6pt}\textbf{276}, 238 (1963).

% 90.
\bibitem{GK97}
O. Gunnarsson, and E. Koch, \emph{Phys. Lett.} A\,\textbf{235}, 530 (1997).

% 91.
\bibitem{HFT59}
H.\,F. Trotter, \emph{Proc. Amer. Math. Soc.} \textbf{10}, 545 (1959).

% 92.
\bibitem{MS76}
M. Suzuki, \emph{Commun. math. Phys.} \textbf{51}, 183 (1976).

% 93.
\bibitem{FFA05}
F.\,F. Assaad, \emph{Phys. Rev.} B\,\textbf{71}, 075103 (2005).

% 94.
\bibitem{AW82}
A. Wiesler, \emph{Phys. Lett.} \textbf{89}\,A, 359 (1982).

% 95.
\bibitem{DRdR83}
H. De Raedt, and B. De Raedt, \emph{Phys. Rev.} A\,\textbf{28}, 3575 (1983).

% 96.
\bibitem{Note1}
An extensive body of texts exists on the subject matter of Monte Carlo techniques, which fall into two main categories of \textsl{classical} \cite{FS02} and \textsl{quantum} Monte Carlo, both of which are rooted in the Metropolis algorithm, due to Metropolis, Rosenbluth, Teller, and Teller \cite{MRRTT53}, and its generalisations, notably the one based on Markov chains \cite{WF68}, by Handscomb \cite{DCH62}, and Hastings. \cite{WKH70} For a short but informative introduction to stochastic methods, the reader is referred to Chapter 8 of Ref.\,\citen{NO98}. Four pedagogical texts regarding the quantum Monte Carlo method are by Creutz and Freedman \cite{CF81}, Thijssen \cite{JMT07}, Wipf \cite{AW13}, and Gattringer and Lang. \cite{GL10} More general texts are by Hammersley and Handscomb \cite{HH75}, Madras \cite{NM02}, Gentle \cite{JEG03}, Ceperley \cite{DMC95,DMC13}, Kalos and Whitlock \cite{KW08}, Binder and Heermann \cite{BH10}, and Landau and Binder. \cite{LB15} For reviews, see De Raedt and Lagendijk \cite{DRL85}, Suzuki \cite{MS87}, Von der Linden \cite{VdL92}, Sandvik \cite{AWS97}, and Trebst and Troyer \cite{TT06}.

% 97.
\bibitem{FS02}
D. Frenkel, and B. Smit, \emph{Understanding Molecular Simulation: From Algorithms to Applications}, 2nd edition (Academic Press, London, 2002).

% 98.
\bibitem{MRRTT53}
N. Metropolis, A.\,W. Rosenbluth, M.\,N. Rosenbluth, A.\,H. Teller, and E. Teller, \emph{J. Chem. Phys.} \textbf{21}, 1087 (1953).

% 99.
\bibitem{WF68}
W. Feller, \emph{An Introduction to Probability Theory and Its Applications},  Vol.\,I, 3rd edition (John Wiley \& Sons, New York, 1968).

% 100.
\bibitem{DCH62}
D.\,C. Handscomb, \emph{Proc. Camb. Phil. Soc.} \textbf{58}, 594 (1962).

% 101.
\bibitem{WKH70}
W.\,K. Hastings, \emph{Biometrika} \textbf{57}, 97 (1970).

% 102.
\bibitem{CF81}
M. Creutz, and B. Freedman, \emph{Ann. Phys.} (N.\,Y.) \textbf{132}, 427 (1981).

% 103.
\bibitem{JMT07}
J.\,M. Thijssen, \emph{Computational Physics}, 2nd edition (Cambridge University Press, 2007).

% 104.
\bibitem{AW13}
A. Wipf, \emph{Statistical Approach to Quantum Field Theory: An Introduction} (Springer, Berlin, 2013).

% 105.
\bibitem{GL10}
C. Gattringer, and C.\,B. Lang, \emph{Quantum Chromodynamics on the Lattice: An Introductory Presentation} (Springer, Berlin, 2010).

% 106.
\bibitem{HH75}
J.\,M. Hammersley, and D.\,C. Handscomb, \emph{Monte Carlo Methods} (Methuen, London, 1975).

% 107.
\bibitem{NM02}
N. Madras, \emph{Lectures on Monte Carlo Methods} (American Mathematical Society, Providence, 2002).

% 108.
\bibitem{JEG03}
J.\,E. Gentle, \emph{Random Number Generation and Monte Carlo Methods}, 2nd edition (Springer, New York, 2003).

% 109.
\bibitem{DMC95}
D.\,M. Ceperley, \emph{Rev. Mod. Phys.} \textbf{67}, 279 (1995).

% 110.
\bibitem{DMC13}
D.\,M. Ceperley, \emph{Path Integral Methods for Continuum Quantum Systems}, Ch.\,14, in \emph{Emergent Phenomena in Correlated Matter Modeling and Simulations}, Vol.\,3, edited by E. Pavarini, E. Koch, and U. Schollw\"{o}ck (Forschungszentrum J\"{u}lich, 2013).

% 111.
\bibitem{KW08}
M.\,H. Kalos, and P.\,A. Whitlock, \emph{Monte Carlo Methods}, 2nd revised and enlarged edition (Wiley-VCH, Weinheim, 2008).

% 112.
\bibitem{BH10}
K. Binder, and D.\,W. Heermann, \emph{Statistical Simulations in Statistical Physics: An Introduction}, 5th edition (Springer, Berlin, 2010).

% 113.
\bibitem{LB15}
D.\,P. Landau, and K. Binder, \emph{A Guide to Monte Carlo Simulations}, 4th edition (Cambridge University Press, 2015).

% 114.
\bibitem{DRL85}
H. De Raedt, and A. Lagendijk, \emph{Phys. Rep.} \textbf{127}, 233 (1985).

% 115.
\bibitem{MS87}
M. Suzuki, editor, \emph{Quantum Monte Carlo Methods in Equilibrium and Nonequilibrium Systems} (Springer, Berlin, 1987).

% 116.
\bibitem{VdL92}
W. von der Linden, \emph{Phys. Rep.} \textbf{220}, 53 (1992).

% 117.
\bibitem{AWS97}
A.\,W. Sandvik, \emph{An Introduction to Quantum Monte Carlo Methods}, pp.\,109-135, in \emph{Strongly Correlated Magnetic and Superconducting Systems}, edited by G. Sierra, and M.\,A. Mart\'{i}n-Delgado (Springer, Berlin, 1997).

% 118.
\bibitem{GBPF93}
G.\,G. Batrouni, and P. de Forcrand, \emph{Phys. Rev.} B\,\textbf{48}, 589 (1993).

% 119.
\bibitem{TW05}
M. Troyer, and U.-J. Wiese, \emph{Phys. Rev. Lett.} \textbf{94}, 170201 (2005).

% 120.
\bibitem{GMLRTW11}
E. Gull, A.\,J. Millis, A.\,I. Lichtenstein, A.\,N. Rubtsov, M. Troyer, and P. Werner, \emph{Rev. Mod. Phys.} \textbf{83}, 349 (2011).

% 121.
\bibitem{SK91}
A.\,W. Sandvik, and J. Kurkij\"{a}rvi, \emph{Phys. Rev.} B\,\textbf{43}, 5950 (1991); A.\,W. Sandvik, \emph{Phys. Rev.} B\,\textbf{59}, R14157 (1999); O.\,F. Sylju{\aa}sen, and A.\,W. Sandvik, \emph{Phys. Rev.} E\,\textbf{66}, 046701 (2002).

% 122.
\bibitem{RGM13}
R.\,G. Melko, \emph{Stochastic Series Expansion Quantum Monte Carlo}, pp.\,185-206, in \emph{Strongly Correlated System: Numerical Methods}, edited by A. Avella, and F. Mancini (Springer, Berlin, 2013).

% 123.
\bibitem{TT06}
S. Trebst, and M. Troyer, \emph{Ensemble Optimization Techniques for Classical and Quantum Systems}, pp.\,591-640, in \emph{Computer Simulations in Condensed Matter: From Materials to Chemical Biology}, Vol.\,1, edited by M. Ferrario, G. Cioccotti, and K. Binder (Springer, Berlin, 2006).

% 124.
\bibitem{BW967}
B.\,B. Beard, and U.-J. Wiese, \emph{Phys. Rev. Lett.} \textbf{77}, 5130 (1996); \emph{Nucl. Phys.} B (Proc. Suppl.) \textbf{53}, 838 (1997).

% 125.
\bibitem{FG92}
E. Farhi, and S. Gutmann, \emph{Ann. Phys.} (N.\,Y.) \textbf{213}, 182 (1992).

%. 126.
\bibitem{ANR03}
A.\,N. Rubtsov, \emph{Quantum Monte Carlo determinantal algorithm without Hubbard-Stratonovich transformation: a general consideration}, \href{http://arxiv.org/abs/cond-mat/0302228}{arXiv:0302228v1}.

% 127.
\bibitem{RL04}
A.\,N. Rubtsov, and A.\,I. Lichtenstein, \emph{JETP Lett.} \textbf{80}, 61 (2004).

% 128.
\bibitem{RSL05}
A.\,N. Rubtsov, V.\,V. Savkin, and A.\,I. Lichtenstein, \emph{Phys. Rev.} B\,\textbf{72}, 035122 (2005).

% 129.
\bibitem{RHJ98}
S.\,[M.\,A.] Rombouts, K. Heyde, and N. Jachowicz, \emph{Phys. Lett.} A\,\textbf{242}, 271 (1998); \emph{Phys. Rev. Lett.} \textbf{82}, 4155 (1999).

% 130.
\bibitem{WCMTM06}
P. Werner, A. Comanac, L. de' Medici, M. Troyer, and A.\,J. Millis, \emph{Phys. Rev. Lett.} \textbf{97}, 076405 (2006).

% 131.
\bibitem{GWPT08}
E. Gull, P. Werner, O. Parcollet, and M. Troyer, \emph{EPL} \textbf{82}, 57003 (2008).

% 132.
\bibitem{GT13}
E. Gull, and M. Troyer, \emph{Fermionic and Continuous Time Monte Carlo}, pp.\,293-319, in \emph{Strongly Correlated System: Numerical Methods}, edited by A. Avella, and F. Mancini (Springer, Berlin, 2013).

% 133.
\bibitem{PS98}
N.\,V. Prokof'ev, and B.\,V. Svistunov, \emph{Phys. Rev. Lett.} \textbf{81}, 2514 (1998).

% 134.
\bibitem{PS08} % Bold diagrams
N.\,V. Prokof'ev, and B.\,V. Svistunov, \emph{Phys. Rev.} B\,\textbf{77}, 125101 (2008).

% 135.
\bibitem{ASM08}
A.\,S. Mishchenko, \emph{Diagrammatic Monte Carlo and Stochastic Optimization Methods for Complex Composite Objects in Macroscopic Baths}, pp.\,367-395, in \emph{Computational Many-Particle Physics}, edited by H. Fehske, R. Schneider, and A. Wei{\ss}e (Springer, Berlin, 2008).

% 136.
\bibitem{vHKPS10}
K. Van Houcke, E. Kozik, N.\,[V.] Prokof'ev, and B.\,[V.] Svistunov, \emph{Diagrammatic Monte Carlo}, \href{http://dx.doi.org/10.1016/j.phpro.2010.09.034}{pp.\,95-105}, in \textsl{Computer Simulations Studies in Condensed Matter Physics XXI}, edited by D.\,P. Landau, S.\,P. Lewis and H.-B. Sch\"{u}ttler, \href{https://www.sciencedirect.com/journal/physics-procedia/vol/6/suppl/C}{\emph{Physics Procedia} \textbf{6}, 1-126 (2010)}.

% 137.
\bibitem{KvHGPPST10}
E. Kozik, K. Van Houcke, E. Gull, L. Pollet, N.\,[V.] Prokof'ev, B.\,[V.] Svistunov, and M. Troyer, \emph{EPL} \textbf{90}, 10004 (2010).

% 138.
\bibitem{GWFSPT11}
E. Gull, P. Werner, S. Fuchs, B. Surer, T. Pruschke, and M. Troyer, \emph{Comput. Phys. Commun.} \textbf{182}, 1078 (2011).

% 139.
\bibitem{vHetal12}
K. Van Houcke, F. Werner, E. Kozik, N.\,[V.] Prokof'ev, B.\,[V.] Svistunov, M.\,J.\,H. Ku, A.\,T. Sommer, L.\,W. Cheuk, A. Schirotzek, and M.\,W. Zwierlein, \emph{Nature Phys.} \textbf{8}, 366 (2012).

% 140.
\bibitem{NP13}
N.\,[V.] Prokof'ev, \emph{Diagrammatic Monte Carlo and Worm Algorithm Techniques}, pp.\,273-292, in \emph{Strongly Correlated System: Numerical Methods}, edited by A. Avella, and F. Mancini (Springer, Berlin, 2013).

% 141.
\bibitem{PS08R}
N.\,[V.] Prokof'ev, and B.\,[V.] Svistunov, \emph{Phys. Rev.} B\,\textbf{77}, 020408(R) (2008).

% 142.
\bibitem{PPS10}
L. Pollet, N.\,V. Prokof'ev, and B.\,V. Svistunov,  \emph{Phys. Rev. Lett.} \textbf{105}, 210601 (2010).

% 143.
\bibitem{AM76}
N.\,W. Ashcroft, and N.\,D. Mermin, \emph{Solid State Physics} (Thompson Learning, London, 1976).

% 144.
\bibitem{GRM10} % Bold diagrams
E. Gull, D.\,R. Reichman, and A.\,J. Millis, \emph{Phys. Rev.} B\,\textbf{82}, 075109 (2010).

% 145.
\bibitem{CGRM14}
G. Cohen, E. Gull, D.\,R. Reichman, and A.\,J. Millis, \emph{Phys. Rev. Lett.} \textbf{112}, 146802 (2014).

% 146.
\bibitem{CRMG14}
G. Cohen, D.\,R. Reichman, A.\,J. Millis, and E. Gull, \emph{Phys. Rev.} B\,\textbf{89}, 115139 (2014).

% 147.
\bibitem{BF07}
B. Farid, \textsl{On the Luttinger theorem concerning number of particles in the ground states of systems of interacting fermions}, \href{http://arxiv.org/abs/0711.0952}{arXiv:0711.0952}.

% 148.
\bibitem{KK71}
H. Keiter, and J.\,C. Kimball, \emph{Intern. J. Magnetism} \textbf{1}, 233 (1971).

% 149.
\bibitem{GK81}
N. Grewe, H. Keiter, \emph{Phys. Rev.} B\,\textbf{24}, 4420 (1981).

% 150.
\bibitem{KM84}
H. Keiter, and G. Morandi, \emph{Phys. Rep.} \textbf{109}, 227 (1984).

% 151.
\bibitem{PG89}
Th. Pruschke, and N. Grewe, \emph{Z. Phys.} B\,\textbf{74}, 439 (1989).

% 152.
\bibitem{NEB87}
N.\,E. Bickers, \emph{Rev. Mod. Phys.} \textbf{59}, 845 (1987).

% 153.
\bibitem{ACH97}
A.\,C. Hewson, \emph{The Kondo Problem to Heavy Fermions}, 1st paperback edition with corrections (Cambridge University Press, 1997).

% 154.
\bibitem{PF02}
P. Fulde, \emph{Electron Correlations in Molecules and Solids}, 3rd enlarged edition, corrected 2nd printing (Springer, Berlin, 2002).

% - The formalism:

% 155.
\bibitem{LPG58}
L.\,P. Gor'kov, \emph{Sov. Phys. JETP} \textbf{34}, 505 (1958).

% 156.
\bibitem{YN60}
Y. Nambu, \emph{Phys. Rev.} \textbf{117}, 648 (1960).

% 157.
\bibitem{JRS99}
J.\,R. Schrieffer, \emph{Theory of Superconductivity}, revised printing (Westview Press, Boulder, Colorado, 1999).

% 158.
\bibitem{KM69}
K. Maki, \emph{Gapless Superconductivity}, Ch.\,18, pp.\,1035-1105, in \emph{Superconductivity}, Vol.\,2, edited by R.\,D. Parks (Marcel Dekker, New York, 1969).

% 159.
\bibitem{PWA58}
P.\,W. Anderson, \emph{Phys. Rev.} \textbf{112}, 1900 (1958).

% 160.
\bibitem{AZ14}
A.\,[M.] Zagoskin, \emph{Quantum Theory of Many-Body Systems: Techniques and Applications}, 2nd edition (Springer, Heidelberg, 2014).

% - Models:

% 161.
\bibitem{MM86}
M. Marinaro, \emph{Phys. Rep.} \textbf{137}, 81 (1986).

% 162.
\bibitem{JS61}
J. Schwinger, \emph{J. Math. Phys.} \textbf{2}, 407 (1961).

% 163.
\bibitem{vBH72}
U. von Barth, and L. Hedin, \emph{J. Phys.} C\,\textbf{5}, 1629 (1972).

% 164.
\bibitem{RC73}
A.\,K. Rajagopal, and J. Callaway, \emph{Phys. Rev.} B\,\textbf{7}, 1912 (1973).

% 165.
\bibitem{BF9799}
B. Farid, \emph{Phil. Mag.} B\,\textbf{76}, 145 (1997); \emph{Solid State Commun.} \textbf{104}, 227 (1997); \emph{Phil. Mag. Lett.} \textbf{79}, 581 (1999).

% 166.
\bibitem{JF09}
J. Fr\"{o}hlich, \emph{Spin, or actually: Spin and Quantum Statistics}, pp.\,1-60, in \emph{The Spin: Poincar\'{e} Seminar 2007}, edited by B. Duplantier, J.-M. Raimond, and V. Rivasseau (Birkh\"{a}user, Basel, 2009).

% 167.
\bibitem{FS97}
J. Fuchs, and C. Schweigert, \emph{Symmetries, Lie Algebras and Representations: A Graduate Course for Physicists} (Cambridge University Press, 1997).

% 168.
\bibitem{RK57}
R. Kubo, \emph{J. Phys. Soc. Japan} \textbf{12}, 570 (1957).

% - The thermo-field dynamics (TFD):

% 169.
\bibitem{BM61}
G. Baym, and N.\,D. Mermin, \emph{J. Math. Phys.} \textbf{2}, 232 (1961).

% 170.
\bibitem{MW58}
E.\,W. Montroll, and J.\,C. Ward, \emph{Phys. Fluids} \textbf{1}, 55 (1958).

% 171.
\bibitem{AGD59}
A.\,A. Abrikosov, L.\,P. Gor'kov, and I.\,E. Dzyaloshinskii, \emph{Sov. Phys. JETP} \textbf{36}, 636 (1959).

% 172.
\bibitem{JSL61}
J.\,S. Langer, \emph{Phys. Rev.} \textbf{124}, 997 (1961).

% 173.
\bibitem{TCL19}
A. Taheridehkordi, S.\,H. Curnoe, and J.\,P.\,F. LeBlanc, \emph{Phys. Rev.} B\,\textbf{99}, 035120 (2019).

% 174.
\bibitem{VF19}
J. Vu\v{c}i\v{c}evi\'{c}, and M. Ferrero, \emph{Phys. Rev.} B\,\textbf{101}, 075113 (2020).

% 175.
\bibitem{HM77}
H. Matsumoto, \emph{Fortsch. Phys.} \textbf{25}, 1 (1977).

% 176.
\bibitem{HM86}
H. Matsumoto, \emph{Z. Phys.} C\,\textbf{33}, 201 (1986).

% 177.
\bibitem{MOU84}
H. Matsumoto, I. Ojima, and H. Umezawa, \emph{Ann. Phys.} (N.\,Y.) \textbf{152}, 348 (1984).

% 178.
\bibitem{SU83}
G.\,W. Semenoff, and H. Umezawa, \emph{Nucl. Phys.} B\,\textbf{220}, 196 (1983).

% 179.
\bibitem{MNUMM83}
H. Matsumoto, Y. Nakano, H. Umezawa, F. Mancini, and M. Marinaro, \emph{Prog. Theor. Phys.} \textbf{70}, 599 (1983).

% 180.
\bibitem{RB94}
M. Revzen, and J.\,L. Birman, \emph{Phys. Rev.} E\,\textbf{49}, 2688 (1994).

% 181.
\bibitem{CU93}
H. Chu, and H. Umezawa, \emph{Phys. Lett.} A\,\textbf{177}, 385 (1993).

% 182.
\bibitem{HM88}
H. Matsumoto, \emph{Prog. Theor. Phys.} \textbf{80}, 57 (1988).

% 183.
\bibitem{TU75}
Y. Takahashi, and H. Umezawa, \emph{Collect. Phenom.} \textbf{2}, 55 (1975). Reprinted: \emph{Int. J. Mod. Phys.} B\,\textbf{10}, 1755 (1996).

% 184.
\bibitem{IO81}
I. Ojima, \emph{Ann. Phys.} (N.\,Y.)  \textbf{137}, 1 (1981).

% 185.
\bibitem{AU85}
T. Arimitsu, and H. Umezawa, \emph{Prog. Theor. Phys.} \textbf{74}, 429 (1985).

% 186.
\bibitem{MS85}
M. Suzuki, \emph{J. Phys. Soc. Japan} \textbf{54}, 4483 (1985).

% 187.
\bibitem{JPW85}
J.\,P. Whitehead, \emph{Prog. Theor. Phys.} \textbf{74}, 1168 (1985).

% 188.
\bibitem{APU86}
T. Arimitsu, J. Pradko, and H. Umezawa, \emph{Physica} \textbf{135}A, 487 (1986).

% 189.
\bibitem{UA86}
H. Umezawa, and T. Arimitsu, \emph{Prog. Theor. Phys. Suppl.} No. 86, 243 (1986).

% 190.
\bibitem{AU87}
T. Arimitsu, and H. Umezawa, \emph{Prog. Theor. Phys.} \textbf{77}, 32 (1987).

% 191.
\bibitem{LvW87}
N.\,P. Landsman, and Ch.\,G. van Weert, \emph{Phys. Rep.} \textbf{145}, 141 (1987).

% 192.
\bibitem{MS78}
M. Schmutz, \emph{Z. Phys.} B\,\textbf{30}, 97 (1978).

% 193.
\bibitem{MNU85}
H. Matsumoto, Y. Nakano, and H. Umezawa, \emph{Phys. Rev.} D\,\textbf{31}, 429 (1985).

% 194.
\bibitem{PAH95}
P.\,A. Henning, \emph{Phys. Rep.} \textbf{253}, 235 (1995).

% 195.
\bibitem{HHW67}
R. Haag, N.\,M. Hugenholtz, and M. Winnink, \emph{Commun. math. Phys.} \textbf{5}, 215 (1967).

% 196.
\bibitem{RH96}
R. Haag, \emph{Local Quantum Physics: Fields, Particles, Algebras}, 2nd revised and enlarged edition (Springer, Berlin, 1996).

% 197.
\bibitem{KL17}
N.\,P. (Klaas) Landsman, \emph{Foundations of Quantum Theory: From Classical Concepts to Operator Algebras} (Springer Open, Cham, Switzerland, 2017). [\href{https://doi.org/10.1007/978-3-319-51777-3}{Open Access}]

% 198.
\bibitem{EHUY92}
T.\,S. Evans, I. Hardman, H. Umezawa, and Y. Yamanaka, \emph{J. Math. Phys.} \textbf{33}, 370 (1992).

% 199.
\bibitem{NS84a}
A.\,J. Niemi, and G.\,W. Semenoff, \emph{Ann. Phys.} (N.\,Y.) \textbf{152}, 105 (1984).

% 200.
\bibitem{NS84b}
A.\,J. Niemi, and G.\,W. Semenoff, \emph{Nucl. Phys.} B\,\textbf{230}, 181 (1984).

% 201.
\bibitem{MNU84}
H. Matsumoto, Y. Nakano, and H. Umezawa, \emph{J. Math. Phys.} \textbf{25}, 3076 (1984).

% 202.
\bibitem{EHUY93}
T.\,S. Evans, I. Hardman, H. Umezawa, and Y. Yamanaka, \emph{Fortschr. Phys.} \textbf{41}, 151 (1993).

% 203.
\bibitem{FG85}
Y. Fujimoto, and R. Grigjanis, \emph{Z. Phys.} C\,\textbf{28}, 395 (1985).

% 204.
\bibitem{TSE87}
T.\,S. Evans, \emph{Z. Phys.} C\,\textbf{36}, 153 (1987).

% 205.
\bibitem{FY88}
Y. Fujimoto, and H. Yamada, \emph{Z. Phys.} C\,\textbf{37}, 265 (1988).

% 206.
\bibitem{KS88}
K. Soutome, \emph{Z. Phys.} C\,\textbf{40}, 479 (1988).

% 207.
\bibitem{KN74}
H.\,J. Kreuzer, and K. Nakamura, \emph{Physica} \textbf{78}, 131 (1974).

% 208.
\bibitem{KK76}
H.\,J. Kreuzer, and C.\,G. Kuper, \emph{J. Phys.} G\,\textbf{2}, 9 (1976).

% 209.
\bibitem{GL51}
M. Gell-Mann, and F. Low, \emph{Phys. Rev.} \textbf{84}, 350 (1951).

% - The lattice model: ...

% 210.
\bibitem{PWA64}
P.\,W. Anderson, \emph{Special Effects in Superconductivity}, pp.\,113-135, in \emph{Lectures on the Many-Body Problem}, Vol.\,2, edited by E.\,R. Caianiello (Academic Press, London, 1964).

% 211.
\bibitem{KBE80}
K.\,B. Efetov, \emph{Sov. Phys. JETP} \textbf{51}, 1015 (1980).

% 212.
\bibitem{FG88}
M.\,P.\,A. Fisher, and G. Grinstein, \emph{Phys. Rev. Lett.} \textbf{60}, 208 (1988).

% 213.
\bibitem{FWGF89}
M.\,P.\,A. Fisher, P.\,B. Weichman, G. Grinstein, and D.\,S. Fisher, \emph{Phys. Rev.} B\,\textbf{40}, 546 (1989).

% 214.
\bibitem{ES94}
E. \u{S}im\'{a}nek, \emph{Inhomogeneous Superconductors: Granular and Quantum Effects} (Oxford University Press, 1994).

% 215.
\bibitem{FM94}
J.\,K. Freericks, and H. Monien, \emph{Europhys. Lett.} \textbf{26}, 545 (1994).

% 216.
\bibitem{KR97}
J. Kisker, and H. Rieger, \emph{Physica} A\,\textbf{246}, 348 (1997).

% 217.
\bibitem{JBCGZ98}
D. Jaksch, C. Bruder, J.\,I. Cirac, C.\,W. Gardiner, and P. Zoller, \emph{Phys. Rev. Lett.} \textbf{81}, 3108 (1998).

% 218.
\bibitem{OSS03}
D. van Oosten, P. van der Straten, and H.\,T.\,C. Stoof, \emph{Phys. Rev.} A\,\textbf{67}, 033606 (2003).

% 219.
\bibitem{WSGY94}
M. Wallin, E.\,S. S{\o}rensen, S.\,M. Girvin, and A.\,P. Young, \emph{Phys. Rev.} B\,\textbf{49}, 12115 (1994).

% 220.
\bibitem{PS04}
N.\,[V.] Prokof'ev, and B.\,[V.] Svistunov, \emph{Phys. Rev. Lett.} \textbf{92}, 015703 (2004).

% 221.
\bibitem{SS11}
S. Sachdev, \emph{Quantum Phase Transitions}, 2nd edition (Cambridge University Press, 2011).

% - The one-particle Green function:

% 222.
\bibitem{YY70I}
K. Yosida, and K. Yamada, \emph{Suppl. Prog. Theor. Phys.}, No.\,46, 244 (1970).

% - The TFD revisited:

% - The Hubbard Hamiltonian revisited:

% 223.
\bibitem{EMH89b}
E. M\"{u}ller-Hartmann, \emph{Z. Phys.} B\,\textbf{76}, 211 (1989).

% 224.
\bibitem{MV89}
W. Metzner, and D. Vollhardt, \emph{Phys. Rev. Lett.} \textbf{62}, 324 (1989). Erratum: \emph{loc. cit.}, 1066 (1989).

% 225.
\bibitem{EMH89a}
E. M\"{u}ller-Hartmann, \emph{Z. Phys.} B\,\textbf{74}, 507 (1989).

% 226.
\bibitem{PF03}
P. Fazekas, \emph{Lecture Notes on Electron Correlation and Magnetism} (World Scientific, Singapore, 2003).

% - The perturbation series expansion of G in terms of (v,G_0):

% - Details:

% 227.
\bibitem{BR91}
R.\,A. Brualdi, and H.\,J. Ryser, \emph{Combinatorial Matrix Theory} (Cambridge University Press, 1991).

% - Discussion:

% - Fredholm integral equation:

% 228.
\bibitem{MF53}
P.\,M. Morse, and H. Feshbach, \emph{Methods of Theoretical Physics}, Parts I and II (McGraw-Hill, New York, 1953).

% 229.
\bibitem{WW62}
E.\,T. Whittaker, and G.\,N. Watson, \emph{A Course of Modern Analysis}, 4th edition (Cambridge University Press, 1962).

% 230.
\bibitem{SG09}
M. Stone, and P. Goldbart, \emph{Mathematics for Physics: A Guided Tour for Graduate Students} (Cambridge University Press, 2009).

% - The self-energy operator Sigma_{00}:

% 231.
\bibitem{RGN80}
R.\,G. Newton, \emph{Ann. Phys.} (N.\,Y.) \textbf{124}, 327 (1980).

% 232.
\bibitem{MB83}
M. Bauer, \emph{Ann. Phys.} (N.\,Y.) \textbf{150}, 1 (1983).

% - The self-energy operator Sigma_{01}:

% - The self-energy operator Sigma_{10}:

% 233.
\bibitem{KKN60}
T. Kato, T. Kobayashi, and M. Namiki, \emph{Suppl. Prog. Theor. Phys.} No. 15, 3 (1960).

% 234.
\bibitem{ED11}
E. Engel, and R.\,M. Dreizler, \emph{Density Functional Theory: An Advanced Course} (Springer, Berlin, 2011).

% 235.
\bibitem{FEDvH91}
B. Farid, G.\,E. Engel, R. Daling, and W. van Haeringen, \emph{Phys. Rev.} B\,\textbf{44}, 13349 (1991); G.\,E. Engel, B. Farid, C.\,M.\,M. Nex, and N.\,H. March, \emph{loc. cit.}, 13356 (1991).

% - The self-energy operator Sigma_{11}:

% 236.
\bibitem{RDM92}
R.\,D. Mattuck, \emph{A guide to Feynman diagrams in the many-body problem}, 2nd edition (Dover, New York, 1992).

% 237.
\bibitem{LH65}
L. Hedin, \emph{Phys. Rev.} \textbf{139}, A\,796 (1965);  L. Hedin, and S. Lundqvist, \emph{Effects of Electron-Electron and Electron-Phonon Interactions on the One-Electron States of Solids}, pp.\,1-181, in \emph{Solid State Physics}, Vol.\,23, edited by F. Seitz, D. Turnbull, and H. Ehrenreich (Academic Press, New York, 1969).

% - The dynamical screened interaction potential W:

% - The two-particle Green function G_2 and its ...:

% 238.
\bibitem{HJ13}
R.\,A. Horn, and C.\,R. Johnson, \emph{Matrix Analysis}, 2nd edition (Cambridge University Press, 2013).

% - The proper polarisation operators ...:

% - The Sequence {P_{01} ...} and ...:

% - The sequence {P_{11} ...}:

% - Summary and concluding remarks:

% 239.
\bibitem{HJR63}
H.\,J. Ryser, \emph{Combinatorial Mathematics} (published for The Mathematical Association of America by John Wiley \& Sons, New York, 1963).

% - Appendix A:

% 240.
\bibitem{BF19}
B. Farid, \emph{On the Wick operator identities and Wick decompositions: A pedagogical review}, to be published.

% 241.
\bibitem{BR86}
J.-P. Blaizot, and G. Ripka, \emph{Quantum Theory of Finite Systems} (The MIT Press, Cambridge, Massachusetts, 1986).

% 242.
\bibitem{HP59}
N.\,M. Hugenholtz, and D. Pines, \emph{Phys. Rev.} \textbf{116}, 489 (1959). For completeness, the details of the references 6 and 7 in this publication, marked on p.\,490 as `unpublished', are: [K. Sawada,] \emph{Phys. Rev.} \textbf{116}, 1344 (1959), and [T.\,T. Wu,] \emph{Phys. Rev.} \textbf{115}, 1390 (1959).

% 243.
\bibitem{STB58}
S.\,T. Beliaev [Belyaev], \emph{Sov. Phys. JETP} \textbf{34}, 289 (1958).

% 244.
\bibitem{NNB47}
N. Bogolubov [N.\,N. Bogoliubov], \emph{J. Phys.} (USSR) \textbf{11}, 23 (1947).

% 245.
\bibitem{EKS98}
T.\,S. Evans, T.\,W.\,B. Kibble, and D.\,A. Steer, \emph{J. Math. Phys.} \textbf{39}, 5726 (1998).

% 246.
\bibitem{JR80}
J.\,M. Jauch, and F. Rohrlich, \emph{The Theory of Photons and Electrons: The Relativistic Quantum Field Theory of Charged Particles with Spin One-half}, 2nd expanded edition, 2nd corrected printing (Springer, Berlin, 1980).

% 247.
\bibitem{SM95}
S. Majid, \textsl{Foundations of Quantum Group Theory} (Cambridge University Press, 1995).

% 248.
\bibitem{CB05}
C. Brouder, \emph{Phys. Rev.} A\,\textbf{72}, 032720 (2005).

% 249.
\bibitem{PMC05}
P.\,M. Cohn, \emph{Basic Algebra: Groups, Rings and Fields}, 2nd printing (Springer, London, 2005).

% 250.
\bibitem{AA14}
M.\,R. Adhikari, and A. Adhikari, \emph{Basic Modern Algebra with Applications} (Springer, New Delhi, 2014).

% 251.
\bibitem{RNFK11}
R. Nederpelt, and F. Kamareddine, \emph{Logical Reasoning: A First Course}, 2nd revised edition (College Publications, London, 2011).

% 252.
\bibitem{BM82}
J.\,A. Bondy, and U.\,S.\,R. Murty, \emph{Graph Theory with Applications}, 5th printing (North-Holland, New York, 1982).

% 253.
\bibitem{NN71}
N. Nakanishi, \emph{Graph Theory and Feynman Integrals} (Gordon \& Breach, New York, 1971).

% 254.
\bibitem{MA07}
M. Aigner, \emph{A Course in Enumeration} (Springer, Berlin, 2007).

% 255.
\bibitem{FH69}
F. Harary, \emph{Graph Theory} (Addison-Wesley, Reading, Massachusetts, 1969).

% 256.
\bibitem{LI18}
L. Isserlis, \emph{Biometrika} \textbf{12}, 134 (1918).

% 257.
\bibitem{SJ97}
S. Janson, \emph{Gaussian Hilbert Spaces} (Cambridge University Press, 1997).

% 258.
\bibitem{JR02}
J. Riordan, \emph{Introduction to Combinatorial Analysis} (Dover, New York, 2002).

% 259.
\bibitem{CHM51}
S. Chowla, I.\,N. Herstein, and W.\,K. Moore, \emph{Canad. J. Math.} \textbf{3}, 328 (1951).

% 260.
\bibitem{GJ87}
J. Glimm, and A. Jaffe, \emph{Quantum Physics: A Functional Integral Point of View}, 2nd edition (Springer, New York, 1987).

% 261.
\bibitem{BS74}
B Simon, \emph{The $P(\Phi)_2$ Euclidean (Quantum) Field Theory} (Princeton University Press, 1974).

% 262.
\bibitem{WB08}
A. Wurm, and M. Berg, \emph{Am. J. Phys.} \textbf{76}, 65 (2008).

% 263.
\bibitem{GS75}
G. Szeg\"{o}, \emph{Orthogonal Polynomials}, 4th edition (American Mathematical Society, Providence, Rhode Island, 1975).

% 264.
\bibitem{PRH74}
P.\,R. Halmos, \emph{Measure Theory} (Springer, New York, 1974).

% 265.
\bibitem{VR91}
V. Rivasseau, \emph{From Perturbative to Constructive Renormalization} (Princeton University Press, 1991).

% 266.
\bibitem{MS99}
M. Salmhofer, \emph{Renormalization: An Introduction} (Springer, Berlin, 1999).

% 267.
\bibitem{FKT02}
J. Feldman, H. Kn\"{o}rrer, and E. Trubowitz, \emph{Fermionic Functional Integrals and the Renormalization
Group} (American Mathematical Society, Providence, Rhode Island, 2002). [\href{https://www.math.ubc.ca/~feldman/aisen.html}{\emph{Aisenstadt Lectures} (August 22-27, 1999)}]

% 268.
\bibitem{GR07}
I.\,S. Gradshteyn, and I.\,M. Ryzhik, \emph{Table of Integrals, Series, and Products}, 7th edition (Elsevier, Amsterdam, 2007).

% 269.
\bibitem{PAH90}
P.\,A. Henning, \emph{Nucl. Phys.} B\,\textbf{337}, 547 (1990).

% 270.
\bibitem{vLS12}
R. van Leeuwen, and G. Stefanucci, \emph{Phys. Rev.} B\,\textbf{85}, 115119 (2012).

% 271.
\bibitem{SF69}
S. Fujita, \emph{J. Phys. Soc. Japan} \textbf{27}, 1096 (1969).

% 272.
\bibitem{SF71}
S. Fujita, \emph{Phys. Rev.} A\,\textbf{4}, 1114 (1971).

% 273.
\bibitem{BB69}
R. Balian, and E. Brezin, \emph{Nuovo Cimento} \textbf{64}\,B, 37 (1969).

% 274.
\bibitem{BF02}
B. Farid, \emph{Phil. Mag.} B\,\textbf{82}, 1413 (2002).

% 275.
\bibitem{JS49}
J. Schwinger, \emph{Phys. Rev.} \textbf{75}, 651 (1949).

% 276.
\bibitem{JLA54}
J.\,L. Anderson, \emph{Phys. Rev.} \textbf{94}, 703 (1954).

% 277.
\bibitem{BdD58}
C. Bloch, and C. de Dominicis, \emph{Nucl. Phys.} \textbf{7}, 459 (1958).

% 278.
\bibitem{MG60}
M. Gaudin, \emph{Nucl. Phys.} \textbf{15}, 89 (1960).

% 279.
\bibitem{DJT57}
D.\,J. Thouless, \emph{Phys. Rev.} \textbf{107}, 1162 (1957).

% 280.
\bibitem{ES96}
T.\,S. Evans, and D.\,A. Steer, \emph{Nucl. Phys.} B \textbf{474}, 481 (1996).

% 281.
\bibitem{Note2}
\emph{Historical note.} Reference \protect\citen{AGD75} (published by Dover), cited in the present paper, is an ``Unabridged republication, with slight corrections, of the revised (1963) English edition.'' Pergamon, Oxford, has published a second edition of this book  (1965) under the modified title ``Quantum Field Theoretical Methods in Statistical Physics''. The text has been translated from the Russian by D.\,E. Brown, and the English translation has been edited by D. ter Haar. On page 108 of both editions, one encounters the \textsl{erroneous} statement (originating from the work by Matsubara \protect\cite{TM55} (1955)) regarding the Wick decomposition theorem for thermal-ensemble averages associated with volume $V$ finite and $V\to\infty$. Interestingly, on page 46 of Ref.\,\protect\citen{TH77} (1977) ter Haar states: ``AGD [Abrikosov \emph{et al.}] discuss \dots, although their proof of Wick's theorem is incorrect.'' As we have just indicated, the ``incorrect'' proof originates from Ref.\,\protect\citen{TM55}. The incorrectness was pointed out and rectified by Thouless in Ref.\,\protect\citen{DJT57} (1957); a later extensive and transparent discussion of this problem is due to Evans and Steer \protect\cite{ES96} (1996). The reference cited by AGD for the Wick theorem, \emph{i.e.} the Wick operator identity, is Ref.\,``A9'' in the Dover edition (p.\,67), and Ref.\,``[25]'' in the Pergamon edition (p.\,65), both referring to the book \emph{Quantum Electrodynamics}, by Akhiezer and Berestetskii \protect\cite{AB65} (1965). We have not detected any trace of an ``incorrect'' proof of the Wick theorem in this book.

% 282.
\bibitem{TH77}
D. ter Haar, \emph{Lectures on Selected Topics in Statistical Mechanics} (Pergamon Press, Oxford, 1977).

% 283.
\bibitem{AB65}
A.\,I. Akhiezer, and V.\,B. Berestetskii, \emph{Quantum Electrodynamics} (Interscience, New York, 1965).

% 284.
\bibitem{XT90}
H.-H. Xu, and C.-H. Tsai, \emph{Phys. Rev.} A\,\textbf{41}, 53 (1990).

% 285.
\bibitem{EHL68}
E.\,H. Lieb, \emph{J. Combin. Theory} \textbf{5}, 313 (1968).

% 286.
\bibitem{DC09}
D. Callan, \emph{A combinatorial survey of identities for the double factorial}, \href{http://arxiv.org/abs/0906.1317v1}{arXiv:0906.1317v1}.

% 287.
\bibitem{YY75III}
K. Yosida, and K. Yamada, \emph{Prog. Theor. Phys.} \textbf{53}, 1286 (1975).

% 288.
\bibitem{KY75II}
K. Yamada, \emph{Prog. Theor. Phys.} \textbf{53}, 970 (1975).

% 289.
\bibitem{JHH66}
J.\,H. Halton, \emph{J. Combin. Theory} \textbf{1}, 224 (1966).

% 290.
\bibitem{LSM61}
E.\,H. Lieb, T. Schultz, and D. Mattis, \emph{Annal. Phys.} (N.\,Y.) \textbf{16}, 407 (1961).

% 291.
\bibitem{SML64}
T.\,D. Schultz, D.\,C. Mattis, and E.\,H. Lieb, \emph{Rev. Mod. Phys.} \textbf{36}, 856 (1964).

% 292.
\bibitem{JW28}
P. Jordan, and E. Wigner, \emph{Z. Physik} \textbf{47}, 631 (1928).

% 293.
\bibitem{HAK64}
H.\,A. Kramers, \emph{Quantum Mechanics}, translated by D. ter Haar (Dover, New York, 1964).

% 294.
\bibitem{DCM88}
D.\,C. Mattis, \emph{The Theory of Magnetism I: Statics and Dynamics}, corrected 2nd printing (Springer, Berlin, 1988).

% 295.
\bibitem{EF13}
E. Fradkin, \emph{Field Theories of Condensed Matter Systems}, 2nd edition (Cambridge University Press, 2013).

% 296.
\bibitem{HP40}
T. Holstein, and H. Primakoff, \emph{Phys. Rev.} \textbf{58}, 1098 (1940).

% 297.
\bibitem{GA59}
M. Girardeau, and R. Arnowitt, \emph{Phys. Rev.} \textbf{113}, 755 (1959).

% 298.
\bibitem{MPW63}
E.\,W. Montroll, R.\,B. Potts, and J.\,C. Ward, \emph{J. Math. Phys.} \textbf{4}, 308 (1963).

% 299.
\bibitem{KO49}
B. Kaufman, and L. Onsager, \emph{Phys. Rev.} \textbf{76}, 1244 (1949).

% - Appendix B:

% 300.
\bibitem{SR12}
S. Roman, \emph{Fundamentals of Group Theory: An Advanced Approach} (Birkh\"{a}user, Boston, 2012).

% 301.
\bibitem{PS03}
S. Pemmaraju, and S. Skiena, \emph{Computational Discrete Mathematics: Combinatorics and Graph Theory with Mathematica} (Cambridge University Press, 2003).

% 302.
\bibitem{JFM72}
J.\,F. Meyer, \emph{Algebraic isomorphism invariants for graphs of automata}, pp.\,123-152, in \emph{Graph Theory and Computing}, edited by R.\,C. Read (Academic Press, New York, 1972).

% 303.
\bibitem{HHW88}
F. Harary, J.\,P. Hayes, and H.-J. Wu, \emph{Comput. Math. Applic.} \textbf{15}, 227 (1988).

% 304.
\bibitem{FH00}
F. Harary, \emph{J. Univ. Comput. Sci.} \textbf{6}, 136 (2000).

% 305.
\bibitem{TH01}
T. Hahn, \emph{Comput. Phys. Commun.} \textbf{140}, 418 (2001).
[\href{http://www.feynarts.de/}{The \textsl{FeynArts} Visitor Center}]

% - Appendix C:

% - Appendix D:

% 306.
\bibitem{SP84}
S. Pissanetzky, \emph{Sparse Matrix Technology} (Academic Press, London, 1984).
$\hfill\Box$

\end{thebibliography}
\end{document}